\documentclass[12pt,a4paper]{book}
\usepackage{amssymb}

\usepackage{tcibkpk}
\usepackage{sw20wspc}
\usepackage{amsmath}
\usepackage{graphicx}
\usepackage{bezier}
\usepackage{curves}
\usepackage{epsfig}

\input{tcilatex}

\begin{document}

\author{CHRISTIAN KORFF}
\title{LIE ALGEBRAIC STRUCTURES IN \\
INTEGRABLE MODELS, \\
AFFINE TODA FIELD THEORY }
\date{im Fachbereich Physik der \\
Freien Universit\"{a}t Berlin \\
eingereichte Dissertation\\
12. Juli 2000}
\maketitle

\newpage

\bigskip

\newpage

{\LARGE Referees:\bigskip }

{\Large Prof. Dr. Robert Schrader}

{\Large Institut f\"{u}r theoretische Physik}

{\Large Freie Universit\"{a}t Berlin}

{\Large D-14195 Berlin}

\bigskip

\bigskip

{\Large Prof. Dr. David I. Olive (FRS)}

{\Large Physics Department}

{\Large University of Wales Swansea}

{\Large Singleton Park}

{\Large Swansea SA2 8PP, Wales UK}

\bigskip

\bigskip

\bigskip

\bigskip

\bigskip

\bigskip

\bigskip

{\LARGE Advisor:\bigskip }

{\Large Dr. Andreas Fring}

{\Large Institut f\"{u}r theoretische Physik}

{\Large Freie Universit\"{a}t Berlin}

{\Large D-14195 Berlin\bigskip }

\newpage

\bigskip

\newpage

\tableofcontents

\chapter{Introduction}

{\small \emph{One cannot accomplish things simply with cleverness. One must
take a broad view. It will not do to make rash judgements concerning good or
evil. However, one should not be sluggish. It is said that one is not truly
a samurai if he does not make his decisions quickly} \emph{and break right
through to completion}.\bigskip }

\qquad \qquad \qquad \qquad \qquad \qquad {\small From 'The Book of the
Samurai, Hagakure'}

\section{Perspective and motivation}

Since the first steps towards a relativistic wave mechanics in the early
1930s up to the ''gauge revolution'' in the 1970s and 1980s leading to the
formulation of the Standard Model, quantum field theory (QFT) has developed
to the most successful physical theory. Its computational power in
comparison with experimental data as well as its conceptual scope in
explaining the subatomic world are met by no other model in theoretical
physics. Yet despite this undeniable success there remains a discrepancy
between the glorious predictive power on the one hand and the scarce
knowledge about the underlying intricate mathematical structure on the other
hand. Indeed, perturbative expansions together with a thicket of detailed
renormalization prescriptions are in most cases the only possible access to
the theory failing to reveal its true nature.

The search for a deeper understanding of the basic physical principles
underlying QFT and for concise mathematical formulations took its main
starting point in the 1950s after the observation that even the
applicability of perturbation theory breaks down in the range of
strong-interactions. The problem of quark confinement is one of the subtle
up-to-day unsolved problems, which impressively demonstrates that
non-perturbative aspects can become dominant and are inevitable in order to
understand the theory to its depth. This initiated the search for
alternative and entirely new techniques going beyond the conventional
approach of calculating numerous Feynman diagrams. Clearly, the exact
solution of \emph{any} non-trivial QFT would yield profound insight and
valuable help in this task, irrespective whether the model under
consideration is linked directly to a concrete physical problem or not.
Solving a QFT exactly is understood as the explicit calculation of all its $%
n $-point or correlation functions, since from the latter the field content
and the physical state space can be recovered by means of \textbf{Wightman's
reconstruction theorem} \cite{Wight,QFT}. It will become important below
that this holds true for Minkowski as well as Euclidean space. In fact, the
correlation functions in Minkowski space can be recovered from those in
Euclidean space as was worked out by Osterwalder and Schrader \cite{Ost}.
Especially important is this connection in the context of \textbf{%
constructive field theory} which by use of functional methods achieved the
mathematical exact construction of a number of field theories in two and
three dimensions, see \cite{Jaffe} and references therein. However, explicit
and closed expressions for the relevant physical quantities are missing in
this approach. It remains the ambitious aim to obtain these in order to
check general assumptions about the structure of QFT for their consistency,
prove certain non-perturbative approaches workable, test new concepts and
learn about concrete physical problems.

\subsubsection{Coleman-Mandula theorem}

Naturally, the most likely candidates of QFT's for which the ambitious aim
of an exact solution seems conceivable are those with powerful symmetries
giving rise to a large number of conservation laws. The latter might ease
the actual computations or even impose such severe constraints that exact
solutions can be constructed. Unfortunately, in 3+1 dimensions\footnote[1]{%
We follow here the standard convention in splitting the number of dimensions
into a sum with the first summand refering to the space and the second to
the time dimension.} any progress in this direction is blocked by the 
\textbf{Coleman-Mandula theorem} \cite{Cole}, which states that additional
symmetries besides Poincar\'{e} invariance and an internal gauge group
describing the degeneracy of the particle spectrum render the scattering
matrix of each massive QFT to be trivial, i.e. it describes non-interacting
quantum particles. With only theses symmetries present it has not been
achieved so far to overcome the conceptual difficulties which one encounters
when looking for exact expressions.

\subsubsection{1+1 dimensional integrable field theories}

The situation greatly improves in 1+1 dimensional systems. Here the blockade
of the Coleman-Mandula theorem is lifted due to the fact that one of the
crucial assumptions in its proof ceases to hold true, the scattering
amplitude does not any longer depend analytically on the scattering angle.
(In one space dimension only forward and backward scattering are possible,
whence the scattering angle can only assume the discrete values $0$ or $\pi
. $) One might therefore look for theories with higher conservation laws, in
fact with infinitely many of them, in which case these theories are called 
\textbf{integrable}. The latter term originates in classical mechanics,
where a system is said to be integrable if there are as many conserved
quantities as degrees of freedom allowing to solve the equations of motion
by integration. In case of a field theory \emph{infinitely} many degrees of
freedom are present, since the field configuration at each point in
space-time has to be specified. In a loose sense one then refers to a field
theory as integrable if it gives rise to an \emph{infinite} set of
conservation laws.

At first sight the specialization to 1+1 dimensions and a class of field
theories with infinitely many conserved charges might appear quite
restrictive. However, the study of low-dimensional QFT has turned out to
reproduce many features which are of interest in higher dimensions, such as
confinement mentioned earlier, duality or gauge anomalies, see e.g. \cite
{Abda} and references therein for further details. Furthermore, integrable
models have found direct physical applications in the off-critical
description of statistical mechanics and condensed matter systems reduced to
two space-dimensions as will be outlined below. In this context
integrability appears naturally as a relict of broken conformal symmetry.

\subsubsection{Exact scattering matrices}

The presence of an infinite set of conservation laws imposes severe
restrictions on the dynamics. In particular, it enforces that the particle
number as well as the individual particle momenta are asymptotically
conserved in a scattering process. Moreover, each scattering process can be
decomposed into two-particle ones, reducing the task to determine the full
scattering matrix (also referred to as S-matrix) to the calculation of the
two-particle amplitude. These powerful constraints together with the idea of 
\textbf{minimal analyticity} originating in the \textbf{S-matrix theory} of
the sixties \cite{Heisb,Chew,Eden}\ allow to construct fully\emph{\ exact
scattering amplitudes.} The basic idea is to regard the scattering
amplitudes as boundary values of analytic functions in the complex plane.
Setting up a set of functional relations reflecting general requirements
like unitarity or crossing symmetry the general form of a two-particle
scattering amplitude can already be written down without relying on a
classical Lagrangian of the field theory. The actual dependence on the
integrable model at hand is then invoked by the famous \textbf{Yang-Baxter } 
\cite{YB} or the \textbf{bootstrap equation}. The former is tightly linked
to the integrability property and describes equivalent ways to factorize a
scattering matrix in two-particle ones, while the latter reflects the
principle of \textbf{nuclear democracy} which states that each bound state
in the theory should also appear in the asymptotic particle spectrum. This
method for constructing exact scattering matrices was pioneered in the
articles \cite{boot} and has become known as the \textbf{bootstrap approach}
in the literature. It has proven extremely successful over the years. In
particular, it lead to the implicit definition of entirely new integrable
models by writing down exact scattering matrices which satisfy all physical
requirements. In a sense this realizes partially the ambitious program of
the S-matrix theory of the sixties in low dimensions.\bigskip

\noindent \emph{The bootstrap construction of scattering matrices will play
a central role in this thesis. Its powerful structure will be in particular
exploited when constructing a class of hitherto unknown scattering matrices
giving rise to new integrable quantum field theories.}

\subsubsection{Off-shell investigations and form factors}

The exact construction of the scattering matrix is not only of interest in
order to obtain a complete knowledge about the on-shell structure of a QFT,
but it also serves as a preliminary step towards the calculation of the
measurable quantities of the system, the $n$-point or correlation functions.
Calculating the latter is the ultimate goal of each theory, since as
mentioned above it amounts to solving the whole model completely via the
reconstruction theorem. One of the most promising approaches in this context
is the \textbf{form factor program} \cite{FF}. Writing down the correlation
function of a local operator $\mathcal{O}$ one inserts a complete set of
states, usually depending on the momentum, giving in a simplifying notation 
\begin{equation*}
\left\langle 0|\mathcal{O}(x)\mathcal{O}(0)|0\right\rangle =\sum_{n}\int
d^{n}p\;|\left\langle 0|\mathcal{O}(0)|\mathbf{p}\right\rangle
|^{2}\,e^{i\tsum\limits_{k=1}^{n}p_{k}\cdot x}\;.
\end{equation*}
Here the sum runs over all possible particle numbers $n$, the vector $%
\mathbf{p}=(p_{1},...,p_{n})$ consists of the individual particle momenta
and the matrix elements $\left\langle 0|\mathcal{O}(0)|\mathbf{p}%
\right\rangle $ are referred to as form factors (actually a slightly
modified version of them). Proceeding conceptually very similar as in the
case of the scattering matrix one then continues analytically the form
factors in the momentum variables. The purpose is to set up recursive
functional\ equations, which require the exact scattering matrix as
prerequisite input. In principle these equations enable one to derive all $n$%
-particle form factors and to calculate the correlation functions by
evaluating the above infinite sum of integrals. Being a highly non-trivial
step this remains an open challenge for almost all theories except the Ising
model. Also the calculation of the complete set of form factors has so far
only been achieved in a few cases, e.g. \cite{FFex,CFK}.

Even though the final and complete construction of correlation functions in
form of explicit analytic expressions is yet outstanding the form factor
program gives reasonable hope that this might be achieved in the near
future. Already now one might exploit the fact that the sum over the
particle number $n$ is rapidly convergent. Hence for many practical purposes
it is sufficient to determine only the first few particle form factors,
which correctly capture the low energy behaviour. The approximative
description of the correlations obtained this way is of high accuracy and
due to the non-perturbative input of the scattering matrix more precise than
any calculation in perturbation theory.

Although the calculation of form factors will not be performed in this
thesis, it has been mentioned since it constitutes nowadays one of the most
interesting techniques in the study of integrable field theory and is
tightly linked to the bootstrap construction of scattering matrices. In
fact, in our article \cite{CFK} the full set of form factors for the $%
su(3)_{2}$-Homogeneous Sine-Gordon model has recently been obtained. The
latter belongs to a class of integrable models which are studied in some
detail in this thesis.\bigskip

\noindent \emph{In conclusion, one might say that from the field theoretic
point of view the motivation to study integrable systems in 1+1 dimensions
is to learn about the structure of QFT by constructing exact solutions and
finding explicit expressions for the relevant physical quantities.}

\subsubsection{Integrability and broken scale invariance}

The interest in integrable models not only resides in their role as
excellent ``testing laboratories'' for exact non-perturbative methods of
QFT, but more recently also in their interpretation as deformed \textbf{%
conformal field theories}\emph{\ }(CFT) \cite{perCFT}. The latter form a
particular class of integrable field theories but in contrast to the cases
mentioned before they are associated with \emph{massless} particles. In this
particular case the infinite set of conversation laws is linked to conformal
space-time symmetry in two-dimensions. An intense research activity in this
area was initiated by the seminal paper of Belavin, Polyakov and
Zamolodchikov \cite{BPZ}. They combined the representation theory of the
Virasoro algebra describing the infinitesimal quantum generators of
conformal transformations with the concept of a local operator algebra in
order to show that a certain class of conformal theories, the so-called
minimal models, constitute particular examples of solvable massless QFT's.
Motivated by the observation of Polyakov that for physical systems with
local interactions conformal symmetry is an immediate extension of scale
invariance \cite{Pol}, the techniques of Euclidean CFT have been applied to
study statistical mechanics and condensed matter systems in two \emph{space}%
-dimensions, which undergo a second order phase transition. The latter
become scale invariant at the critical point and fall into different
universality classes fixed by the critical exponents, which describe the
power law behaviour of the correlations in the system. One of the
motivations to study CFT is the classification of all these universality
classes.

A simple example for this picture is provided by the two-dimensional Ising
model. In the continuum limit it can be described by a Euclidean field
theory of free Majorana fermions whose mass is proportional to $%
|1/T-1/T_{c}| $ with $T$ being the temperature and $T_{c}$ its special value
at the critical point. Away from criticality the fermions in the system are
massive and the correlations fall off over a finite length scale fixed by
the Compton wave length. If the system approaches the critical point at $%
T=T_{c}$ the particles become massless and the associated correlation length
diverges. In particular, the theory loses its dependence on the only
dimensionful parameter and becomes therefore scale invariant.

This scenario can be generalized to more complicated cases. Given a
conformal field theory at the critical point one might in particular ask
what happens to the infinite conservation laws linked to conformal
invariance when the system becomes off-critical and scale invariance is
lost. As Zamolodchikov pointed out \cite{perCFT} an infinite set of these
conserved charges -- even though they get deformed -- might survive the
breaking of conformal symmetry and render also the perturbed theory
integrable. Provided the particle spectrum of the perturbed theory is purely
massive one might now exploit the above non-perturbative techniques of the
bootstrap program to obtain information about the off-critical behaviour of
the system. This point of view has been supported by numerous concrete
examples, starting with the study of the Ising model in an external magnetic
field \cite{Zper}. In fact, this interplay between field theoretic
considerations and phase transitions in statistical mechanics renewed the
interest in integrable field theories and made the subject flourish in the
last years. To name a few examples in the area of condensed matter theory,
non-perturbative methods of integrable field theory have been discussed in
the context of quantum impurity problems, see \cite{Sal} and references
therein. Other possible applications have been investigated in the context
of two leg Hubbard ladders and Carbon nanotubes \cite{Lud}. There is also a
series of papers which apply the theory of so called $W$-algebras, closely
connected to affine or Kac-Moody algebras, to the quantum Hall effect \cite
{Capp}.

\subsubsection{Recovering conformal invariance}

In contrast to the picture just described one might proceed in reverse order
and start with a massive integrable field theory and ask how to recover the
associated conformal model. Scale and therefore conformal invariance will be
approximately restored in the high-energy regime, where the masses of the
particles become negligible. The technique which will be applied in this
thesis to investigate the high-energy limit of integrable quantum field
theories is the \textbf{thermodynamic Bethe ansatz} (TBA) \cite{Yang,TBAZam1}%
. Using the exact scattering matrix as the only input, this approach allows
to calculate the free energy of the integrable field theory on an infinite
cylinder after performing a Wick rotation and interpreting the imaginary
time axis as temperature. When the latter reaches an energy scale which is
large compared to the one set by the particle masses in the spectrum,
numerous characteristic quantities of the CFT governing the ultraviolet
behaviour can be extracted from the free energy, as for instance the
(effective) \textbf{central charge} $c$ playing the role of a Casimir
energy. In this way the TBA forms an important interface between the
conformal and the massive integrable model. In particular, it can be used to
test scattering matrices constructed via the bootstrap approach for
consistency and to relate massive to conformal spectra with the ultimate
goal of obtaining a deeper understanding of the origin of mass. Moreover,
regarding the applications to the off-critical behaviour of statistical
mechanics or condensed matter systems one might assign by means of the TBA
to each scattering matrix a conformal field theory or a universality class.

Additional motivation for the investigation of the intimate relation between
integrable and conformal models also comes from \textbf{string theory},
whose objective is the unification of all forces in nature. Here the time
evolution of a one-dimensional object, the string, sweeps a two-dimensional
world-sheet in space-time and the shape of the string is described by fields
which live on this world-sheet. In case of the vacuum state the world-sheet
is assumed to be invariant under reparametrizations, what implies conformal
symmetry for the field content. Another concept which is reminiscent of
string theory is the notion of dual models or duality, which in an
elementary and simple form will also be encountered in the context of the
integrable models investigated in this thesis.

After this outline of the general perspective and techniques of
two-dimensional integrable quantum field theory, the different aspects
mentioned will now be elaborated with particular hindsight to \textbf{affine
Toda field theories} (ATFT) \cite{ATFT}, which constitute the most
prominent, best studied and largest class of integrable models.

\section{Affine Toda field theory}

The simplest and best known examples of affine Toda field theories (ATFT)
are the Sine-Gordon and the Sinh-Gordon model for imaginary and real values
of the coupling constant, respectively. In general, they are associated with
the following classical Lagrangian 
\begin{equation}
\mathcal{L}_{\text{ATFT}}(\frak{g})=\dfrac{1}{2}\left\langle \partial _{\mu
}\phi ,\partial ^{\mu }\phi \right\rangle -\frac{m^{2}}{\beta ^{2}}%
\sum_{i=0}^{n}n_{i}e^{\beta \left\langle \alpha _{i},\phi \right\rangle }\;.
\label{La}
\end{equation}
Here $\phi =(\phi _{1},...,\phi _{n})$ are $n$-component fields transforming
as scalars under the Lorentz group and $m,\beta $ define a classical mass
scale and coupling constant, respectively. The latter are classically
unimportant but enter the quantum theory associated with the above
Lagrangian. The integer constants $n_{i}$ and the constant vectors $\alpha
_{i}\in \mathbb{R}^{n}$ are restricted to special values in order to
guarantee that the resulting field theory is integrable. In fact, it turns
out that the allowed set of external parameters is in general linked to an 
\textbf{affine Lie algebra} $\frak{\hat{g}}$ of rank $n$ \cite{Kac}. In many
cases, however, it turns out that the structure of a simple
finite-dimensional Lie algebra $\frak{g}$, whose affine extension is $\frak{%
\hat{g}}$, is sufficient for the description. The integers $n_{i}$ are then
interpreted as its Kac labels, the $\alpha _{i}$'s constitute its simple
roots with $i=1,...n$ and $-\alpha _{0}$ is the highest root with $n_{0}=1$.
The latter mathematical objects will be explained in more detail in course
of the thesis. Choosing $\frak{g}=su(2)$ we recover the Sine-Gordon or
Sinh-Gordon Lagrangian upon noting that then one has $n=n_{0}=n_{1}=1$ and $%
\alpha _{1}=-\alpha _{0}$, whence the above potential acquires the form of a 
$\cos $ or a $\cosh $-function for $\beta $ purely imaginary or real,
respectively.

In this thesis the discussion will exclusively deal with the models
associated with a real coupling constant, i.e. with the generalizations of
the Sinh-Gordon model. While in the latter case the above Lagrangian is
manifestly real, this property is in general lost for imaginary $\beta $.
However, due to the soliton solutions to which they give rise, also the
latter models starting with the Sine-Gordon theory have been studied in
detail in the literature. (For a relatively recent review of ATFT and
references see \cite{Corrigan}.)

The outstanding property common to all these models is the rich underlying
Lie algebraic structure encoded in $\frak{g}$, which allows for the
application of powerful mathematical concepts. Classically it can be used to
show integrability by a Lax pair construction and to determine the solutions
to the classical equations of motion \cite{ATFT}. Remarkably, also on the
quantum level physical quantities like the mass spectrum, the fusing
processes of particles as well as the S-matrix reflect the Lie algebraic
structure.\medskip

\noindent \emph{One of the central aims of this thesis is to exploit this
Lie algebraic structure in order to obtain generic and concise formulas for
all relevant quantities, such that all models are encompassed at once. These
universal expressions will in particular allow to separate model dependent
features from more general ones and unify numerous case-by-case discussions
found in the literature.}

\subsubsection{The search for the universal scattering matrix of ATFT}

The scattering matrices of affine Toda models associated with simply-laced
Lie algebras, the so-called $ADE$ series, were the first to be studied by
standard perturbative methods as well as by the bootstrap approach,
beginning with the paper by Arinshtein et al. about the $\frak{g}%
=A_{n}\equiv su(n+1)$ theories and followed by articles from Mussardo and
Christe as well as Braden et al., who considered the remaining cases \cite
{TodaS}. Their results were put into a universal form in \cite{PD,FO}
describing the scattering matrices of all $ADE$ models in a unique and\
generic formula. The key feature they exploited is the Coxeter geometry
naturally assigned to each simple Lie algebra $\frak{g}$. This had been
noticed to be crucial in the description of the three-point couplings of the
theory \cite{FLO}. The latter determine the bound state structure (so-called
fusing processes of the particles), an information required to perform the
bootstrap construction of the scattering matrix.

Similar attempts failed for the remaining ATFT involving non simply-laced
Lie algebras (the $BCFG$ series) due to the fact that their renormalization
behaviour turned out to be quite different form the $ADE$ series, where the
classical mass ratios survive quantization up to one loop order in
perturbation theory. For several years different proposals for the
scattering matrices were put forward. They were plagued by mysterious
higher-order poles which were coupling dependent and resisted a consistent
physical interpretation.

The breakthrough in the understanding of the non simply-laced models started
with the paper of Delius et al. \cite{Gust}. Based on their perturbative
calculations it was suggested \cite{Do1} that these theories are governed by
two classical Lagrangians belonging to a pair of ''dual'' algebras, one
describing the system in the weak and the other in the strong coupling
regime.\ Another crucial step made by Corrigan et al. was the formulation of
the generalized bootstrap principle \cite{nons} giving a consistent
prescription how to identify those poles in the physical sheet which are
relevant to the bootstrap approach. However, the construction of the various
non simply-laced scattering matrices was performed separately for the
different models and a concise Lie algebraic formulation was lacking.

First steps towards this direction were made in the work by Chari and
Pressley \cite{CP}, who managed to reproduce the allowed fusing processes in
terms of Coxeter geometry associated with the two dual algebras, and the
article by Khastgir \cite{Khast}, who employed the idea of folding to
reproduce the scattering matrices found in \cite{nons}. However, it was Oota
who finally succeeded in writing down a closed universal expression for the
scattering matrices by introducing $q$-deformed Coxeter elements \cite{Oota}%
. The latter allow to link the fusing rules directly to the scattering
matrices and to accommodate the coupling dependence of the theories by a
special choice of the deformation parameter.

Various formulas found by Oota, which were until then only claimed on the
base of a case-by-case analysis, have been rigorously derived in our paper 
\cite{FKS2} together with numerous entirely new identities. In particular,
the precise relation between the different versions of fusing rules has been
obtained therein and their consistency with the formulation of the mass
spectrum as null vector of a $q$-deformed Cartan matrix demonstrated.
Additional results of our work \cite{FKS2} include the systematic discussion
of the bootstrap properties, the rigorous derivation of a generic integral
representation for the scattering matrix found in \cite{Oota} as well as the
proof of new S-matrix equations, so-called combined bootstrap equations,
which are intimately linked to the underlying Lie algebraic structure. The
discussion of ATFT in this thesis will closely follow the arguments provided
in \cite{FKS2}.\medskip

\noindent The motivation for extracting as much of the Lie algebraic
structures underlying ATFT as possible is twofold. First they provide a
concise mathematical framework which eases the investigation of quantum
integrable models. Having universal expressions for the characteristic
physical quantities of the theory at hand, general claims about the
structure of QFT might be checked for the infinite class of affine Toda
models at once instead of only a few cases. Moreover, in future
applications, e.g. the form factor program, calculations might be performed
in a generic Lie algebraic framework avoiding tedious case-by-case studies.
Second, once the interplay between the powerful mathematical structures and
physical quantities has been understood, the Lie algebraic concepts might be
employed to construct entirely new integrable models with similar features.

\subsubsection{Colour valued scattering matrices and a new class of
integrable models}

In our article \cite{FKcol} a general construction principle has been
suggested leading to new scattering matrices associated with integrable
quantum models. Given the mass spectrum of an integrable model one might
multiply it by assigning to each particle additional quantum numbers,
so-called colours. Provided the scattering matrix of the original theory is
explicitly coupling dependent one might then let particles of different
colours act at different values of the coupling. A slightly more complicated
version of this principle can be employed if the coupling dependence of the
original scattering matrix can be absorbed in a separate factor. This is the
case for the ATFT S-matrix $S^{ADE}$ associated with simply-laced algebras.
Explicitly one has a decomposition of the form 
\begin{equation}
S^{ADE}=S^{\min }\,S^{\text{CDD}}\;.  \label{sepp}
\end{equation}
The so-called minimal factor $S^{\min }$ incorporates all the physical
relevant information about the bound state structure and is independent of
the coupling, while $S^{\text{CDD}}$ is a so called CDD-factor \cite{CDD},
which only introduces poles outside the physical sheet and displays the full
coupling dependence. Note that the particular feature (\ref{sepp}) is
characteristic of the $ADE$ series and ceases to hold for non-simply-laced
algebras \cite{Gust}. Letting particles of the same colour interact through $%
S^{\min }$ and those of different colours through $S^{\text{CDD}}$ one might
invoke the same Lie algebraic concepts as in the case of ATFT leading to a
class of $\frak{g}|\frak{\tilde{g}}$-theories \cite{FKcol} associated with
two simply-laced simple Lie algebras. One describes the bound state
structure and the other the colour degrees of freedom. In total this yields
a class containing as many exact S-matrices as possible pairs $(\frak{g},%
\frak{\tilde{g}})\in ADE\times ADE$. There are certain scattering matrices
obtained earlier in the literature which are contained as a subclass in the $%
\frak{g}|\frak{\tilde{g}}$-theories, namely the scaling models or minimal
ATFT associated with $S^{\min }$ (see e.g. \cite{Zper}) and the Homogeneous
Sine-Gordon models \cite{HSGS}. (Both theories are explained in more detail
below.) Additional motivation for the definition of this particular class of
scattering matrices becomes apparent when one discusses the high-energy
behaviour of the associated integrable quantum field theories by means of
the TBA and determines the underlying conformal field theories as explained
above.

\subsubsection{Affine Toda field theories as perturbed conformal models}

As outlined in the first section of the introduction, integrable quantum
field theories have a natural interpretation in terms of perturbed conformal
field theories. In this context one decomposes the associated classical
action functional of the integrable model into two parts, 
\begin{equation*}
S=S_{\text{CFT}}+\lambda \int d^{2}x\;\Phi (x,t)\;,
\end{equation*}
where $S_{\text{CFT}}$ denotes the action of the conformal model describing
the system at the critical point, $\Phi $ is a relevant spinless field
operator of the unperturbed theory and $\lambda $ is the coupling constant
of the perturbation term. On dimensional grounds $\lambda $ is proportional
to the mass scale of the perturbed theory, $\lambda \propto m^{2-d_{\Phi }}$
with $d_{\Phi }<2$ being the anomalous scaling dimension of $\Phi $ in the
conformal limit. This writing of the classical action functional appeals to
a renormalization group point of view originating in the study of critical
phenomena. At $\lambda =0$ the system is massless and constitutes a fixed
point w.r.t. renormalization group transformations. Adding a relevant
perturbation term, $\lambda \neq 0$, it is dragged away from the critical
point giving rise to a renormalization group flow towards a massive
theory.\medskip

\noindent \emph{The second part of this thesis is concerned with reversing
this renormalization group flow. Employing the thermodynamic Bethe ansatz
the high-energy regime of the mentioned integrable quantum field theories is
analyzed in detail. In this approach the mass scale is eventually sent to
zero, i.e. }$\lambda \rightarrow 0$\emph{, what implies in the above picture
that the massive system floats back into the fixed point. This in particular
will assign to each of the exact scattering matrices the central charge of
the underlying ultraviolet conformal field theory and associate them with an
off-critical model. \medskip }

The affine Toda models constitute particular examples for perturbed
conformal field theories: Omitting the term associated with the affine or
highest root $\alpha _{0}$ in (\ref{La}) one obtains the so-called \textbf{%
Toda field theories} (TFT), which are conformally invariant \cite{cToda}.
For instance the Sinh-Gordon theory can be viewed as perturbation of the
ubiquitous Liouville theory, which is the simplest and best known example of
TFT. In particular, it appears in applications related to string theory and
two-dimensional quantum gravity. All Toda models have central charge $c\geq
25$ and belong to a class of conformal field theories about the structure of
which not much is known. Having the scattering matrix of the off-critical
theory at hand, namely the affine Toda S-matrix, one might investigate the
high-energy regime by means of the TBA and perform certain consistency
checks on conjectures concerning the structure of the unperturbed conformal
model. In fact, this has first been done for the simplest examples,
Liouville theory and the Sinh-Gordon model, see \cite{Zamref}. Therein a
numerical TBA calculation was used to supplement semi-classical
considerations giving support to a proposal for the explicit form of the
three and four-point function of Liouville theory on a sphere.

Focussing on the massive side a detailed TBA analysis of all ATFT has been
performed in our articles \cite{FKS1} and \cite{FKtba}, the results of which
are presented in this thesis. Combining extensive numerical investigations
with approximate analytical considerations the off-critical Casimir energy
is determined to leading order. Based on the Lie algebraic framework
developed in the context of ATFT it is demonstrated that the latter can be
obtained in a universal formula involving only basic Lie algebraic data like
the rank of the algebra or the Coxeter number \cite{FKtba}. The behaviour
found matches with the results in \cite{Zamref} and is also in agreement
with the findings in \cite{Fateev,Fateev2}. The latter articles rely also on
semi-classical considerations and are similar in spirit to the work on the
Liouville model by Zamolodchikov and Zamolodchikov, but treat general cases
of TFT and ATFT.

In view of this comparison between data obtained from the conformal model
and the perturbed massive theory the TBA analysis also yields a consistency
check for the bootstrap construction of the ATFT S-matrix. This application
of the TBA to test scattering matrices is of particular importance when mass
spectra and the scattering amplitudes have been derived by semiclassical
arguments which need to be verified on the quantum level as it is the case
for the Homogeneous Sine-Gordon (HSG) models \cite{HSG} mentioned above as
particular subclass of the $\frak{g}|\frak{\tilde{g}}$-theories.

\subsubsection{Colour valued scattering matrices, WZNW cosets and
off-critical models}

The interest in the class of colour valued scattering matrices described
above stems from the fact that many of them can be related in the
ultraviolet limit to Wess-Zumino-Novikov-Witten (WZNW) cosets \cite{Witten}.
WZNW theories constitute the best known and possibly best understood
conformal models due to an underlying Lie algebraic structure similar to the
case of ATFT. In addition, the coset construction of these models allows one
to construct nearly every other conformal field theory. Hence, WZNW models
can be viewed as the basic building blocks in constructing conformally
invariant theories. Concrete examples are the minimal conformal models
listed in Table 1.1. They can be described by cosets of the form 
\begin{equation*}
\text{coset:}\quad \frak{g}_{1}\otimes \frak{g}_{1}/\frak{g}_{2}\quad \quad 
\text{central charge:}\quad c=\frac{2\dim \frak{g}}{(h+1)(h+2)}\;
\end{equation*}
Here $h=\sum_{i=0}^{n}n_{i}$ denotes the Coxeter number of the Lie algebra $%
\frak{g}$ and the lower index indicates the so called level of the
representation of the associated affine Lie algebra \cite{Kac}. These cosets
are the most prominent representatives, since they can be directly linked to
statistical mechanics systems in two-dimensions. Initiated by the paper of
Zamolodchikov on the Ising model in an external magnetic field, it was
realized by extensive TBA studies, that the off-critical behaviour of these
systems under a perturbation by the relevant field operator with conformal
weights $\Delta =\bar{\Delta}=2/(h+2)$ is described by so-called \textbf{%
scaling models} or \textbf{minimal affine Toda theories }\cite
{Zper,PT,TBAZam1,TBAZamun}, already mentioned above. The name indicates that
the massive field theory describing the perturbed system away from
criticality is associated with the minimal factor $S^{\min }$ of the ATFT
S-matrix \cite{TodaS} only (compare the separation property (\ref{sepp}) for
simply-laced algebras). Historically, these were the first examples of exact
scattering matrices considered in context of affine Toda field theory. They
provide concrete examples for the application of integrable quantum field
theory to statistical mechanics. This observation had important impact on
the subsequent investigations of ATFT in the literature. \medskip

\begin{center}
$
\begin{tabular}{|l|l|l|l|}
\hline\hline
$\frak{g}$ & minimal model & $c$ & perturbation \\ \hline\hline
$A_{1}$ & critical Ising & $1/2$ & $\Delta =1/2$ \\ \hline\hline
$A_{2}$ & three state Potts & $4/5$ & $\Delta =2/5$ \\ \hline\hline
$E_{6}$ & tricritical Potts & $6/7$ & $\Delta =1/7$ \\ \hline\hline
$E_{7}$ & tricritical Ising & $7/10$ & $\Delta =1/10$ \\ \hline\hline
$E_{8}$ & critical Ising & $1/2$ & $\Delta =1/16$ \\ \hline\hline
\end{tabular}
\medskip $
\end{center}

\noindent {\small Table 1.1: Minimal conformal field theories as }$\frak{g}%
_{1}\otimes \frak{g}_{1}/\frak{g}_{2}${\small \ WZNW cosets and the
conformal weights of the perturbing operator corresponding to\ minimal
affine Toda field theory.}\vspace{0.4cm}

\noindent Notice that the Ising model can be realized in two different ways
involving either the Lie algebra $A_{1}\equiv su(2)$ or the exceptional Lie
algebra $E_{8}$. This can also be seen from the Virasoro characters \cite
{KM,Vch}. While $A_{1}$ represents the simplest example of a simple Lie
algebra, $E_{8}$ is the most complex and intricate one. The two different
realizations also play an important role for the perturbed theories given by
the different field operators in Table 1.1. In case of $A_{1}$ the system is
disturbed thermally and the corresponding minimal ATFT S-matrix describes a
single massive free Majorana fermion, i.e. it is almost trivial $S^{\min
}=-1 $. In contrast, the perturbation in case of $E_{8}$ corresponds to an
external magnetic field and the associated off-critical field theory
contains now 8 different species of quantum particles, which interact with
each other in a highly non-trivial fashion. This is reflected by a
complicated scattering matrix, which involves up to 12$^{\text{th}}$ order
poles. This shows that already for the easiest example of a conformal field
theory, the free fermion, complex integrable structures might appear, when
the system becomes off-critical. Moreover, it motivates the study of the
general Lie algebraic structures, since a broad range of different simple
Lie algebras might occur in the applications as the above list of examples
impressively demonstrates.

The close connection between the Lie algebraic structures of ATFT and of
WZNW models is further supported by the class of $\frak{g}|\frak{\tilde{g}}$%
-theories. In our paper \cite{FKcol} a universal formula for the central
charge was obtained from the exact scattering matrices by means of a TBA
analysis which is also presented in this thesis, 
\begin{equation*}
c=\frac{\tilde{h}}{h+\tilde{h}}\frak{\,}\limfunc{rank}\frak{g}\cdot \limfunc{%
rank}\frak{\tilde{g}}
\end{equation*}
with $h,\tilde{h}$ being the Coxeter numbers of the algebra $\frak{g}$ and $%
\frak{\tilde{g}}$, respectively. Remarkably, this formula has also been
stated in a different context involving conformal parafermionic theories 
\cite{Dunne}. Specializing this general result to particular choices of the
related simply-laced Lie algebras, as for instance keeping $\frak{g}=ADE$
generic and restricting $\frak{\tilde{g}}$ to $A_{k-1}\equiv su(k),\,k>1$
the above central charge matches with the one of the following series of
WZNW coset models for simply-laced algebras, 
\begin{equation*}
\underset{\text{\quad \quad \quad \quad \quad }k}{\text{coset:\quad }%
\underbrace{\frak{g}_{1}\otimes \cdots \otimes \frak{g}_{1}}}/\frak{g}%
_{k}\quad \quad \quad \text{central charge:}\quad c=k\frak{\,}\frac{\dim 
\frak{g}}{h+1}-\frac{k\dim \frak{g}}{h+k}\;.
\end{equation*}
Choosing $k=2$, i.e. $\frak{\tilde{g}}=A_{1}\equiv su(2)$, one obtains in
particular the minimal affine Toda theories in accordance with the
considerations outlined above. From this point of view the latter class of
scattering matrices can be viewed as an extension of the scaling models.
Exchanging now the role of both algebras, i.e. choosing $\frak{\tilde{g}}%
=ADE $ generic and imposing this time the restriction $\frak{g}=A_{k-1},$ $%
k>1$, one obtains the central charges belonging to WZNW cosets of the form 
\begin{equation*}
\text{coset:\quad }\frak{\tilde{g}}_{k}/u(1)^{\times \limfunc{rank}\frak{%
\tilde{g}}}\quad \quad \quad \text{central charge:\quad }c=\frac{k\dim \frak{%
\tilde{g}}}{k+h}-\limfunc{rank}\frak{\tilde{g}\;.}
\end{equation*}
For this last choice of the Lie algebraic structure the associated exact
scattering matrices coincide with the ones proposed in \cite{HSGS} in the
context of the Homogeneous Sine-Gordon models \cite{HSG}. These integrable
models were explicitly constructed from the above WZNW cosets by perturbing
with an operator of conformal weight $\Delta =\bar{\Delta}=h/(k+h)$. In \cite
{HSG2} it was argued that the particular choice of the coset then ensures
that these theories are purely massive and quantum integrable, ensuring that
the scattering matrix can be constructed by the bootstrap approach.
Additional distinguished features of these integrable models are parity
violation and soliton solutions. The latter allow for a semi-classical
quantization of the mass spectrum by means of the Bohr-Sommerfeld rule \cite
{HSGsol}.

With regard to these considerations a detailed TBA analysis of the HSG
models has then been performed in our work \cite{CFKM}, the results of which
are presented in this thesis. The latter show in particular that the
conjectured S-matrix is consistent with the semi-classical picture and in
the ultraviolet limit gives rise to the correct coset central charge as
mentioned above. Moreover, extensive numerical calculations for the $\frak{%
\tilde{g}}=su(3)$ case together with analytical findings elucidate the role
of resonance poles in the proposed HSG scattering matrix. The latter have
been introduced in \cite{HSGS} in order to accommodate unstable particles in
the semi-classical spectrum. In the TBA investigation the resonance poles
give rise to a so-called \textbf{staircase pattern} in the free energy of
the system. That is, the free energy exhibits plateaus of constant height
corresponding to a flow of the perturbed theory towards different conformal
field theories, which is controlled by the external parameters, such as the
temperature and the mass scale of the unstable particles. The physical
picture one then recovers for the $\frak{\tilde{g}}=su(3)$ example is that
depending on whether the formation of unstable particles has set in or not
the HSG model in the ultraviolet regime approaches one of the two cosets 
\cite{CFKM}, 
\begin{eqnarray*}
\text{unstable particle formation} &:&\text{\quad \quad \quad \quad }\frak{%
\tilde{g}}_{k}/u(1)^{\times \limfunc{rank}\frak{\tilde{g}}} \\
\text{no unstable particle formation} &:&\text{\quad \quad \quad \quad }%
\limfunc{rank}\frak{\tilde{g}}\times su(2)/u(1)\;.
\end{eqnarray*}
In the second case it is understood that one has $\limfunc{rank}\frak{\tilde{%
g}}$ copies of the $su(2)/u(1)$ theory which do not interact with each
other. In addition, the TBA analysis of the $\frak{\tilde{g}}=su(3)$ case
strongly suggests that in an alternative formulation one might describe the
interpolation of the HSG theory between these two cosets by a massless
ultraviolet-infrared flow. For the case at hand one recovers as a subsystem
the flow between the tricritical and the critical Ising model \cite{triZam}.

Similar observations of resonance poles, staircase patterns and massless
flows have occurred in the literature before \cite{ZamoR,Stair} in the
context of the so-called \textbf{affine Toda resonance} or \textbf{staircase
models}. Here the ATFT scattering matrix for simply-laced algebras is
evaluated at complex values of the effective coupling constant. This
introduces in particular resonance poles. Similarly to the situation above
they generate a staircase pattern in the free energy of the system which
interpolates between the WZNW\ coset models 
\begin{equation*}
\text{coset:\quad }\frak{g}_{1}\otimes \frak{g}_{k}/\frak{g}_{k+1}\quad 
\text{central charge:\quad }c=\dim \frak{g}\left( \frac{1}{1+h}+\frac{k}{k+h}%
-\frac{k+1}{k+1+h}\right)
\end{equation*}
Also in this case the resonance poles of the scattering matrix have been
suggested to belong to unstable particles. However, in contrast to the HSG
models a concrete Lagrangian description of the resonance models, where this
relation would become manifest is still lacking. It is this particular
advantage of a classical Lagrangian allowing for a semi-classical approach
together with the new feature of parity violation, which makes the HSG
theories particularly interesting candidates for further investigations on
the quantum level.\medskip

\noindent In summary, the above outline demonstrates that the Lie algebraic
structures encountered in context of affine Toda models are relevant beyond
these theories, since they appear at numerous other occasions in the
discussion of integrable quantum field theories. In particular, this Lie
algebraic framework is important for the connection between exact scattering
matrices of affine Toda type and conformal field theories related to WZNW
models. An overview of the different relations is given in Figure 1.1.

Starting form a conformal WZNW theory associated with a compact simple Lie
group $G$ and an associated simple Lie algebra $\frak{g}$ upon suitable
manipulation one might obtain different other conformal theories, as for
example WZNW cosets or TFT. (The latter can be obtained by a decomposition
of the WZNW fields and a suitable reductio procedure as was worked out in 
\cite{WZWToda}.) Perturbing these conformal theories by introducing a mass
scale yields different integrable models, whose S-matrices can be obtained
by the bootstrap approach. The hinge in linking the different models is the
splitting of the ATFT scattering matrix in a minimal part and a CDD factor
for simply-laced algebras. In contrast, the non-diagonal affine Toda models
belonging to purely imaginary coupling, e.g. the Sine-Gordon theory, need to
be understood in more detail. Here many scattering matrices have not even
been constructed yet. It is to be expected that the study of the diagonal
ATFT with real coupling will also be of great help in this study. For
instance, it is well known that the scattering matrix of the breather
spectrum in the Sine-Gordon model (i.e. the bound state spectrum of the
fundamental particles) coincides with the minimal scattering matrix of $%
A_{n}^{(2)}$-ATFT for real coupling.\newpage


\begin{center}
\includegraphics[width=5.9404in,height=8.2792in,angle=0]{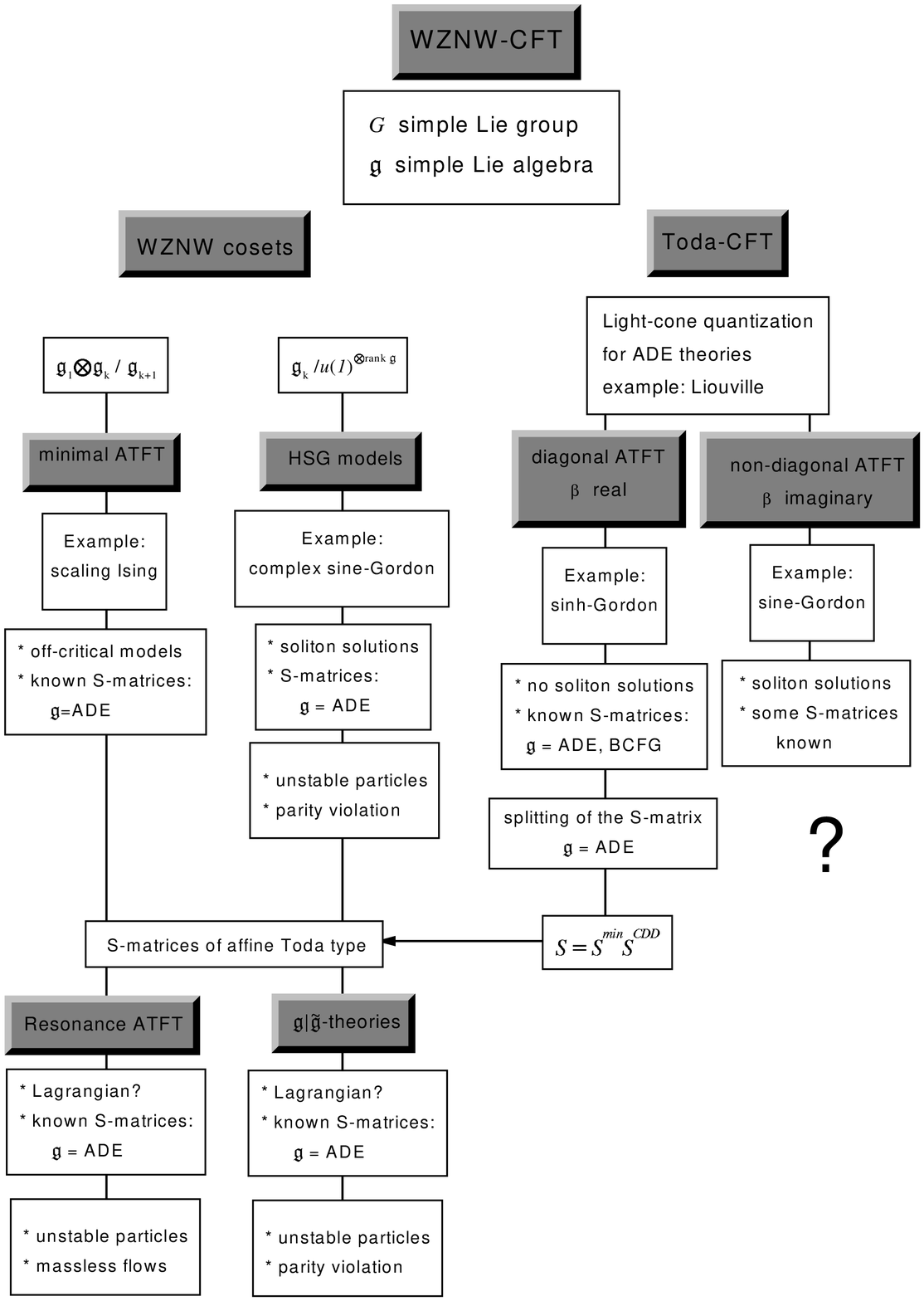}
\end{center}

$\medskip $

\begin{center}
\noindent {\small Figure 1.1: Overview of conformal and integrable field
theories off affine Toda type.}
\end{center}

\section{Outline of the content}

The description of the content is restricted to a few key-words, since a
more detailed account is given at the beginning of each chapter.
Introductory surveys in the first sections of the different chapters have
the purpose to keep the discussion self-contained.

\begin{description}
\item  \textbf{Chapter 2} is concerned with introducing those mathematical
concepts which will be used subsequently in the thesis. The presentation
starts with the theory of simple Lie algebras and their affine extensions
focussing on those aspects which are important for the physical applications
in due course. Of particular importance are the sections on Coxeter
geometry, which prepare the subsequent introduction of $q$-deformed Coxeter
elements and Cartan matrices needed to describe the renormalized mass-flow
in ATFT.

\item  \textbf{Chapter 3} is devoted to the bootstrap analysis of exact
scattering matrices, starting with a survey of the general analytic
structure of the two-particle scattering amplitude in integrable models.
Particular emphasis is given to ATFT and their Lie algebraic structure.
Fusing rules, mass spectrum and scattering matrix are discussed in a generic
Lie algebraic setting for all models at once and universal formulas are
derived. Specializing the general case to simply-laced Lie algebras the
separation property of the scattering matrix in a minimal and a CDD factor
is presented. This serves as a preliminary step towards the definition of
the colour valued S-matrices. The discussion of their bootstrap properties
together with the feature of parity violation inherent to the $\frak{g}|%
\frak{\tilde{g}}$-theories closes this chapter.

\item  \textbf{Chapter 4} contains the thermodynamic Bethe ansatz of the
integrable models discussed in the previous chapter. Since the link between
conformal and integrable quantum field theories is at the center of interest
in this presentation, some basic notions of conformal invariance are
recalled. Different numerical and analytical methods in the context of the
TBA are discussed in full generality and then applied to concrete examples,
starting with ATFT. The central charge and the free energy up to leading
order are determined, employing once more the generic Lie algebraic
apparatus to find universal formulas. The discussion continues with a
detailed analysis of the Homogeneous Sine-Gordon models with special focus
on the semi-classical predictions made in the literature. Using the Lie
algebraic structures underlying also these models, the methods of the TBA
analysis are then extended to the more general class of $\frak{g}|\frak{%
\tilde{g}}$-theories and the associated colour valued scattering matrices.
The main result is a generic formula depending only on the rank and the
Coxeter numbers of the Lie algebras involved and which associates with each
of these scattering matrices a central charge of the underlying conformal
model in the ultraviolet limit.

\item  \textbf{Chapter 5} summarizes the results and presents the
conclusions. In addition, an outlook is given on related problems and future
investigations.\bigskip
\end{description}

\noindent Throughout this thesis natural units will be used, i.e. Planck's
constant and the velocity of light are set to one, $\hbar =c=1.$\bigskip
\bigskip

\noindent As has been indicated in the introduction, parts of this thesis
have already been published. The main results presented here are contained
in the following articles:

\begin{enumerate}
\item  A. Fring, C. Korff and B.J. Schulz, \emph{The ultraviolet behaviour
of integrable field theory, affine Toda field theory}, Nucl. Phys. \textbf{%
B549} (1999) 579-612.

\item  A. Fring, C. Korff and B.J. Schulz, \emph{On the universal
representation of the scattering matrix of affine Toda field theory}, Nucl.
Phys. \textbf{B567} (2000) 409-453.

\item  A. Fring and C. Korff, \emph{Colour valued S-matrices}, Phys. Lett. 
\textbf{B477} (2000) 380-386.

\item  O.A. Castro-Alvaredo, A. Fring, C. Korff and J.L. Miramontes, \emph{%
Thermodynamic Bethe ansatz of the Homogeneous Sine-Gordon models}, Nucl.
Phys. \textbf{B573} (2000) 535-560.

\item  A. Fring and C. Korff, \emph{Large and small density approximations
to the thermodynamic Bethe ansatz}, Nucl. Phys. \textbf{B579} (2000) 617-631

\item  O.A. Castro-Alvaredo, A. Fring and C. Korff, \emph{Form factors of
the homogeneous Sine-Gordon models}, Phys. Lett. \textbf{B484} (2000)
167-176.
\end{enumerate}

\chapter{Mathematical Preliminaries}

{\small \emph{11.15. Restate my assumptions.}}

{\small \emph{1. Mathematics is the language of nature.}}

{\small \emph{2. Everything around us can be represented and understood
through numbers.}}

{\small \emph{3. If you graph these numbers patterns emerge. }}

{\small \emph{Therefore: There are patterns everywhere in nature.}\bigskip }

\qquad \qquad \qquad \qquad \qquad \qquad \qquad {\small Max Cohen in the
film }$\pi ${\small .\bigskip }

This first chapter prepares the discussion of affine Toda field theory by
introducing the mathematical framework used in the subsequent chapters to
treat all affine Toda models at once in a complete and generic way. As
pointed out in the introduction Toda and affine Toda field theories come
naturally equipped with a Lie algebraic structure. Consider the classical
Toda field equations (see e.g. \cite{LS}) which read 
\begin{equation}
\partial _{\mu }\partial ^{\mu }\varphi _{i}+\frac{m^{2}}{\beta }%
\sum_{j=1}^{n}A_{ij}\,e^{\beta \varphi _{j}}=0\;,\quad \quad i=1,...,n
\label{toda}
\end{equation}
where the field $\varphi =(\varphi _{1},...,\varphi _{n})$ consists of $n$%
-components and transforms as a scalar under the Lorentz group. The
constants $m,\beta $ represent an overall mass scale and the coupling,
respectively. Being classically unimportant they become relevant when the
system is quantized. The field equations are characterized by the so-called 
\textbf{Cartan matrix} $A=(A_{ij})$ which has only integral entries and is
non-singular, i.e. $\det A\neq 0$. Moreover it is subject to the following
restrictions:

\begin{itemize}
\item[(A1)]  Its diagonal elements take the value $A_{ii}=2$.

\item[(A2)]  The off-diagonal elements are zero or negative, $A_{ij}\leq 0$.

\item[(A3)]  For $i\neq j$, $A_{ij}=0$ implies $A_{ji}=0$.
\end{itemize}

Every such matrix can be shown to code the structure of a finite dimensional
simple Lie algebra as will be explained below. Similar, the field equations
of affine Toda field theory \cite{ATFT}, which can be viewed as perturbation
of Toda theory, read 
\begin{equation}
\partial _{\mu }\partial ^{\mu }\varphi _{i}+\frac{m^{2}}{\beta }%
\sum_{j=0}^{n}\hat{A}_{ij}\,e^{\beta \varphi _{j}}=0\;,\quad \quad
i=0,1,...,n\;,  \label{atoda}
\end{equation}
where the set of fields has been extended by a component ($i=0$) and $\hat{A}%
=(\hat{A}_{ij})$ is said to be the \textbf{extended }or \textbf{affine
Cartan matrix}. The latter can be constructed from $A$ by adding a row and a
column in a specific way explained below. It satisfies the same properties
(A1),(A2) and (A3) as the ordinary Cartan matrix but is singular\footnote{%
More precisely, it is degenerate positive definite. That is, there exists a
diagonal matrix $D$ such that $DA$ is symmetric and positive semidefinite.},
i.e. $\det \hat{A}=0$. Like in the non affine case, a Lie algebra can be
linked to the matrix $\hat{A}$, but this time it is infinite dimensional.
However, as it will turn out in the chapters to follow, the structure of the
finite dimensional simple Lie algebra is sufficient to describe most of
affine Toda field theory. Emphasis is therefore given to finite simple Lie
algebras and their geometry, which will appear in its full complexity and
elegance when we discuss the classical and quantum mass spectrum of affine
Toda theory, its fusing processes of particles and the two-particle S-matrix.

In preparation to this discussion we recapitulate in the first section of
this chapter how the Lie algebraic structure can be extracted from the
Cartan matrix $A$ appearing in the Toda equations. Starting point is the
classification of all possible Cartan matrices which gives the complete set
of possible Toda and affine Toda models with regard to (\ref{toda}) and (\ref
{atoda}). In the next step simple roots, Weyl groups and abstract root
systems are constructed. They constitute the central objects in describing
the structure of simple Lie algebras. The latter are introduced by means of
the Chevalley-Serre relations and it is shown how the root systems naturally
arise in the context of the adjoint representation.

In Section 2.2 the theory of Coxeter and twisted Coxeter elements of a Weyl
group is reviewed. Their properties are stated and their action on the root
system discussed. The formulas presented are preparatory for the subsequent
section.

Section 2.3 contains the key results of the chapter. After introducing the
concept of dual algebras $q$-deformed Coxeter and twisted Coxeter elements
are defined. The $q$-deformation will play a crucial role in exhibiting the
coupling dependence of affine Toda field theory as well as in the
formulation of its fusing rules and its scattering matrix. Furthermore, it
is shown how the information encoded in the root orbits of the $q$-deformed
Coxeter elements is contained in so-called $q$-deformed Cartan matrices. The
latter are of special importance for the integral representation of the ATFT
S-matrix which will be exploited in Chapter 4 when we investigate the high
energy limit of the associated integrable quantum field theories.

\section{Simple Lie algebras}

In this section we follow the Cartan-Killing classification of simple Lie
algebras in reverse order. We start with the most compact and reduced object
coding all the necessary information, the Cartan matrix $A$ specified above.
Step by step we then introduce more complex structures, fundamental systems,
the Weyl group and root systems. The motivation for this procedure will
become apparent when relating these objects to simple Lie algebras and their
adjoint representation in the final step. The material presented can be
found in more detail in several text books, e.g. \cite{Sam,Kac}, however,
the following summary ought to keep the discussion self-contained and is
focussed on relations and quantities which are relevant for our physical
application.

\subsection{Fundamental systems and Dynkin diagrams}

Suppose we are a given an $n\times n$ Cartan matrix $A$, then we can always
assign to it an $n$-dimensional Euclidean vector space $\frak{E}$ on which
it naturally acts. That is, we understand in the following $\frak{E}$ to be
a real vector space with positive definite inner product $\left\langle \cdot
,\cdot \right\rangle :\frak{E}\times \frak{E}\rightarrow \mathbb{R}$. The
action of the Cartan matrix is then defined w.r.t. some special basis $%
\{\alpha _{1},...,\alpha _{n}\}$ which spans $\frak{E}$ and whose scalar
products give the Cartan matrix elements, 
\begin{equation}
A_{ij}=\left\langle \alpha _{i}^{\vee },\alpha _{j}\right\rangle \;,\quad
\quad \alpha _{i}^{\vee }:=\frac{2\alpha _{i}}{\left\langle \alpha
_{i},\alpha _{i}\right\rangle }\;.  \label{Cmatrix}
\end{equation}
The elements $\alpha _{i}\,$of such a basis are called \textbf{simple roots}
and the vector $\alpha _{i}^{\vee }$ introduced above is named a \textbf{%
simple coroot}. The real vector space $\frak{E}$ with its inner product $%
\left\langle \cdot ,\cdot \right\rangle $ and the basis $\{\alpha
_{1},...,\alpha _{n}\}$ together are said to form a \textbf{simple system}.
Note that the set of simple roots is linear independent but in general not
orthonormal. In fact, exploiting the relation (\ref{Cmatrix}) one
immediately sees that 
\begin{equation*}
A_{ij}A_{ji}=4\left\langle \alpha _{i},\alpha _{j}\right\rangle ^{2}/|\alpha
_{i}|^{2}|\alpha _{j}|^{2}=4\cos ^{2}\delta _{ij}
\end{equation*}
where $0\leq \delta _{ij}\leq \pi $ is the angle between the simple roots $%
\alpha _{i},\alpha _{j}$. Since by its definition the Cartan matrix has only
integral values, the possible values for the product $A_{ij}A_{ji}$ are
restricted to $0,1,2,3,4$. The latter is excluded by linear independence of
the simple roots, because it implies $\delta _{ij}=0$ or $\pi $.
Furthermore, from the properties (A2) and (A3) we infer that the allowed off
diagonal entries in $A$ are $0,-1,-2,-3$ corresponding to the angles $\delta
_{ij}=\pi /2,2\pi /3,3\pi /4,5\pi /6$. Thus, by virtue of introducing the
Euclidean space $\frak{E}$ it follows from simple geometric arguments that
the set of all possible Cartan matrices is fairly restricted. This
observation turns out to be crucial for their classification, which we will
now perform with the help of so-called Dynkin diagrams.

A \textbf{Dynkin diagram} $\Gamma $ is a connected graph consisting of
vertices and links encoding a given Cartan matrix and its assigned simple
system in the following way. To each of the simple roots there corresponds a
vertex and the vertices of two simple roots $\alpha _{i},\alpha _{j}$ are
connected by $A_{ij}A_{ji}=0,1,2$ or $3$ lines. In particular, if $\alpha
_{i},\alpha _{j}$ are orthogonal they are not connected at all. In case that 
$A_{ij}A_{ji}=2$ or $3$ we must have that $\alpha _{i}^{2}<\alpha _{j}^{2}$
or $\alpha _{i}^{2}>\alpha _{j}^{2}$ according to (\ref{Cmatrix}) and the
fact that all matrix entries are integral. We then choose the convention to
draw an arrow pointing towards the shorter root on the line connecting the
corresponding vertices, i.e. $A_{ij}=-1$ and $A_{ji}=-2$ or $-3$ when $%
\alpha _{i}^{2}<\alpha _{j}^{2}$. In this manner, to each simple system
there exists a Dynkin diagram. Vice versa, we can now construct all such
diagrams graphically where the number of links between vertices is at most
three and interpret them in the above manner as Cartan matrices or simple
systems. Then many combinations drop out by contradicting the positive
definiteness of the inner product in $\frak{E}$. The set of allowed diagrams
is depicted in Figure 2.1 and Figure 2.2 together with its nomenclature.
They are separated in two classes namely those where only simple links occur
(the $ADE$ series) and those which allow for multiple links (the $BCFG$
series). For obvious reasons, they are called \textbf{simply-laced} and 
\textbf{non simply-laced}. Clearly, in the former case all simple roots have
the same length, $\alpha _{i}^{2}=\alpha _{j}^{2}$, while in the latter they
can be different as discussed above. The numeration of the vertices and
their colour, black or white, will become important in due course. The
dotted lines connecting two vertices in the Dynkin diagrams of the $A,D$
series and $E_{6}$ indicate possible permutations of the vertices under
which the inner product evaluated on the linear span of the simple roots
stays invariant. They are called \textbf{Dynkin diagram automorphisms} and
will become important later on when discussing dual algebras.


\begin{center}
\includegraphics[width=4.6043in,height=6.2275in,angle=0]{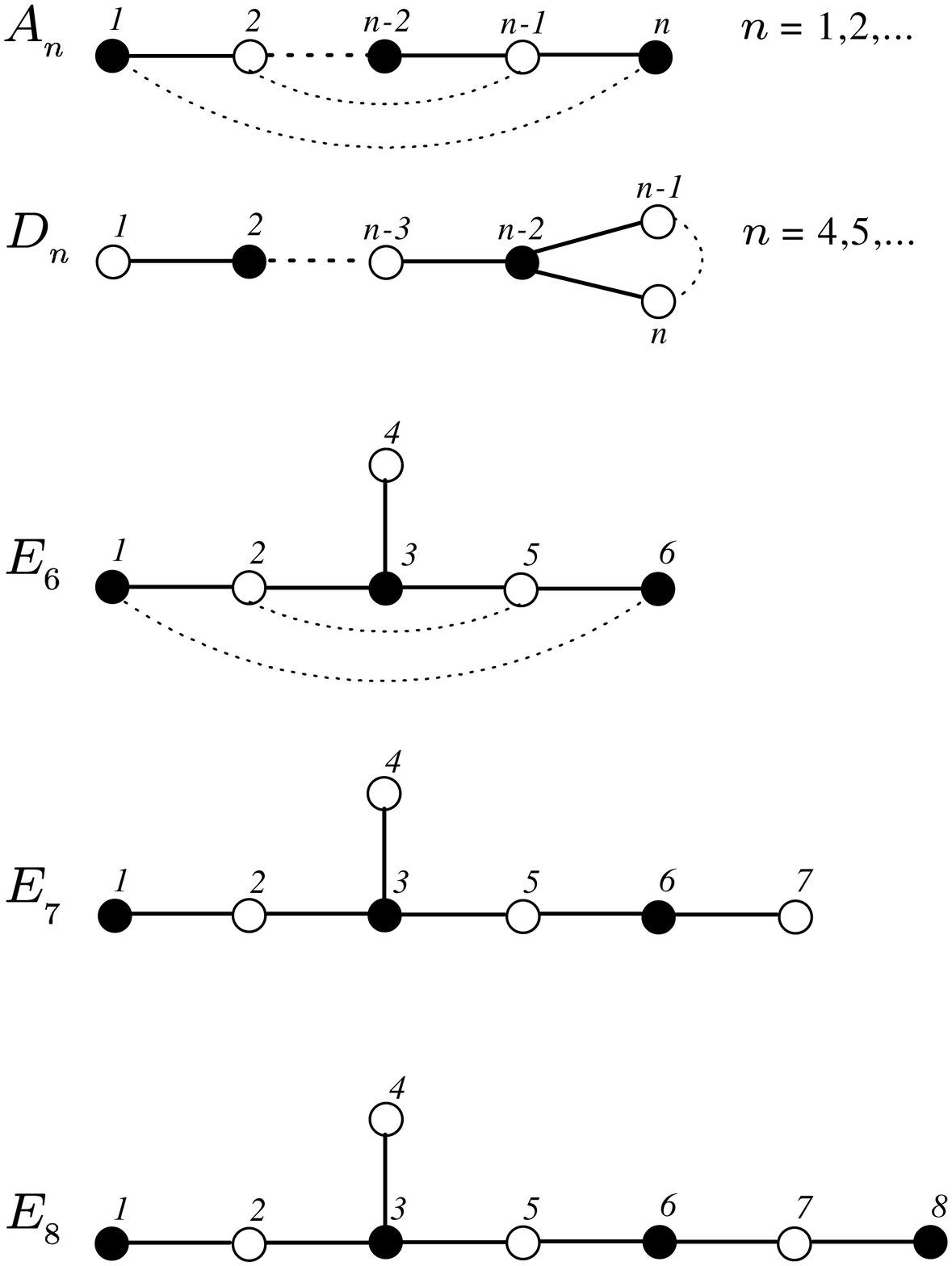}

{\small Figure 2.1: Simply-laced Dynkin diagrams\bigskip }
\end{center}


\begin{center}
\includegraphics[width=4.1857in,height=3.1367in,angle=0]{bcfg.epsi}

{\small Figure 2.2: Non simply-laced Dynkin diagrams.\bigskip }
\end{center}

\subsection{Simple Weyl reflections, the Weyl group and root systems}

In this subsection we show how to generate a group of isometries $\frak{E}%
\rightarrow \frak{E}$ from the simple system of a Cartan matrix $A$. Define
the mapping $\sigma _{i}:\frak{E}\rightarrow \frak{E}$ to be the reflection
w.r.t. to the hyperplane orthogonal to $\alpha _{i}$, 
\begin{equation}
v\rightarrow \sigma _{i}v:=v-\left\langle v,\alpha _{i}^{\vee }\right\rangle
\alpha _{i}\;.  \label{sweyl}
\end{equation}
These reflections associated with each simple root are called \textbf{simple
Weyl reflections}. They generate the so-called Weyl group $W$ via
composition. The latter belongs to the class of Coxeter groups, i.e. its
structure is completely determined by a set of relations of the form 
\begin{equation}
(\sigma _{i}\sigma _{j})^{x_{ij}}=1\;  \label{weylg}
\end{equation}
where $x_{ij}$ is always a finite integer number depending on the angle
between the two roots $\alpha _{i},\alpha _{j}$. In particular, $x_{ii}=1$.
From the generating relations (\ref{weylg}) it follows that the Weyl group
is finite, since only a limited number of composite simple Weyl reflections
are different from the identity. Thus, the action of the simple Weyl
reflections on the set of simple roots yields a finite subset $R$ of $\frak{E%
}$ which is called a \textbf{root system}, 
\begin{equation}
R=W\{\alpha _{1},...,\alpha _{n}\}\;.  \label{roots}
\end{equation}
The elements in $R$ are called \textbf{roots} and will be denoted by Greek
letters $\alpha ,\beta ,$etc. in the following. As before for simple roots
we define also so-called \textbf{coroots} by setting $\alpha ^{\vee
}:=2\alpha /|\alpha |^{2}$. Even though, we have introduced the root system
as a set generated from the simple roots by the Weyl group, $R$ can be
defined independently of them by its properties:\medskip

\noindent \textbf{Root systems.} \emph{Let }$\frak{E}$ \emph{be a finite
dimensional Euclidean space of dimension }$n$\emph{\ with inner product }$%
\left\langle \cdot ,\cdot \right\rangle $\emph{. Then a finite subset }$%
R\subset \frak{E}$\emph{\ spanning }$\frak{E}$\emph{\ and not containing }$0$%
\emph{\ is called an (abstract) root system if the following conditions are
satisfied:}

\begin{itemize}
\item[(R1)]  \emph{For any two elements }$\alpha ,\beta \in R$\emph{\ the
scalar product }$\left\langle \alpha ,\beta ^{\vee }\right\rangle $\emph{\
is an integer.}

\item[(R2)]  \emph{If }$\alpha ,\beta \in R$\emph{\ the element }$\beta
-\left\langle \alpha ,\beta ^{\vee }\right\rangle \alpha $\emph{\ also lies
in }$R$\emph{.}

\item[(R3)]  \emph{The only multiple of }$\alpha \in R$\emph{\ lying also in 
}$R$\emph{\ is }$-\alpha $\emph{.}
\end{itemize}

It needs to be emphasized that any root system characterized by the
properties (R1),(R2) and (R3) can be obtained from a simple system in the
way described. Thus, Cartan matrices, simple systems and root systems
constitute in a loose sense equivalent objects and all of them are
classified by the Dynkin diagrams depicted in Figures 2.1 and 2.2. In
particular, we might also define the Weyl group solely in terms of the root
system by associating to each $\alpha \in R$ a Weyl reflection $\sigma
_{\alpha }:\frak{E}\rightarrow \frak{E}$ completely analogous to (\ref{sweyl}%
). The latter then generate the same Weyl group $W$. Clearly, the condition
(R2) just states the invariance of $R$ under $W$.

From (\ref{roots}) we infer that every root $\alpha \in R$ can be written as
an integral linear combination of simple roots by observing that the simple
Weyl reflections act on simple roots via the Cartan matrix, $\sigma
_{i}\alpha _{j}=\alpha _{j}-A_{ji}\alpha _{i}$. This naturally induces a $%
\mathbb{Z}$-gradation of $R$ when associating to each root its height ht$%
:R\rightarrow \mathbb{Z}$, 
\begin{equation}
\text{ht\thinspace }(\alpha ):=\sum_{i=1}^{n}k_{i}\quad \text{with}\quad
\alpha =\sum_{i=1}^{n}k_{i}\alpha _{i}\;.  \label{ht}
\end{equation}
The height functional can be used to define a partial ordering $\succ $ of
the root system writing symbolically $\alpha \succ \beta ,$ whenever ht$%
(\alpha )>$ht$(\beta )$ for $\alpha ,\beta \in R$. One then calls $\alpha
\in R$ a \textbf{positive root}, in symbols $\alpha \succ 0$, if ht$(\alpha
)>0$. The set of all positive roots will be denoted by $R_{+}$. In the
obvious way, the set of negative roots $R_{-}$ is defined and its elements
are called \textbf{negative roots}, in symbols $\alpha \prec 0$. Clearly, $%
R_{-}=-R_{+}$ and more importantly $R=R_{+}\cup R_{-}$. For the classified
root systems the partial ordering singles out a unique \textbf{highest }or 
\textbf{maximal root} $\theta $ satisfying ht$(\theta )>$ht$(\alpha )$ for
all roots $\alpha $. Throughout this thesis the normalization of the inner
product is chosen such that 
\begin{equation}
\left\langle \theta ,\theta \right\rangle =2\;.  \label{convention}
\end{equation}
The linear coefficients $n_{i},n_{i}^{\vee }\in \mathbb{N}$ of the highest
root w.r.t. the basis of simple roots and coroots are called \textbf{Coxeter}
and \textbf{dual Coxeter labels}, respectively. From the definition of the
coroots we see that the Coxeter labels are related to their dual
counterparts by $n_{i}=|\alpha _{i}|^{2}n_{i}^{\vee }/2$. Their sums define
two important and characteristic constants of a simple Lie algebra, the
so-called \textbf{Coxeter }and \textbf{dual Coxeter number}, both of which
we will frequently encounter in the following, 
\begin{equation}
h:=1+\sum_{i=1}^{n}n_{i}\quad \text{and}\quad h^{\vee
}:=1+\sum_{i=1}^{n}n_{i}^{\vee }\;.  \label{Coxnumbers}
\end{equation}
For later purposes it is important to note that the height functional can be
implemented by an element of the vector space $\frak{E}$. Let $\{\lambda
_{1},...,\lambda _{n}\}$ and $\{\lambda _{1}^{\vee },...,\lambda _{n}^{\vee
}\}\ $denote the dual basis of the simple coroots and simple roots,
respectively, i.e. 
\begin{equation}
\left\langle \lambda _{i},\alpha _{j}^{\vee }\right\rangle =\delta
_{ij}\quad \text{and}\quad \left\langle \lambda _{i}^{\vee },\alpha
_{j}\right\rangle =\delta _{ij}\;.  \label{weights}
\end{equation}
\ The elements $\lambda _{i}$ are called \textbf{fundamental weights} and
the elements $\lambda _{i}^{\vee }$ \textbf{fundamental coweights}. Their
sums define the so-called \textbf{Weyl }and \textbf{dual Weyl vector }which
equivalently can also be expressed in terms of the positive roots and
coroots, respectively, 
\begin{equation}
\rho =\sum_{i=1}^{n}\lambda _{i}=\frac{1}{2}\sum_{\alpha \in R_{+}}\alpha
\quad \text{and}\quad \rho ^{\vee }=\sum_{i=1}^{n}\lambda _{i}^{\vee }=\frac{%
1}{2}\sum_{\alpha \in R_{+}}\alpha ^{\vee }\;.  \label{weylv}
\end{equation}
The Weyl vectors are uniquely determined by the properties $\left\langle
\rho ^{\vee },\alpha _{i}\right\rangle =\left\langle \rho ,\alpha _{i}^{\vee
}\right\rangle =1$ which is immediate by their definition in terms of
fundamental (co)weights. The equivalent expressions in terms of (co)roots
can be derived by use of the Weyl group. The height functional can now be
expressed by the scalar product ht$(\alpha )=\left\langle \rho ^{\vee
},\alpha \right\rangle $. In particular, Coxeter and dual Coxeter number
read 
\begin{equation}
h=1+\left\langle \rho ^{\vee },\theta \right\rangle \quad \quad \quad
\,h^{\vee }=1+\left\langle \rho ,\theta \right\rangle \;.
\end{equation}
This closes the discussion of abstract root systems and we now move on from
linear algebra to the theory of simple Lie algebras by associating to the
elements of each root system a set of Lie algebra generators. However, most
of the actual calculations will only use the framework of linear algebra
introduced in this and the preceding subsection.

\subsection{Lie algebras and the Chevalley-Serre relations}

The concept of a Lie algebra naturally arises in physics when one considers
the infinitesimal generators of symmetry operations mathematically linked to
a Lie group $G$. For instance, if we consider the group $G=SU(2)$ their
infinitesimal generators are given by the angular momentum operators $%
\{J_{+},J_{-},J_{3}\}$ which form the well known Lie algebra $su(2)$. The
group structure is encoded in their commutation relations, 
\begin{equation}
\lbrack J_{+},J_{-}]=2J_{3}\;,\quad \quad \lbrack J_{3},J_{\pm }]=\pm J_{\pm
}\;  \label{su2}
\end{equation}
together with the property that $J_{+}$ is the Hermitian conjugate of $J_{-}$
and $J_{3}$ is self-conjugate. The Lie group is then regained by
''exponentiation'' of the generators in an appropriate way. However, from a
mathematical point of view the concept of a Lie algebra can be treated
separately from the one of a Lie group. We thus start with the abstract
definition of a simple Lie algebra and then comment on its realizations in
terms of Cartan matrices and simple systems.\medskip

\noindent \textbf{Lie algebra. }\emph{A vector space }$\frak{g}$\emph{\
(real or complex) is called a Lie algebra if it is equipped with a bilinear
map }$[\cdot ,\cdot ]:\frak{g}\times \frak{g}\rightarrow \frak{g},$\emph{\
the so-called \textbf{Lie product} or \textbf{bracket}, satisfying
skew-symmetry and the Jacobi identity, i.e., } 
\begin{equation}
\lbrack x,y]=-[y,x]\quad \text{\ and\quad }[x,[y,z]]+[y,[z,x]]+[z,[x,y]]=0
\label{jacobi}
\end{equation}
\emph{for all }$x,y,z\in \frak{g}$\emph{.}\medskip

The Lie algebras we are going to consider in this section are all finite
dimensional, meaning that $\frak{g}$ viewed as a vector space has finite
dimension. Furthermore, we restrict ourselves to \textbf{simple} \textbf{Lie
algebras}. The latter are characterized by the property that they are not
abelian, i.e. $[\frak{g},\frak{g}]\neq \{0\}$, and contain no proper ideal,
i.e. there exists no subalgebra $\frak{i}$ which obeys the relation $[\frak{g%
},\frak{i}]\subset \frak{i}$. This particular class of Lie algebras is
distinguished by the fact that all its elements are classified through the
Cartan matrices or equivalently the root systems presented in the preceding
subsections.

Suppose we are given a Cartan matrix $A$ together with an assigned simple
system. The most direct way to associate a Lie algebra $\frak{g}=\frak{g}(A)$
to them is to define to each simple root $\alpha _{i}$ a subalgebra $%
\{e_{i},f_{i},h_{i}/2\}$ isomorphic to $su(2)$ specified in (\ref{su2}).
Here $e_{i},f_{i}$ are identified with the step operators $J_{+},J_{-}$,
respectively, and $h_{i}/2$ plays the role of $J_{3}$. In order to determine
the Lie algebra $\frak{g}(A)$ generated by $\{e_{i},f_{i},h_{i}/2%
\}_{i=1}^{n} $ completely we need to define in addition the bracket
relations between generators belonging to different simple roots. These are
known as \textbf{Chevalley-Serre relations} and read, 
\begin{equation*}
\lbrack h_{i},h_{j}]=0\,,\quad \lbrack h_{i},e_{j}]=A_{ij}e_{j}\,,\quad
\lbrack h_{i},f_{j}]=-A_{ij}f_{j}\,,\quad \lbrack e_{i},f_{j}]=\delta
_{ij}h_{j}\,,
\end{equation*}
\begin{equation}
(\limfunc{ad}\nolimits_{e_{i}})^{1-A_{ij}}e_{j}=0\quad \text{and}\quad (%
\limfunc{ad}\nolimits_{f_{i}})^{1-A_{ij}}f_{j}=0  \label{serre}
\end{equation}
where in the second set of relations the symbol $\limfunc{ad}$ stands for
the \textbf{adjoint action}, i.e. $\limfunc{ad}_{x}y:=[x,y]$. Now, the
important result is that the constructed Lie algebra $\frak{g}(A)$ is simple
and that any simple Lie algebra can be obtained in this way, i.e. there is a
bijective mapping $A\rightarrow \frak{g}(A)$ from the set of Cartan matrices
into the set of simple Lie algebras. Hence, the classification scheme for
Cartan matrices via Dynkin diagrams carries through to simple Lie algebras.
To make the abstract definition of simple Lie algebras more explicit, some
examples which can be realized in terms of matrix algebras are stated in the
following table,\medskip

\begin{center}
$
\begin{tabular}{|c|c|}
\hline\hline
Lie algebra & matrix algebra \\ \hline\hline
$A_{n}$ & $su(n+1)$ \\ \hline\hline
$B_{n}$ & $so(2n+1)$ \\ \hline\hline
$C_{n}$ & $sp(2n)$ \\ \hline\hline
$D_{n}$ & $so(2n)$ \\ \hline\hline
\end{tabular}
$\medskip

\vspace{0.4cm} {\small Table 2.1: Matrix representations of simple Lie
algebras.}
\end{center}

\noindent Here $su(n)$ denotes the Lie algebra of the unitary $n\times n$
matrices with unit determinant, $so(n)$ the Lie algebra of the orthogonal $%
n\times n$ matrices with unit determinant and $sp(n)$ the $n$-dimensional
symplectic algebra.

Having constructed all simple Lie algebras via the above identification
scheme concludes the Cartan-Killing classification program. In the next
subsection we briefly comment on how the simple Lie algebras give naturally
rise to root systems by means of the adjoint representation. To simplify
this discussion we henceforth assume the Lie algebra to be defined over the
complex numbers.\footnote{%
So far we have constructed $\frak{g}(A)$ as a real Lie algebra using an
underlying real Euclidean vector space $\frak{E}$, however, the abstract
definition of a Lie algebra allows also for vector spaces over the complex
numbers. In fact, we might associate to $\frak{g}(A)$ the complex Lie
algebra $\frak{g}(A)\otimes \mathbb{C=}\frak{g}(A)\oplus \sqrt{-1}\frak{g}%
(A) $. On the contrary, every complex Lie algebra can be reduced to a real
Lie algebra by restriction to real linear combinations. Under this mutual
relation the classification scheme presented above carries over to complex
simple Lie algebras as well, that is, all complex simple Lie algebras are
obtained in this manner.}

\subsection{The adjoint representations of simple Lie algebras}

Every simple Lie algebra $\frak{g}$ might be represented as an operator
algebra acting on itself. Regard $\frak{g}$ as a vector space then each
element in it can be interpreted as operator by means of the adjoint action
defined above, 
\begin{equation}
\frak{g}\ni x\mapsto \limfunc{ad}\nolimits_{x}\;.  \label{adjoint}
\end{equation}
Then as an immediate consequence of the relations (\ref{jacobi}) the Lie
bracket might be identified with the commutator, $[x,y]\sim \limfunc{ad}%
\nolimits_{x}\circ \limfunc{ad}\nolimits_{y}-\limfunc{ad}\nolimits_{y}\circ 
\limfunc{ad}\nolimits_{x}$. This realization of a simple Lie algebra is
called its \textbf{adjoint representation} and suggests in a natural way the
introduction of a metric on the algebra by setting 
\begin{equation}
(x,y)\mapsto \kappa (x,y):=\frac{1}{2h^{\vee }}\limfunc{Tr}(\limfunc{ad}%
\nolimits_{x}\circ \limfunc{ad}\nolimits_{y})\;.  \label{killing}
\end{equation}
Here the normalization constant is the so-called Dynkin index of the adjoint
representation (see e.g. \cite{Sam}) which due to our convention (\ref
{convention}) equals twice the dual Coxeter number. The metric (\ref{killing}%
) is known as \textbf{Killing form}. Clearly, $\kappa $ is symmetric,
bilinear and invariant w.r.t. the bracket in the sense that the following
relation holds, $\kappa ([x,y],z)=\kappa (x,[y,z])$ for all $x,y,z\in \frak{g%
}$. Moreover, for $\frak{g}$ simple it is non-degenerate, i.e. $\kappa
(x,y)=0$ for all $x\in \frak{g}$ implies $y=0$. In the following we now
outline how the abstract root system defined above naturally appears in the
adjoint representation.

From the Serre relations (\ref{serre}) we infer that we can write $\frak{g}$
as the following direct sum in the sense of vector spaces, 
\begin{equation*}
\frak{g}(A)=\frak{g}_{+}\oplus \frak{h}\oplus \frak{g}_{-}
\end{equation*}
where $\frak{g}_{+},\frak{g}_{-}$ are the subspaces generated by the step
operators $e_{i}$ and $f_{i}$, respectively, and $\frak{h}$ denotes the
subalgebra spanned by the $h_{i}$'s. According to (\ref{serre}) it is
maximal abelian and such a subalgebra $\frak{h}$ is said to be a \textbf{%
Cartan subalgebra}.\footnote{%
In fact, for any complex simple Lie algebra we might speak of \emph{the}
Cartan subalgebra, since one can show that independent of the choice of the
fundamental system all Cartan subalgebras of $\frak{g}$ are conjugate to
each other with respect to the group of inner automorphisms. That is, $g%
\frak{h}g^{-1}$ yields another Cartan subalgebra where $g$ is an element of
the Lie group $G$ associated with $\frak{g}$ and every Cartan subalgebra may
be obtained this way.} Its dimension as a vector space is called the \textbf{%
rank} of $\frak{g}$ and by construction it coincides with the dimension of
the Euclidean space $\frak{E}$ introduced in the context of fundamental and
root systems, 
\begin{equation}
\limfunc{rank}\frak{g}:=\dim \frak{h}=n\;.  \label{rank}
\end{equation}
Because $\frak{h}$ is maximal abelian the adjoint representation restricted
to $\frak{h}$ must decompose into a direct sum of one-dimensional
representations, i.e. all the elements in $\frak{h}$ can be diagonalized
simultaneously. Thus, we may write $\frak{g}$ as a direct sum of $\frak{h}$
and one-dimensional subspaces $\frak{g}_{\alpha }$ invariant under the
action of $\limfunc{ad}\frak{h}$, 
\begin{equation}
\frak{g}=\frak{h\oplus }\bigoplus_{\alpha \in R}\frak{g}_{\alpha }\quad 
\text{with\quad }[h,\QTR{frak}{g}_{\alpha }]=\alpha (h)\frak{g}_{\alpha
},\quad h\in \frak{h}.  \label{decomp}
\end{equation}
Here the labels $\alpha $ are elements in the dual space $\frak{h}^{\ast }$
of the Cartan subalgebra defined by $\alpha :h\rightarrow \alpha (h)$, where
the scalar factors $\alpha (h)$ are the eigenvalues of the operators $%
\limfunc{ad}\nolimits_{h},h\in \frak{h}$ when restricted to the eigenspace $%
\frak{g}_{\alpha }$. As the notation indicates the functionals $\alpha \in 
\frak{h}^{\ast }$ constitute a root system $R$ by identifying the Euclidean
space $\frak{E}$ as the subspace of $\frak{h}^{\ast }$ obtained when
restricting the linear combinations to real numbers. Conversely, given an
abstract root system $R$ in an Euclidean space $\frak{E}$ there always
exists a simple Lie algebra $\frak{g}(R)$ determined by the direct sum (\ref
{decomp}). The latter is then called \textbf{root space decomposition}. In
fact, from the above decomposition we immediately derive that the number of
roots is related to the dimension of the simple Lie algebra by 
\begin{equation}
\dim \frak{g}=|R|+\limfunc{rank}\frak{g}=|R|+n  \label{rnumber}
\end{equation}
The remaining structure which needs to be identified is the inner product $%
\left\langle \cdot ,\cdot \right\rangle $ in $\frak{E}$. It is related to
the Killing form of the adjoint representation by setting 
\begin{equation*}
\kappa (h_{\alpha },h_{\beta })=\left\langle \alpha ,\beta \right\rangle
\end{equation*}
where the element $h_{\alpha }\in \frak{h}$ is uniquely defined by the
relation $\alpha (h)=\kappa (h_{\alpha },h)$. Similar, like in the case of
simple roots one finds to each $\alpha \in R$ a triplet $\{e_{\alpha
},e_{-\alpha },h_{\alpha ^{\vee }}/2\}$ of generators which forms a
subalgebra isomorphic to $su(2)$ with $e_{\alpha },e_{-\alpha }$ playing the
role of the ladder operators. The latter span the one-dimensional subspaces $%
\frak{g}_{\alpha },\frak{g}_{-\alpha }$ in (\ref{decomp}) and are obtained
from the Chevalley generators $e_{i}=e_{\alpha _{i}},f_{i}=e_{-\alpha _{i}}$
as multiple commutators via the Chevalley-Serre relations (\ref{serre}).
These describe how the whole root system $R$ can be constructed from simple
roots in terms of Lie algebra generators. The latter represents the
decomposition of the Lie algebra $\frak{g}(A)$ into $su(2)$ subalgebras
whose mutual commutation rules are reflected by the linear structure of $R$.
In particular, one infers by means of (\ref{jacobi}) that the subspaces obey
the relation 
\begin{equation*}
\lbrack \frak{g}_{\alpha },\frak{g}_{\beta }]\subset \frak{g}_{\alpha +\beta
}\;,
\end{equation*}
where it is understood that $\frak{g}_{\alpha }=\{0\}$ if $\alpha $ is not a
root.

\section{Affine Lie Algebras}

We now come to an extension of the finite simple Lie algebras studied so
far. As mentioned in the introduction to this chapter affine Toda field
theories are parametrized by an extended or affine Cartan matrix. The latter
describes in a similar fashion a Lie algebra $\frak{\hat{g}},$ which is
infinite dimensional and closely related to a simple finite dimensional Lie
algebra $\frak{g}$. In this section we will first turn to the Lie algebraic
structure and then see at the end how the affine Cartan matrix emerges from
the affine root system. However, throughout the subsequent chapters we will
only use the structure of the finite Lie algebras but as the name of the
integrable theories we are going to investigate indicates there is naturally
associated an affine structure. The latter is particularly important in the
context of classical affine Toda field theory as for example in the
construction of solutions to the classical equations of motion (\ref{atoda}%
), see e.g. \cite{LS,OTU,KO}. For completeness we therefore present a short
survey on the construction of affine Lie algebras, but this exposition will
be less detailed than the discussion of simple Lie algebras. For a more
profound treatise of affine Lie algebras and their important role in physics
see e.g. \cite{Kac,GO}.

Let $\frak{g}$ be a simple finite Lie algebra and $\left\langle \cdot ,\cdot
\right\rangle :\frak{g}\times \frak{g}\rightarrow \mathbb{C}$ its Killing
form (\ref{killing}). We now assign to this pair an infinite-dimensional
graded Lie algebra $\frak{\hat{g}}$ which is called (untwisted) \textbf{%
affine Lie algebra}. At the heart of this construction lies the following
geometrical picture in terms of the associated groups. Suppose that $G$ is a
compact Lie group associated with $\frak{g}$. Then we might consider the set
of diffeomorphisms from the unit circle into the group, $\mathbb{S}%
^{1}\rightarrow G$. Defining the group multiplication on this set by
pointwise multiplication one obtains an infinite-dimensional Lie group, the
so-called \textbf{loop group} of $G$ whose infinitesimal generators give
rise to the \textbf{loop algebra}, 
\begin{equation}
L(\frak{g}):=\frak{g}\otimes \mathbb{C}[t,t^{-1}]\;.  \label{loop}
\end{equation}
The above definition as a tensor product is understood in the sense of
vector spaces and $\mathbb{C}[t,t^{-1}]$ denotes the commutative algebra of
all Laurent polynomials over the field $\mathbb{C}$, i.e. an element $f\in 
\mathbb{C}[t,t^{-1}]$ in this algebra is of the form 
\begin{equation*}
f=\sum_{n\in \mathbb{Z}}f_{n}\,t^{n},\quad f_{n}\in \mathbb{C},
\end{equation*}
where only a finite number of the coefficients $f_{n}$ is nonzero. While the
first factor in (\ref{loop}) represents the generators of $G$, the second is
associated with an infinitesimal smooth mapping defined on the unit circle,
which always allows for an expansion in a Laurent polynomial when setting $%
t=e^{i\tau },\tau \in \mathbb{R}$. To determine the loop algebra completely
we need to specify the Lie bracket structure, which is given by 
\begin{equation}
\lbrack x\otimes t^{m},y\otimes t^{n}]=[x,y]\otimes t^{m+n},
\end{equation}
where $x,y\in \frak{g}$. In context of quantum theory central extensions of
Lie algebras play an immanent role. Their occurrence can be understood in
the sense of anomalies which arise when classical symmetries get quantized.
In Chapter 4 this will be discussed in more detail in the context of
conformal field theory. The loop algebra $L(\frak{g})$ has a unique
nontrivial central extension $\frak{\hat{g}}$ defined by the exact sequence 
\begin{equation}
0\rightarrow \mathbb{C}\,k\rightarrow \frak{\hat{g}}\rightarrow L(\frak{g}%
)\rightarrow 0,
\end{equation}
where $\mathbb{C}\,k$ is a one dimensional space spanned by the so-called 
\textbf{central element} $k$. The latter is characterized by the property
that it commutes with all other generators. Explicitly, $\frak{\hat{g}}$ as
vector space is given by 
\begin{equation}
\frak{\hat{g}}=\frak{g}\otimes \mathbb{C}[t,t^{-1}]\oplus \mathbb{C}\,k
\end{equation}
and the bracket structure $[\cdot ,\cdot ]:\frak{\hat{g}}\times \frak{\hat{g}%
}\rightarrow \frak{\hat{g}}$ takes the following form 
\begin{equation}
\lbrack \frak{\hat{g}},k]=0\quad \text{and}\quad \lbrack x\otimes
t^{m},y\otimes t^{n}]=[x,y]\otimes t^{m+n}+\left\langle x,y\right\rangle
\,m\,\delta _{m+n,0}\,k\;.
\end{equation}
As a rule we will call $\frak{\hat{g}}$ the \textbf{affine Lie algebra}
associated with $\frak{g}$. Analogous to the case of simple Lie algebras one
might consider the adjoint representation in order to extract the root
system as the eigenvalues of the generators in the Cartan subalgebra.
However, the latter turn out to be infinitely degenerate, whence one usually
introduces a grading of the affine Lie algebra via the derivation $d\equiv t%
\frac{d}{dt}$ acting on the polynomial part of the Lie algebra elements
setting 
\begin{equation}
\lbrack d,k]=0\quad \;\text{and\ }\quad \lbrack d,x\otimes f]=x\otimes df\;
\end{equation}
with $f\in \mathbb{C}[t,t^{-1}]$. Thus, there is a natural gradation of $%
\frak{\hat{g}}$ obtained by considering the eigenspaces of the degree
operator $d$, 
\begin{equation}
\frak{\hat{g}}=\bigoplus_{n\in \mathbb{Z}}\frak{\hat{g}}_{n},\quad \frak{%
\hat{g}}_{n}:=\{x\in \frak{\hat{g}}:[d,x]=nx\}.
\end{equation}
Clearly, these eigenspaces are spanned by the monomials $f(t)=t^{n}$. Adding
this degree operator to the affine Lie algebra gives rise to what is called
the \textbf{extended affine Lie algebra}, 
\begin{equation}
\frak{\tilde{g}}=\frak{\hat{g}}\oplus \mathbb{C}d=\frak{g}\otimes \mathbb{C}%
[t,t^{-1}]\oplus \mathbb{C}\,k\oplus \mathbb{C}d\;.
\end{equation}
In order to obtain the same structures as in the finite-dimensional,
non-affine case it remains to extend the Killing form $\left\langle \cdot
,\cdot \right\rangle $ to $\frak{\tilde{g}}$. This is done by setting 
\begin{equation}
\left\langle x\otimes f,y\otimes g\right\rangle :=\frac{\left\langle
x,y\right\rangle }{2\pi i}\oint_{\mathbb{S}^{1}}\frac{dt}{t}%
\,f(t)g(t)\;,\quad \quad f,g\in \mathbb{C}[t,t^{-1}]
\end{equation}
and 
\begin{equation}
\left\langle k,k\right\rangle =\left\langle d,d\right\rangle =\left\langle
x\otimes f,k\right\rangle =\left\langle x\otimes f,d\right\rangle =0\quad
\quad \left\langle k,d\right\rangle =1\;.
\end{equation}
One might verify that this definition makes $\left\langle \cdot ,\cdot
\right\rangle $ non-degenerate and invariant on the extended affine Lie
algebra, while its restriction to the affine Lie algebra $\frak{\hat{g}}$ is
degenerate.

\subsection{Affine roots and the generalized Cartan matrix}

Similar to the non-affine case discussed in the previous section the root
system can be recovered from the dual space to the Cartan subalgebra. In
analogy to the finite-dimensional case one therefore defines the so-called 
\textbf{affine Cartan subalgebras}, 
\begin{equation}
\frak{\hat{h}}=\frak{h}\oplus \mathbb{C}k\subset \frak{\hat{g}\quad }\text{%
and}\quad \frak{\tilde{h}}=\frak{h}\oplus \mathbb{C}k\oplus \mathbb{C}%
d\subset \frak{\tilde{g}}\text{\ }
\end{equation}
with $\frak{h}$ being the Cartan subalgebra of the non-affine finite simple
Lie algebra $\frak{g}$. The dual spaces are then given by 
\begin{equation}
\frak{\hat{h}}^{\ast }=\frak{h}^{\ast }\oplus \mathbb{C}\lambda _{0}\frak{%
\quad }\text{and}\quad \frak{\tilde{h}}^{\ast }=\frak{h}^{\ast }\oplus 
\mathbb{C}\lambda _{0}\oplus \mathbb{C}\delta \text{,}
\end{equation}
respectively. The new extra elements $\lambda _{0},\delta $ are the linear
functionals corresponding to the central element and the degree operator and
are defined by 
\begin{equation*}
\lambda _{0}(k)=1,\quad \lambda _{0}(d)=\lambda _{0}(h)=0\quad \text{%
and\quad }\delta (d)=1,\quad \delta (k)=\delta (h)=0,
\end{equation*}
for all $h\in \frak{h}$. As in the finite dimensional case (\ref{decomp})
one has now the following affine root space decomposition, 
\begin{equation}
\frak{\tilde{g}}=\frak{\tilde{h}}\oplus \bigoplus_{\hat{\alpha}\in \hat{R}}%
\frak{\tilde{g}}_{\hat{\alpha}}\,,  \label{affdecomp}
\end{equation}
where the set of \textbf{affine roots} is obtained to 
\begin{equation}
\hat{R}=\left\{ \alpha +n\delta :\alpha \in R,\,n\in \mathbb{Z}\}\cup
\{n\delta :\,n\in \mathbb{Z}\backslash \{0\}\right\}  \label{affRoot}
\end{equation}
and the subspaces $\frak{\tilde{g}}_{\hat{\alpha}}:=\{x\in \frak{\tilde{g}}%
:[h,x]=\hat{\alpha}(h)x,\,h\in \frak{\tilde{h}}\}$ invariant under the
adjoint action of the affine Cartan subalgebras can be explicitly written
down as 
\begin{equation}
\frak{\tilde{g}}_{\alpha +n\delta }=\{e_{\alpha }\otimes t^{n}\}\quad \;%
\text{and\ }\quad \frak{\tilde{g}}_{n\delta }=\frak{h}\otimes t^{n}\;.
\end{equation}
Note that the set of affine roots contains also the roots of the non-affine
Lie algebra $\frak{g}$, i.e. $R\subset \hat{R}$. In particular, for $n=0$ we
can identify $\frak{\tilde{g}}_{\alpha +n\delta }$ with the non-affine
subspace $\frak{g}_{\alpha }$ in (\ref{decomp}).

The next step is to find a basis for set of affine roots. Looking at the
structure of the affine root set (\ref{affRoot}) one immediately infers,
that the set of simple roots $\{\alpha _{1},...,\alpha _{n}\}\subset R$ of
the non-affine Lie algebra must be supplemented by an additional element
generating the dependence on the linear functional $\delta $. This element
is called the \textbf{affine root} and is defined as 
\begin{equation*}
\alpha _{0}\equiv -\theta +\delta \;.
\end{equation*}

Recall from the non-affine case that there is to each of the simple roots a $%
su(2)$ subalgebra associated, the so-called Chevalley generators. From the
latter the simple Lie algebra can be constructed by means of the
Chevalley-Serre relations (\ref{serre}) encoded in the Cartan matrix $A$.
The same is true for affine Lie algebras w.r.t. the generalized Cartan
matrix $\hat{A}$ which is defined as 
\begin{equation*}
\hat{A}_{ij}=\left\langle \alpha _{i}^{\vee },\alpha _{j}\right\rangle
\;,\quad \quad i,j=0,1,...,n\;.
\end{equation*}
Defining the Chevalley generators for the affine root as 
\begin{equation*}
e_{0}\equiv e_{-\theta }\otimes t\;,\quad f_{0}\equiv e_{\theta }\otimes
t^{-1}\;,\quad h_{0}\equiv k+h_{-\theta }\;
\end{equation*}
the set $\{e_{i},f_{i},h_{i}\}_{i=0}^{n}$ generates the affine Lie algebra $%
\frak{\hat{g}}$ upon invoking the affine version of the Chevalley-Serre
relations, 
\begin{equation*}
\lbrack h_{i},h_{j}]=0\,,\quad \lbrack h_{i},e_{j}]=\hat{A}%
_{ij}e_{j}\,,\quad \lbrack h_{i},f_{j}]=-\hat{A}_{ij}f_{j}\,,\quad \lbrack
e_{i},f_{j}]=\delta _{ij}h_{j}\,,
\end{equation*}
\begin{equation}
(\limfunc{ad}\nolimits_{e_{i}})^{1-\hat{A}_{ij}}e_{j}=0\quad \;\text{and\ }%
\quad (\limfunc{ad}\nolimits_{f_{i}})^{1-\hat{A}_{ij}}f_{j}=0\;.
\end{equation}
The remarkable fact that the above algebraic relations lead to same Lie
algebra $\frak{\hat{g}}$ as the geometrical construction in terms of the
loop algebra was proven in \cite{Kac}. The interplay between these two
aspects gives rise to a rich mathematical structure and is the deeper reason
why affine Lie algebras play an important role in physics \cite{Kac,GO}.
After this short digression we now turn back to simple Lie algebras and
their finite-dimensional root spaces.

\section{Coxeter and twisted Coxeter elements}

In the preceding sections we saw how to each simple Lie algebra a Euclidean
space and a root system is naturally assigned. In this section we explore
some of their intrinsic geometry which we already encountered when
constructing the root system by means of simple Weyl reflections. Their
collection forms the Weyl group which belongs to the class of Coxeter
groups. These kind of groups are distinguished by the existence of special
group elements, so-called Coxeter elements, whose properties we are going to
exploit in Chapter 3 when discussing the Lie algebraic structure of affine
Toda theory.

\subsection{Bicolouration and Coxeter elements}

Consider the Weyl group $W$ of a simple Lie algebra $\frak{g}$. All elements
in the Weyl group are generated from simple Weyl reflections $\sigma _{i}$,
i.e. for any $w\in W$ there is a decomposition $w=\sigma _{i_{1}}\cdots
\sigma _{i_{\ell }}$ in simple Weyl reflections. There exists a longest
element in the sense that it is built up from a maximal number of simple
Weyl reflections. It is called \textbf{Coxeter element} or \textbf{%
transformation }and defined by the product over all simple Weyl reflections, 
\begin{equation*}
\sigma =\sigma _{1}\sigma _{2}\cdots \sigma _{n}\;.
\end{equation*}
Clearly, this definition depends on the particular choice of simple roots,
which define the simple Weyl reflections. Also the ordering of the
reflections is only a matter of choice. The Coxeter element is therefore
only defined up to conjugacy, i.e., given a longest element $\sigma $ we get
another one by the adjoint action of the Weyl group, $\sigma \rightarrow
w\sigma w^{-1}$. In fact, this adjoint action is exhaustive on the set of
possible Coxeter elements. However, the geometric properties of the Coxeter
element we are interested in are shared by all representatives of this
conjugacy class \cite{Kostant}:

\begin{itemize}
\item[(C1)]  \emph{The Coxeter element fixes no non-zero vector.}

\item[(C2)]  \emph{It is of finite order, }$\sigma ^{h}=1,$\emph{\ where }$h$%
\emph{\ is the Coxeter number defined in (\ref{Coxnumbers}). This means that
the Coxeter element permutes the roots in orbits of length }$h$\emph{.}

\item[(C3)]  \emph{The eigenvalues of }$\sigma $\emph{\ are of the form } 
\begin{equation*}
\exp \frac{i\pi s_{j}}{h}\;,\quad j=1,...,n
\end{equation*}
\emph{where the characteristic set of integers }$1=s_{1}\leq s_{2}\leq
...\leq s_{n}=h-1$\emph{\ are called the exponents of the Lie algebra }$%
\frak{g}$\emph{\ satisfying the relation }$s_{n+1-i}=h-s_{i}$\emph{.}
\end{itemize}

Below we will calculate concrete physical quantities in terms of Coxeter
elements, whence we need a unique prescription to determine which of the
possible Coxeter elements we are going to use. This is achieved by
introducing the concept of \textbf{bicolouration} for Dynkin diagrams. To
every vertex in the Dynkin diagram $\Gamma (\frak{g})$ we assign a colour,
black or white, such that two vertices linked to each other are differently
coloured. See Figure 2.1 and 2.2 for our conventions. This bicolouration
polarizes the index set $\Delta =\{1,...,n\}$ into two subsets $\Delta _{+}$
and $\Delta _{-}$ corresponding to white and black coloured vertices,
respectively. Grouping all simple Weyl reflections according to this
polarization a Coxeter element is unambiguously specified by setting \cite
{FLO} 
\begin{equation}
\sigma :=\sigma _{-}\sigma _{+},\quad \quad \sigma _{\pm }:=\prod_{i\in
\Delta _{\pm }}\sigma _{i}\;.  \label{Cox}
\end{equation}
This particular definition fixes the Coxeter element uniquely for the
following reason. By the definition of the bicolouration the sub-elements $%
\sigma _{\pm }$ only contain simple Weyl reflections corresponding to roots
which are orthogonal to each other. Therefore, all of them commute in
accordance with (\ref{weylg}) and their relative order in $\sigma _{\pm }$
does not matter. Moreover, as an immediate consequence of this construction
one has the following identities between the sub-elements $\sigma _{\pm }$
and the Coxeter element, 
\begin{equation}
\sigma _{\pm }^{2}=1\,,\quad \sigma ^{-1}=\sigma _{+}\sigma _{-}\,,\quad
\sigma _{\pm }\sigma ^{x}=\sigma ^{-x}\sigma _{\pm }\;.  \label{sigpm}
\end{equation}
Having these formulas at hand facilitates to calculate the action of the
Coxeter element on the sets of simple roots and fundamental weights.

\subsubsection{Coxeter orbits and coloured simple roots}

In the course of our argumentation \textbf{Coxeter orbits} denoted by $%
\Omega _{i}$ will play an essential role. Following the conventions in \cite
{FLO,FO} they are generated by the successive action of the above Coxeter
element (\ref{Cox}) on a \textbf{``coloured'' simple root} $\gamma
_{i}=c_{i}\alpha _{i}$ with $c_{i}=+1$ or $-1$ when the $i^{\text{th}}$ node
in the Dynkin diagram $\Gamma (\frak{g})$ is white or black, respectively.
The associated orbit reads explicitly 
\begin{equation}
\Omega _{i}:=\{\sigma ^{x}\gamma _{i}:1\leq x\leq h\}\;.  \label{orbit}
\end{equation}
Note that we have used the period (C2) of $\sigma $ in this definition. The
motivation to define the Coxeter orbits $\Omega _{i}$ via the coloured roots
is that they do not intersect, i.e. $\Omega _{i}\cap \Omega _{j}=\emptyset $%
, and are exhaustive on the set of roots. Moreover, all $\gamma _{i}$'s lie
in different orbits and all elements in one orbit are linear independent 
\cite{FLO}. This let the coloured simple roots appear as natural entities in
the context of the bicolouration of Dynkin diagrams and the definition (\ref
{Cox}), since they constitute a complete set of representatives for the
Coxeter orbits $\Omega _{i}$.

From these facts we can now easily derive the number of roots of the Lie
algebra $\frak{g}$ and its dimension. Every coloured simple root gives rise
to an orbit with $h$ elements when the Coxeter element acts on it. The set
of all these orbits gives the total root system $R$ whence we deduce $|R|=nh$%
. Together with formula (\ref{rnumber}) derived from the root space
decomposition we obtain the dimension of the Lie algebra in terms of the
Coxeter number 
\begin{equation}
\dim \frak{g}=n(h+1)\;.  \label{dim}
\end{equation}

\subsubsection{Action on simple roots and fundamental weights}

Exploiting that the simple Weyl reflections in $\sigma _{\pm }$ all commute
we obtain via the definition of simple Weyl reflections (\ref{sweyl}) the
following simple relation for the action of $\sigma $ on coloured roots 
\begin{equation}
\sigma _{c_{i}}\gamma _{i}=-\gamma _{i}\quad \text{and\quad }\sigma
_{-c_{i}}\gamma _{i}=\gamma _{i}-\sum_{j\in \Delta _{-c_{i}}}I_{ij}\gamma
_{j}\;.  \label{action}
\end{equation}
Here the notation $\sigma _{c_{i}}$ is understood in the obvious sense, that 
$\sigma _{c_{i}}=\sigma _{\pm }$ for $c_{i}=\pm 1$, and we have defined the
so-called \textbf{incidence matrix} 
\begin{equation}
I:=2-A\;,  \label{inc}
\end{equation}
which will frequently be used in the course of our argumentation. Adding
both identities (\ref{action}) yields the relation 
\begin{equation}
(\sigma _{+}+\sigma _{-})\gamma _{i}=-\sum_{j\in \Delta
_{-c_{i}}}I_{ij}\gamma _{j}  \label{eigenAsig}
\end{equation}
which upon squaring is seen to relate the eigenvalues of the Cartan matrix
with the eigenvalues of the Coxeter element defined in (C3). Indeed, the
former can be determined to be $4\sin ^{2}\pi s_{k}/h$, where $%
s_{k},\,k=1,...,n$ are the exponents.

The action of simple Weyl reflections can be extended to weights. Of
particular interest are the fundamental weights (\ref{weights}) which form
the dual basis of the simple coroots. The action of the special elements is
then derived to 
\begin{equation}
\sigma _{c_{i}}\lambda _{i}=\lambda _{i}-\alpha _{i}\quad \text{and\quad }%
\sigma _{-c_{i}}\lambda _{i}=\lambda _{i}\;.  \label{waction}
\end{equation}
Using the identities (\ref{action}) and (\ref{waction}) the Coxeter element
can be shown to relate simple roots and fundamental weights which is of
practical importance, 
\begin{equation*}
\gamma _{i}=(\sigma _{-}-\sigma _{+})\lambda _{i}=(1-\sigma ^{-1})\sigma ^{%
\frac{1-c_{i}}{2}}\lambda _{i}\;.
\end{equation*}
Note that $(1-\sigma ^{-1})$ never vanishes because of (C1) and the fact
that $\sigma ^{-1}=\sigma _{-}\sigma \sigma _{-}$ is also an Coxeter
element. Hence, it can be inverted and by exploiting (C2) we then obtain 
\begin{equation}
\lambda _{i}=\sum_{x=1}^{h}\frac{x}{h}\,\sigma ^{x+\frac{c_{i}-1}{2}}\gamma
_{i}  \label{weightroot}
\end{equation}
which supplements the former equation giving $\gamma _{i}$ in terms of $%
\lambda _{i}$. This last identity will play a central role for our further
discussion since it allows to relate the information combined in a Coxeter
orbit of a simple root to the Cartan matrix via the variant, 
\begin{equation}
A_{ji}^{-1}\frac{|\alpha _{i}|^{2}}{2}=\left\langle \lambda _{j},\lambda
_{i}\right\rangle =\sum_{x=1}^{h}\frac{x}{h}\,\left\langle \lambda
_{j},\sigma ^{x+\frac{c_{i}-1}{2}}\gamma _{i}\right\rangle \;,
\label{classM}
\end{equation}
where the first equation follows from the definition of the Cartan matrix (%
\ref{Cmatrix}) and the fundamental weights (\ref{weights}). Below $q$%
-deformed versions of the relations presented here will be introduced and
turn out to be the hinge for revealing the Lie algebraic structure behind
affine Toda theory.

\subsection{Twisted Coxeter elements}

In case the Dynkin diagram $\Gamma $ of the simple Lie algebra $\frak{g}$
allows for symmetry operations leaving $\Gamma $ invariant, there is a
generalized notion of Coxeter elements \cite{Springer,Kac}. From Figure 2.1
we infer that such symmetries are present in the $A$ and $D$ series of
simple algebras as well as in $E_{6}$. That is, for any of these algebras
there exists an automorphism $\frak{g}\rightarrow \frak{g}$ induced by a
reordering of the labels in the Dynkin diagram, $\omega
:\{1,...,n\}\rightarrow \{\omega (1),...,\omega (n)\},$ such that all inner
products, i.e. the Killing form, are invariant. A natural action of $\omega $
on the root system is given by 
\begin{equation*}
\omega (\alpha _{i}):=\alpha _{\omega (i)}\;.
\end{equation*}
Note that we use the same symbol $\omega $ for the mapping defined on the
index set and for the one defined on the simple roots. Necessarily, the
automorphism $\omega $ has finite order, i.e. there is a finite integer $%
\ell $ such that $\omega ^{\ell }=1$, and from Figure 2.1 we infer that the
latter is given by $\ell =1,2$ or $3$. While $\ell =1$ represents the case
when no symmetry is present, the last possibility $\ell =3$ only occurs for $%
\frak{g}=D_{4}$ which has an enhanced symmetry compared to the other Dynkin
diagrams of the $D$ series.

Exploiting the Dynkin diagram automorphism $\omega $ we might now define
twisted Coxeter elements which share similar properties as the ordinary ones
defined in the preceding subsection. Choose an index set $\Delta ^{\omega
}\subset \{1,...,n\}$ containing a representative of each orbit of $\omega $
such that for any pair $i,j\in \Delta ^{\omega }$ the roots $\gamma
_{i},\gamma _{j}$ lie in different orbits. Set $\Delta _{\pm }^{\omega
}:=\Delta ^{\omega }\cap \Delta _{\pm }$ where $\Delta _{\pm }$ are the
white and black vertices defined in the preceding subsection and note that
the Dynkin diagram automorphism is colour preserving, i.e. $\gamma _{\omega
(i)}\in \Delta _{\pm }$ if $\gamma _{i}\in \Delta _{\pm }$. Then the
corresponding twisted Coxeter element is defined by 
\begin{equation}
\hat{\sigma}:=\omega ^{-1}\hat{\sigma}_{-}\hat{\sigma}_{+},\quad \hat{\sigma}%
_{\pm }:=\prod_{i\in \Delta _{\pm }^{\omega }}\sigma _{i}\,.  \label{tCox}
\end{equation}
The above definition of the twisted Coxeter element differs from the
original one given in \cite{Springer} by conjugation with the automorphism $%
\omega ^{-1}$. However, this does not alter any of its properties but rather
amounts to a different choice of the set of representatives, $\Delta
^{\omega }\rightarrow \omega ^{-1}(\Delta ^{\omega })$. This is immediate to
see by deriving the transformation properties of the simple Weyl reflections
from which one then infers that 
\begin{equation*}
\omega \hat{\sigma}_{\pm }\omega ^{-1}=\prod_{i\in \omega (\Delta _{\pm
}^{\omega })}\sigma _{i}\,.
\end{equation*}
Hence, the adjoint action of $\omega $ on $\hat{\sigma}$ gives another
twisted Coxeter element associated with the index set $\omega (\Delta
^{\omega })$. Note that this action is not exhaustive on the set of possible
twisted Coxeter elements, that is, there exist index sets which can not be
obtained from $\Delta ^{\omega }$ by successive actions of $\omega $. We
list further properties of twisted Coxeter elements \cite{Springer}:

\begin{itemize}
\item[(TC1)]  \emph{The twisted Coxeter element fixes no non-zero vector.}

\item[(TC2)]  \emph{It has finite order, }$\hat{\sigma}^{H}=1$\emph{\ with }$%
H=\ell h$\emph{\ being the so-called }$\ell ^{\text{\emph{th}}}$\emph{\
Coxeter number \cite{Kac}. This means in particular that }$\hat{\sigma}$%
\emph{\ permutes the roots in orbits of length }$H$\emph{.}

\item[(TC3)]  \emph{The eigenvalues of }$\sigma $\emph{\ are of the form } 
\begin{equation*}
\varepsilon _{j}^{-1}\,\exp \frac{2\pi i\,s_{j}}{H}\,,\quad j=1,...,n\,.
\end{equation*}
\emph{Here }$\varepsilon _{i}$\emph{\ denote the eigenvalues of }$\omega $%
\emph{\ and }$s_{i}$\emph{\ are the exponents defined in context of the
ordinary Coxeter element, see (C3).}
\end{itemize}

Analogous to the non-twisted Coxeter element we now investigate the action
on simple roots. Since the twisted Coxeter element only contains the simple
Weyl reflections associated with the choice of representatives $\Delta
^{\omega }$ it is convenient to define the integers 
\begin{equation}
\hat{t}_{i}=\left\{ 
\begin{array}{l}
1\qquad \qquad \text{for\thinspace \thinspace \thinspace }\alpha _{i}\in
\Delta ^{\omega } \\ 
0\qquad \qquad \text{for\thinspace \thinspace \thinspace }\alpha _{i}\notin
\Delta ^{\omega }
\end{array}
\right. \,\,.  \label{tint}
\end{equation}
The action of the special elements $\hat{\sigma}_{\pm }$ on a simple root
can now conveniently be written in the following form 
\begin{equation}
\hat{\sigma}_{c_{i}}\gamma _{i}=(-1)^{\hat{t}_{i}}\gamma _{i}\qquad \text{%
and\qquad }\hat{\sigma}_{-c_{i}}\gamma _{i}=\gamma _{i}-\sum\limits_{j\in 
\hat{\Delta}_{-c_{i}}^{\omega }}I_{ij}\gamma _{j}\,\,\,,  \label{taction}
\end{equation}
where the sum involving the incidence matrix $I=2-A$ is restricted to the
representatives $\hat{\Delta}_{\pm }^{\omega }$. Analogously to the
untwisted case we also define twisted Coxeter orbits by setting 
\begin{equation*}
\hat{\Omega}_{i}:=\{\hat{\sigma}^{x}\gamma _{i}:1\leq x\leq H\}
\end{equation*}
where in the definition the finite period of the twisted Coxeter element has
been employed, see (TC2).

\section{Dual algebras and $q$-deformation}

For the discussion of the ATFT S-matrix it will become necessary to
introduce the concept of a pair of \textbf{dual affine Lie algebras} $(\frak{%
g},\frak{g}^{\vee })$ which are related to each other by exchanging short
and long roots. In other words the identification of roots and coroots, $%
\alpha \rightarrow \alpha ^{\vee }$, maps the root system of the one algebra
onto the other. This is sometimes referred to as Langlands duality. Clearly,
simply-laced Lie algebras are left invariant under this duality
transformation since all roots are of the same length, i.e. $\frak{g}=\frak{g%
}^{\vee }$. The remaining pairs of algebras which are not self-dual must
therefore involve non simply-laced algebras. It turns out that one factor,
say $\frak{g}$, is given by an untwisted affine Lie algebra denoted by $%
X_{n}^{(1)},$ while the second factor $\frak{g}^{\vee }$ involves a twisted
affine Lie algebra denoted by $\hat{X}_{\hat{n}}^{(\ell )}$. Here we adopted
the notation of \cite{Kac}. The lower indices $n,\hat{n}$ state the rank of
the algebras while the upper index $\ell =1,2$ or $3$ refers to the order of
a Dynkin diagram automorphism $\omega $ used in the construction of the
twisted algebra, see \cite{Kac} for details. We list all non self-dual pairs
in the following table,

\begin{center}
$
\begin{tabular}{|c|c|}
\hline\hline
untwisted & twisted \  \\ \hline\hline
$B_{n}^{(1)}$ & $A_{2n-1}^{(2)}$ \\ \hline\hline
$C_{n}^{(1)}$ & $D_{n+1}^{(2)}$ \\ \hline\hline
$F_{4}^{(1)}$ & $E_{6}^{(2)}$ \\ \hline\hline
$G_{2}^{(1)}$ & $D_{4}^{(3)}$ \\ \hline\hline
\end{tabular}
$

\vspace{0.4cm} {\small Table 2.2: Pairs of non self-dual affine Lie algebras.%
}
\end{center}

In terms of the associated affine Dynkin diagrams these pairs of Lie
algebras are related by reversing the arrow between vertices associated with
short and long roots. However, in the following we will not use the affine
structure, since it will turn out that all the necessary information is
already contained in the finite dimensional algebras $X_{n}$ and $\hat{X}_{%
\hat{n}}$ whose relation to each other can be easily understood via folding.

\subsection{Pairs of finite algebras and folding}

From the above table we infer that $\hat{X}_{\hat{n}}$ is simply-laced and
equipped with a Dynkin diagram automorphism $\omega $ of order $\ell =2$
except for $\hat{X}_{\hat{n}}=D_{4}$, where $\ell =3$. Folding the algebra $%
\hat{X}_{\hat{n}}$ w.r.t. to this automorphism, i.e. fixing the subalgebra
invariant under $\omega $, we obtain the Langlands dual of the finite
non-simply laced algebra $X_{n}$. Following the procedure outlined in \cite
{TO} this means that given a Chevalley-basis $\{\hat{h}_{i},\hat{e}_{i},\hat{%
f}_{i}\}$ of $\hat{X}_{\hat{n}}$ we choose an index set of representatives $%
\hat{\Delta}^{\omega }\subset \{1,...,\hat{n}\}$ (as explained in the
context of twisted Coxeter elements in subsection 2.3.2) and define the
generators of the ``folded'' non simply-laced Lie algebra by setting 
\begin{equation*}
h_{i}^{\omega }:=\sum_{n=1}^{\ell _{i}}\hat{h}_{\omega (i)},\quad
e_{i}^{\omega }:=\sum_{n=1}^{\ell _{i}}\hat{e}_{\omega (i)}\quad \text{and}%
\quad f_{i}^{\omega }:=\sum_{n=1}^{\ell _{i}}\hat{f}_{\omega (i)}\,
\end{equation*}
where $i\in \hat{\Delta}^{\omega }$ runs over the representatives of the
different orbits and $\ell _{i}$ denotes the length of the orbit of the $i^{%
\text{th}}$ simple root, i.e. $\omega ^{\ell _{i}}\hat{\alpha}_{i}=\hat{%
\alpha}_{i}$. It is straightforward to verify that these generators satisfy
the Chevalley-Serre relations (\ref{serre}) w.r.t. the \textbf{''folded''
Cartan matrix}, 
\begin{equation}
A_{ij}^{\omega }=\sum_{n=1}^{\ell _{i}}\hat{A}_{\omega (i)j}\;,\quad \quad
i,j\in \hat{\Delta}^{\omega }\;.  \label{foldedA}
\end{equation}
By the Cartan-Killing classification scheme this determines the folded Lie
algebra uniquely and one immediately verifies that the folded Cartan matrix $%
A^{\omega }$ just coincides with the Cartan matrix $A$ of the non
simply-laced algebra $X_{n}$ given in Table 2.2. In particular, $A^{\omega }$
is a $n\times n$ matrix and the rank $n$ of $X_{n}$ equals the number of
orbits $|\hat{\Delta}^{\omega }|$ of the automorphism $\omega $.

For convenience the numeration of the vertices in the Dynkin diagrams of the
dual Lie algebras in Table 2.2 and the set of representatives $\hat{\Delta}%
^{\omega }$ will be chosen such that the orbits of the first $n$ simple
roots of $\hat{X}_{\hat{n}}$ are identified under folding with the simple
roots of $X_{n}$, i.e. 
\begin{equation*}
\hat{\Delta}^{\omega }:=\{1,...,n\}\subset \{1,...,\hat{n}\}\;.
\end{equation*}
Below this will enable us to identify particles in the affine Toda theories
associated with the dual algebras without relabeling. Furthermore, the
bicolouration of the Dynkin diagrams of the dual algebras will be fixed by
requiring that in $\Gamma (X_{n})$ the unique vertex of the short root which
is connected to a long root is black. Upon the above identification between
both dual algebras this determines also the bicolouration of $\Gamma (\hat{X}%
_{\hat{n}})$.

Another feature of the duality between both algebras $(X_{n}^{(1)},\hat{X}_{%
\hat{n}}^{(\ell )})$ we will exploit in the following is the close
relationship between their Coxeter numbers $h$ and $\hat{h}$. The
interchange of roots and coroots leads to an interchange of the Kac labels $%
n_{i}$ and the dual Kac labels $n_{i}^{\vee }$, whence one must have that 
\begin{equation}
h=\hat{h}^{\vee }\qquad \text{and\qquad }h^{\vee }=\hat{h}\,\,,
\label{dualCox}
\end{equation}
where $h^{\vee }$,$\hat{h}^{\vee }$ are the dual Coxeter numbers of $X_{n}$
and $\hat{X}_{\hat{n}}$, compare (\ref{Coxnumbers}). Thus, we might express
the Coxeter numbers of the one algebra completely in terms of the other and
vice versa. This will become important in the following subsections where we
are going to introduce $q$-deformed Coxeter and twisted Coxeter elements on
the two dual algebras $(X_{n}^{(1)},\hat{X}_{\hat{n}}^{(\ell )})$. As we
already saw in the non-deformed case the Coxeter numbers determine the
periods of the Coxeter elements and the length of the orbits. We will
encounter this mutual dependence of quantities belonging to the two algebras 
$X_{n}$ and $\hat{X}_{\hat{n}}$ more often in due course. In fact, the close
relationship between the dual algebras builds the cornerstone for the
definition of generalized Coxeter elements via $q$-deformation. The latter
allows to combine the structures of both algebras in the orbits of one
Coxeter transformation.

\subsection{q-deformed Coxeter element of $X_{n}^{(1)}$}

The idea of introducing $q$-deformed Coxeter elements was first put forward
by Oota \cite{Oota}. In his work many formulas were supported solely by a
case-by-case study. The discussion of the following subsections will provide
the necessary proofs in a rigorous, generic setting and includes also
numerous entirely new relations \cite{FKS2}.

\subsubsection{Definitions}

From the defining equation (\ref{Cmatrix}) of the Cartan matrix we know that
all its entries are integers. This allows in a natural way to introduce $q$%
-deformation via so-called $q$\textbf{-deformed integers} notated in the
standard fashion by 
\begin{equation*}
\lbrack n]_{q}=\frac{q^{n}-q^{-n}}{q^{1}-q^{-1}}\,,\quad q\in \mathbb{C}\;.
\end{equation*}
In this chapter the $q$-deformation will be discussed as a purely
mathematical operation only and $q$ will have the status of a formal
parameter. That is, for the time being we assume the deformation parameter $%
q $ to be completely generic. However, in Chapter 3 when discussing the
S-matrix of ATFT we will specify $q$ to be a root of unity and also
introduce a particular parameterization $q(\beta ),$ where $\beta $ is a
coupling constant. In this situation the limit\footnote[1]{%
In the literature this limit is often referred to as ''classical'' in the
sense that $\ln q\propto \hbar $ with $\hbar \rightarrow 0$ being the Planck
constant.} $q\rightarrow 1,$ which we are going to perform at several
occasions in this chapter as a consistency check, will correspond to the
weak or strong coupling limit.

As we know from the preceding discussion the action of simple Weyl
reflections $\sigma _{i}$ on simple roots $\alpha _{i}$ can be expressed in
terms of the Cartan matrix\thinspace $A$. We now define a $q$-\textbf{%
deformed Weyl reflection} $\sigma _{i}^{q}$ by deforming the off-diagonal
entries of $A$, 
\begin{equation}
\sigma _{i}^{q}(\alpha _{j}):=\alpha _{j}-\left( 2\delta _{ij}-\left[ I_{ji}%
\right] _{q}\right) \alpha _{i}\,\,.  \label{qweyl}
\end{equation}
Here $I=2-A$ denotes the incidence matrix related to the non simply-laced
Lie algebra $X_{n}$. As motivation for the above definition we observe that
the $q$-deformed Weyl reflection shares two essential features of the
non-deformed one, namely it is idempotent and reflections belonging to
orthogonal roots commute, i.e., 
\begin{equation}
(\sigma _{i}^{q})^{2}=1\qquad \;\text{and\ }\qquad (\sigma _{i}^{q}\sigma
_{j}^{q})^{2}=1\quad \text{for\ \ }\left\langle \alpha _{i},\alpha
_{j}\right\rangle =0\;.  \label{qweylg}
\end{equation}
In particular, in the classical limit $q\rightarrow 1$ we obtain the usual
Weyl reflection defined in (\ref{sweyl}). In general, however, the Weyl
reflections (\ref{qweyl}) do not preserve the inner product.

We now proceed analogously as in the non-deformed case to define $q$%
-deformed Coxeter elements, i.e. we introduce the bicolouration of the
Dynkin diagram with colour values $c_{i}=\pm 1$ as defined above and order
the simple roots in two sets $\Delta _{\pm }$ whose elements are mutually
orthogonal (see Subsection 2.3.1). This allows to define the two special
elements 
\begin{equation}
\sigma _{\pm }^{q}:=\prod\limits_{i\in \Delta _{\pm }}\sigma _{i}^{q}\;,
\label{spm}
\end{equation}
uniquely which analogously to (\ref{sigpm}) share the property $(\sigma
_{\pm }^{q})^{2}=1$. So far the definition has closely followed the one of
ordinary Coxeter elements. As a new feature in the context of $q$%
-deformation we will now take into account the different length of the roots
for the definition of the Coxeter element. By our convention (\ref
{convention}) the long roots have squared length equal to two and we
therefore define integers $t_{i}$ by 
\begin{equation}
t_{i}=\ell \,\frac{\left\langle \alpha _{i},\alpha _{i}\right\rangle }{2}
\label{symm}
\end{equation}
The ratio is indeed an integer as follows directly from the definition of
the Cartan matrix $A$. Here $\ell =1,2$ or $3$ equals the order of the
Dynkin diagram automorphism, since the dual of $X_{n}$ is obtained from $%
\hat{X}_{\hat{n}}$ as folded algebra w.r.t. $\omega $ (see our discussion
above). In particular, these integers symmetrize the Cartan matrix $A$ when
combined into a diagonal matrix $D$, 
\begin{equation}
AD=DA^{t}\;,\quad \quad \quad D_{ij}=\delta _{ij}t_{i}\;.  \label{D}
\end{equation}
In fact, the last property fixes the integers $t_{i}$ up to a normalization
constant. In the present context we need the $t_{i}$'s for introducing the
map 
\begin{equation}
\alpha _{i}\rightarrow \tau (\alpha _{i}):=q^{t_{i}}\alpha _{i}\;
\label{tau}
\end{equation}
in terms of which the $q$\textbf{-deformed Coxeter element} is then defined
by setting \cite{Oota} 
\begin{equation}
\sigma _{q}:=\sigma _{-}^{q}\,\tau \,\sigma _{+}^{q}\,\tau \,\,.
\label{qCox}
\end{equation}
Note that $\sigma _{q}$ is defined unambiguously by our choice of the
bicolouration of the Dynkin diagram of $X_{n}$. The introduction of the map $%
\tau $ and the particular ordering of the maps in (\ref{qCox}) will be
motivated in retrospect by their actions on simple roots and the identities
to which they give rise. Before we start to investigate the latter, notice
that the above definition (\ref{qCox}) is consistent with the ''classical''
limit, i.e. sending the deformation parameter to one we recover the ordinary
Coxeter element (\ref{Cox}), $\lim_{q\rightarrow 1}\sigma _{q}=\sigma \;.$

\subsubsection{Action of $\protect\sigma _{q}$ in the root space}

From the definition of the simple Weyl reflections (\ref{qweyl}) it is
straightforward to derive the action on coloured simple roots $\gamma
_{i}=c_{i}\alpha _{i}$ yielding the $q$-deformed version of equation (\ref
{action}) 
\begin{equation}
\sigma _{c_{i}}^{q}\gamma _{i}=-\gamma _{i}\quad \text{and}\quad \sigma
_{-c_{i}}^{q}\gamma _{i}=\gamma _{i}-\sum\limits_{j\in \Delta _{-c_{i}}}%
\left[ I_{ij}\right] _{q}\gamma _{j}\,\,.  \label{qaction}
\end{equation}
From this relation together with the definition of the $q$-deformed Coxeter
element we see that $\sigma _{q}$ generates polynomials in the variable $q$
and simple roots\footnote{%
Note that the possible nonzero entries in the incidence matrix are $%
I_{ij}=1,2,3$ which upon $q$-deformation give rise to the polynomials $%
[1]_{q}=1,\,[2]_{q}=q+q^{-1}$ and $[3]_{q}=1+q^{2}+q^{-2}$.}. Thus, denoting
by $\Omega _{i}^{q}$ the orbit of the simple root $\gamma _{i}$ a typical
element will be of the form 
\begin{equation*}
(\sigma _{q})^{x}\gamma _{i}=a_{1}q^{y_{1}}\gamma
_{i_{1}}+a_{2}q^{y_{2}}\gamma _{i_{2}}+...+a_{k}q^{y_{k}}\gamma _{i_{k}}\;
\end{equation*}
where $x$ and $a_{1},...,a_{k},y_{1},...,y_{k}$ are some integers depending
on $i$ and $i_{1},...,i_{k}$. Now, the most important feature which
motivates the definition (\ref{qCox}) is the remarkable identity 
\begin{equation}
q^{-2H}\left( \sigma _{q}\right) ^{h}=1\,\,.  \label{quasi}
\end{equation}
A general proof of (\ref{quasi}) is not known so far, but it is expected to
be quite involved when considering the analogue of the non-deformed case.
Therefore, we shall here be content with confirming it by means of a
case-by-case analysis in the appendix. Clearly, the property (\ref{quasi})
generalizes condition (C2) of the usual Coxeter element. It states, that the
roots up to multiplication by the factor $q^{2H}$ are permuted in orbits of
length $h$ rendering its action on the root space ``finite''. It needs to be
stressed that the length of the orbit is determined by the Coxeter number $h$
of the algebra $X_{n}$, while the quasi-periodicity in the deformation
parameter is determined by the $\ell ^{\text{th}}$ Coxeter number $H$ of the
dual algebra $\hat{X}_{\hat{n}}$. This indicates that by means of the chosen 
$q$-deformation the characteristics of \emph{both} dual algebras are
combined in the structure of the orbits $\Omega _{i}^{q}$ generated by $%
\sigma _{q}$.

In anticipation of the discussion of the ATFT S-matrix we mention that each
of the orbits $\Omega _{i}^{q}$ will be associated with a particle species.
The operation of charge conjugation yielding the antiparticle is then
identified by the transition to another orbit, $C:\Omega _{i}^{q}\rightarrow
\Omega _{\bar{\imath}}^{q}$, which contains the element \cite{FKS2} 
\begin{equation}
-q^{-H+\frac{c_{\bar{\imath}}-c_{i}}{2}t_{i}}\,\sigma _{q}^{\frac{h}{2}+%
\frac{c_{i}-c_{\bar{\imath}}}{4}}\gamma _{i}\in \Omega _{\bar{\imath}%
}^{q}\,\,.  \label{anti}
\end{equation}
Note that despite the first impression the power of the Coxeter element in
this relation is integral due to the property $c_{i}c_{\bar{\imath}%
}=(-1)^{h} $. Repeating the charge conjugation in (\ref{anti}) leads to (\ref
{quasi}), provided $t_{i}=t_{\bar{\imath}}$. Thus, the property $C^{2}=1$
required on physical grounds is satisfied for simply-laced algebras. For the
non-simply laced algebras, all particles will turn out to be self-conjugate
whence the relation (\ref{anti}) reduces to 
\begin{equation}
\,\sigma _{q}^{\frac{h}{2}}\gamma _{i}\,\,=-q^{H}\gamma _{\bar{\imath}}\,\,.
\label{self}
\end{equation}

\noindent \textbf{Remark. }\emph{The motivation of the definition (\ref{anti}%
) is analogue to the one known from the simply laced case \cite{FLO}. This
means complex conjugating the field which creates the particle of type }$i$%
\emph{\ in the classical theory corresponds to the creation of the
anti-particle }$\bar{\imath}$\emph{, suggesting to associate }$-\gamma _{i}$%
\emph{\ to the anti-particle. However, in context of the quantum theory it
will turn out that the classical theory is only recovered in the extreme
weak or extreme strong limit of the coupling constant. That is, for }$%
q\rightarrow 1$\emph{\ we recover the known identity \cite{FO} for the
simply-laced case }$\sigma ^{\frac{h}{2}+\frac{c_{i}-c_{\bar{\imath}}}{4}%
}\gamma _{i}\,\,=\gamma _{\bar{\imath}}$\emph{\ which relates particles and
anti-particles.\medskip }

For the time being all further information about the action of the $q$%
-deformed Coxeter element we need to proceed is the relation 
\begin{equation}
\left( q^{-c_{i}t_{i}}(\sigma _{q})^{c_{i}}+q^{c_{i}t_{i}}\right) \gamma
_{i}=\sum\limits_{j\in \Delta _{-c_{i}}}q^{\frac{1+c_{i}}{2}t_{i}-\frac{%
1+c_{j}}{2}t_{j}}\left[ I_{ij}\right] _{q}\gamma _{j}\,\,  \label{bootp}
\end{equation}
which will turn out to be crucial subsequently. It is derived from (\ref
{qaction}) by acting with $\sigma _{-c_{i}}^{q}$ on the first identity and
then using the second. Note that for $c_{i}=-1$ this amounts to the action
of the inverse Coxeter element\footnote{%
We differ here from the definition of the inverse in \cite{Oota}.} $(\sigma
_{q})^{-1}=\tau ^{-1}\sigma _{+}^{q}\,\tau \,^{-1}\sigma _{-}^{q}$. We are
now going to exploit (\ref{bootp}) to relate the orbits of the Coxeter
element $\sigma _{q}$ to a $q$-deformed Cartan matrix.

\subsection{q-deformed Cartan matrix of $X_{n}^{(1)}$}

In this subsection we derive the analogue of identity (\ref{classM}) for the 
$q$-deformed Coxeter element. This will allow us to relate the action of $%
\sigma _{q}$ on $X_{n}$ to the action of the $q$-deformed twisted Coxeter
element on $\hat{X}_{\hat{n}}$ defined below. In particular, the close
interplay between the dual algebras will become manifest.

We start by defining a $n\times n$ matrix $M$\ depending on \emph{two}
deformation parameters $q,\hat{q}\in \mathbb{C}$, one of which will be
related to $X_{n}$ and the other to $\hat{X}_{\hat{n}}$, 
\begin{equation}
M_{ij}(q,\hat{q}):=-\frac{[t_{j}]_{\hat{q}}}{2}\,\hat{q}^{\frac{1-c_{j}}{2}%
t_{j}-\frac{1+c_{i}}{2}t_{i}}\sum_{x=1}^{h}\left\langle \lambda _{j}^{\vee
},(\sigma _{\hat{q}})^{x+\frac{c_{i}-1}{2}}\gamma _{i}\right\rangle q^{2x+%
\frac{c_{i}-c_{j}}{2}-1}.  \label{M0}
\end{equation}
Here the $\lambda _{j}^{\vee }$'s denote the coweights (\ref{weights}) of $%
X_{n}$, the $t_{i}$'s are the integers defined in (\ref{symm}) and $%
\left\langle \cdot ,\cdot \right\rangle $ is the inner product induced by
the Killing form. The reason for introducing the matrix $M$ will become
apparent momentarily. Multiplying now equation (\ref{bootp}) by $q^{2x+\frac{%
c_{i}-c_{j}}{2}-1}\sigma _{\hat{q}}^{x}$ and performing the sum over the
powers in the range $1\leq x\leq h$ we obtain by taking the inner product
with $\lambda _{j}^{\vee }$ and exploiting the periodicity property (\ref
{quasi}) the determining equation 
\begin{equation}
(q^{-1}\hat{q}^{-t_{i}}+q\hat{q}^{t_{i}})M_{ij}(q,\hat{q})-%
\sum_{k=1}^{n}[I_{ik}]_{\hat{q}}M_{kj}(q,\hat{q})=\tfrac{1-q^{2h}\hat{q}^{2H}%
}{2}\,[t_{i}]_{\hat{q}}\delta _{ij}\,.  \label{M1}
\end{equation}
Note that the dependence on the colour values of the roots has vanished and
that the last equation only involves elementary data of the algebra $X_{n}$
such as the incidence matrix $I=2-A$, the entries of the diagonal matrix $D$
defined in (\ref{D}) and the Coxeter numbers $h,H$. More importantly, (\ref
{M1}) can be solved for the matrix $M$ to give the alternative expression 
\begin{equation}
M(q,\hat{q})=\frac{1-q^{2h}\hat{q}^{2H}}{2}\,A(q,\hat{q})^{-1}[D]_{\hat{q}%
}\;.  \label{M2}
\end{equation}
Here we have defined two new objects, the $q$-deformed symmetrizer $%
([D]_{q})_{ij}:=[D_{ij}]_{q}$ and the \textbf{doubly }$q$\textbf{-deformed
Cartan matrix}\footnote{%
In a different context not involving Coxeter geometry a similar expression
was obtained in \cite{FR}.} 
\begin{equation}
A(q,\hat{q}):=q^{-1}\hat{q}^{\,-D}+q\,\hat{q}^{D}-[\,I\,]_{\hat{q}}\;,
\label{qA}
\end{equation}
where $([\,I\,]_{q})_{ij}:=[I_{ij}]_{q}$ and $q^{D}\equiv \exp (\ln q\cdot
D) $ is diagonal with entries $(q^{D})_{ij}=\delta _{ij}q^{t_{i}}$. Note
that the classical limit $q,\hat{q}\rightarrow 1$ yields again the
non-deformed objects $A$ and $D$. In particular, after dividing by $(1-q^{2h}%
\hat{q}^{2H}) $ in (\ref{M1}) we recover the identity (\ref{classM}) in this
limit.

Some comments are due to fully appreciate the equivalence of the expressions
(\ref{M0}) and (\ref{M2}). Looking at the defining relation (\ref{M0}) we
know from the action of the $q$-deformed Coxeter element described in the
previous subsection that the matrix elements $M_{ij}(q,\hat{q})$ consist of
polynomials in the variables $q,\hat{q}$, 
\begin{equation}
M(q,\hat{q})=\sum_{x=1}^{2h}\sum_{y=1}^{2H}\mu (x,y)q^{x}\hat{q}^{y}\;.
\label{Mpoly}
\end{equation}
Looking at (\ref{M2}) this is far from obvious, since the expression
contains the inverse of the deformed Cartan matrix (\ref{qA}). The latter is
only well-defined for $q,\hat{q}$ away from roots of unity, since then $A(q,%
\hat{q})$ might become singular. The singular values are given by the zeroes
of the determinant of $A(q,\hat{q}),$ which on a case-by-case basis can be
established to 
\begin{equation}
\det A(q,\hat{q})=\prod_{k=1}^{n}(q\,\hat{q}^{H/h}+q^{-1}\hat{q}%
^{\,-H/h}-2\cos \tfrac{\pi s_{k}}{h})\;.  \label{detA}
\end{equation}
Here $\{s_{1},...,s_{n}\}$ is the set of exponents$\,$of the Lie algebra $%
X_{n}$ defined by (C3) in Section 2.3.1. However, the matrix $M(q,\hat{q})$
is defined for arbitrary complex values of the deformation parameters, since
the determinant $\det A(q,\hat{q})$ is always contained as a factor in $%
(1-q^{2h}\hat{q}^{\,2H})$ standing in front of the Cartan matrix in (\ref{M2}%
). Hence, possible poles of the determinant are cancelled. Later on we will
see that the column vectors of the $M$-matrix evaluated at the zeroes$\,$of
the determinant (\ref{detA}) can be directly linked to conserved quantities
in ATFT. In particular, they will yield the mass spectrum.

\subsection{Inner product identities}

Having established the equivalence between the expressions (\ref{M0}) and (%
\ref{M2}) we might now turn the picture around and ask what we can learn
from the matrix expression about the action of the $q$-deformed Coxeter
element. In fact, we might plug in (\ref{M2}) into the l.h.s. of (\ref{M0})
and then solve for the inner product $\left\langle \lambda _{j}^{\vee
},\sigma _{q}^{x}\gamma _{i}\right\rangle $ by discrete Fourier
transformation in the variable $q$, 
\begin{equation}
-\frac{[t_{j}]_{\hat{q}}}{2}\,\hat{q}^{\frac{1-c_{j}}{2}t_{j}-\frac{1+c_{i}}{%
2}t_{i}}\left\langle \lambda _{j}^{\vee },(\sigma _{\hat{q}})^{x}\gamma
_{i}\right\rangle =\frac{1}{2h}\sum_{n=1}^{2h}M_{ij}(\tau ^{n},\hat{q})\tau
^{n(2x-\frac{c_{i}+c_{j}}{2})}  \label{F1}
\end{equation}
where $\tau $ is any \thinspace root of unity of order $2h$. Here we have
used that the powers of the variable $q$ run over the range $1,2,...,2h$
when the powers of the Coxeter element are restricted to $1,2,...,h$. For
practical purposes in calculating the orbits of the Coxeter element $\sigma
_{q}$ formula (\ref{F1}) might not yield an advantage. However, we can use
it to prove some inner product identities which are immediate consequences
of the matrix properties of (\ref{M2}). For example, noting that equation (%
\ref{D}) also holds for the $q$-deformed quantities, 
\begin{equation}
A(q,\hat{q})[D]_{\hat{q}}=[D]_{\hat{q}}A(q,\hat{q})^{t}\;,  \label{qD}
\end{equation}
we infer that the $M$-matrix is symmetric. Upon using the formula (\ref{F1})
we end up with the non-trivial identity \cite{FKS2} 
\begin{equation}
(q^{2t_{j}}-1)\left\langle \lambda _{j}^{\vee },(\sigma _{q})^{x}\gamma
_{i}\right\rangle =(q^{2t_{i}}-1)\,\left\langle \lambda _{i}^{\vee },(\sigma
_{q})^{x}\gamma _{j}\right\rangle \,\,.  \label{spari}
\end{equation}
This particular equality has special significance, since it will be used in
Chapter 3 to show parity invariance of the ATFT scattering matrix. Another
inner product equality we are going to exploit in the discussion of the
S-matrix is \cite{FKS2} 
\begin{equation}
q^{\frac{(1-c_{j})t_{j}-(1+c_{i})t_{i}}{2}}\,\left\langle \lambda _{j}^{\vee
},\sigma _{q}^{x}\gamma _{i}\right\rangle +\,q^{2H+\frac{%
(c_{j}-1)t_{j}+(1+c_{i})t_{i}}{2}}\,\left\langle \lambda _{j}^{\vee }\right.
,\sigma _{q^{-1}}^{h-x+\frac{c_{i}+c_{j}}{2}}\bigl. \gamma _{i}\bigr\rangle %
=0\,\,,  \label{sm}
\end{equation}
which will be linked to analyticity in the physical sheet. Similar like
before, it follows by means of (\ref{M2}) and (\ref{F1}) from the relation 
\begin{equation}
q^{2h}\hat{q}^{2H}\,M(q^{-1},\hat{q}^{-1})=-M(q,\hat{q})  \label{Mmero}
\end{equation}
which is immediate to verify upon noting that $A(q^{-1},\hat{q}^{-1})=A(q,%
\hat{q})$ and $[D]_{q^{-1}}=[D]_{q}$. The simple derivation of both formulas
(\ref{spari}) and (\ref{sm}) shows the usefulness of formula (\ref{M2}). It
should, however, be noted that both inner product identities can also be
proven directly \cite{FKS2}.

\subsection{q-deformed twisted Coxeter element of $\hat{X}_{\hat{n}}^{(\ell
)}$}

In this subsection we perform a discussion similar to the previous one but
now focussing on the dual algebra $\hat{X}_{\hat{n}}$. We introduce the $q$%
-deformed twisted Coxeter element and afterwards show a matrix identity
leading to the $q$-deformed folded Cartan matrix associated with (\ref
{foldedA}). As is the non-deformed case the folded Cartan matrix will prove
to be identical to the Cartan matrix (\ref{qA}) of $X_{n}$ and, thus, relate
the twisted and non-twisted $q$-deformed Coxeter elements of the dual
algebras.

\subsubsection{Definitions}

We shortly recall the definitions from the beginning of this section. The
dual algebra $\hat{X}_{\hat{n}}$ is always simply-laced and equipped with a
Dynkin diagram automorphism $\omega $ of order $\ell $. The simple roots $%
\hat{\alpha}_{i}$ fall in different orbits w.r.t. $\omega $ whose length is
given by the integers $\ell _{i}$ such that $\omega ^{\ell _{i}}\hat{\alpha}%
_{i}=\hat{\alpha}_{i}$. The largest value of $\ell _{i}$ corresponds to $%
\ell $. Furthermore, we choose an index set of representatives $\hat{\Delta}%
^{\omega }\subset \{1,...,\hat{n}\}$ and separate it in two subsets $\hat{%
\Delta}_{\pm }^{\omega }$ according to the bicolouration of the Dynkin
diagram associated to $\hat{X}_{\hat{n}}$. Now, a $q$-deformation of the
simple Weyl reflections 
\begin{equation*}
\hat{\sigma}_{i}(\hat{\alpha}_{j})=\hat{\alpha}_{j}-\hat{A}_{ji}\hat{\alpha}%
_{i}\,\,\,,
\end{equation*}
analogously to (\ref{qweyl}) would not lead to a new structure since the
incidence matrix $\hat{I}=2-\hat{A}$ only contains the entries $0$ or $1$.
One possible way to introduce a non-trivial deformation is by defining the
analogue of the map $\tau $ in (\ref{tau}) but this time using the integers $%
\hat{t}_{i}$ defined in (\ref{tint}) which single out the representatives, 
\begin{equation}
\hat{\tau}(\hat{\alpha}_{i}):=q^{2\hat{t}_{i}}\hat{\alpha}_{i}\,\,\,.
\label{ttau}
\end{equation}
Using the non-deformed elements $\hat{\sigma}_{\pm }=\prod_{i\in \hat{\Delta}%
_{\pm }^{\omega }}\hat{\sigma}_{i}$ of Subsection 2.3.2 a $q$-\textbf{%
deformed twisted Coxeter element} is defined as \cite{Oota}, 
\begin{equation}
\hat{\sigma}_{q}:=\omega ^{-1}\,\hat{\sigma}_{-}\,\hat{\tau}\,\hat{\sigma}%
_{+}\,\,\,.  \label{qtCox}
\end{equation}
Once again the bicolouration ensures that $\hat{\sigma}_{q}$ is uniquely
defined. The particular ordering of the maps will prove important for the
characteristics of $\hat{\sigma}_{q}$. For $q\rightarrow 1$ we obtain the
standard twisted Coxeter element (\ref{tCox}).

\subsubsection{Action of $\hat{\protect\sigma}_{q}$ in the root space}

Due to the occurrence of the automorphism $\omega $ in the definition (\ref
{qtCox}) it turns out to be convenient to regard the objects $\hat{\gamma}%
_{i}^{\omega }:=\omega ^{\frac{c_{i}-1}{2}}\hat{\gamma}_{i}$ instead of the
usual coloured roots. From (\ref{taction}) and (\ref{qtCox}) the action of $%
\hat{\sigma}_{q}$ is then obtained to 
\begin{equation}
(-q^{-2c_{i}})^{\hat{t}_{i}}\hat{\sigma}_{q}^{c_{i}}\hat{\gamma}_{i}^{\omega
}=\omega ^{-c_{i}}\hat{\gamma}_{i}^{\omega }-\sum\limits_{j\in \hat{\Delta}%
_{-c_{i}}^{\omega }}\!I_{ij}\hat{\gamma}_{j}^{\omega }\;.  \label{qtaction}
\end{equation}
This tells us that the orbits $\hat{\Omega}_{i}^{q}$ generated by $\hat{%
\sigma}_{q}$ consist of polynomials in the deformation parameter $q$ and
simple roots analogous to the untwisted case. The structure of $\hat{\Omega}%
_{i}^{q}$ is periodic because of the crucial relation 
\begin{equation}
q^{-2h}\hat{\sigma}_{q}^{H}=1\,\,.  \label{quasi2}
\end{equation}
This on the one hand reflects the property (TC2) of the usual twisted
Coxeter element (\ref{tCox}) and on the other hand it shows the duality
relation between the algebras $X_{n}$ and $\hat{X}_{\hat{n}}$ by comparison
with (\ref{quasi}). We note that the roles of $h$ and $H$ are just
interchanged. Like before we do not give a generic proof of this periodicity
property, but verify it case-by-case in the appendix.

For later purposes we define the charge conjugation operation as done in
Subsection 2.4.2 for the untwisted Coxeter element. Assign the anti-particle
to the orbit $\hat{\Omega}_{\bar{\imath}}^{q}$ in which we find the element 
\cite{FKS2} 
\begin{equation}
-q^{-h+\frac{c_{\bar{\imath}}-c_{i}}{2}\hat{t}_{i}}\,\hat{\sigma}_{q}^{\frac{%
H}{2}+\frac{c_{i}-c_{\bar{\imath}}}{4}(2-\ell _{i})}\hat{\gamma}_{i}^{\omega
}\in \hat{\Omega}_{\bar{\imath}}^{q}\;,\quad i\in \hat{\Delta}^{\omega }\;.
\label{antit}
\end{equation}
Here $i$ is assumed to belong to the set of representatives $\hat{\Delta}%
^{\omega }$ since only these will correspond to particle species later on.
Repeating the conjugation yields again $\hat{\Omega}_{i}^{q}$ by use of (\ref
{quasi2}) provided that $\ell _{i}=\ell _{\bar{\imath}}$. Assuming
self-conjugation for the non-simply laced algebras as in (\ref{anti}), the
latter relation reduces to 
\begin{equation}
\,\hat{\sigma}_{q}^{\frac{H}{2}}\hat{\gamma}_{i}^{\omega }\,\,=-q^{h}\hat{%
\gamma}_{\bar{\imath}}^{\omega }\,\,.
\end{equation}
In the limit $q\rightarrow 1$ we obtain $\hat{\sigma}^{\frac{H}{2}+\frac{%
c_{i}-c_{\bar{\imath}}}{4}(2-\ell _{i})}\hat{\gamma}_{i}^{\omega }\,\,=\hat{%
\gamma}_{\bar{\imath}}^{\omega }$, which relates particles and
anti-particles in twisted algebras.

\subsection{q-deformed folded Cartan matrix of $\hat{X}_{\hat{n}}^{(\ell )}$}

In order to relate the action of $\hat{\sigma}_{q}$ to a matrix expression
we now perform an analogous calculation as in Subsection 2.4.3 for the
untwisted algebra. However, due to the appearance of the Dynkin diagram
automorphism $\omega $ the computation turns out to be a bit more involved,
whence we first state the result and then present the main steps of the
derivation. Let $\,\lambda _{i}^{\omega }$ denote the fundamental weight
which is dual to all elements inside one $\omega $-orbit, i.e. 
\begin{equation}
\lambda _{i}^{\omega }:=\sum\limits_{k=1}^{\ell _{i}}\hat{\lambda}_{\omega
^{k}(i)}\;,\quad i\in \hat{\Delta}^{\omega }\;.  \label{foldedw}
\end{equation}
Then we will show below that the $n\times n$ matrix depending on two
deformation parameters $q,\hat{q}$ and defined by 
\begin{equation}
N_{ij}(q,\hat{q}):=-\frac{q^{-\frac{c_{i}+c_{j}}{2}}}{2}\sum_{y=1}^{H}\left%
\langle \lambda _{j}^{\omega },(\hat{\sigma}_{q})^{y+\frac{c_{i}-1}{2}}\hat{%
\gamma}_{i}^{\omega }\right\rangle \hat{q}^{2y+\frac{c_{i}-1}{2}\ell _{i}-%
\frac{c_{j}-1}{2}\ell _{j}-1},\quad i,j\in \hat{\Delta}^{\omega }  \label{N0}
\end{equation}
can be equivalently expressed in terms of the following $q$-deformed
matrices, 
\begin{equation}
N(q,\hat{q})=\frac{1-q^{2h}\hat{q}^{2H}}{2}\,A^{\omega }(q,\hat{q})^{-1}[%
\hat{D}]_{\hat{q}}\;.  \label{N2}
\end{equation}
The objects appearing in the last equation will be explained step by step.
Consider the diagonal matrix $D_{ij}:=\ell _{i}\delta _{ij}$ and the folded
incidence matrix $I^{\omega }=2-A^{\omega }$ associated with (\ref{foldedA}%
), then we define 
\begin{equation}
([\hat{D}]_{q})_{ij}:=[\ell _{i}]_{q}\delta _{ij},\quad \quad ([I^{\omega
}]_{q})_{ij}:=\left[ \sum_{n=1}^{\ell _{i}}\hat{I}_{\omega ^{n}(i)j}\right]
_{q}\;,\quad i,j\in \hat{\Delta}^{\omega }
\end{equation}
to be their $q$-deformed counterparts. Setting now analogously to (\ref{qA}) 
\begin{equation}
A^{\omega }(q,\hat{q})=q^{-1}\hat{q}^{\,-\hat{D}}+q\,\hat{q}^{\hat{D}%
}-[I^{\omega }]_{\hat{q}}  \label{qfA}
\end{equation}
we obtain a $q$\textbf{-deformed version of the folded Cartan matrix} (\ref
{foldedA}) \cite{FKS2}. The restriction of the indices in (\ref{N0}) to the
subset of representatives, which was chosen to be $\hat{\Delta}^{\omega
}=\{1,...,n\},$ appears naturally due to the folding procedure explained at
the beginning of this section. In fact, studying the latter we saw that $%
A^{\omega }=A$. Furthermore, one easily deduces that the length of the roots
in $X_{n}$ is related to the length of the $\omega $-orbits in $\hat{X}_{%
\hat{n}}$, namely $\ell _{i}=t_{i},\;i=1,...,n$. Thus, we conclude that $%
A^{\omega }(q,\hat{q})=A(q,\hat{q})$ holds for generic values of the
deformation parameters, whence we have proven \cite{FKS2} 
\begin{equation}
N(q,\hat{q})=M(q,\hat{q})\;,\quad q,\hat{q}\in \mathbb{C}\;.  \label{N=M}
\end{equation}
This establishes the previously mentioned equivalence of the roles played by
the $q$-deformed Coxeter and twisted Coxeter elements on the dual algebras.
It should be emphasized that the identities (\ref{M0}), (\ref{M2}) and (\ref
{N0}), (\ref{N2}) together with the equality of the two $q$-deformed Cartan
matrices (\ref{qA}) and (\ref{qfA}) are the key results of this chapter.
They will allow for a universal treatment of ATFT upon choosing the
deformation parameters in a special coupling dependent way to be specified
later. Moreover, they provide an important mathematical tool in which the
duality relation between the two dual algebras becomes manifest.

Before we now turn to the proof of (\ref{N2}) we state another neat formula
for the determinant of the $q$-deformed folded Cartan matrix \cite{FKS2}, 
\begin{equation}
\det A^{\omega }(q,\hat{q})=\prod_{i=1}^{\hat{s}}(q^{h/H}\,\hat{q}+q^{-h/H}%
\hat{q}^{\,-1}-2\cos \tfrac{\pi \hat{s}_{i}}{H})  \label{detfA}
\end{equation}
where the product runs over the exponents of the algebra $\hat{X}_{\hat{n}}$%
. In view of (\ref{detA}) this is a different specification of the singular
values of the Cartan matrix $A(q,\hat{q})$. However, since the exponents of
the dual algebras only differ by multiples of the Coxeter number $h$ \cite
{Kac} both formulas are consistent. Note that (\ref{detfA}) is established
on a case-by-case study.

We now turn to the proof of the identity (\ref{N2}). Let us first define the
following auxiliary $\hat{n}\times \hat{n}$ matrix, 
\begin{equation}
\hat{N}_{ij}(q,\hat{q}):=-\frac{1}{2}\,q^{-\frac{c_{i}+c_{j}}{2}%
}\sum_{y=1}^{H}\left\langle \hat{\lambda}_{j},(\hat{\sigma}_{q})^{y+\frac{%
c_{i}-1}{2}}\hat{\gamma}_{i}^{\omega }\right\rangle \hat{q}^{2y+\frac{c_{i}-1%
}{2}\ell _{i}-\frac{c_{j}-1}{2}\ell _{j}-1}.  \label{Na}
\end{equation}
Here $\hat{\lambda}_{j}$ denotes a fundamental weight of the algebra $\hat{X}%
_{\hat{n}}$ and the indices run now over the full index set $\hat{\Delta}%
=\{1,...,\hat{n}\}$. Multiplying equation (\ref{qtaction}) from the left
with $\hat{q}^{2y+\frac{c_{i}-1}{2}\ell _{i}-\frac{c_{j}-1}{2}\ell _{j}-1}%
\hat{\sigma}_{q}^{y}$ and performing the sum over the powers $y$ in the
appropriate range yields upon using the periodicity (\ref{quasi2}) the
determining equation 
\begin{multline*}
(-1)^{\hat{t}_{i}+1}(q^{\hat{t}_{i}}\hat{q})^{-2c_{i}}\hat{N}_{ij}+\hat{N}%
_{\omega ^{-c_{i}}(i)j}= \\
\sum_{l\in \hat{\Delta}_{-c_{i}}^{\omega }}q^{-c_{i}}\hat{q}^{-2c_{i}+\frac{%
c_{i}-1}{2}\ell _{i}+\frac{c_{i}+1}{2}\ell _{l}}I_{il}\hat{N}_{lj}+(q\hat{q}%
)^{-c_{i}}\frac{(1-q^{2h}\hat{q}^{2H})}{2}\delta _{i\omega ^{\frac{1+c_{i}}{2%
}}(j)}\,\,.
\end{multline*}
In contrast to (\ref{M1}) it can not be directly solved for $\hat{N}$
because of the indices transformed by $\omega $ appearing in the equation.
Therefore, we successively replace $i\rightarrow \omega ^{-c_{i}}(i)$ until
the order $\ell $ of the automorphism is reached. The resulting set of
iterated equations then allows to set up an equation for the restricted $%
n\times n$ matrix (\ref{N0}) obtained from $\hat{N}$ by the prescription $%
N_{ij}:=\sum_{n=1}^{\ell _{j}}\hat{N}_{i\omega ^{n}(j)}$ with $i,j\in \hat{%
\Delta}^{\omega }=\{1,...,n\}$. The determining equation for $N$ then reads 
\begin{multline*}
q^{-c_{i}}\hat{q}^{\,-\ell _{i}c_{i}}N_{ij}+q^{c_{i}}\hat{q}^{\ell
_{i}c_{i}}(-1)^{\ell -\ell _{i}}(q\hat{q})^{2c_{i}(\ell -\ell _{i})}N_{ij}=
\\
\sum_{n=0}^{\ell -1}I_{\omega ^{n}(i)k}(-q^{2c_{i}})^{n\delta _{1,\ell
_{i}}}\,\hat{q}^{2nc_{i}}\hat{q}^{\frac{c_{i}+1}{2}(\ell _{k}-\ell
_{i})}N_{kj}+ \\
+\hat{q}^{c_{i}(1-\ell _{i})}\sum_{n=0}^{\ell -1}(-q^{2c_{i}})^{n\delta
_{1,\ell _{i}}}\,\hat{q}^{2nc_{i}}\,\frac{1-q^{2h}\hat{q}^{2H}}{2}\,\delta
_{ij}\,.
\end{multline*}
This expression can be simplified by discussing the cases $\ell _{i}=1$ and $%
\ell _{i}=\ell $ separately and upon noting that $I_{\omega (i)j}=0$ for $%
\ell _{i}=\ell $ and $I_{\omega (i)j}=I_{ij}$ for $\ell _{i}=1$. In
addition, using the identity 
\begin{equation*}
(q\hat{q})^{\ell }+(-1)^{\ell -1}(q\hat{q})^{\ell }=(q\hat{q}+q^{-1}\hat{q}%
^{\,-1})(q\hat{q})^{\ell -1}\sum_{n=0}^{\ell -1}(-1)^{n}(q\hat{q})^{2n}
\end{equation*}
the sums in the last expression can be simplified and we obtain the
stringent formula (\ref{N2}).

\subsection{Inner Product Identities}

We close the section by stating the equivalent relations for the inner
product identities (\ref{spari}) and (\ref{sm}) for the dual algebra $\hat{X}%
_{\hat{n}}$. They can be proven by the same procedure as in the non-twisted
case. The identity linked to parity invariance is given by \cite{FKS2} 
\begin{equation}
\left\langle \lambda _{j}^{\omega },\hat{\sigma}_{q}^{x}\hat{\gamma}%
_{i}^{\omega }\right\rangle =\left\langle \lambda _{i_{{}}}^{\omega
}\!\right. ,\hat{\sigma}_{q}^{x+\frac{c_{j}-c_{i}}{2}+\frac{c_{i}-1}{2}\ell
_{i}+\frac{1-c_{j}}{2}\ell _{j}}\left. \hat{\gamma}_{j}^{\omega
}\right\rangle  \label{paritt}
\end{equation}
and the one related to analyticity of the S-matrix by \cite{FKS2} 
\begin{equation}
\left\langle \lambda _{j}^{\omega },\hat{\sigma}_{q}^{x}\hat{\gamma}%
_{i}^{\omega }\right\rangle =-q^{2h+c_{i}+c_{j}}\,\,\left\langle \lambda
_{j}^{\omega }\right. ,\hat{\sigma}_{q^{-1}}^{H-x+c_{i}+\frac{1-c_{i}}{2}%
\ell _{i}+\frac{c_{j}-1}{2}\ell _{j}}\left. \hat{\gamma}_{i_{{}}}^{\omega
}\right\rangle  \label{mm}
\end{equation}
It should be mentioned that the proof of both relations in terms of the
Coxeter elements turns out to be more complicated as in the non-twisted
case. However, it would be desirable to formulate these, since this might
provide more profound insight of the Lie algebraic duality between the
algebra $X_{n}$ and $\hat{X}_{\hat{n}}$ and the action of the twisted
Coxeter element.

\chapter{Exact S-matrices}

{\small \emph{When meeting calamities or difficult situations, it is not
enough to simply say that one is not at all flustered. When meeting
difficult situations, one should dash forward bravely and with joy. It is
the crossing of a single barrier and is like the saying, ``The more the
water, the higher the boat.''}}

\qquad \qquad \qquad \qquad \qquad \qquad {\small From 'The Book of the
Samurai, Hagakure'\medskip }

One of the central objects in quantum field theory is the scattering matrix,
henceforth also referred to as S-matrix, which determines the on-shell
structure of the model and describes the collision of quantum particles.
While in general the S-matrix can only be calculated in the framework of
perturbation theory by means of Feynman diagrams, one might pursue in case
of integrable models an alternative approach originating in the early works
of Heisenberg \cite{Heisb} and Chew \cite{Chew}. The basic idea is that the
scattering matrix by itself has to obey a number of physically motivated
constraints which might be restrictive enough to calculate it directly
without relying on the field content of the theory. Some of the general
properties an S-matrix should satisfy are:

\begin{itemize}
\item[(S1)]  conservation of probability

\item[(S2)]  Lorentz invariance

\item[(S3)]  analyticity in the energy variables

\item[(S4)]  crossing symmetry
\end{itemize}

In higher dimensions 3+1 the application of this method has led to the
famous dispersion relations yielding rigid constraints on cross sections.
However, a systematic way to fully construct an S-matrix is not known. In
contrast, the approach becomes extremely powerful in 1+1 dimensions in
context of integrable field theories. Their key feature is the presence of
an infinite set of conserved charges (a pair of higher spin charges is
actually sufficient, see \cite{Parke}) implying the following severe
restrictions on a scattering process,

\begin{itemize}
\item  absence of particle production

\item  conservation of the individual particle momenta

\item  factorization of the scattering matrix into two-particle amplitudes
\end{itemize}

Especially the last property is of importance since it reduces the problem
to determining the two-particle S-matrix. Furthermore, it gives rise to two
sets of consistency conditions, the Yang-Baxter and the bootstrap equations.
While the former describe equivalent ways to factorize a three particle
scattering process when reflection is present, the latter ensure consistency
when an intermediate bound state occurs. Provided the particle content and
the bound state structure of the model are known the two-particle S-matrix
can then be systematically constructed by exploiting the above restrictions
(S1)-(S4) and the results obtained ought to be ``exact'', i.e. they should
hold to all orders of perturbation theory.

Besides giving an exact answer this ``bootstrap'' approach, as it is called
in the literature, supersedes the conventional perturbative one in two
aspects. First it elegantly circumvents the messy and tedious computations
of higher order perturbation and renormalization theory. Second, since the
bootstrap does not rely on the field content, it might be performed whether
or not an underlying classical Lagrangian formulation of the theory is
known. Nevertheless, it is of advantage to have a classical Lagrangian at
one's disposal since it serves physical intuition and also enables
perturbative consistency checks of the bootstrap approach which are useful
from time to time. In the subsequent chapter we will, however, encounter a
different method to check the scattering matrices for consistency which also
does not use the field quantities and yields valuable insight in the high
energy regime of the associated quantum field theories.

This chapter starts with a review of the ideas underlying the bootstrap
approach in 1+1 dimensional integrable quantum field theories in Section
3.1. The factorization of the scattering matrix is motivated followed by a
discussion of the analytic structure of the two-particle S-matrix which
becomes the central object of interest. The physical constraints on the
scattering matrix stated above are translated to concrete functional
relations for the two-particle amplitude and the general form of the
solutions satisfying the properties (S1)-(S4) is stated. In order to extract
from the possible solutions the S-matrix of a concrete quantum field theory
one has to relate them to the particle content in a second step by invoking
the bootstrap equation.

In Section 3.2 this is done for affine Toda field theory. As a starting
point the classical Lagrangian, the classical mass spectrum and the
classical fusing processes present in this class of theories are recalled.
This serves as motivation for the subsequent discussion of the corresponding
structures on the quantum level and the universal formulation of the
S-matrix. This discussion will be given for all Toda models at once by
exploiting the Lie algebraic techniques introduced in the previous chapter.
In particular, we will see that the $q$-deformation of Chapter 2 has a
concrete physical meaning in describing the renormalized quantum mass flow
dependent on the coupling constant and it will become apparent how the weak
and strong coupling regime of affine Toda theories is governed by different
Lie algebraic structures. After having stated the quantum particle content
of affine Toda theories two universal expressions of the two-particle
scattering matrix are given which in their most elegant form only involve
the $q$-deformed Cartan matrices discussed before.

One of the main observations we will recover from the universal treatment of
affine Toda theory is the natural splitting of the models in two subclasses
corresponding to simply-laced and non-simply laced algebras. The former are
known to have simpler renormalization properties and the associated
two-particle S-matrix can be separated in two factors one containing all the
physical information about the particle content of the model and the other
displaying the coupling dependence. This separation of the affine Toda
S-matrix will then be used to define new integrable models by constructing
new solutions to the functional equations of the bootstrap approach in
Section 3.3.

\section{Analyticity, crossing and the bootstrap equations}

Since the S-matrix ought to describe a scattering experiment the forces are
assumed to be sufficiently short ranged and the particles should become free
at sufficiently large times ($t\rightarrow \pm \infty $). The space of
physical states should therefore asymptotically be spanned by incoming or
outgoing momentum eigenstates obeying the mass-shell condition of free
particles, i.e. $p_{\mu }p^{\mu }=m^{2}$ with $p$ being the relativistic
two-momentum. This statement is called \textbf{asymptotic completeness} and
it motivates the introduction of momentum creation and annihilation
operators $a_{in}(p),a_{in}^{\ast }(p),a_{out}(p),a_{out}^{\ast }(p)$ which
generate the Fock space of the $in$ and $out$-states upon acting on the
vacuum state $|0\rangle $, 
\begin{equation*}
|p_{1},\ldots ,p_{n}\rangle _{{\QATOP{in }{out}}}=a_{{\QATOP{in }{out}}%
}^{\ast }(p_{1})\cdots a_{{\QATOP{in }{out}}}^{\ast }(p_{n})|0\rangle \;.
\end{equation*}
The asymptotic states might depend in addition on internal quantum numbers
but for simplicity these are suppressed in the notation. Both sets of these
states are assumed to be complete and orthonormal whence they have to be
mapped onto each other by a unitary operator, $\mathcal{S}^{\ast }\mathcal{S}%
=\mathcal{SS}^{\ast }=1,$ which is just the S-matrix, 
\begin{eqnarray}
\,_{out}\left\langle p_{1}^{\prime },\ldots ,p_{m}^{\prime }|\,p_{1},\ldots
,p_{n}\right\rangle _{in} &=&\,_{in}\left\langle p_{1}^{\prime },\ldots
,p_{m}^{\prime }|\,\mathcal{S}\,|\,p_{1},\ldots ,p_{n}\right\rangle _{in}
\label{S1} \\
&=&\,_{out}\left\langle p_{1}^{\prime },\ldots ,p_{m}^{\prime }|\,\mathcal{S}%
\,|\,p_{1},\ldots ,p_{n}\right\rangle _{out}\;.  \notag
\end{eqnarray}
The above equations show that we might drop the $in$ and $out$ labels. Thus,
the first constraint (S1) is a direct consequence of postulating asymptotic
completeness for the set of $in$ and $out$ states.

To satisfy the second condition of Lorentz invariance we require that the
S-matrix elements are invariant under the action of the Lorentz group, e.g.
the two particle amplitude should obey 
\begin{equation}
\left\langle p_{1},p_{2}|\,\mathcal{S}\,|\,p_{3},p_{4}\right\rangle
=\left\langle p_{1},p_{2}|\,\Lambda \mathcal{S}\Lambda
^{-1}|\,p_{3},p_{4}\right\rangle  \label{S2}
\end{equation}
for every proper Lorentz transformation $\Lambda $. Here we followed the
standard convention (see e.g. \cite{Eden}) and specified Lorentz invariance
for the whole matrix elements instead for their absolute values only. This
fixes the phase uniquely. As a consequence the two-particle amplitude
depends only on the so-called \textbf{Mandelstam variables} 
\begin{equation*}
s=(p_{1}+p_{2})^{2}\,,\quad t=(p_{1}-p_{3})^{2}\quad \text{and\quad }%
u=(p_{1}-p_{4})^{2}\,
\end{equation*}
up to an overall factor $\delta ^{(4)}(p_{1}+p_{2}-p_{3}-p_{4})$ reflecting
energy-momentum conservation. The Mandelstam variables satisfy the relation $%
s+t+u=\sum_{i=1}^{4}m_{i}^{2}$ from which it can be deduced that in 1+1
dimensions only one of them is independent. Before discussing the
outstanding properties (S3) and (S4) we now turn to further restrictions
imposed by integrability.

\subsection{Conserved charges and factorization}

As mentioned before integrable field theories are distinguished by the
presence of an infinite set of conversation laws which are in involution and
transform as higher rank tensors under the Lorentz group. Let $\mathcal{Q}%
^{(s)}$ denote such a conserved charge with Lorentz spin $\pm s>1$ then we
can choose the one-particle momentum eigenstates in the $in$ and $out$ basis
such that they are simultaneously also eigenstates of $\mathcal{Q}^{(s)}$, 
\begin{equation*}
\mathcal{Q}^{(s)}\left| \,p\right\rangle =Q^{(s)}(p^{\pm })^{s}\left|
\,p\right\rangle \;.
\end{equation*}
Here $p^{\pm }=p^{0}\pm p^{1}$ are the light cone components of the
two-momentum, $Q^{(s)}$ is a scalar and the upper or lower sign applies if $%
s $ is either positive or negative, respectively. This particular form of
the eigenvalue is required by the Lorentz transformation property of the
conserved charge $\mathcal{Q}^{(s)}$. If we assume the charge to be local,
i.e. to be an integral of a local charge density, then its action on a
multiparticle state must be additive, 
\begin{equation*}
\mathcal{Q}^{(s)}\left| \,p_{1},...,p_{n}\right\rangle =\left\{
Q_{1}^{(s)}(p_{1}^{\pm })^{s}+\cdots +Q_{n}^{(s)}(p_{n}^{\pm })^{s}\right\}
\left| \,p_{1},...,p_{n}\right\rangle \;.
\end{equation*}
Now conservation of $\mathcal{Q}^{(s)}$ means that for a generic scattering
process $\left| \,p_{1},...,p_{n}\right\rangle ^{in}\rightarrow \left|
\,p_{1}^{\prime },...,p_{m}^{\prime }\right\rangle ^{out}$ the following
sums must be equal 
\begin{equation*}
\sum_{i=1}^{n}Q_{i}^{(s)}(p_{i}^{\pm
})^{s}=\sum_{i=1}^{m}Q_{i}^{(s)}(p_{i}^{\prime \pm })^{s}\;.
\end{equation*}
For an infinite set of higher spin charges this results in an infinite set
of equations which allow only for the trivial solution namely that the sets
of incoming and outgoing particle momenta are equal, 
\begin{equation}
\{p_{1},\ldots ,p_{n}\}=\{p_{1}^{\prime },\ldots ,p_{m}^{\prime }\}\;.
\label{A}
\end{equation}
Note that this in particular implies the absence of particle production. To
see the third condition imposed by integrability, i.e. the factorization of
the S-matrix, we now follow a line of reasoning given by Shankar and Witten 
\cite{SW}.

Let us assume for simplicity that one of the higher spin conserved charges $%
\mathcal{Q}^{(s)}$ acts on the momentum eigenstates just as $(p^{1})^{s}$
with $p^{1}$ being the spatial component of the two-momentum $p$. Then a
localized one-particle state described by a wave function of the form 
\begin{equation*}
\psi (x)=N\int dp\,e^{-a(p-p^{1})^{2}+ip(x-\xi )}
\end{equation*}
is transformed into 
\begin{equation*}
e^{i\lambda \mathcal{Q}^{(s)}}\psi (x)=N\int dp\,e^{-a(p-p^{1})^{2}+ip(x-\xi
)+i\lambda p^{s}}\;.
\end{equation*}
Here $N$ is a normalization constant, $\lambda $ an arbitrary real shift
parameter and $\xi $ the center of the localized wave packet. Thus, the
action of the conserved charge has shifted the center of the wave packet
dependent on the particle momentum, 
\begin{equation*}
\xi \rightarrow \xi ^{\prime }=\xi -s\lambda (p^{1})^{s-1}\;.
\end{equation*}
Consider now a multiparticle state with wave packets of different momenta
and localized in position just like the wave function above. Then the
application of the operator $e^{i\lambda \mathcal{Q}^{(s)}},s>1$ will
displace the wave-packets relative to each other. Explicitly, consider a
three particle collision process with $p_{1}<p_{2}<p_{3}$ as depicted in
Figure 3.1. By successive action with the conserved charge for different
values of $\lambda $ one might change the impact parameters of the colliding
particles and produce all three different space-time diagrams shown.
Clearly, the last two diagrams correspond to successive two-particle
collisions showing that a three-particle scattering process factorizes in
two-particle ones. Moreover, by conservation of $\mathcal{Q}^{(s)}$ both
possible decompositions of the three-particle scattering amplitude must
coincide, 
\begin{equation}
S^{(2)}(p_{2},p_{3})S^{(2)}(p_{3},p_{1})S^{(2)}(p_{1},p_{2})=S^{(2)}(p_{1},p_{2})S^{(2)}(p_{1},p_{3})S^{(2)}(p_{2},p_{3})\;,
\label{YB}
\end{equation}
where $S^{(2)}(p_{i},p_{j})=\left\langle p_{i},p_{j}|\,\mathcal{S}%
\,|\,p_{i},p_{j}\right\rangle $ denotes the two-particle amplitude. This
factorization identity is the celebrated \textbf{Yang-Baxter equation}\emph{%
\ }\cite{YB}. Note that the two-particle amplitude might depend on internal
quantum numbers whence the above equation is to be understood in a matrix
notation. In the presence of such internal symmetries the Yang-Baxter
equation imposes a surprisingly powerful constraint which often suffices to
construct the S-matrix explicitly \cite{boot}.

\begin{center}
\includegraphics[width=8cm,height=12cm,angle=-90]{YB.epsi}
\end{center}


\begin{center}
\noindent {\small Figure 3.1: Depiction of the Yang-Baxter equation.\bigskip 
}
\end{center}

However, in due course we will only encounter theories where the particle
spectrum is non-degenerate. Either because all the particle masses are
different or since the particles can be distinguished by the eigenvalues of
one of the conserved charges. This excludes a possible redistribution of the
particles with the same quantum numbers in a scattering process and there is
no reflection. One is left with a set of diagonal S-matrices, i.e. a set of
phases, which trivially satisfy the Yang-Baxter equation. Thus, for the
construction of diagonal scattering matrices (\ref{YB}) does not impose an
additional constraint to (S1)-(S4). However, by the same argument one now
motivates more generally that the $n$-particle scattering amplitude should
factorize into two-particle ones, 
\begin{equation}
S^{(n)}(p_{1},\ldots ,p_{n}):=\,^{out}\left\langle p_{1},\ldots
,p_{n}|\,p_{1},\ldots ,p_{n}\right\rangle
^{in}=\prod_{i<j}S^{(2)}(p_{i},p_{j})  \label{B}
\end{equation}
meaning that every scattering process can be decomposed into $n(n-1)/2$
subprocesses involving only a pair of particles. Note that we have omitted a
pair of $\delta $-function in the above definition of the scattering
amplitude. The properties (\ref{A}) and (\ref{B}) are characteristic for
integrable field theories and constitute the cornerstones for the successful
application of the bootstrap approach in constructing exact S-matrices.

It should be mentioned that the assumption of an infinite number of
conserved charges is actually too strong. As was shown by Parke \cite{Parke}
a pair of higher spin charges is already sufficient to deduce the crucial
properties (\ref{A}) and (\ref{B}). For theories invariant under a parity
transformation even \emph{one} charge of spin $s>1$ is sufficient because
upon a parity transformation one obtains another one of spin $-s$. However,
for the integrable models we are going to consider it is believed that an
infinite set of higher spin charges is present.

\subsection{The analytic structure of the two-particle S-matrix}

So far we have discussed the restrictions on the scattering matrix imposed
by the general constraints (S1), (S2) and integrability. From the latter we
have learned that for integrable models one only needs to regard the
two-particle amplitude which must be a function of one of the Mandelstam
variables, say $s$. For the following discussion of the remaining general
constraints (S3), (S4) and the formulation of functional equations
reflecting them, it is convenient to introduce a different variable, the
rapidity $\theta $. The latter is defined by parametrizing the two-momentum
in the following way, 
\begin{equation}
p_{i}=m_{i}(\cosh \theta _{i},\sinh \theta _{i})\;.  \label{theta}
\end{equation}
Clearly, rewriting $p$ in the above manner has the advantage that the
on-shell condition $p_{\mu }p^{\mu }=m^{2}$ is satisfied automatically. The
parametrization of the Mandelstam variable is then determined by 
\begin{equation}
s=m_{i}^{2}+m_{j}^{2}+2m_{i}m_{j}\cosh \theta _{ij}  \label{stheta}
\end{equation}
where $\theta _{ij}=\theta _{i}-\theta _{j}$ is the rapidity difference.
Henceforth, the two-particle scattering amplitude will therefore be written
in either one of the following variants, 
\begin{equation}
S^{(2)}(p_{i},p_{j})=:S_{ij}(s)\quad \text{or}\quad
S^{(2)}(p_{i},p_{j})=:S_{ij}(\theta _{ij})\;.  \label{2S}
\end{equation}
The sole dependence either on the Mandelstam variable $s$ or the rapidity
difference $\theta _{ij}$ incorporates relativistic invariance according to
requirement (S2). The ultimate reason for introducing the rapidity variable
is that it simplifies the discussion of the analytic structure (S3)\ of the
two-particle S-matrix to which we now turn. However, for sake of
completeness we will relate the analytic properties in the variable $\theta $
to the ones in the variable $s$ and vice versa.

\subsubsection{The analytic domains}

At the heart of the bootstrap principle lies the idea to view the physical
scattering amplitudes as real-boundary values of analytic functions defined
on domains of the complex plane. This property has been motivated to be
linked to macroscopic causality properties, see e.g. \cite{Eden}. We
therefore interpret the two-particle amplitude (\ref{2S}) now as function of
complex variables $s,\theta \in \mathbb{C}$ defined on the domains specified
in Figure 3.2. Starting with $S_{ij}(s)$ one can deduce from unitarity
arguments that the two-particle amplitude must have a square root branch
point in the $s$-channel at the two-particle threshold $s=(m_{i}+m_{j})^{2}$%
. This branching point signals the least energy required that inelastic
processes can take place like the production of extra particles. A second
one arises by crossing symmetry -- which will be explained below -- in the $%
t $-channel and is located at $s=(m_{i}-m_{j})^{2}$. Since in the context of
integrable models one only deals with two-particle unitarity, these are
assumed to be the only branching points implying that the two particle
scattering amplitude is meromorphic. The resulting left and right branch
cuts in the $s$-plane are defined to lie in the ranges $s\leq
(m_{i}-m_{j})^{2}$ and $(m_{i}+m_{j})^{2}\leq s$ .Via the parametrization (%
\ref{stheta}) these branch cuts are unfolded in the complex $\theta $-plane
and mapped onto the axes $\limfunc{Im}\theta =\pi $ and $\limfunc{Im}\theta
=0$, respectively. They enclose the strip $0<\limfunc{Im}\theta <\pi $ which
is referred to as \textbf{physical sheet}\emph{,} since only the analytic
continuation into this region is assumed to be related to concrete physical
processes.

\subsubsection{Hermitian analyticity and unitarity}

Having introduced real branch cuts in the $s$-plane we need to specify how
to obtain the physical values of the scattering amplitude, since for
on-shell processes the Mandelstam variable $s$ is real. Comparison with
perturbation theory shows that these are recovered by the limit onto the
branch cut from the upper-half plane \cite{Eden}, 
\begin{equation*}
\text{physical values:}\quad S_{ij}(s)\equiv \lim_{\varepsilon \rightarrow
0^{+}}S_{ij}(s+i\varepsilon )\;,\quad s\in \mathbb{R}\;.
\end{equation*}
This limit reflects Feynman's prescription in perturbation theory to add to
each particle mass a small \emph{negative} imaginary part $-i\varepsilon $
in order to make the integration over internal lines of the relevant Feynman
diagrams well defined \cite{Eden}. However, one might also consider the 
\emph{unphysical} limit on the branch cut from the lower-half plane. The
complex conjugate of this value can be linked to the parity transformed 
\emph{physical} scattering amplitude 
\begin{equation*}
\lim_{\varepsilon \rightarrow 0^{+}}S_{ij}(s-i\varepsilon )^{\ast
}=\lim_{\varepsilon \rightarrow 0^{+}}S_{ji}(s+i\varepsilon )\;,\quad s\in 
\mathbb{R}\;.
\end{equation*}
Thus, two different scattering processes are linked to each other across the
branch cut as boundary values of the same analytic function. This property
is known as \textbf{Hermitian analyticity} \cite{HERMAN}. Analytic
continuation to complex arguments then results in the following functional
equation in terms of the rapidity variable, 
\begin{equation}
S_{ij}(\theta )=S_{ji}(-\theta ^{\ast })^{\ast }\;.  \label{ha}
\end{equation}
Note that for parity invariant theories one has $S_{ij}=S_{ji}$, whence
Hermitian analyticity then amounts to \textbf{real analyticity}. Combining (%
\ref{ha}) with the unitarity requirement $S_{ij}(\theta )S_{ij}(\theta
)^{\ast }=1,\;\theta \in \mathbb{R}$ for the two-particle amplitude yields
upon analytic continuation the functional equation 
\begin{equation}
S_{ij}(\theta )S_{ji}(-\theta )=1\quad ,  \label{uni}
\end{equation}
which is assumed to hold for all complex arguments $\theta \in \mathbb{C}$.
The concept that different boundary values of one analytic function relate
different scattering processes is also exploited in the context of crossing
symmetry.

\subsubsection{Crossing symmetry}

In a relativistic theory a crossing transformation amounts to the
replacement of an incoming particle of momentum $p$ by an outgoing
antiparticle with momentum $-p$ and has its origin in the
Lehmann-Symanzik-Zimmermann formalism, see any standard textbook on quantum
field theory. In terms of the Mandelstam variables this corresponds to the
transition from the $s$-channel to the $t$-channel 
\begin{equation}
s=(p_{i}+p_{j})^{2}\;\rightarrow
\;t=(p_{i}-p_{j})^{2}=2m_{i}^{2}+2m_{j}^{2}-s\;.  \label{tscross}
\end{equation}
Note that we have used here the conservation of the individual particle
momenta in integrable theories. Crossing symmetry now requires that the
scattering amplitudes corresponding to the two collision processes 
\begin{equation*}
\left| \,p_{i},p_{j}\right\rangle ^{in}\rightarrow \left|
\,p_{i},p_{j}\right\rangle ^{out}\quad \text{and}\quad \left| \,p_{j},-p_{_{%
\bar{\imath}}}\right\rangle ^{in}\rightarrow \left| \,p_{j},-p_{\bar{\imath}%
}\right\rangle ^{out}\;
\end{equation*}
can be obtained from each other by analytic continuation. Here $\bar{\imath}$
denotes the anti-particle of the particles species $i$. The path of analytic
continuation is depicted in Figure 3.2 and connects the upper side of the
right branch cut with the lower side of the left branch cut according to (%
\ref{tscross}). Upon noting that the right and left cut in the $s$-plane
correspond to the axes $\limfunc{Im}\theta =0$ and $\limfunc{Im}\theta =\pi $
in terms of the rapidity variable, crossing symmetry then simplifies to the
functional equation 
\begin{equation}
S_{ij}(i\pi -\theta )=S_{j\bar{\imath}}(\theta )\;.  \label{cross}
\end{equation}
The requirements of unitarity, Hermitian analyticity and crossing symmetry
form powerful constraints on the possible form of a solution to the
functional equations (\ref{uni}) and (\ref{cross}). It was shown in \cite
{Mitra} that the most general form of a scattering amplitude obeying the
stated restrictions must consist of ratios of hyperbolic functions%
\footnote[2]{%
In non-diagonal theories one might encounter als infinite products of $%
\Gamma $-functions or elliptic functions.}, 
\begin{equation}
S_{ij}(\theta )=\prod_{x\in X_{ij}}\frac{\sinh \frac{1}{2}(\theta +i\pi x)}{%
\sinh \frac{1}{2}(\theta -i\pi x)}  \label{gensol}
\end{equation}
where the set of real numbers $X_{ij}$ incorporates the information
characteristic to the specific quantum field theory under consideration. To
determine the latter one needs an additional equation for the S-matrix
displaying the particle content and the structure of the interaction.

\begin{center}
\includegraphics[width=4.3431in,height=6.2855in,angle=0]{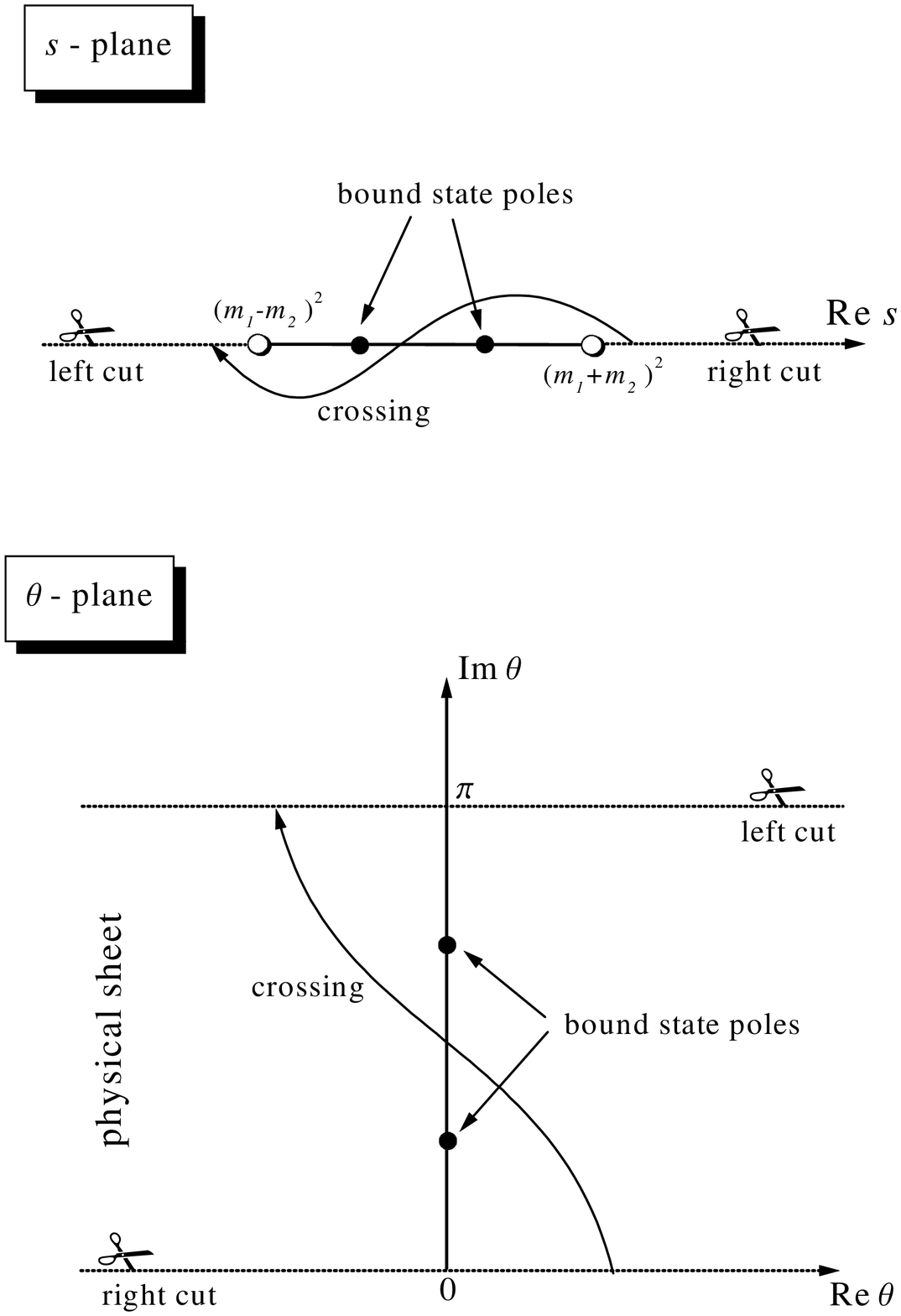}
\end{center}


\begin{center}
{\small Figure 3.2: The analytic domains of the two-particle
S-matrix.\bigskip }
\end{center}

\subsubsection{Bound state poles}

The general solution (\ref{gensol}) resulting from (\ref{uni}) and (\ref
{cross}) is obviously meromorphic and exhibits poles in the complex rapidity
plane whenever the denominator vanishes. The unspecified numbers $x\in
X_{ij} $ determine therefore the singularity structure of the S-matrix. The
general discussion of this singularity structure is a delicate issue, here
we will restrict ourselves to simple poles which lie in the physical strip
and are associated with stable bound states. Explicitly, consider a simple
pole in the two-particle amplitude of the particles $i,j$ which are assumed
to form a bound state labelled by $\bar{k}$. In the vicinity of the
singularity the S-matrix must be of the form 
\begin{equation}
S_{ij}(\theta )\sim \frac{iR_{ij}^{k}}{(\theta -iu_{ij}^{k})}\;,\quad
0<u_{ij}^{k}<\pi \;.  \label{residue}
\end{equation}
Comparison with perturbation theory motivates that these kind of poles are
linked to the propagation of an intermediate bound state in the $s$-channel
for $R_{ij}^{k}>0$ and in the $t$-channel for $R_{ij}^{k}<0$. That is, the
two-particle state of a scattering process becomes dominated by a one
particle state at purely imaginary rapidity difference $\theta
_{ij}=iu_{ij}^{k}$ and the Mandelstam variable $s$ ought to satisfy 
\begin{equation}
s=m_{\bar{k}}^{2}=m_{i}^{2}+m_{j}^{2}+2m_{i}m_{j}\cos u_{ij}^{k}\;.
\label{fuseangle}
\end{equation}
Such a process is called \textbf{fusing} $i+j\rightarrow \bar{k}$ in the
literature and the rapidity difference $u_{ij}^{k}$ fixing the singularity
is referred to as \textbf{fusing angle}.\medskip

\begin{picture}(240.00,200.00)(-50.00,0.00)
\put(188.00,182.00){$\bar u_{ji}^{k}$}
\put(236.00,106.00){$\bar u_{kj}^{i}$}
\put(38.00,106.00){$\bar u_{ik}^{j}$}
\put(92.00,10.00){$m_{\bar \imath}$}
\put(92.00,190.00){$m_{\bar \imath}$}
\put(18.00,40.00){$m_{\bar \jmath}$}
\put(18.00,160.00){$m_{\bar\jmath}$}
\put(158.00,10.00){$m_i$}
\put(158.00,190.00){$m_i$}
\put(230.00,40.00){$m_j$}
\put(230.00,160.00){$m_j$}
\put(130.00,90.00){$m_k$}
\qbezier(245.00,124.00)(227.00,121.00)(223.00,100.00)
\qbezier(172.00,185.00)(190.00,166.00)(214.00,176.00)
\qbezier(47.00,123.00)(58.00,116.00)(56.00,100.00)
\put(60.00,0.00){\line(2,1){200.00}}
\put(0.00,100.00){\line(3,-5){60.00}}
\put(60.00,200.00){\line(-3,-5){60.00}}
\put(260.00,100.00){\line(-2,1){200.00}}
\put(200.00,0.00){\line(3,5){60.00}}
\put(0.00,100.00){\line(2,-1){200.00}}
\put(10.00,100.00){\line(-1,0){10.00}}
\put(0.00,100.00){\line(1,0){10.00}}
\put(200.00,200.00){\line(-2,-1){200.00}}
\put(260.00,100.00){\line(-3,5){60.00}}
\put(10.00,100.00){\line(1,0){250.00}}
\end{picture}

{\small \noindent Figure 3.3: Mass triangles in the complex velocity plane.
The fusing angles are defined as $\bar{u}_{ij}^{k}=\pi -u_{ij}^{k}.$ }%
\medskip\ 

\noindent Assuming \textbf{nuclear democracy} the intermediate bound state
labelled by $\bar{k}$ is supposed to belong to the asymptotic particle
spectrum and compatibility with crossing symmetry then requires that also
the fusing processes $j+k\rightarrow \bar{\imath}$ and $i+k\rightarrow \bar{j%
}$ are present in the theory. Hence, one must have 
\begin{equation}
u_{ij}^{k}+u_{ki}^{j}+u_{jk}^{i}=2\pi \;.  \label{2pi}
\end{equation}
The fusing angles and the fusing condition (\ref{fuseangle}) can be
geometrically visualized by drawing a triangle whose sides have length equal
to the masses of the three particles, see Figure 3.3.

The crucial assumption is now that the intermediate state is present at
macroscopic times which allows to formulate the following consistency
equation known as \textbf{bootstrap identity}, 
\begin{equation}
S_{li}(\theta -i\bar{u}_{ki}^{j})S_{lj}(\theta +i\bar{u}_{jk}^{i})=S_{l\bar{k%
}}(\theta )\;,\quad \bar{u}_{ij}^{k}=\pi -u_{ij}^{k}\;.  \label{boot}
\end{equation}
This functional relation for the S-matrix states the equivalence of the two
possibilities that either the scattering with a particle $l\,$takes place
before the fusing $i+j\rightarrow \bar{k}$ occurs or afterwards. For a
graphical depiction see Figure 3.4. Note that analytic continuation is also
here crucial, since the S-matrices are evaluated at unphysical values. For
later purposes we write the bootstrap equation in a more symmetric variant
exploiting (\ref{uni}) and (\ref{cross}), 
\begin{equation}
S_{li}(\theta )S_{lj}(\theta +iu_{ij}^{k})S_{lk}(\theta
+iu_{ij}^{k}+iu_{jk}^{i})=1\;.  \label{boot1}
\end{equation}
The bootstrap identity plays the key role in the construction of diagonal
S-matrices. It connects the on-shell scattering processes with the bound
state structure characteristic for the field theory under consideration by
treating the bound states at the same footing as the asymptotic particle
states. The construction of an S-matrix can now be summarized in the
following steps.

\begin{itemize}
\item  Write down the general solution (\ref{gensol}) satisfying unitarity,
Hermitian analyticity and crossing symmetry.

\item  Given some information about the bound state structure, i.e. either
the mass spectrum or the fusing processes and angles, introduce a minimum
number of physical poles in the general solution (\ref{gensol}) by fixing
the numbers $x\in X_{ij}$.

\item  Check this solution for consistency by means of the bootstrap
identity (\ref{boot}).

\item  Interpret the complete singularity structure inside the physical
sheet.
\end{itemize}

Even though equation (\ref{boot}) is extremely powerful in the construction
of exact S-matrices it turns out, that it only provides a consistency check
and does not determine the solution uniquely. Once a ''minimal'' S-matrix
obeying the functional equations is constructed one might multiply it by a
factor possessing the same properties except that its poles lie exclusively
outside of the physical sheet. Such a solution is called a \textbf{CDD factor%
} \cite{CDD} and does not alter the bound state structure of the S-matrix.
In order to remove this ambiguity an additional consistency check is
required, either by perturbation theory or by the thermodynamic Bethe
ansatz, which we will study in Chapter 4.

\begin{center}
\includegraphics[width=9cm,height=12cm,angle=-90]{boot.epsi}
\end{center}


\begin{center}
{\small Figure 3.4: Depiction of the bootstrap equation.}
\end{center}

\section{The S-matrix of affine Toda theory}

After the comments on the general structure of the two-particle S-matrix in
1+1 dimensional integrable field theories and the introduction of the
bootstrap principle we now start to consider a class of concrete examples,
affine Toda field theories. The latter have been under investigation for a
long time and because of their appealing Lie algebraic structure belong to
the most prominent and best studied examples by now. Due to their simple
renormalization properties the set of S-matrices associated with
simply-laced algebras was the first one to be completely constructed on the
level of a case-by-case study \cite{TodaS}. Exploiting the Lie algebraic
symmetry present in ATFT these S-matrices were put into a universal form 
\cite{PD,FO}. in particular, Fring and Olive demonstrated how the bootstrap
properties of the S-matrix could be derived from generic Lie algebraic
expressions involving Coxeter geometry \cite{FO}. \emph{To push a theory to
this level of universality is not only for economic and aesthetic reasons,
since all models can be treated at once, but it also allows to distinguish
model specific properties from more general ones. In this manner, it is the
first step towards the possible discovery of more powerful mathematical
concepts in constructing field theoretic models}.

The aim of this section is to achieve a similar formulation for the set of
affine Toda field theories connected with non simply-laced Lie algebras by
applying the method of $q$-deformation discussed at the end of the previous
chapter. In fact, it will turn out that the latter provides the appropriate
framework to give a concise treatment not only for non simply-laced affine
Toda models but \emph{all} of them. In order to stress this point the
generic case is discussed first and it is then shown at the end how the
special case of simply-laced affine Toda models can be extracted from the
general formulas.

Following the spirit of the bootstrap program we could directly write down
the S-matrix of affine Toda field theory and then discuss its bootstrap
properties. However, it turns out to be more instructive to discuss some
aspects of the underlying classical theory in advance for two reasons.
First, many of the classical properties of ATFT survive the quantization
when the underlying Lie algebra is simply-laced. Second, the classical
discussion will prepare the formulation of universal formulas at the quantum
level by showing that the transition \emph{classical\thinspace }$\rightarrow
\,$\emph{quantum} can in many cases achieved by replacing ordinary Lie
algebraic objects by $q$-deformed ones.

\subsection{The classical regime}

The study of classical affine Toda field theory has been performed over
years and is a subject on its own \cite{ATFT} not to mention the huge area
of associated discrete models, so-called Toda chains \cite{Todachain}.
Therefore, the following presentation is limited to those classical formulas
which form the counterparts of the quantum formulas presented in due course.
In particular, emphasis will be given to the mass spectrum and the fusing
rules, since these are the only informations needed as an input for
constructing the S-matrix via the bootstrap equations (\ref{boot}).
Comparison between classical and quantum results will then demonstrate how $%
q $-deformation naturally fits into the subject.

We start with the classical affine Toda equations (\ref{atoda}). The first
step in making contact to a field theory is to find a classical action
functional from which the equations follow under a small variation of the
fields. For this purpose it is convenient to introduce new field variables $%
\phi =(\phi _{1},...,\phi _{n})$ defined through the relation, 
\begin{equation*}
\varphi _{i}=\left\langle \alpha _{i},\phi \right\rangle +\beta ^{-1}\ln 
\frac{2n_{i}}{\left\langle \alpha _{i},\alpha _{i}\right\rangle }\;,\quad
i=0,1,...,n\;.
\end{equation*}
The equations of motion are then rewritten in the variant 
\begin{equation}
\partial _{\mu }\partial ^{\mu }\phi +\frac{m^{2}}{\beta }%
\sum_{i=0}^{n}n_{i}\alpha _{i}e^{\beta \left\langle \alpha _{i},\phi
\right\rangle }=0\;.  \label{motion}
\end{equation}
The latter can be shown to coincide with the Euler-Lagrange equations w.r.t.
the following action functional 
\begin{equation}
S_{\text{ATFT}}(\frak{g})=\int \dfrac{1}{2}\left\langle \partial _{\mu }\phi
,\partial ^{\mu }\phi \right\rangle -\frac{m^{2}}{\beta ^{2}}%
\sum_{i=0}^{n}n_{i}e^{\beta \left\langle \alpha _{i},\phi \right\rangle
}d^{2}x\;.  \label{ADEaction}
\end{equation}
Here $\frak{g}$ denotes the simple Lie algebra of rank $n$ associated with
the Cartan matrix in (\ref{atoda}), $m$ the bare mass scale, $\beta $ the
classical coupling constant and $\{\alpha _{1},...,\alpha _{n}\}$ a set of
simple roots, $n_{i}$ the Coxeter labels and $\alpha _{0}=-\theta $ the
negative highest root with $n_{0}=1$. All these Lie algebraic quantities
have been discussed in the previous chapter and determine the Lie algebraic
structure of affine Toda field theory. In order to extract the mass spectrum
and the possible fusing processes from the above functional we expand the
potential in powers of the field variables, 
\begin{eqnarray*}
V(\phi ) &=&\frac{m^{2}}{\beta ^{2}}\sum_{i=0}^{n}n_{i}e^{\beta \left\langle
\alpha _{i},\phi \right\rangle } \\
&=&\frac{m^{2}}{\beta ^{2}}\sum_{i=0}^{n}n_{i}+\frac{1}{2}(M^{2})_{ij}\phi
^{i}\phi ^{j}+\,\frac{1}{3!}\,C_{ijk}\phi ^{i}\phi ^{j}\phi ^{k}+...
\end{eqnarray*}
The constant term just shifts the vacuum energy and can be disregarded. The
coefficient of the quadratic term is interpreted as the mass matrix 
\begin{equation*}
(M^{2})_{ij}=m^{2}\sum_{k=0}^{n}n_{k}(\alpha _{k})^{i}(\alpha _{k})^{j}
\end{equation*}
whose eigenvalues $\mathbf{m}=(m_{1},...,m_{n})$ give the classical masses
of the fundamental particles in the spectrum. One of the striking results in
classical ATFT is that for \emph{untwisted} affine algebras these coincide
with the components of the Perron-Frobenius eigenvector of the Cartan matrix 
$A$, i.e. the eigenvector to the smallest eigenvalue, associated with the
Lie algebra $\frak{g}$ \cite{mass,FLO}, 
\begin{equation}
A\cdot \mathbf{m}=4\sin ^{2}\tfrac{\pi }{2h}\,\mathbf{m}\;,\quad \mathbf{m}%
=(m_{1},...,m_{n})\;.  \label{Perron}
\end{equation}
Note that this definition of the masses is compatible with physical
requirements since the components of the Perron-Frobenius vector can be
shown to be always positive. Furthermore, (\ref{Perron}) motivates a
one-to-one correspondence between particle labels and vertices in the Dynkin
diagram $\Gamma (\frak{g})$ of the associated Lie algebra. For later
purposes we rewrite the above eigenvector equation in terms of the incidence
matrix, 
\begin{equation}
\left( 2\cos \tfrac{\pi }{h}-I\right) \cdot \mathbf{m}=0\;,\quad I=2-A\;.
\label{Perron1}
\end{equation}
The observation (\ref{Perron}) can be generalized to all eigenvectors of the
Cartan matrix whose components yield the eigenvalues of higher spin
conserved quantities 
\begin{equation}
A\cdot \mathbf{Q}^{(s)}=4\sin ^{2}\tfrac{\pi s}{2h}\,\mathbf{Q}%
^{(s)}\;,\quad \mathbf{Q}^{(s)}=(Q_{1}^{(s)},...,Q_{n}^{(s)})\;,
\label{ADEeigen}
\end{equation}
\begin{equation}
\left( 2\cos \tfrac{\pi s}{h}-I\right) \cdot \mathbf{Q}^{(s)}=0\;,
\end{equation}
where $s$ runs over the exponents of the algebra specified in Section 2.3.1.
For $s=1$ we recover the masses. For $s>1$ the physical interpretation of
the charges (\ref{ADEeigen}) is less direct. Their relevance lies in the
preservation of the fusing relations to which we now turn.

A fusing process of the kind $i+j\rightarrow \bar{k}$ in terms of particles
is related to a non-vanishing three point coupling of the associated fields
defined in (\ref{ADEaction}), 
\begin{equation}
C_{ijk}\neq 0\quad \Rightarrow \quad i+j\rightarrow \bar{k}\;.
\label{fusing}
\end{equation}
Here $\bar{k}$ labels the anti-particle of $k$. From the above expansion of
the potential the three-point couplings are read off as 
\begin{equation*}
C_{ijk}=\beta \,m^{2}\sum_{l=0}^{n}n_{l}(\alpha _{l})^{i}(\alpha
_{l})^{j}(\alpha _{l})^{k}\;.
\end{equation*}
Exploiting this explicit form the fusing condition (\ref{fusing}) can be
translated in terms of Coxeter geometry (again we have to restrict ourselves
to untwisted algebras). This was first observed by Dorey \cite{PD} on a
case-by-case basis and later rigorously derived by Fring, Liao and Olive 
\cite{FLO}\ making use of the classical Lagrangian (\ref{ADEaction}).\medskip

\noindent \emph{The three-point coupling }$C_{ijk}$\emph{\ does not vanish
if and only if there exist three representatives in the orbits }$\Omega
_{i},\Omega _{j},\Omega _{k}$\emph{\ of the Coxeter element (\ref{Cox})
which sum up to zero.\medskip }

More explicitly, whenever $C_{ijk}\neq 0$ there is a triple of integers $%
(\xi _{i},\xi _{j},\xi _{k})$ such that 
\begin{equation}
\sum_{l=i,j,k}\sigma ^{\xi _{l}}\gamma _{l}=0\;.  \label{ADEfuse}
\end{equation}
Vice versa the existence of such a triple implies that particles labelled by 
$i,j,k\ $couple to each other. Equation (\ref{ADEfuse}) is known as \textbf{%
fusing rule} and has been the starting point for numerous attempts to derive
more general constraints in terms of representation theory of the associated
algebra being analogues of the well-known Clebsch-Gordon rules, e.g. \cite
{Braden}. However, so far such formulations turned out to be less
restrictive than the ones presented here.

Notice that the particular form of the fusing rule allows for a whole set of
solutions, simply by multiplying with an arbitrary power $\sigma ^{x}$ of
the Coxeter element from the left. Thus, the integers are only defined up to
equivalence w.r.t. the transformation $\xi _{l}\rightarrow \xi _{l}+x$.
However, it is important to note that besides these equivalent solutions
there exits always one which can not be obtained by a simultaneous shift in
the integers. This second version of the fusing rule reads \cite{FO}, 
\begin{equation}
\sum_{l=i,j,k}\sigma ^{\xi _{l}^{\prime }}\gamma _{l}=0\quad \text{with\quad 
}\xi _{l}^{\prime }=-\xi _{l}+\frac{c_{l}-1}{2}\;.  \label{ADEfuse1}
\end{equation}
The existence of a second solution is important for two reasons. Classically
it is needed for proving what is known as \textbf{area law} in the
literature. The latter states that the magnitude of the three-point coupling 
$|C_{ijk}|$ is proportional to the area of the fusing triangle depicted in
Figure 3.3. In formulas \cite{FLO}, 
\begin{equation*}
|C_{ijk}|=\frac{2\beta }{\sqrt{h}}m_{i}m_{j}\sin u_{ij}^{k}\;.
\end{equation*}
On the quantum level, however, we will need two non-equivalent solutions of
the fusing rule in order to solve the bootstrap equation (\ref{boot}).

Having stated the fusing rule, one might ask how to relate the fusing angles
defined in (\ref{fuseangle}) to it. Exploiting that the eigenvalues of the
Cartan matrix and the Coxeter element are related to each other by (\ref
{eigenAsig}) it was derived in \cite{FO} that the components of the
eigenvectors (\ref{ADEeigen}) satisfy the relations 
\begin{equation}
\sum_{l=i,j,k}e^{i\frac{\pi s}{h}\eta _{l}}Q_{l}^{(s)}=0\;,\quad \eta
_{l}=-2\xi _{l}+\frac{c_{l}-1}{2}  \label{ADEcharge}
\end{equation}
We will see in a subsequent sections how to derive this formula as a special
case from the $q$-deformed Cartan matrices. Complex conjugation in (\ref
{ADEcharge}) gives the relation w.r.t. to the second solution of the fusing
rule, $\eta _{l}\rightarrow -\eta _{l}=-2\xi _{l}^{\prime }+\frac{c_{l}-1}{2}
$. Relation (\ref{ADEcharge}) can be interpreted geometrically as fixing a
triangle in the euclidean plane upon identifying the complex numbers as
euclidean vectors whose sum is zero. Specializing to $s=1$ the length of the
sides of the triangle are just proportional to the masses and the triangle
is the one depicted in Figure 3.3. We then find from the above equation with
the help of (\ref{fuseangle}) that the fusing angles are given by 
\begin{equation}
u_{ij}^{k}=\frac{\pi }{h}\left( \eta _{j}-\eta _{i}\right)  \label{ADEangles}
\end{equation}
Note that there are two possible triangles corresponding to the two
non-equivalent solutions of the fusing rule. At the same time it can be seen
on geometrical grounds that these are the only two possible solutions \cite
{FO}. Analogously one might construct similar triangles for the higher spin
conserved charges $Q_{i}^{(s)}$ in (\ref{ADEeigen})\ which shows that they
are preserved in the fusing processes determined by (\ref{ADEfuse}).

\subsection{Renormalization properties}

Having stated the mass spectrum and the fusing processes of classical ATFT
in a universal form for all models associated with untwisted algebras, we
are now prepared to turn to the quantization of them. This usually requires
renormalization theory and in the following two paragraphs a short and
qualitative summary of the renormalization properties of ATFT is presented.
For details the reader is referred to the original literature.

In general it is to be expected that each of the classical masses will
renormalize differently w.r.t. the coupling constant $\beta $. However, in
case of ATFT the following important distinction can be made. Those theories
which belong to simply-laced algebras, i.e. the $ADE$ series, were shown to
renormalize up to one loop order in a uniform manner: The quantum masses are
the same as the classical ones up to an overall ''re''-normalization factor 
\cite{TodaS}, 
\begin{equation}
\frac{\delta m_{i}}{m_{i}}=\frac{\beta ^{2}}{2h}\cot \frac{\pi }{h}\,\,.
\label{ADEmass}
\end{equation}
That this factor can be expressed by the above universal formula was derived
by Braden et al. This simple renormalization behaviour has several important
consequences. We immediately infer from (\ref{ADEmass}) that the classical
mass ratios are preserved in the quantum theory and therefore also the
classical fusing angles (\ref{ADEangles}) hold true on the quantum level. In
view of the analytic structure of the two-particle S-matrix this fixes the
bound state poles and one might start to derive generic Lie algebraic
expressions for the scattering amplitude using (\ref{ADEfuse}) and (\ref
{ADEangles}) as it was done in \cite{FO}.

In sharp contrast to these properties the mass spectra of the $BCFG$ series
of ATFT models renormalize in a totally different way. By use of
perturbation theory it was realized that for these kind of theories the
classical mass ratios and fusing angles are not preserved in the quantum
theory. First trial S-matrices with poles at the classical values were shown
to incorporate extra singularities which could not be backed up by
perturbation theory. In particular, it was shown to low orders in the
perturbation expansion that the simple relation (\ref{ADEmass}) ceases in
general to be valid and that the renormalized mass ratios flow between
different classical values dependent on the coupling constant \cite{Gust,G2}%
. This lead to the suggestion \cite{Do1} that the quantum field theories are
dominated by different classical ATFT in the weak and strong coupling regime
and one might have to consider pairs of Lie algebras $(\frak{g,g}^{\vee })$
-- as introduced in Chapter 2, Table 2.2 -- to formulate a consistent
quantum field theory. To be more explicit consider the following example.
Let $(G_{2}^{(1)},D_{4}^{(3)})$ be the pair of algebras then the classical
mass ratios of the associated ATFT (\ref{ADEaction}) can be derived to 
\begin{equation}
\left( \frac{m_{1}}{m_{2}}\right) _{G_{2}^{(1)}}=\frac{\sin \frac{\pi }{6}}{%
\sin \frac{\pi }{3}}\;\quad \text{and\quad\ }\left( \frac{m_{1}}{m_{2}}%
\right) _{D_{4}^{(3)}}=\frac{\sin \frac{\pi }{12}}{\sin \frac{\pi }{6}}
\label{exmass}
\end{equation}
Based on the above assumption the corresponding quantum mass ratios are
expected to asymptotically approximate these different classical values in
the weak ($\beta \rightarrow 0$) and in the strong ($\beta \rightarrow
\infty $) coupling regime, respectively. A transformation from small
coupling values to large ones should therefore amount to an exchange of the
dual Lie algebras and the associated classical ATFT. In formulas this can be
summarized in a loose sense by the following equivalence of Olive-Montonen 
\cite{OM} and Langlands duality, 
\begin{equation}
\beta \rightarrow 4\pi /\beta \quad \simeq \quad \alpha \rightarrow \alpha
^{\vee }=\frac{2\,\alpha }{\left\langle \alpha ,\alpha \right\rangle }\;.
\label{duality}
\end{equation}
This off course only describes the qualitative picture and one now has to
make the quantitative relation of the mass flow explicit. Note that this
also effects the fusing angles and the statement of the fusing rules in the
corresponding theories. In fact, these look quite different in the dual
classical theories. In the above example $(G_{2}^{(1)},D_{4}^{(3)})$ the
non-vanishing three point couplings can be derived to be 
\begin{equation}
G_{2}^{(1)}:\quad C_{111},C_{112},C_{222}\quad \text{and\quad }%
D_{4}^{(3)}:\quad C_{111},C_{112},C_{222},C_{221}\;.  \label{exfuse}
\end{equation}
Therefore, one needs also a criterion to select those fusing processes which
survive quantization and are consistent with the mass flow. Below it is
demonstrated how this can be achieved in a generic and universal manner by
means of $q$-deformation.

\begin{center}
\includegraphics[width=6cm,height=12cm,angle=-90]{renorm.epsi}

\medskip {\small Figure 3.5: Schematic description of the renormalization
flow in affine Toda models.\bigskip }
\end{center}

We conclude by pointing out that as far as this qualitative renormalization
picture is concerned also the simply-laced algebras fit into the scheme
expressed by (\ref{duality}). In the $ADE$ case an interchange of roots and
coroots does not alter the Lie algebra, i.e. both dual algebras coincide, $%
\frak{g}=\frak{g}^{\vee }$. According to (\ref{duality}) this amounts to a
self-duality in the coupling constant and the coupling dependent mass flow
is then just the trivial one, namely the mass ratios stay constant at their
classical values.

\subsection{Quantum fusing rules}

Looking at (\ref{exfuse}) giving the classical fusing rules of two dual
algebras one immediately infers that the classical fusing rule (\ref{ADEfuse}%
) has to be modified on the quantum level in order to accommodate the
renormalization flow of ATFT. The first solution which comes to mind is to
take the intersection of the non-vanishing three-point couplings in both
dual theories upon identifying the particles in a suitable manner as
explained in 2.4.1. In fact, this remedy to the problem of obviously
different fusing structures of the dual partners was first suggested by
Chari and Pressley \cite{CP}.\medskip

\noindent \textbf{Fusing rule in }$\Omega ,\hat{\Omega}$. \emph{The }quantum%
\emph{\ three-point coupling does not vanish if and only if there exist
three representatives in the orbits }$\Omega _{i},\Omega _{j},\Omega _{k}$%
\emph{\ of the Coxeter element (\ref{Cox}) of the Lie algebra }$\frak{g}$%
\emph{\ and three representatives in the orbits }$\hat{\Omega}_{i},\hat{%
\Omega}_{j},\hat{\Omega}_{k}$\emph{\ of the twisted Coxeter element (\ref
{tCox}) of }$\frak{g}^{\vee }$\emph{\ which separately sum up to zero.}%
\medskip

In formulas this means that in contrast to the classical case (\ref{ADEfuse}%
) there exist now \emph{two} triples of integers $(\xi _{i},\xi _{j},\xi
_{k})$ and $(\hat{\xi}_{i},\hat{\xi}_{j},\hat{\xi}_{k})$ such that 
\begin{equation}
\tsum\limits_{l=i,j,k}\sigma ^{\xi _{l}}\,\gamma _{l}=0\,\quad \text{and}%
\quad \tsum\limits_{l=i,j,k}\hat{\sigma}^{\hat{\xi}_{l}}\,\hat{\gamma}%
_{l}^{\omega }=0\,\,  \label{CPfuse}
\end{equation}
holds. Vice versa the existence of such triples implies that the quantum
particles labelled by $i,j,k\ $couple to each other. Like before there
always exists a second solution which is non-equivalent in the sense that it
cannot be obtained by a simple shift. Namely, there are integers 
\begin{equation}
\xi _{l}^{\prime }=-\xi _{l}+\frac{c_{l}-1}{2}\quad \text{and\quad }\hat{\xi}%
_{l}^{\prime }=-\hat{\xi}_{l}+\frac{1-c_{l}}{2}\ell _{l}+c_{l}+1,\quad
l=i,j,k.  \label{neCP}
\end{equation}
such that (\ref{CPfuse}) holds with the replacement $\xi _{l}\rightarrow \xi
_{l}^{\prime }$ and $\hat{\xi}_{l}\rightarrow \hat{\xi}_{l}^{\prime }$. The
argument why this is the only non-equivalent solution is similar to the one
in the classical case and will be given below. The identification of the
particles in the dual theories follows the prescription of Chapter 2, i.e.
we identify the orbits of the first $n$ roots in the twisted algebra with
the roots of the untwisted algebra without relabelling. Note that there is a
slight change in this fusing rule to the one stated in \cite{CP} due to
differently chosen conventions for the twisted Coxeter element (\ref{tCox})
as explained in Section 2.3.2.

Even though (\ref{CPfuse}) correctly displays the fusing processes for the
quantum ATFT it has the disadvantage to treat the dual algebras separately.
In order to describe the coupling dependent mass flow and to derive a
universal expression of the S-matrix it will become apparent below that it
is essential to combine the information of both algebras in one setting. It
was Oota who first observed that this might be done by means of $q$%
-deformation \cite{Oota}.

Using the definitions of Chapter 2 we state now two other fusing rules in
terms of ordinary and twisted $q$-deformed Coxeter elements. Afterwards the
precise relation among them and to (\ref{CPfuse}) will be derived and all
three fusing rules will turn out to be equivalent. Let us start with the
non-twisted algebra $\frak{g}$.\medskip

\noindent \textbf{Fusing rule in }$\Omega ^{q}$. \emph{The }quantum\emph{\
three-point coupling does not vanish if and only if there exist three
representatives in the orbits }$\Omega _{i}^{q},\Omega _{j}^{q},\Omega
_{k}^{q}$\emph{\ of the }$q$\emph{-deformed Coxeter element (\ref{qCox}) of
the Lie algebra} $\frak{g}$ \emph{which sum up to zero.}\medskip

In concrete terms this fusing rule is expressed as follows. The existence of
the mentioned three representatives implies for the $q$-deformed case that
there are two triples of integers $(\xi _{i},\xi _{j},\xi _{k})$, $(\zeta
_{i},\zeta _{j},\zeta _{k})$ such that 
\begin{equation}
\tsum\limits_{l=i,j,k}q^{\zeta _{l}}\,\sigma _{q}^{\xi _{l}}\,\gamma
_{l}=0\quad .  \label{Ootafuse1}
\end{equation}
As we will see below the powers of the deformation parameter $q$ now
incorporate the information of the dual algebra. Again, we find equivalent
solutions to (\ref{Ootafuse1}) by acting with $q^{y}\sigma _{q}^{x}$ from
the right on the above equation. A second \emph{non}-equivalent solution,
however, is obtained when replacing 
\begin{equation}
\zeta _{l}\rightarrow \zeta _{l}^{\prime }=-\zeta _{l}-(1+c_{l})t_{l}\quad 
\text{and\quad }\xi _{l}\rightarrow \xi _{l}^{\prime }=-\xi _{l}+\frac{%
c_{l}-1}{2},\quad l=i,j,k\;.  \label{O1}
\end{equation}
This second solution can be constructed directly from (\ref{Ootafuse1}), see 
\cite{FKS2}, and in addition it is the only one. This will be proven below
when rewriting the fusing structure in terms of $q$-deformed matrices.

The third and last fusing rule to be stated in terms of Coxeter geometry
involves the data of the twisted or dual algebra $\frak{g}^{\vee }$
only.\medskip

\noindent \textbf{Fusing rule in }$\hat{\Omega}^{q}$. \emph{The }quantum%
\emph{\ three-point coupling does not vanish if and only if there exist
three representatives in the orbits }$\hat{\Omega}_{i}^{q},\hat{\Omega}%
_{j}^{q},\hat{\Omega}_{k}^{q}$\emph{\ of the }$q$\emph{-deformed twisted
Coxeter element (\ref{qtCox}) of the Lie algebra} $\frak{g}^{\vee }$ \emph{%
which sum up to zero.\medskip }

This fusing rule implies that there are again two triples of integers $(\hat{%
\xi}_{i},\hat{\xi}_{j},\hat{\xi}_{k})$, $(\hat{\zeta}_{i},\hat{\zeta}_{j},%
\hat{\zeta}_{k})$ such that the following equation holds, 
\begin{equation}
\tsum\limits_{l=i,j,k}\,q^{\hat{\zeta}_{l}}\hat{\sigma}_{q}^{\hat{\xi}_{l}}\,%
\hat{\gamma}_{l}^{\omega }=0\;.  \label{Ootafuse2}
\end{equation}
Analogously to the cases considered before this fusing rule is complemented
by a second solution when replacing 
\begin{equation}
\hat{\zeta}_{l}\rightarrow \hat{\zeta}_{l}^{\prime }=-\hat{\zeta}%
_{l}+1-c_{l}\quad \text{and\quad }\hat{\xi}_{l}\rightarrow \hat{\xi}%
_{l}^{\prime }=-\hat{\xi}_{l}+\frac{1-c_{l}}{2}\ell _{l}+c_{l}+1\;,\quad
l=i,j,k\;.  \label{O2}
\end{equation}
This second solution can be derived from the given one (\ref{Ootafuse2}) and
as in the untwisted case it is unique, whence the statement of only one is
sufficient as existence criterion.

Having stated three versions of fusing rules we need to clarify whether they
are equivalent and how they are related to each other in order to obtain a
consistent picture. One conclusion can be drawn immediately. From the
definition of the $q$-deformed Coxeter and twisted Coxeter element we see
that in the ''classical'' limit $q\rightarrow 1$ one recovers from the two $%
q $-deformed versions (\ref{Ootafuse1}) and (\ref{Ootafuse2}) the
non-deformed fusing rule (\ref{CPfuse}). This is a first hint that the three
versions are in some sense compatible to each other. It remains, however, to
give the precise relation between the integers appearing in the different
versions and to prove that one of them is sufficient to imply all the other.
This discussion is postponed after the next subsection in which the
connection between the fusing rules and the conserved quantities, in
particular the masses and their fusing angles, will be worked out.

\subsection{Quantum mass spectrum and conserved charges}

For the moment let us assume as a working hypothesis that the three variants
(\ref{CPfuse}), (\ref{Ootafuse1}) and (\ref{Ootafuse2}) of the fusing rule
are equivalent. Then the next step towards the construction of the S-matrix
is the determination of the mass spectrum and the corresponding fusing
angles. As the classical relations hold true for the $ADE$ subclass of ATFT
one expects to find close analogues for the eigenvalue equations (\ref
{Perron}) and more generally (\ref{ADEeigen}). After comparing the classical
fusing rule with the $q$-deformed ones a possible conjecture which comes to
mind is to replace the ordinary Cartan matrix by the $q$-deformed one (\ref
{qA}) or equivalently (\ref{qfA}) defined in Chapter 2. However, the
relation turns out to be slightly different and in particular one has to
specify the coupling dependence of the masses first by a specific choice of
the deformation parameters $q,\hat{q}$, which hitherto have been kept
completely generic.\medskip

\noindent \textbf{Quantum charges}. \emph{The quantum analogues to the
charges (\ref{ADEcharge}) preserved by the fusing processes of the theory
are given by the following null vector of the }$q$\emph{-deformed Cartan
matrix as specified in (\ref{qA}), } 
\begin{equation}
\sum_{j=1}^{n}A_{ij}(q=e^{i\pi s\vartheta _{h}},\hat{q}=e^{i\pi s\vartheta
_{H}})Q_{j}^{(s)}=0\,\,.  \label{null}
\end{equation}
\emph{Here the deformation parameters are chosen in terms of the angles } 
\begin{equation}
\vartheta _{h}:=\frac{2-B}{2h}\text{\quad \quad }\vartheta _{H}:=\frac{B}{2H}
\label{Bangle}
\end{equation}
\emph{which determine the coupling dependence of the charges via the
function } 
\begin{equation}
B(\beta )=\frac{2\beta ^{2}}{4\pi h/h^{\vee }+\beta ^{2}}\;.  \label{effB}
\end{equation}
\emph{As in the classical case the spin }$s$\emph{\ runs over the exponents
of the Lie algebra} $\frak{g}$.\medskip

Admittedly, this definition of the quantum conserved charges and their
coupling dependence appears to be a bit \emph{ad hoc}. The particular
parametrization of the \textbf{effective coupling constant} $B$ was first
suggested in \cite{Donons} and will allow to express the coupling dependence
of the fusing angles in linear terms. Note that its range lies in the
interval $0\leq B\leq 2$ where the boundary values correspond to the weak
and strong coupling limit, respectively. Under a duality transformation in
the coupling constant $\beta \rightarrow 4\pi /\beta $ \emph{and} a
simultaneous exchange of the Coxeter numbers one has $B\rightarrow 2-B$. In
this sense it reflects the renormalization behaviour (\ref{duality}).

The null vector equation (\ref{null}) can be motivated by the following
arguments. According to the renormalization behaviour (\ref{ADEmass}) we
ought to recover the classical relation (\ref{Perron}) for the $ADE$ series
of ATFT. If the algebras are simply-laced then $\frak{g}=\frak{g}^{\vee }$
and all the roots have same length implying the integers (\ref{symm}) to be
all one. Since also the entries of the incidence matrix $I=2-A$ are zero or
one the $q$-deformed Cartan matrix (\ref{qA}) reduces to 
\begin{equation}
ADE:\quad A(q,\hat{q})=q\hat{q}+q^{-1}\hat{q}^{-1}-I
\end{equation}
which upon inserting in (\ref{null}) and noting that $h=h^{\vee }$ reduces
to the classical relation (\ref{Perron1}) which can be transformed into the
eigenvalue equation (\ref{Perron}). In the generic case, however, a genuine
eigenvalue equation cannot be regained from (\ref{null}). In particular for $%
s=1$ giving the quantum masses it leads to 
\begin{equation}
\sum_{j=1}^{n}[I_{ij}]_{\hat{q}}\,m_{j}=2\cos \pi \left( \vartheta
_{h}+t_{i}\vartheta _{H}\right) \,m_{i}\;,\quad \hat{q}=e^{i\pi s\vartheta
_{H}}\;.  \label{ATFTmass}
\end{equation}
The scalar factor corresponding to the eigenvalue in the classical equation (%
\ref{Perron1}) now depends via the symmetrizer $t_{i}$ on the particle index
spoiling the eigenvalue property. Nevertheless, (\ref{ATFTmass}) exhibits
nicely the Lie algebraic structure present and provides us with a universal
formula of the quantum mass spectrum for \emph{all} models of ATFT.
Moreover, in the limit $\beta \rightarrow 0$ the classical equation is
recovered once more, which demonstrates compatibility with the
renormalization properties outlined in Subsection 3.2.2.

Despite these compatibility checks which leave little doubt about the
correctness of the above assertion a generic Lie algebraic proof is still
missing and requires more profound insight in the structure of the
associated quantum field theory. On a case-by-case study, however, it has
been established in \cite{Oota} and \cite{FKS2}, see also the appendix.

\subsection{The quantum fusing angles}

After having stated how to obtain the charges preserved by a fusing process
we need to check it for consistency with the three fusing rules (\ref
{Ootafuse1}), (\ref{Ootafuse2}) and (\ref{CPfuse}). This will give
additional support to the mass spectrum formula (\ref{ATFTmass}) and provide
us not only with the fusing angles needed for the bootstrap equation but
also yield an intrinsic proof of the equivalence of all three versions of
the fusing rule \cite{FKS2}. From the discussion of the classical regime we
infer that an equation analogous to (\ref{ADEcharge}) for the quantum
conserved charges is required in order to determine the fusing triangle
(Figure 3.3) and, thus, the fusing angles.\medskip

\noindent \textbf{Fusing angles}. \emph{The quantities defined through the
null vector (\ref{null}) are conserved in the fusing processes specified by
the rules (\ref{Ootafuse1}), (\ref{Ootafuse2}) and (\ref{CPfuse}). That is,
they satisfy the equation } 
\begin{equation}
\tsum\limits_{l=i,j,k}e^{i\pi s(\eta _{l}\vartheta _{h}+\hat{\eta}%
_{l}\vartheta _{H})}\,Q_{l}^{(s)}=0\;  \label{ATFTcharge}
\end{equation}
\emph{whenever the particles labelled by }$i,j,k$\emph{\ couple to each
other. In particular, the integers }$\eta _{l},\hat{\eta}_{l}$\emph{\
determine the (quantum) fusing angles to be } 
\begin{equation}
u_{ij}^{k}=(\eta _{j}-\eta _{i})\pi \vartheta _{h}+(\hat{\eta}_{j}-\hat{\eta}%
_{i})\pi \vartheta _{H}  \label{ATFTangles}
\end{equation}
\emph{The angle coefficients }$\eta _{l},\hat{\eta}_{l}$\emph{\ are related
to the integers }$\zeta _{l},\xi _{l}$\emph{\ and }$\hat{\zeta}_{l},\hat{\xi}%
_{l}$\emph{\ defined in (\ref{Ootafuse1}) and (\ref{Ootafuse2}) via the
equations } 
\begin{equation}
\eta _{l}=-2\xi _{l}+\frac{c_{l}-1}{2}\quad \quad \hat{\eta}_{l}=\zeta _{l}+%
\frac{1+c_{l}}{2}\,t_{l}\;,  \label{1eta}
\end{equation}
\begin{equation}
\eta _{l}=\hat{\zeta}_{l}+\frac{c_{l}-1}{2}\quad \quad \hat{\eta}_{l}=-2\hat{%
\xi}_{l}+\frac{1-c_{l}}{2}\,\ell _{l}+c_{l}+1\;,  \label{2eta}
\end{equation}
\emph{for }$l=i,j,k$\emph{. Moreover, the identity (\ref{ATFTcharge}) is
equivalent to the fusing rules (\ref{Ootafuse1}) and (\ref{Ootafuse2}%
).\medskip }

The proof of these assertion is given momentarily. First notice that the
fusing angles (\ref{ATFTangles}) are now coupling dependent via the defining
relations (\ref{Bangle}) and (\ref{effB}). Moreover, while the first summand 
$(\eta _{j}-\eta _{i})$ determines the \emph{classical} fusing angle of the
non-twisted algebra via (\ref{ADEangles}), the second $(\hat{\eta}_{j}-\hat{%
\eta}_{i})$ yields the classical value for the dual, twisted algebra. This
is in accordance with the renormalization picture outlined in 3.2.2.
Specializing to $s=1$ we deduce from (\ref{ATFTcharge}) the following
``floating'' mass ratios 
\begin{equation}
\frac{m_{i}}{m_{j}}=\frac{\sin \left[ (\eta _{k}-\eta _{j})\pi \vartheta
_{h}+(\hat{\eta}_{k}-\hat{\eta}_{j})\pi \vartheta _{H}\right] }{\sin \left[
(\eta _{i}-\eta _{k})\pi \vartheta _{h}+(\hat{\eta}_{i}-\hat{\eta}_{k})\pi
\vartheta _{H}\right] }\;.  \label{mratio}
\end{equation}
This especially implies that the fusing triangle in Figure 3.3 which is
``static'' in the classical case now starts to vary its shape when the
coupling constant $\beta $ is tuned between the weak and strong coupling
regime. It then interpolates between the classical values at $\beta
\rightarrow 0$ and $\beta \rightarrow \infty $. More generally we have the
relation 
\begin{equation}
Q_{i}^{(s)}/Q_{j}^{(s)}=\frac{\sin \left[ (\eta _{k}-\eta _{j})\pi
s\vartheta _{h}+(\hat{\eta}_{k}-\hat{\eta}_{j})\pi s\vartheta _{H}\right] }{%
\sin \left[ (\eta _{i}-\eta _{k})\pi s\vartheta _{h}+(\hat{\eta}_{i}-\hat{%
\eta}_{k})\pi s\vartheta _{H}\right] }\;.  \label{Qratio}
\end{equation}
Both relations can be interpreted in the complex velocity plane as explained
in the $ADE$ case \cite{FO}. Together with the last two equations all the
structures presented in the context of classical ATFT have been
``translated'' to the quantum level yielding all the necessary ingredients
for the bootstrap construction of the two-particle S-matrix. However, before
constructing the latter, we need to prove the above assertions (\ref
{ATFTcharge}),(\ref{1eta}),(\ref{2eta}) and to verify the equivalence
between the fusing rules. To do this we exploit the matrix structure present
in the theory.

\subsection{The relation between the fusing rules}

We start by identifying the matrix elements of the $M$-matrix defined in (%
\ref{M0}) as the quantum conserved charges (\ref{null}). In the determining
equation (\ref{M1}) for the $M$-matrix it is immediate to see that 
\begin{equation*}
A(q,\hat{q})M(q,\hat{q})=0
\end{equation*}
whenever $q^{2h}\hat{q}^{2H}=1$. The last property is easily verified for
the special choice of the deformation parameters in (\ref{null}). Moreover,
from the discussion of Chapter 2, and the particular form of the determinant
(\ref{detA}) we infer that for $q=e^{i\pi s\vartheta _{h}},\hat{q}=e^{i\pi
s\vartheta _{H}}$ the matrix $M$ is only non-zero when $s$ is an exponent of
the Lie algebra $\frak{g}$. Recalling in addition that $M$ is symmetric we
are led to the conclusion 
\begin{equation}
M_{ij}(e^{i\pi s\vartheta _{h}},e^{i\pi s\vartheta _{H}})\propto
Q_{i}^{(s)}Q_{j}^{(s)}  \label{Mcharge}
\end{equation}
by comparison with (\ref{null}). Note that the proportionality factor in (%
\ref{Mcharge}) does not depend on the particle indices $i$ or $j$. The
expected fusing equation (\ref{ATFTcharge}) in terms of conserved quantities
can therefore be translated into a fusing equation of the matrix $M$, 
\begin{equation}
\tsum\limits_{l=i,j,k}q^{\eta _{l}}\hat{q}^{\hat{\eta}_{l}}M_{ml}(q,\hat{q}%
)=0\;,\quad q=e^{i\pi s\vartheta _{h}},\hat{q}=e^{i\pi s\vartheta _{H}}\;,
\label{Mfuse}
\end{equation}
for all indices $1\leq m\leq n$. The crucial step is now to relate (\ref
{Mfuse}) to one of the ''quantum'' fusing rules, say (\ref{Ootafuse1}), and
to derive the explicit expressions for $\eta _{l},\hat{\eta}_{l}$ in terms
of the integers $\zeta _{l},\xi _{l}$.

Let us assume that the fusing rule (\ref{Ootafuse1}) in terms of the $q$%
-deformed Coxeter element holds for some integers $(\zeta _{i},\zeta
_{j},\zeta _{k})$ and $(\xi _{i},\xi _{j},\xi _{k})$. Since $\hat{q}%
\,^{-2H}\sigma _{\hat{q}}^{h}=1$, we may assume $0<\xi _{l}<h$ without loss
of generality. Acting with $[t_{m}]_{\hat{q}}/2\,\,\hat{q}^{\frac{1-c_{m}}{2}%
\,t_{m}}q^{2x-\frac{c_{m}}{2}-1}\sigma _{\hat{q}}^{x}$ on the above equation
and summing over $x$ afterwards yields, 
\begin{equation*}
0=-\frac{[t_{m}]_{\hat{q}}}{2}\sum_{l=i,j,k}\hat{q}^{\zeta _{l}+\frac{1-c_{m}%
}{2}\,t_{m}}\sum_{x=1}^{h}\,\left\langle \lambda _{m}^{\vee },\sigma _{\hat{q%
}}{}^{x+\xi _{l}}\,\gamma _{l}\right\rangle \,q^{2x-\frac{c_{m}}{2}-1}.
\end{equation*}
Now splitting up the sum into two parts, 
\begin{multline*}
\sum_{x=1}^{h}\,\left\langle \lambda _{m}^{\vee },\sigma _{\hat{q}}{}^{x+\xi
_{l}}\,\gamma _{l}\right\rangle \,q^{2x-\frac{c_{m}}{2}-1}=q^{-2\xi
_{l}}\sum_{x=\xi _{l}+1}^{h+\frac{c_{l}-1}{2}}\,\left\langle \lambda
_{m}^{\vee },\sigma _{\hat{q}}{}^{x}\,\gamma _{l}\right\rangle \,q^{2x-\frac{%
c_{m}}{2}-1} \\
+q^{-2\xi _{l}}\sum_{x=h+\frac{c_{l}+1}{2}}^{h+\xi _{l}}\left\langle \lambda
_{m}^{\vee },\sigma _{\hat{q}}{}^{x}\,\gamma _{l}\right\rangle \,q^{2x-\frac{%
c_{m}}{2}-1}
\end{multline*}
and remembering that $q^{2h}\hat{q}^{2H}=1$ we have by means of equation (%
\ref{M0}) that 
\begin{equation*}
\sum_{l=i,j,k}q^{-2\xi _{l}+\frac{c_{l}-1}{2}}\,\hat{q}^{\zeta _{l}+\frac{%
1+c_{l}}{2}\,t_{l}}\,M_{lm}(q,\hat{q})=0,
\end{equation*}
which is the desired reformulation of the fusing rule with $\eta _{l}=-2\xi
_{l}+\frac{c_{l}-1}{2}$ and $\hat{\eta}_{l}=\zeta _{l}+\frac{1+c_{l}}{2}%
\,t_{l}$. To prove the complementary assertion we assume that (\ref{Mfuse})
holds and define $\eta _{l}=:-2\xi _{l}+\frac{c_{l}-1}{2}$. Then by use of (%
\ref{M0}) we see that 
\begin{equation*}
0=\sum_{l=i,j,k}\hat{q}^{\bar{\eta}_{l}}\sum_{n=1}^{2h}M_{lm}(\tau ^{n},\hat{%
q})\tau ^{n(\eta _{l}+\frac{c_{m}+1}{2})}=-\frac{[t_{m}]_{\hat{q}}}{2}\hat{q}%
^{\frac{1-c_{m}}{2}\,t_{m}}\sum_{l=i,j,k}\hat{q}^{\hat{\eta}_{l}-\frac{%
1+c_{l}}{2}\,t_{l}}\left\langle \lambda _{m}^{\vee },\sigma _{\hat{q}%
}{}^{\xi _{l}}\,\gamma _{l}\right\rangle
\end{equation*}
for all $1\leq m\leq n$. But since the fundamental co-weights form a basis,
this implies the fusing rule (\ref{Ootafuse1}) with $\eta _{l}=-2\xi _{l}+%
\frac{c_{l}-1}{2}$ and $\hat{\eta}_{l}=\zeta _{l}+\frac{1+c_{l}}{2}\,t_{l}$.

Completely, analogous one proves the equivalence between (\ref{Ootafuse2})
and (\ref{Mfuse}) for the twisted algebra by means of the $N$-matrix and its
definition in terms of the $q$-deformed twisted Coxeter element (\ref{N0})
and the identity (\ref{N=M}). Note that the identification of the $N$ and
the $M$-matrix is crucial in this step. In particular, one has as an
immediate consequence also the equivalence of the two fusing rules (\ref
{Ootafuse1}) and (\ref{Ootafuse2}) in terms of the $q$-deformed Coxeter
elements \cite{FKS2}. Having established the equivalence between the latter
fusing rules the one in terms of the non-deformed Coxeter elements now also
follows in the classical limit $q\rightarrow 1$ as we already have seen. A
Lie algebraic proof how to derive the $q$-deformed versions from the
non-deformed ones is outstanding, but it can be verified case-by-case. This
shows that all three different formulations are consistent and that any of
them implies all the others. In particular, we also proved that the
definition of the quantum charges (\ref{ATFTcharge}) is consistent with the
fusing rules \cite{FKS2}, which makes their definition as null-vector of the 
$q$-deformed Cartan matrix very suggestive. To close the picture we
summarize the relations between the various integers used in the different
equations. \medskip

\noindent \textbf{Summary}. \emph{The fusing rules involving the }$q$\emph{%
-deformed Coxeter and twisted Coxeter element are linked to each other by 
\cite{FKS2} } 
\begin{equation}
-2\xi _{l}=\hat{\zeta}_{l},\quad \text{and\quad }\zeta _{l}=-2\hat{\xi}_{l}+%
\frac{1-c_{l}}{2}\ell _{l}-\frac{1+c_{l}}{2}t_{l}+c_{l}+1\;,\quad l=i,j,k\;.
\label{fus}
\end{equation}
$\,$\emph{Notice that as in the classical case the non-equivalent solutions (%
\ref{O1}) and (\ref{O2}) are reflected in (\ref{ATFTcharge}) by complex
conjugation, i.e. }$\eta _{l}\rightarrow -\eta _{l},\,\hat{\eta}%
_{l}\rightarrow -\hat{\eta}_{l}$\emph{\ when either }$\xi _{l}\rightarrow
\xi _{l}^{\prime },\,\zeta _{l}\rightarrow \zeta _{l}^{\prime }$\emph{\ or }$%
\hat{\xi}_{l}\rightarrow \hat{\xi}_{l}^{\prime },\,\hat{\zeta}%
_{l}\rightarrow \hat{\zeta}^{\prime }$\emph{. In particular, this proves the
existence of the non-equivalent solutions. In terms of them the relations
between the powers appearing in the fusing rules can be simplified \cite
{FKS2}, } 
\begin{equation}
\eta _{l}=\xi _{l}^{\prime }-\xi _{l}=\frac{\hat{\zeta}_{l}-\hat{\zeta}%
_{l}^{\prime }}{2}\quad \text{and\quad }\hat{\eta}_{l}=\frac{\zeta
_{l}-\zeta _{l}^{\prime }}{2}=\hat{\xi}_{l}^{\prime }-\hat{\xi}_{l}\;,\quad
l=i,j,k\;.  \label{3eta}
\end{equation}
Furthermore, interpreting (\ref{ATFTcharge}) geometrically we conclude by
the same argumentation as in the classical or $ADE$\ case \cite{FO} that
only two non-equivalent solutions corresponding to the two fusing triangles
shown in Figure 3.3 exist.

\subsection{The S-matrix in blocks of meromorphic functions}

The universal expressions for the fusing rules, the mass spectrum and the
fusing angles allow to write down a generic formula for the ATFT S-matrix
instead of constructing the scattering amplitudes for each model separately,

\begin{equation}
S_{ij}(\theta )=\prod\limits_{x=1}^{2h}\prod\limits_{y=1}^{2H}\left\{
x,y\right\} _{\theta }^{\mu
_{ij}(x,y)}=\prod\limits_{x=1}^{h}\prod\limits_{y=1}^{H}\left\{ x,y\right\}
_{\theta }^{2\mu _{ij}(x,y)}\;.  \label{blockS}
\end{equation}
Here $\left\{ x,y\right\} _{\theta }$ denotes a ratio of hyperbolic
functions similar to (\ref{gensol}) mentioned in the context of the
functional equations and which will be specified momentarily. The power
function $\mu =\mu (x,y)$ takes its values in the half integers $\frac{1}{2}%
\mathbb{Z}$ and is positive for the range of arguments stated in (\ref
{blockS}). In the course of the argumentation both writings of the product
will be used and their equivalence derived. Moreover, it will be
demonstrated that $\mu $ can be expressed in universal Lie algebraic terms
and the structure developed in the preceding paragraphs will then serve to
prove the correct bootstrap properties of (\ref{blockS}).

\subsubsection{Blocks of meromorphic functions}

We already have seen that the analytic requirements on the two-particle
scattering amplitude together with unitarity (\ref{uni}) restricted severely
the form of solutions to the functional equations. For the discussion of the
ATFT matrix it was first observed by Dorey \cite{Donons} that the
combination of hyperbolic functions into the following building blocks is
most suitable, 
\begin{equation}
\left\{ x,y\right\} _{\theta }:=\frac{\left[ x,y\right] _{\theta }}{\left[
x,y\right] _{-\theta }}\quad \lbrack x,y]_{\theta }:=\frac{\left\langle
x-1,y-1\right\rangle _{\theta }\left\langle x+1,y+1\right\rangle _{\theta }}{%
\left\langle x-1,y+1\right\rangle _{\theta }\left\langle
x+1,y-1\right\rangle _{\theta }}\quad  \label{block}
\end{equation}
and 
\begin{equation}
\left\langle x,y\right\rangle _{\theta }:=\sinh \tfrac{1}{2}\left( \theta
+x\theta _{h}+y\theta _{H}\right) \,.
\end{equation}
This is of the general form (\ref{gensol}) discussed in 3.1.2. The shifts
depending on the integer entries $x,y$ are defined in terms of the angles (%
\ref{Bangle}) introduced in context of the fusing rule and the conserved
charges, 
\begin{equation*}
\theta _{h}:=i\pi \vartheta _{h}\quad \quad \text{and\quad }\quad \theta
_{H}:=i\pi \vartheta _{H}\;.
\end{equation*}
It should be emphasized that this definition of the building blocks as
innocent as it might look at first sight is essential in displaying the Lie
algebraic structure. Indeed, as it will turn out below the integer entries
will be related to the powers of the $q$-deformed Coxeter elements and the
deformation parameters. Already from the defining relation involving the
Coxeter numbers $h,H$ we can deduce that the first entry $x$ will be related
to the structure of the untwisted algebra $\frak{g}$ while the second is
connected to the twisted one $\frak{g}^{\vee }$. Moreover, the block
structure together with the effective coupling constant $B$ defined in (\ref
{effB}) nicely incorporates the renormalization properties: Upon a
strong-weak duality transformation in the classical coupling constant, $%
\beta \rightarrow 4\pi /\beta $, and a simultaneous exchange of the
algebras, $\frak{g}\leftrightarrow \frak{g}^{\vee }$, the effective coupling
transforms as $B\rightarrow 2-B$ and the S-matrix stays invariant.

For later purposes it is important to note that the blocks (\ref{block}) can
alternatively be expressed in terms of Fourier integrals of the form 
\begin{gather}
\left\{ x,y\right\} _{\theta }=\exp \dint\limits_{0}^{\infty }\frac{dt}{%
t\sinh t}\,\,f_{x,y}^{h,H}(t,B)\sinh \left( \frac{\theta t}{i\pi }\right) \;,
\label{iblock} \\
f_{x,y}^{h,H}(t,B):=8\sinh t\vartheta _{h}\sinh t\vartheta _{H}\sinh
t(1-x\vartheta _{h}-y\vartheta _{H})\;.  \notag
\end{gather}
$\,$However, for proving the bootstrap properties of the S-matrix (\ref
{blockS}) the block form in terms of hyperbolic functions is more
convenient. Especially, easy to derive are the subsequent functional
relations which will become important in due course, 
\begin{eqnarray}
\left\{ x,y\right\} _{\theta } &=&\left\{ x+2h,y+2H\right\} _{\theta
}=\left\{ -x,-y\right\} _{\theta }^{-1}\quad  \label{shif1} \\
\left\{ x,y\right\} _{\theta +x^{\prime }\theta _{h}+y^{\prime }\theta _{H}}
&=&\frac{\left[ x+x^{\prime },y+y^{\prime }\right] _{\theta }}{\left[
x-x^{\prime },y-y^{\prime }\right] _{-\theta }}  \label{shif2} \\
\left\{ x,y\right\} _{\theta +p\theta _{h}+q\theta _{H}}\left\{ x,y\right\}
_{\theta -p\theta _{h}-q\theta _{H}} &=&\left\{ x+p,y+q\right\} _{\theta
}\left\{ x-p,y-q\right\} _{\theta }\,\,.  \label{shif}
\end{eqnarray}
The equivalent integral expression requires some comments about convergence
when one analytically continues into the complex plane w.r.t. the rapidity.
Performing the shift $\theta \rightarrow \theta +x^{\prime }\theta
_{h}+y^{\prime }\theta _{H}$ convergence is maintained if 
\begin{equation*}
0\leq (x-x^{\prime }-1)\vartheta _{h}+(y-y^{\prime }-1)\vartheta _{H}\leq
2(1-(1+x^{\prime })\vartheta _{h}-(1+y^{\prime })\vartheta _{H})\,\,.\,
\end{equation*}
An additional aspect which deserves careful attention is the
non-commutativity of certain limits when $\theta $ tends to zero. While we
infer that in general $\left\{ x,y\right\} _{\theta =0}=1$, we take the
convention to set $\left\{ 1,1\right\} _{\theta =0}=-1$ meaning that one
first should set $x=y=1$ and then take the limit $\theta \rightarrow 0$.
Similarly, the integral representation (\ref{iblock}) requires also to set $%
x=y=1$ first, to integrate thereafter and finally to take the limit $\theta
\rightarrow 0$.

Having specified the building blocks and their analytic properties the
second step in the derivation of the universal expression (\ref{blockS}) is
the definition of the power function. There are three equivalent ways to
define $\mu (x,y)$ and all of them have different advantages for displaying
the various Lie algebraic structures connected to the pair of dual algebras $%
(\frak{g},\frak{g}^{\vee })$.

\subsubsection{The power function in terms of $q$-deformed matrices}

In regard to the building blocks just discussed we start with the ``matrix
representation'' of the power function $\mu $ since this is the most
convenient one to show how the analytic properties (\ref{shif1}) of a single
block are reflected in the powers of the S-matrix (\ref{blockS}). In Chapter
2 it was argued that the $M$-matrix (\ref{M0}) consists of a polynomial in
the deformation parameters $q,\hat{q}$ and the power function $\mu (x,y)$
was implicitly defined in 2.4.3 as the coefficient of the monomials $q^{x}%
\hat{q}^{y}$ in the expansion (\ref{Mpoly}). Formally, the latter equation
might be inverted by discrete Fourier transformation, 
\begin{equation}
\mu (x,y)=\frac{1}{2h}\sum_{r=1}^{2h}\frac{1}{2H}\sum_{s=1}^{2H}M(\tau ,\hat{%
\tau})\tau ^{rx}\hat{\tau}^{sy}\;,  \label{muM}
\end{equation}
where $\tau ,\hat{\tau}$ are roots of unity of order $2h$ and $2H$,
respectively. This equality just states that up to a factor one half the
power of the block $\{x,y\}$ appearing in (\ref{blockS}) equals the number
of times the monomial $q^{x}\hat{q}^{y}$ occurs in the expansion of $M(q,%
\hat{q})$. The factor $1/2$ originates in the defining relation (\ref{M0}).
By the restriction of the deformation parameters to roots of unity, $q=\tau ,%
\hat{q}=\hat{\tau}$, in (\ref{muM}) and the matrix identity (\ref{M2}) it is
now straightforward to verify \cite{FKS2} the following properties of the
power function , 
\begin{equation}
\mu (x,y)=\mu (x+2h,y+2H)=-\mu (2h-x,2H-y)=\mu (x,y)^{t}  \label{mu}
\end{equation}
The first property is obvious since $\tau ^{2h}=\hat{\tau}^{2H}=1$, the
second follows from (\ref{Mmero}) and the last one just reflects that the $M$%
-matrix is symmetric due to (\ref{qD}), i.e. $M(q,\hat{q})=M(q,\hat{q})^{t}$%
. Recall that the manipulations of the $M$-matrix at roots of unity in terms
of the identity (\ref{M2}) require some care. As discussed in Chapter 2 the
determinant of the $q$-deformed Cartan matrix (\ref{detA}) has to be
canceled against the prefactor in (\ref{M2}) first and then one might safely
set $q=\tau ,\hat{q}=\hat{\tau}$. Clearly, the first two properties (\ref{mu}%
) are analogues of the block behaviour (\ref{shif1}). In particular we infer
from (\ref{mu}) that the expansion of the $M$-matrix as polynomial has to be
of the form 
\begin{equation}
M(q,\hat{q})=\sum_{x=1}^{h}\sum_{y=1}^{H}\mu (x,y)\left( q^{x}\hat{q}%
^{y}-q^{2h-x}\hat{q}^{2H-y}\right)  \label{polyM}
\end{equation}
from which the second equality in (\ref{blockS}) is deduced. The latter
rewriting is important, because it ensures that the building blocks $\{x,y\}$
appear with integral powers in the S-matrix. This in turn guarantees that $%
S_{ij}(\theta )$ is a meromorphic function of the rapidity meeting the
general requirement (S3).

There are two more important identities needed for the discussion of the
bootstrap properties of the S-matrix (\ref{blockS}). The first one is
connected to crossing symmetry and is a direct consequence of the definition
of the anti-particle (\ref{anti}), 
\begin{equation}
\mu _{j\bar{\imath}}(x,y)=\mu _{ij}(h-x,H-y)\;.  \label{muCross}
\end{equation}
The second one is linked to a fusing process $i+j\rightarrow \bar{k}$ and
can be derived from the matrix fusing rule (\ref{Mfuse}), 
\begin{equation}
\sum\limits_{l=i,j,k}\mu _{ml}\left( x\pm \eta _{l},y\pm \hat{\eta}%
_{l}\right) =0\,.  \label{muBoot}
\end{equation}
An additional identity for the power function can be deduced from the
determining equation for the $M$-matrix (\ref{M1}) which played a central
role in the discussion of the fusing rules and the conserved quantities, 
\begin{equation}
\mu _{ij}(x+1,y+t_{i})+\mu
_{ij}(x-1,y-t_{i})=\sum\limits_{n=1}^{I_{il}}\sum\limits_{l\in \Delta }\mu
_{lj}(x,y+2n-1-I_{il})\,\,  \label{idd}
\end{equation}
where it is understood that the sum gives zero when $I_{il}=0$. Its
derivation is immediate when inverting (\ref{M1}) by discrete Fourier
transformation. The identity (\ref{idd}) for the power function was first
mentioned in \cite{Oota} where it was used as recursive relation to generate
the powers in (\ref{blockS}). When discussing the corresponding identities
for the S-matrix below we will see that (\ref{idd}) should, however,
regarded as a consequence of the bootstrap equations (\ref{muBoot}) which
turn out to be more fundamental \cite{FKS2}.\medskip

\noindent \textbf{Remark}. \emph{Notice that instead of the }$M$\emph{%
-matrix one might also use the structure of the twisted algebra }$\frak{g}%
^{\vee }$\emph{\ and define everything in terms of the }$N$\emph{-matrix
together with the identity (\ref{N2}). But since both matrices were shown to
be equivalent in (\ref{N=M}) the same structure emerges \cite{FKS2}.\medskip 
}

Furthermore, it should be emphasized that the matrix representation of the
power function requires only a minimum of Lie algebraic data and is
certainly most convenient in deriving the relations (\ref{mu}) needed for
verifying the bootstrap equations of the S-matrix in the next subsection.
Despite these obvious advantages, however, it should kept in mind that the
matrix identities (\ref{M2}) and (\ref{N2}) were derived by means of Coxeter
geometry. Also the expansion of the $M$-matrix or $N$-matrix in a polynomial
-- necessary for determining $\mu $ -- might turn out to be rather
complicated for higher rank algebras, since it involves the inversion of the 
$q$-deformed Cartan matrix. Another point is the dual structure of the two
Lie algebra $(\frak{g},\frak{g}^{\vee })$ which remains quite hidden in this
formalism. Therefore, it is advisable to present the power function also in
an alternative manner by use of Coxeter geometry.

\subsubsection{The power function in terms of $q$-deformed Coxeter elements}

In the preceding paragraphs the logic of the $q$-deformation as introduced
in Chapter 2 was in a certain sense ''reversed''. Everything was expressed
in terms of the most sophisticated structure derived in the end, the $q$%
-deformed Cartan matrices. We now return to Coxeter geometry and instead of
calculating the power function by first extracting polynomials from the $M$
or $N$-matrix we directly read it off from the orbit structure of $\sigma
_{q}$ and $\hat{\sigma}_{q}$ as defined in (\ref{qCox}) and (\ref{qtCox}),
respectively.

For the untwisted algebra $X_{n}$ the powers in the representation (\ref
{blockS}) can be defined in terms of the $q$-deformed Coxeter element as the
generating function 
\begin{equation}
\sum\limits_{y}\mu _{ij}\left( 2x-\frac{c_{i}+c_{j}}{2},y\right) q^{y}=-%
\frac{[t_{j}]_{q}}{2}q^{\frac{(1-c_{j})t_{j}-(1+c_{i})t_{i}}{2}%
}\,\left\langle \lambda _{j}^{\vee },\sigma _{q}^{x}\gamma _{i}\right\rangle
\,\,,  \label{muCox}
\end{equation}
for fixed integer $x$. Taking $x$ in the range $(3-c_{i})/2\leq x\leq
h+(1-c_{i})/2$ ensures that the first argument of $\mu $ is between $1$ and $%
2h$. This definition of the power function makes it especially evident that
the definition of the S-matrix (\ref{blockS}) is a clear generalization of
the one of the simply-laced case found in \cite{FO}. The $q$-deformation
reflects the presence of the second dual algebra, since the powers of the
deformation parameter determine the second entries for which $\mu (x,y)$ is
non-vanishing. The properties (\ref{mu}) might now be derived directly from (%
\ref{quasi}), (\ref{sm}) and (\ref{spari}) in Section 2.4.4. Also the
crossing (\ref{muCross}) and bootstrap property (\ref{muBoot}) can be
verified by means of (\ref{anti}) and (\ref{Ootafuse1}), respectively \cite
{FKS2}.

Alternatively, one might use the twisted algebra and the associated $q$%
-deformed twisted Coxeter element to define $\mu (x,y)$ as the generating
function 
\begin{equation}
\sum\limits_{x}\mu _{ij}\left( x,2y-c_{i}+\frac{c_{i}-1}{2}\ell _{i}-\frac{%
c_{j}-1}{2}\ell _{j}\right) q^{x}=-\frac{q^{-\frac{c_{i}+c_{j}}{2}}}{2}%
\sum_{k=1}^{\ell _{j}}\,\left\langle \lambda _{\omega ^{k}(j)},\hat{\sigma}%
_{q}^{y}\hat{\gamma}_{i}^{\omega }\right\rangle \,\,.  \label{mutCox}
\end{equation}
This time the second entry $y$ is kept fixed while $x$ varies and now the
structure of the untwisted algebra is incorporated in the deformation
parameter. Using the identities (\ref{quasi2}), (\ref{mm}), (\ref{paritt})
together with the definition of the anti-particle (\ref{antit}) and the
fusing rule (\ref{Ootafuse2}) for the twisted algebra one again verifies all
the desired properties of $\mu $ \cite{FKS2}.\medskip

\noindent \textbf{Remark}. \emph{The three definitions (\ref{muM}), (\ref
{muCox}) and (\ref{mutCox}) of the power function are equivalent by means of
the identities (\ref{M0}), (\ref{N0}) and (\ref{N=M}) proved in Chapter 2 
\cite{FKS2}. This means in particular that the representation (\ref{blockS})
of the ATFT S-matrix might be computed in three different ways, via the
orbits of the }$q$\emph{-deformed Coxeter element, the orbits of the }$q$%
\emph{-deformed twisted Coxeter element and the }$q$\emph{-deformed Cartan
matrix upon inversion.\medskip }

Which of the mentioned methods is preferable in order to construct
explicitly the powers of the building blocks might depend on the particular
example at hand. In the appendix all three of them are demonstrated for
numerous examples. The next step is to establish that the scattering
amplitudes (\ref{blockS}) satisfy the bootstrap functional relations proving
them to be consistent solutions.

\subsection{The bootstrap properties}

Since ATFT is assumed to be invariant under parity transformation we expect
that the two-particle scattering amplitude is symmetric in the particle
indices. Indeed, $S_{ij}(\theta )=S_{ji}(\theta )$ follows for our choice (%
\ref{blockS}) from the last equality in (\ref{mu}). This in particular
implies that Hermitian analyticity is replaced by the stronger requirement
of real analyticity and the unitarity-analyticity equation (\ref{uni})
reduces to 
\begin{equation*}
S_{ij}(\theta )S_{ij}(-\theta )=1\;.
\end{equation*}
That the latter functional equation is fulfilled is immediate to verify from
the property $\left\{ x,y\right\} _{\theta }\left\{ x,y\right\} _{-\theta
}=1 $ of each individual building block in (\ref{blockS}).

\subsubsection{Crossing symmetry}

In contrast, the crossing relation $S_{ij}(\theta )=S_{\bar{\imath}j}(i\pi
-\theta )$ requires in general a little bit more effort, 
\begin{eqnarray*}
S_{ij}(i\pi -\theta ) &=&\prod_{x=1}^{2h}\prod_{y=1}^{2H}\{h-x,H-y\}_{\theta
}^{\mu _{ij}(x,y)}=\prod_{x=1}^{2h}\prod_{y=1}^{2H}\{x+h,y+H\}_{\theta
}^{-\mu _{ij}(x,y)} \\
&=&\prod_{x=1}^{h}\prod_{y=1}^{H}\{x+2h,y+2H\}_{\theta }^{-\mu
_{ij}(x+h,y+H)}\prod_{x=h+1}^{2h}\prod_{y=H+1}^{2H}\{x,y\}_{\theta }^{-\mu
_{ij}(x-h,y-H)} \\
&=&\prod_{x=1}^{2h}\prod_{y=1}^{2H}\{x,y\}_{\theta }^{\mu
_{ij}(h-x,H-y)}=S_{j\bar{\imath}}(\theta )\;.
\end{eqnarray*}
Here we have used first the shifting property (\ref{shif1}) of the building
blocks and second the corresponding relation for the power function (\ref{mu}%
) as well as the definition of the anti-particle (\ref{muCross}).

\subsubsection{Bootstrap equation}

For our purposes it is convenient to use the second variant (\ref{boot1}) of
the bootstrap equations which in terms of the fusing angles (\ref{ATFTangles}%
) reads 
\begin{equation}
\prod\limits_{l=i,j,k}S_{ml}(\theta +\eta _{l}\theta _{h}+\hat{\eta}%
_{l}\theta _{H})=1\,.  \label{ATFTboot}
\end{equation}
This might be verified by similar shifting techniques in the building blocks
as in the case of crossing symmetry. In view of the property (\ref{shif2})
it turns out to be necessary to separate the building blocks in two
sub-blocks of hyperbolic functions, 
\begin{equation*}
S_{ml}(\theta +\eta _{l}\theta _{h}+\hat{\eta}_{l}\theta
_{H})=\prod_{x=1}^{2h}\prod_{y=1}^{2H}\left( \frac{[x+\eta _{l},y+\hat{\eta}%
_{l}]_{\theta }}{[x-\eta _{l},y-\hat{\eta}_{l}]_{-\theta }}\right) ^{\mu
_{ml}(x,y)}\;.
\end{equation*}
Now numerator and denominator may be treated separately. One of which
requires the equation (\ref{muBoot}) for the upper and the other for the
lower sign showing that both non-equivalent solutions to the fusing rules
are needed as has been mentioned earlier. Proceeding similar like in the
case of crossing symmetry one ends up with the expression 
\begin{equation*}
S_{ml}(\theta +\eta _{l}\theta _{h}+\hat{\eta}_{l}\theta
_{H})=\prod_{x=1}^{2h}\prod_{y=1}^{2H}\frac{[x,y]_{\theta }^{\mu
_{ml}(x-\eta _{l},y-\hat{\eta}_{l})}}{[x,y]_{-\theta }^{\mu _{ml}(x+\eta
_{l},y+\hat{\eta}_{l})}}
\end{equation*}
which upon using (\ref{muBoot}) gives the desired bootstrap identity (\ref
{ATFTboot}). Thus, the claim of the universal ATFT S-matrix (\ref{blockS})
is proven to be consistent with the formulated fusing rules of Subsection
3.2.3 \cite{FKS2}.

\subsubsection{Combined bootstrap identities}

Besides the bootstrap identities describing the consistency between
factorization of the S-matrix and the fusing structure, there are additional
product identities special for the affine Toda models. In the course of the
presented argumentation equation (\ref{M0}) turned out to be the backbone in
analyzing the Lie algebraic structure, since it provides the link between $q$%
-deformed Coxeter geometry and $q$-deformed Cartan matrices. It is therefore
remarkable that the same equation translated in terms of the power function (%
\ref{idd}) has also a direct interpretation in terms of S-matrices. Namely,
using the shifting property (\ref{shif}) of the building blocks one obtains 
\cite{FKS2} 
\begin{equation}
S_{ij}\left( \theta +\theta _{h}+t_{i}\theta _{H}\right) S_{ij}\left( \theta
-\theta _{h}-t_{i}\theta _{H}\right)
=\tprod\limits_{l=1}^{n}\tprod\limits_{k=1}^{I_{il}}S_{lj}\left( \theta
+(2k-1-I_{il})\theta _{H}\right) \;,  \label{cb}
\end{equation}
where it is understood that the product contributes $1$ if $I_{il}=0$. In
context of simply-laced algebras such identities were first obtained in \cite
{Rava} but by different methods. Here they follow from a purely geometrical
context (\ref{M0}) and it was first noted in \cite{FKS2} that they can be
built up from the elementary bootstrap equations (\ref{ATFTboot}), whence
they are referred to as ``combined'' bootstrap equations. Note that only
direct Lie algebraic quantities like the length of the roots and the
incidence matrix enter the above equation. Moreover, in contrast to the $ADE$
case the r.h.s. in (\ref{cb}) involves multiple products of S-matrices for
non simply-laced algebras. Explicit examples are worked out in the appendix.

\subsubsection{Singularities and the generalized bootstrap criterion}

From the bootstrap equation (\ref{ATFTboot}) for the ATFT S-matrix we infer
that the bound state poles depend on the coupling via the fusing angles (\ref
{ATFTangles}) which is in accordance with the renormalization properties of
the mass spectrum. On general grounds \cite{Eden} connected to the analytic
structure of the amplitude $S_{ij}(\theta )$ the bound state poles ought to
be of odd order or even simple and to move in the physical sheet when $0\leq
\beta \leq \infty $ or $0\leq B\leq 2$ is varied, 
\begin{equation}
\phi =\pm (\eta _{i}-\eta _{j})\theta _{h}\pm (\hat{\eta}_{i}-\hat{\eta}%
_{j})\theta _{H}\,\,.
\end{equation}
The two signs correspond to the two non-equivalent solutions of the fusing
rules. Furthermore, at a bound state pole the S-matrix is expected to have
up to a factor $\sqrt{-1}$ a residue of definite sign, see the general
discussion in Section 3.1.2 While in the case of simply-laced Lie algebras
all simple poles in the physical sheet could be shown to have this kind of
behaviour and could also be traced to the fusing processes described by (\ref
{ADEfuse}), once more matters turn out to be more involved in the non
simply-laced case. Here additional simple or odd order poles appear in the
physical sheet which cannot consistently be interpreted in terms of bound
states as was shown in perturbation theory. Corrigan et al. noted that these
poles can be characterized by the property that their residue $%
R_{ij}^{k}(\beta )$ eventually changes its sign when $\beta $ sweeps through
the allowed range and suggested the following ``generalized'' bootstrap
principle \cite{nons}:\medskip

\noindent \emph{Only odd order poles which have residue of definite sign
over the whole range of the coupling constant participate in the bootstrap,
while those of varying sign together with even order poles are excluded.}%
\medskip

It are exactly the poles singled out by the above prescription, which are
described by the fusing rules (\ref{CPfuse}), (\ref{Ootafuse1}), and (\ref
{Ootafuse2}). To see this directly for the constructed S-matrices in the
appendix it is convenient to derive a simple criterion in terms of the
building blocks (\ref{block}) in (\ref{blockS}) to decide whether or not an
odd order pole has the physical interpretation of a fusing process.

For this purpose the building blocks (\ref{block}) are not well suited since
numerous cancellations of zeroes and poles take place when there occur
specific combinations of them. This motivates the definition of the combined
block 
\begin{eqnarray}
\left\{ x,y_{n}\right\} _{\theta } &=&\prod\limits_{l=0}^{n-1}\left\{
x,y+2l\right\} _{\theta }  \label{comblock} \\
&=&\frac{\left\langle x-1,y-1\right\rangle _{\theta }\left\langle
x+1,y-1+2n\right\rangle _{\theta }}{\left\langle x+1,y-1\right\rangle
_{\theta }\left\langle x-1,y-1+2n\right\rangle _{\theta }}\times (\theta
\rightarrow -\theta )^{-1}\,\,,  \notag
\end{eqnarray}
together with the angles 
\begin{equation}
\theta _{x,y,n}^{\pm }=(x\pm 1)\theta _{h}+(2n+y-1)\theta _{H}\;.
\label{poles}
\end{equation}
One sees easily that there are four zeros of $\{x,y_{n}\}_{\theta }$ located
at $\pm \theta _{x,y,0}^{\pm },\mp \theta _{x,y,n}^{\pm }$ and four simple
poles at $\pm \theta _{x,y,n}^{\pm },\pm \theta _{x,y,0}^{\mp },$
respectively. Blocks of the above type can be seen to occur gradually in the
S-matrix and the obvious advantage of the definition (\ref{comblock}) is
that it takes the cancellation of several hyperbolic functions in the
product into account. Note that we recover the basic building blocks (\ref
{block}) for $n=1$.

Since only the poles in the physical strip are of interest, one needs a
first criterion to restrict the imaginary part of the angles (\ref{poles})
to the range $0\leq \func{Im}\theta \leq \pi $. From the second equation in (%
\ref{blockS}) we infer that the block entries are limited to the values $%
0<x<h$, $0<y<H$. In addition, the effective coupling lies in the interval $%
0\leq B\leq 2$ whence one derives the condition 
\begin{equation}
0\leq \func{Im}(\theta _{x,y,n}^{\pm })\leq \pi \qquad \text{for }B\leq 
\frac{2H(h-x\mp 1)}{\left| h(2n+y-1)-H(x\pm 1)\right| }\,.  \label{64}
\end{equation}
In order to check the generalized bootstrap principle one needs to evaluate
the corresponding residues \cite{FKS2}, 
\begin{eqnarray}
&& 
\begin{array}{c}
\limfunc{Res}\limits_{\theta =\theta _{x,y,0}^{-}}\left\{ x,y_{n}\right\} =\,%
\frac{-2\sinh \theta _{h}\sinh (n\theta _{H})\sinh \left( x\theta
_{h}+(n+y-1)\theta _{H}\right) \sinh \left( \theta _{x,y,0}^{-}\right) }{%
\sinh \left( \theta _{h}+n\theta _{H}\right) \sinh \left( x\theta
_{h}+(y-1)\theta _{H}\right) \sinh \left( (x-1)\theta _{h}+(y+n-1)\theta
_{H}\right) }
\end{array}
\\
&& 
\begin{array}{c}
\limfunc{Res}\limits_{\theta =\theta _{x,y,n}^{+}}\!\left\{ x,y_{n}\right\} =%
\frac{2\sinh \theta _{h}\sinh (n\theta _{H})\sinh \left( x\theta
_{h}+(n-1+y)\theta _{H}\right) \sinh \left( \theta _{x,y,n}^{+}\right) }{%
\sinh \left( \theta _{h}+n\theta _{H}\right) \sinh \left( (1+x)\theta
_{h}+(n+y-1)\theta _{H}\right) \sinh \left( x\theta _{h}+(2n+y-1)\theta
_{H}\right) }\,\,.\qquad
\end{array}
\end{eqnarray}
For the stated range of $x,y,B,n$ together with the restriction (\ref{64})
it is now tedious but straightforward to check that 
\begin{equation}
\func{Im}\left( \limfunc{Res}_{\theta =\theta _{x,y,0}^{-}}\!\left\{
x,y_{n}\right\} _{\theta }\right) <0\quad \quad \text{and \qquad }\func{Im}%
\left( \limfunc{Res}_{\theta =\theta _{x,y,n}^{+}}\!\left\{ x,y_{n}\right\}
_{\theta }\right) >0\,\,,
\end{equation}
indicating that $\theta _{x,y,n}^{+}$ is a possible candidate for a direct
channel pole obeying the generalized bootstrap prescription. However, there
are additional contributions from other blocks which might change their sign
when the coupling is tuned from the weak to the strong limit. (Note that
this is in contrast to the simply-laced case where the above knowledge is
sufficient to determine the sign of the total residue of the whole S-matrix,
e.g. \cite{FO}). Thus, one has also to investigate the behaviour of a second
independent block at the particular pole $\theta _{x,y,n}^{+}$.

Take this block to be $\left\{ x^{\prime },y_{n^{\prime }}^{\prime }\right\}
_{\theta _{x,y,n}^{+}}$ then a clear indication for a sign change are
different signs of the values in the extreme weak and extreme strong
coupling regime. In general, $\lim_{\beta \rightarrow 0,\infty }\left\{
x^{\prime },y_{n^{\prime }}^{\prime }\right\} _{\theta _{x,y,n}^{+}}$ $=1$
with the exception when $x^{\prime }=x$, where one has 
\begin{eqnarray}
\lim_{\beta \rightarrow 0}\left\{ x,y_{n^{\prime }}^{\prime }\right\}
_{\theta _{x,y,n}^{\pm }} &=&\left( \frac{y^{\prime }-y-2n}{y^{\prime
}-y+2n^{\prime }-2n}\right) ^{\pm 1}  \label{res1} \\
\lim_{\beta \rightarrow \infty }\left\{ x,y_{n^{\prime }}^{\prime }\right\}
_{\theta _{x,y,n}^{\pm }} &=&1\,\,.
\end{eqnarray}
Provided that the right hand side of (\ref{res1}) is negative, the imaginary
parts of the possible additional blocks 
\begin{equation}
\left\{ x,y_{n^{\prime }}^{\prime }\right\} _{\theta _{x,y,n}^{+}}\text{%
\qquad and\qquad }\left\{ x+2,y_{n^{\prime \prime }}^{\prime \prime
}\right\} _{\theta _{x,y,n}^{+}}  \label{kii}
\end{equation}
both change their sign while $\beta $ runs from zero to infinity. Therefore
the pole $\theta _{x,y,n}^{+}$ is excluded from the bootstrap, whenever the
scattering matrix contains in addition the blocks (\ref{kii}) to an odd
power and these do not cross the real axis w.r.t. $\beta $ at the same
point. Given the S-matrix explicitly in block form (\ref{blockS}) the
condition on $y,y^{\prime },n,n^{\prime }$ by which the l.h.s. of (\ref{res1}%
) becomes negative, together with the occurrence of blocks like (\ref{kii})
yields a simple criterion \cite{FKS2}\ which proves to be sufficient for
practical purposes as can be checked at the examples listed in the appendix.
Nevertheless, a direct Lie algebraic formula is desirable.

\subsection{The S-matrix in integral form}

After having established that the two-particle scattering amplitude
possesses all the desired bootstrap properties one might look for
alternative expressions of it. Although the block form (\ref{blockS}) is
most suitable for exhibiting the bootstrap properties in a simple manner as
just demonstrated, it is convenient to rewrite the S-matrix in a different
variant in regard of several applications, e.g. the thermodynamic Bethe
ansatz (see Chapter 4) or the form factor program \cite{FF}. Explicitly, we
are looking for an expression of the type 
\begin{equation}
S_{ij}(\theta )=\,\exp \int\limits_{0}^{\infty }\frac{dt}{t}\,\,\phi
_{ij}(t)\sinh \left( \frac{\theta t}{i\pi }\right) \,\,,  \label{SPPi}
\end{equation}
where the matrix valued integral kernel $\phi $ has to be specified. In
fact, as will be proven below for ATFT it is determined by 
\begin{equation}
\phi \left( t\right) =8\sinh (\vartheta _{h}t)\sinh (t_{j}\vartheta
_{H}t)A(e^{t\vartheta _{h}},e^{t\vartheta _{H}})^{-1}\,\,\,.  \label{Oota}
\end{equation}
Notice that only the inverse $q$-deformed Cartan matrix (\ref{qA}) and the
length of the roots (\ref{symm}) enter the expression. The deformation
parameters have been chosen to be $q=\exp (t\vartheta _{h})$ and $\hat{q}%
=\exp (t\vartheta _{H})$, where the angles $\vartheta _{h},\vartheta _{H}$
are the same as defined in (\ref{Bangle}). That the S-matrix can be cast
into this neat and universal form with a minimum of Lie algebraic
information was first noticed in \cite{Oota} on a case-by-case basis. We now
turn to the proof of this formula \cite{FKS2} which will be an immediate
consequence of the Lie algebraic structure developed in Chapter 2 and the
preceding sections.

Rewriting each of the building blocks occurring in (\ref{blockS}) in the
integral form (\ref{iblock}) the product in (\ref{blockS}) is transformed
into a sum in the exponent. Then the natural question arises whether this
sum may be performed to give a more compact expression. Specifying the
deformation parameters $q,\hat{q}$ as stated above one infers the following
identity by means of (\ref{polyM}), 
\begin{eqnarray}
\phi (t) &=&\frac{1}{\sinh t}\sum_{x=1}^{h}\sum_{y=1}^{H}2\mu
(x,y)\,f_{x,y}^{h,H}(t,B)\;,  \label{phi} \\
&=&-\frac{8\sinh (\vartheta _{h}t)\sinh (\vartheta _{H}t)}{\sinh t}%
e^{-t}M(e^{t\vartheta _{h}},e^{t\vartheta _{H}})  \notag
\end{eqnarray}
upon noting that $q^{2h}\hat{q}^{2H}=e^{2t}$. Applying now the matrix
identity (\ref{M2}) and using some elementary identity for hyperbolic
functions gives the integral representation. Notice that this easy
derivation builds once more on the identification of the link between $q$%
-deformed Coxeter geometry and the $q$-deformed Cartan matrix. In \cite{FKS2}
also an alternative derivation in terms of contour integrals in the complex
rapidity plane can be found. For simply-laced algebras see also \cite{FKS1}.

To conclude it should be emphasized that the $q$-deformed Cartan matrix now
appears clearly to be the central object in the theory, since one can
formulate in terms of it not only the quantum mass spectrum (\ref{ATFTmass})
and the fusing rules (\ref{ATFTcharge}) but also the S-matrix. Each of these
informations can be written down in a single formula for \emph{all} affine
Toda models, whose behaviour differs greatly depending on the algebras in
question. This level of universality is rarely achieved for any other class
of field theories except for conformal WZNW theories, which have similar
underlying Lie algebraic structures. \emph{As was claimed before, such
sophisticated mathematical structures do not only neatly reorganize known
structures in a general setting, but they can also be extrapolated to find
entirely new integrable models as is demonstrated in the next section}.
Before coming to this point we extract from the block form (\ref{blockS})
and the integral formula (\ref{SPPi}) the known universal S-matrix of
simply-laced ATFT \cite{FO} as final consistency check. At the same time
this will prepare the subsequent definition of colour valued S-matrices.

\subsection{Reduction to simply-laced affine Toda}

To make the difference between the two classes of affine Toda models of
simply-laced and non simply-laced Lie algebras also explicit in terms of the
S-matrix, the general formulas (\ref{blockS}) and (\ref{SPPi}) are now
evaluated for the $ADE$ series. In particular, it will turn out that the
simple renormalization properties (\ref{ADEmass}) of the latter ATFT allow
to separate the two-particle scattering amplitude in two factors, one
containing the bound state poles and another one displaying the coupling
dependence. This feature has consequences beyond ATFT, since it might be
used to relate the techniques developed for constructing the ATFT S-matrix
to other integrable models of similar type.

For the following we choose $\frak{g}\equiv ADE,$ which implies as mentioned
before that both dual algebras coincide, $\frak{g}=\frak{g}^{\vee }$, as the
associated root system of $\frak{g}$ is self-dual under the exchange of
roots and coroots. This has a number of consequences. First of all for
simply-laced algebras all the roots are of the same length, whence the
symmetrizer (\ref{symm}) is trivial $t_{i}=1$ and the incidence matrix $%
I=2-A $ has only entries zero or one. Thus, the $q$-deformation of the
simple Weyl reflections (\ref{qweyl}) vanishes and the action of the $q$%
-deformed Coxeter element (\ref{qCox}) simplifies to 
\begin{equation*}
ADE:\quad \sigma _{q}=q^{2}\sigma \;,
\end{equation*}
where $\sigma $ is just the ordinary non-deformed Coxeter transformation (%
\ref{Cox}). To recover the universal formula for the $ADE$ S-matrix \cite{FO}
we express the power function $\mu $ in the variant (\ref{muCox}) and find
for this special case 
\begin{equation}
ADE:\quad \sum\limits_{y}\mu _{ij}\left( 2x-\frac{c_{i}+c_{j}}{2},y\right)
q^{y}=-\frac{1}{2}q^{2x-\frac{c_{i}+c_{j}}{2}}\left\langle \lambda
_{j},\sigma ^{x}\gamma _{i}\right\rangle \;.  \label{possim}
\end{equation}
Hence $y=2x-\frac{c_{i}+c_{j}}{2}$ and the first and second entry of the
building blocks in (\ref{blockS}) are the same, i.e. only blocks of the type 
$\{x,x\}\,$occur\footnote{%
The block $\left\{ x,x\right\} _{\theta }$ corresponds to the block $\left\{
x\right\} _{\theta }$ as defined in \cite{FO} or \cite{FKS1}.}. The block
form (\ref{blockS}) of the scattering matrix then reduces to 
\begin{equation}
ADE:\quad S_{ij}\left( \theta \right) =\prod\limits_{x=1}^{h}\left\{ 2x-%
\frac{c_{i}+c_{j}}{2},2x-\frac{c_{i}+c_{j}}{2}\right\} _{\theta }^{-\frac{1}{%
2}\left\langle \lambda _{j},\sigma ^{x}\gamma _{i}\right\rangle },
\label{sADE}
\end{equation}
which coincides with the expression found by Fring and Olive \cite{FO} upon
noting that for the self-dual case the two relevant Coxeter numbers are the
same, $h=H=\ell h^{\vee }$. In fact, since both dual algebras coincide the
twist of $\frak{g}^{\vee }$ is trivial, $\omega =1,\ell =1,$ meaning that
all orbits have length $\ell _{i}=1$ and expression (\ref{sADE}) follows by
analogous steps.

Reorganizing the building blocks (\ref{block}) in the $ADE$ S-matrix in the
following manner 
\begin{equation}
\{x,x\}=\frac{\left\langle x+1,x+1\right\rangle _{\theta }\left\langle
x-1,x-1\right\rangle _{\theta }}{\left\langle x+1,x+1\right\rangle _{-\theta
}\left\langle x-1,x-1\right\rangle _{-\theta }}\times \frac{\left\langle
x+1,x-1\right\rangle _{-\theta }\left\langle x-1,x+1\right\rangle _{-\theta }%
}{\left\langle x+1,x-1\right\rangle _{\theta }\left\langle
x-1,x+1\right\rangle _{\theta }}  \label{sb}
\end{equation}
one realizes that (\ref{sADE}) can be decomposed into a so-called
''minimal'' and a coupling dependent part \cite{TodaS}, 
\begin{equation}
S_{ij}(\theta )=S_{ij}^{\min }(\theta )F_{ij}(\theta ,B)\;,  \label{split}
\end{equation}
where $S_{ij}^{\min }$ contains only the first and $F_{ij}(\theta ,B)$ the
second factor of the building block as decomposed in (\ref{sb}). Observing
that for simply-laced algebras the definition of the building blocks (\ref
{block}) is simplified by the relations $\vartheta _{h}+\vartheta
_{H}=h^{-1} $ and $\vartheta _{h}-\vartheta _{H}=(1-B)/h$ one immediately
verifies that the coupling dependence in the first factor drops out giving
rise to the factorization (\ref{split}). Moreover, by the same techniques as
before one verifies that each factor obeys the bootstrap equations
separately. However, from the defining relation (\ref{block}) and (\ref{sb})
one immediately infers that the poles of the coupling dependent factor lie
outside of the physical strip. Thus, the bound state poles and therefore the
associated fusing processes are incorporated exclusively in the minimal
part. With regard to the general remarks at the beginning of this chapter, $%
F_{ij}$ is therefore a particular example for a CDD factor, which can be
multiplied to the minimal solution $S_{ij}^{\min }$ without violating the
bootstrap properties.

It needs to be stressed, that this splitting property of the S-matrix is a
characteristic feature of the $ADE$ series and is not shared by the non
simply-laced affine Toda models, as was first pointed out by Delius et al. 
\cite{Gust}.

For the sake of completeness and later purposes we now also discuss the
separation of the $ADE$ S-matrix (\ref{sADE}) in the integral
representation. We already saw in context of the conserved charges that the $%
q$-deformed Cartan matrix simplifies for $ADE$ algebras, $A(q,\hat{q})=q\hat{%
q}+q^{-1}\hat{q}^{-1}-I$ (compare Section 3.2.4). Thus, the universal
integral kernel (\ref{Oota}) specializes to 
\begin{eqnarray}
\phi \left( t\right) &=&8\sinh \tfrac{Bt}{2h}\sinh \tfrac{(2-B)t}{2h}\left(
2\cosh \tfrac{t}{h}-I\right) ^{-1}  \notag \\
&=&4\left( \cosh \tfrac{t}{h}-\cosh \tfrac{t(1-B)}{h}\right) \left( 2\cosh 
\tfrac{t}{h}-I\right) ^{-1}\,\,\,.
\end{eqnarray}
Here the equality of the Coxeter number $h,H$ and the explicit definition of
the angles (\ref{Bangle}) as well as a simple trigonometric identity has
been used. Thus, the integral kernel $\phi $ can be written as the sum of
two terms implying via (\ref{SPPi}) the factorization (\ref{split}).
Explicitly the expressions read 
\begin{equation}
S_{ij}^{\min }(\theta )=\exp \dint\limits_{0}^{\infty }\frac{dt}{t}\,4\cosh 
\frac{t}{h}\left( 2\cosh \frac{t}{h}-I\right) _{ij}^{-1}\sinh \frac{t\theta 
}{i\pi }  \label{iSmin}
\end{equation}
and 
\begin{equation}
F_{ij}(\theta ,B)=\exp -\dint\limits_{0}^{\infty }\frac{dt}{t}\,4\cosh \frac{%
t(1-B)}{h}\left( 2\cosh \frac{t}{h}-I\right) _{ij}^{-1}\sinh \frac{t\theta }{%
i\pi }\;.  \label{iCDD}
\end{equation}
From the integral representation of the CDD factor several properties
concerning the coupling dependence of the $ADE$ S-matrix are immediate to
verify. First of all, in the weak coupling limit one has that 
\begin{equation*}
F_{ij}(\theta ,B=0)=S_{ij}^{\min }(\theta )^{-1}\;.
\end{equation*}
The last property is to be expected on physical grounds, since the particles
should not interact, $S_{ij}=1$, when $\beta \rightarrow 0$. The second
property which can be easily deduced from (\ref{iCDD}) is the strong-weak
self-duality in the coupling constant which is mirrored by the self-duality
in terms of the root system, compare (\ref{duality}). Since the CDD factor $%
F_{ij}$ contains the whole coupling dependence, it ought to obey the
relation 
\begin{equation*}
F_{ij}(\theta ,B)=F_{ij}(\theta ,2-B)\;,\quad B_{ADE}=\frac{2\beta ^{2}}{%
4\pi +\beta ^{2}}
\end{equation*}
implying the invariance of the S-matrix under the mapping $\beta \rightarrow
4\pi /\beta $. Notice that the effective coupling constant $B$ is the one
defined in (\ref{effB}) for the special case of simply-laced algebras and
that the strong-weak self-duality then amounts to the invariance under the
transformation $B\rightarrow 2-B$. This self-duality is obvious from (\ref
{iCDD}) and we conclude that the S-matrix (\ref{sADE}) derived from the
general expression is in accordance with the known behaviour of simply-laced
ATFT.

\section{Colour valued S-matrices}

In this section a general construction principle \cite{FKcol} is proposed to
equip coupling dependent scattering matrices of integrable models with
colour degrees of freedom. This means that for a given mass spectrum of an
integrable model an additional quantum number is assigned to the associated
particles and in this manner the original model is ``multiplied''. At the
heart of this method lies the splitting of the original S-matrix into a
minimal and coupling dependent part as we just encountered in the case of
simply-laced ATFT. It will allow to generate in an easy manner new
S-matrices with underlying Lie algebraic structure and which have the novel
feature to violate parity invariance. To emphasize the usefulness of the
presented construction method the class of generated S-matrices will be
shown for special choices of the colour structure to contain examples
already known in the literature.

To define new integrable models via writing down consistent S-matrices might
at first seem a bit abstract, since in general the associated classical
Lagrangian to these quantum field theories is not known. However, it is in
full accordance with the spirit of the S-matrix theory and the bootstrap
approach which declare the S-matrix as the central and determining object of
each quantum field theory. The latter is supposed to be more fundamental
than any classical description. Nevertheless, valuable insight is gained
when determining the corresponding classical field theories. A first step
towards this direction is postponed to the next chapter, where the new
S-matrices will be linked to perturbed conformal field theories providing
additional motivation for their definition.

\subsection{Construction principle}

Suppose that we are given an integrable model whose two-particle S-matrix
obeys a separation property analogous to the one of simply-laced ATFT (\ref
{split}), i.e. it separates into one minimal factor containing the bound
state poles and one of CDD type, for instance $F_{ij}$ in equation (\ref
{split}), displaying the coupling dependence only, $S_{ij}(\theta
)=S_{ij}^{\min }(\theta )S_{ij}^{\text{CDD}}(\theta ,B)$. Then one possible
way to introduce a ``colour'' dependence in the S-matrix is given by
choosing the coupling to be colour dependent, i.e. one performs the
replacement \cite{FKcol} 
\begin{equation}
S_{ij}(\theta )=S_{ij}^{\min }(\theta )S_{ij}^{\text{CDD}}(\theta
,B)\rightarrow \hat{S}_{ij}^{ab}(\theta )=S_{ij}^{\min }(\theta )S_{ij}^{%
\text{CDD}}(\theta ,B_{ab})\,\,.  \label{cSa}
\end{equation}
Here the indices $i,j$ refer to the (original) particle masses and the
indices $a,b$ to the colours, i.e. each particle carries now two quantum
numbers $(i,a)$. The corresponding ranges may be chosen differently, $1\leq
i\leq n$ and $1\leq a\leq \tilde{n}$, and consequently one obtains a new
mass spectrum of $n\times \tilde{n}$ different particle types. Clearly, the
bootstrap properties of the original S-matrix are not changed by this
prescription, whence the relevant functional equations for the S-matrix are
satisfied. In particular, the original fusing structure for particles of the 
\emph{same} colour is preserved and particles of \emph{different} colours
only interact at different values of the effective coupling. This motivates
to refer to the second type of indices as colour degrees of freedom.
Alternatively, one might take the splitting of the S-matrix more seriously
and even separate both factors to define an S-matrix of the type \cite{FKcol}
\begin{equation}
S_{ij}^{ab}(\theta )=\left\{ 
\begin{array}{ll}
S_{ij}^{\min }(\theta )=(S_{ij}^{\text{CDD}}(\theta ,B_{aa}=0))^{-1} & 
\qquad \qquad \text{for }i=j\medskip \\ 
S_{ij}^{\text{CDD}}(\theta ,B_{ab}) & \qquad \qquad\text{for }i\neq j\,\,\, 
\label{cSb}
\end{array}
\right. \,\,.
\end{equation}
This means whenever $a=b$ we simply have $\tilde{n}$ copies of theories
which interact via a minimal scattering matrix and for $a\neq b$ the
particles interact purely via a CDD-factor. Clearly by construction also (%
\ref{cSb}) satisfies the bootstrap equations. It should be noted here that (%
\ref{cSa}) and (\ref{cSb}) still describe scattering processes for which
backscattering is absent. Hence, these type of colour values play a
different role as those which occur for instance in S-matrices related to
affine Toda field theories \cite{ATFT} with purely imaginary coupling
constant, e.g. \cite{nondiag}. Despite the fact that the relative mass
spectra related to (\ref{cSb}) are degenerate, this is consistent when we
encounter $\tilde{n}$ different overall mass scales dependent on the colour
or the particles have different charges. To provide a concrete realization
for $S_{ij}^{ab}(\theta )$, which is of affine Toda field theory type, the
concrete separation (\ref{split}) for ATFT will now be exploited and main
and colour quantum numbers be linked to a pair of simply-laced Lie algebras.
It is, however, clear from the previous comments that the forms (\ref{cSa})
and (\ref{cSb}) are of a more general nature.

\subsection{The $\frak{g}$\textbf{\TEXTsymbol{\vert}}$\frak{\tilde{g}}$
S-matrix in integral form}

Associate the main quantum numbers $i,j$ as in the context of ATFT to the
vertices of the Dynkin diagram of a simply-laced Lie algebra $\frak{g}$%
\textbf{\ }of rank $n$ and the colour quantum numbers $a,b$ to the vertices
of the Dynkin diagram of a simply-laced Lie algebra $\frak{\tilde{g}}$ of
rank $\tilde{n}$. To each fixed colour quantum number $a$ let there be a
tower of particles whose mass ratios are the same as in the corresponding
ATFT connected with $\frak{g}$, i.e. $m_{i}^{a}/m_{j}^{a}:=\left(
m_{i}/m_{j}\right) _{\text{ATFT,}\frak{g}}$. To define the interaction
between the particles we use now slightly modified version of the second
scheme (\ref{cSb}) outlined above. Let $A$ and $\tilde{A}$ denote the Cartan
matrices associated with $\frak{g}$ and $\frak{\tilde{g}}$, then the
S-matrix in its integral form is chosen to be \cite{FKcol} 
\begin{equation}
S_{ij}^{ab}(\theta )=e^{i\pi \varepsilon _{ab}A_{ij}^{-1}}\exp
\int\limits_{0}^{\infty }\frac{dt}{t}\,2\left( 2\cosh \frac{t}{h}-\tilde{I}%
\right) _{ab}\left( 2\cosh \frac{t}{h}-I\right) _{ij}^{-1}\sinh \frac{%
t\theta }{i\pi }\;.  \label{iSnew}
\end{equation}
Here $I=2-A,\tilde{I}=2-\tilde{A}$ are the corresponding incidence matrices, 
$h$ is the Coxeter number of $\frak{g}$ and $\varepsilon _{ab}$ is the
antisymmetric tensor, i.e. $\varepsilon _{ab}=-\varepsilon _{ba}$. Notice
that due to the appearance of the latter the above expressions break parity
invariance, $S_{ij}^{ab}(\theta )\neq S_{ji}^{ba}(\theta )$. Some comments
are due to clarify how this definition fits into the general prescription (%
\ref{cSb}). Choosing the colour quantum numbers to be equal we obtain in
accordance with (\ref{cSb}) the minimal ATFT S-matrix (\ref{iSmin})
associated with the first Lie algebra $\frak{g},$ since the diagonal
elements in the incidence matrix $\tilde{I}=2-\tilde{A}$ are zero, 
\begin{equation*}
S_{ij}^{aa}(\theta )=S_{ij}^{\min }(\theta )\;.
\end{equation*}
If on the other hand the colours are different, there are now two cases to
distinguish. Whenever the vertices $a$ and $b$ are not linked on the $\frak{%
\tilde{g}}$--Dynkin diagram the S-matrix becomes trivial, i.e. $%
S_{ij}^{ab}=1,$ and the particles do not interact. In contrast, when $a$ and 
$b$ are linked on the $\frak{\tilde{g}}$\textbf{--}Dynkin diagram, we have $%
\tilde{I}_{ab}=1$ from which in comparison with (\ref{iCDD}) it follows that 
\begin{equation*}
\tilde{I}_{ab}=1:\quad S_{ij}^{ab}(\theta )=e^{i\pi \varepsilon
_{ab}A_{ij}^{-1}}F_{ij}(\theta ,B=1)^{\frac{1}{2}}\;.
\end{equation*}
Analogously to (\ref{cSb}) the interaction between particles of different
colours is thus defined via the special CDD factor (\ref{iCDD}) with the
slight change that the square root at effective coupling $B=1$ is taken
first. This, however, changes the bootstrap properties and explains the
occurrence of the ominous looking phase factor as will be demonstrated in
the next subsection when discussing (\ref{iSnew}) in its block form of
meromorphic functions.


\noindent At the moment the neat Lie algebraic structure of (\ref{iSnew})
shall be sufficient as motivation for its definition. In the next chapter we
will then see in retrospect that this particular combination of Lie
algebraic structures can be traced back to WZNW models in the high energy
limit.

Henceforth, the quantum field theories described by (\ref{iSnew}) are
referred to as $\frak{g}$\textbf{\TEXTsymbol{\vert}}$\frak{\tilde{g}}$ models%
\footnote{%
The notation should of course not be understood as a coset.} \cite{FKcol}.
It should be clear from the discussion that this pairing of Lie algebras in
the present context is conceptually not related to the pairing of Lie
algebras encountered in the previous section. Furthermore, these new
theories only involve simply-laced algebras.

\begin{center}
\includegraphics[width=8.5cm,height=8.5cm,angle=-90]{colour.epsi}
\end{center}

\noindent {\small Figure 3.6: The structure of }$\frak{g}|\frak{\tilde{g}}$%
{\small -theories: To each simple root of }$\alpha _{a}${\small \ of }$\frak{%
\tilde{g}}${\small \ there is an tower of }$n=\,${\small rank\thinspace }$%
\frak{g}${\small \thinspace\ particles whose mass ratios are identical to
those of ATFT. Particles inside the same tower scatter via the minimal ATFT
S-matrix associated with }$\frak{g}${\small , while particles in different
towers interact via a CDD-factor.} \bigskip

\subsection{The $\frak{g}$\textbf{\TEXTsymbol{\vert}}$\frak{\tilde{g}}$
S-matrix in block form}

In order to trace down the origin of the phase factor and for analyzing the
pole structure it is convenient to rewrite (\ref{iSnew}) in blocks of
meromorphic functions. Looking at the block expression for the CDD factor (%
\ref{iCDD}) at $B=1$ one observes that $F_{ij}(\theta ,B=1)^{\frac{1}{2}}$
is built up in terms of meromorphic functions of the type 
\begin{equation}
F_{ij}(\theta ,B=1)^{\frac{1}{2}}=\prod_{x}\sqrt{\frac{\left\langle
x+1,x-1\right\rangle _{-\theta }\left\langle x-1,x+1\right\rangle _{-\theta }%
}{\left\langle x+1,x-1\right\rangle _{\theta }\left\langle
x-1,x+1\right\rangle _{\theta }}}=\prod_{x}\frac{\sinh \frac{1}{2}(\theta -i%
\frac{\pi }{h}x)}{\sinh \frac{1}{2}(\theta +i\frac{\pi }{h}x)}
\label{CDDnew}
\end{equation}
from which one immediately deduces that taking the square root does not
affect the desired analytic properties and minimizes the power of the poles
in the unphysical sheet. This shows the operation of taking the square root
in a more natural light. However, the behaviour of the CDD factor under a
crossing transformation is changed by the possible appearance of a minus
sign. This motivates the following definition of the new building blocks 
\cite{FKcol} 
\begin{equation}
\left[ x,B\right] _{\theta ,ab}=e^{\frac{i\pi x}{h}\varepsilon _{ab}}\left( 
\frac{\sinh \tfrac{1}{2}(\theta +i\pi \frac{x-1+B}{h})\sinh \tfrac{1}{2}%
(\theta +i\pi \frac{x+1-B}{h})}{\sinh \tfrac{1}{2}(\theta -i\pi \frac{x-1+B}{%
h})\sinh \tfrac{1}{2}(\theta -i\pi \frac{x+1-B}{h})}\right) ^{\frac{1}{2}%
}\,\,.  \label{bl}
\end{equation}
This block has the obvious properties 
\begin{equation*}
\left[ x,B\right] _{\theta ,ab}\left[ x,B\right] _{-\theta ,ba}=1\quad \text{%
and\quad }\left[ h-x,B=1\right] _{\theta ,ab}=\left[ x,B=1\right] _{i\pi
-\theta ,ba}\,\,.
\end{equation*}
In a slightly loose notation it is understood that in the second equality
one first takes the square root and thereafter performs the shifts in the
arguments. Note further that the order of the colour values is relevant for
crossing symmetry. The appearance of a minus sign in the original building
blocks is now interpreted in terms of parity violation via a colour
dependent phase factor. From (\ref{bl}) we can now construct the block form
of the $\frak{g}$\textbf{\TEXTsymbol{\vert}}$\frak{\tilde{g}}$--scattering
matrix \cite{FKcol}\ by exploiting the equivalence between the analogous
expressions of simply-laced ATFT in 3.2.10, 
\begin{equation}
S_{ij}^{ab}(\theta )=\prod\limits_{x=1}^{h}\left[ 2x-\tfrac{c_{i}+c_{j}}{2},%
\tilde{I}_{ab}\right] _{\theta ,ab}^{-\frac{\tilde{A}_{ab}}{2}\left\langle
\lambda _{j},\sigma ^{x}\gamma _{i}\right\rangle }\,\,.  \label{Snew}
\end{equation}
Here the $\lambda _{i}$'s are again the fundamental weights, the $\gamma
_{i} $'s are simple roots times a colour value $c_{i}=\pm 1$, $h$ is the
Coxeter number and $\sigma $ is the Coxeter element related to the Lie
algebra $\frak{g}$. The remaining step to establish the equivalence between (%
\ref{Snew}) and (\ref{iSnew}) is to prove that the parity breaking phase
factor in (\ref{bl}) combines to the factor in front of the integral
representation, i.e. 
\begin{equation*}
e^{i\pi A_{ij}^{-1}}=\prod\limits_{x=1}^{h}\left[ e^{i\frac{\pi }{h}\left(
2x-\frac{c_{i}+c_{j}}{2}\right) }\right] ^{-\frac{1}{2}\left\langle \lambda
_{j},\sigma ^{x}\gamma _{i}\right\rangle }\,.
\end{equation*}
However, this is immediate from the Lie algebraic identity (\ref{classM}) of
Chapter 2.\medskip

\noindent \textbf{Remark}. \emph{At first sight the power }$1/2$\emph{\ in
the definition of the building block (\ref{bl}) seems to suggest the
presence of square root branch cuts in the S-matrix (\ref{Snew}). A careful
analysis of the cases }$a=b$\emph{\ and }$a\neq b$\emph{\ shows, however,
that one recovers the minimal S-matrix and the CDD factor in (\ref{split}),
respectively. Both are meromorphic functions, in particular for }$a=b$\emph{%
\ one has }$\tilde{A}_{ab}=2$\emph{\ and for }$a\neq b,\,B=\tilde{I}_{ab}=1$%
\emph{\ the square root can be taken directly in (\ref{bl}). The remaining
power }$1/2$\emph{\ stemming from the exponent in (\ref{Snew}) is
compensated by the same mechanism as encountered in the context of ATFT, see
Section 3.2.7, i.e. the combination of various equal blocks.\medskip }

It should be emphasized that there is no need to introduce the phase to
satisfy the unitarity equation (\ref{uni}). It is further clear that (\ref
{Snew}) is Hermitian analytic. However, as already mentioned above the
introduction of the phase factor is crucial in order to satisfy the crossing
relation. \emph{Whence the violation of parity invariance is a direct
consequence of the functional equation (\ref{cross}) in the bootstrap
approach.} Assuming the validity of the $ADE$-fusing rules (\ref{ADEfuse})
one may now verify by the same shifting arguments as in the previous section
that the fusing bootstrap equations (\ref{boot}) are satisfied. This
establishes (\ref{Snew}) or equivalently (\ref{iSnew}) to be a consistent
S-matrix and hence implicitly defines a new class of integrable quantum
field theories.

\subsubsection{Relation to known models}

For special choices of the algebras one recovers from the general class of $%
\frak{g}$\textbf{\TEXTsymbol{\vert}}$\frak{\tilde{g}}$-models two subclasses
of already known S-matrices. Choose $\frak{\tilde{g}}$ to be $A_{1}$ then
the colour structure is removed (since there is only one possible colour
value) and the system reduces to the one described by $S_{ij}^{\min }(\theta
)$. The first examples of these S-matrices were constructed in \cite{Zper}
for $\frak{g}=A_{n,}E_{8}$ and describe so-called scaling models mentioned
in the introduction. This class is at the same time the only example for
which (\ref{Snew}), (\ref{iSnew}) do not violate parity invariance. Choosing
instead $\frak{g}$\textbf{\ }to be $A_{n}$ and $\frak{\tilde{g}}=ADE$ we
recover the S-matrices of the Homogeneous Sine-Gordon models for vanishing
resonance parameter at level $(n+1)$ \cite{HSGS,CFKM}. The latter S-matrices
were just recently formulated in context of massive perturbations of WZNW
models. A detailed discussion of their properties and the possibility to
include resonance poles in their definition will be postponed to the next
chapter, where the high energy behaviour of the integrable models of this
chapter is discussed.

\chapter{From Massive to Massless Models}

{\small \emph{Our bodies are given life from the midst of nothingness.
Existing where there is nothing is the meaning of the phrase, ``Form is
emptiness.'' That all things are provided for by nothingness is the meaning
of the phrase, ``Emptiness is form.'' One should not think that these are
two separate things.}}

\qquad \qquad \qquad \qquad \qquad \qquad {\small From 'The Book of the
Samurai, Hagakure'\medskip }

In the last chapter we saw that integrability of 1+1 dimensional quantum
field theories leads to severe restrictions on the scattering processes
allowing to determine the scattering matrix exactly via the bootstrap
principle. The question which then comes to mind is, how natural is the
concept of integrability? In particular, can one understand the set of
infinite conserved charges in terms of a higher symmetry? The answer to
these questions was first put forward in the work by Zamolodchikov \cite
{perCFT}, who suggested to view integrability as relict of broken conformal
symmetry and to interpret integrable models as \emph{perturbed conformal
field theories}.

Conformal field theories (CFT) are characterized by scale invariance, i.e.
they describe massless relativistic particles or (in Euclidean space)
statistical mechanics systems at a critical point. As was shown in the
seminal paper by Belavin, Polyakov and Zamolodchikov \cite{BPZ} conformal
symmetry becomes extremely powerful in two-dimensions\footnote{%
There exist earlier considerations of field theories in 1+1 dimensions which
focus on the aspect of conformal invariance, e.g. \cite{SCH}. However, the
key feature, i.e. the role played by the Virasoro algebra, which lead to a
more universal formulation and allowed to find their solution was first
realized and exploited in \cite{BPZ}.}. There an infinite number of
conserved currents arises due to the chiral splitting of the theory into an
holomorphic and an anti-holomorphic part (see below for an explanation).
Upon a perturbation by a relevant operator of the conformal theory this
splitting is in general lost and the system dragged away from the critical
point exhibits a finite correlation length, i.e. it becomes massive.
However, in \cite{perCFT} an argument was provided that an infinite set of
the original conserved currents, even though they get deformed, might
survive the breaking of conformal invariance.

To be more precise, consider an action functional in Euclidean space of the
form 
\begin{equation}
S=S_{\text{CFT}}-\lambda \int d^{2}x\,\Phi (x,t)\quad \text{with\quad }%
\Delta _{\Phi }=\bar{\Delta}_{\Phi }<1\;.  \label{perCFT}
\end{equation}
Here $S_{\text{CFT}}$ is the action of the unperturbed conformal field
theory, $\lambda \sim m^{2-d_{\Phi }}$ is a coupling constant of the
perturbation term proportional to the overall mass scale $m$ and $\Phi $ is
assumed to be a relevant spinless field operator, i.e. in the conformal
limit $\lambda =0$ it has anomalous scaling dimension $d_{\Phi }=2\Delta
_{\Phi }$ with conformal weight $\Delta _{\Phi }<1$. See below for an
explanation of the various quantities. Now introducing complex coordinates $%
z=x^{0}+ix^{1},\bar{z}=x^{0}-ix^{1}$ the theory at the conformal point $%
\lambda =0$ exhibits the characteristic chiral splitting, i.e. there are
infinitely many currents $J,\bar{J}$ which depend either on $z$ or $\bar{z}$
only, 
\begin{equation*}
\lambda =0:\quad \bar{\partial}J=0\quad \text{and\quad }\partial \bar{J}=0\;.
\end{equation*}
In the perturbed theory this splitting is in general lost and $J,\bar{J}$
acquire an additional $\bar{z},z$-dependence, respectively. In fact, up to
first order in the coupling constant the change of the above conversation
law changes to \cite{perCFT} 
\begin{equation*}
\lambda \neq 0:\quad \bar{\partial}J(z,\bar{z})=\lambda \oint_{z}\frac{%
d\zeta }{2\pi i}\;\Phi (\zeta ,\bar{z})J(z)+...
\end{equation*}
Now, if there exist another current, say $\bar{J}^{\prime }$, such that the
r.h.s. in the above equation can be expressed as the derivative $-\partial 
\bar{J}^{\prime }$, one obtains a conservation law for the perturbed theory, 
$\bar{\partial}J+\partial \bar{J}^{\prime }=0$. Under the assumption that
this holds true for infinitely many cases and if the perturbed theory is
purely massive one then ends up with an integrable quantum field theory
whose S-matrix should be tractable by the bootstrap approach.

In this chapter we will adopt this point of view and regard the quantum
field theories implicitly defined by the construction of the exact
S-matrices in Chapter 3 as perturbed conformal field theories. In fact, one
can reverse the picture and recover to each massive quantum field theory the
corresponding conformal one by taking the high-energy limit, where the
masses become negligible and the system loses its scale dependence. To
perform the high-energy limit and to regain the conformal field content from
the massive one is in general a highly non-trivial task, which requires
renormalization group techniques and a profound insight into the structure
of the field theory under consideration.

In case of factorizable S-matrices, however, there is an alternative method
originating in the early work of Yang and Yang \cite{Yang}, the \emph{%
thermodynamic Bethe ansatz} (TBA). First formulated in the context of the
non-relativistic Bose gas it was extended about ten years ago to systems of
particles which interact in a relativistic manner through a factorizable
scattering matrix \cite{TBAZam1}.\ Nowadays it is established as one of the
most important techniques in exploring the close relationship between
conformal and integrable field theories. Provided the two-particle
scattering amplitude of the massive quantum field theory is known the TBA
allows to calculate the most characteristic quantities of the underlying UV
conformal model, as for instance the effective central charge or conformal
anomaly (see below for its definition). The latter information is often
sufficient to determine the field content at the critical point and once the
perturbing operator has been identified the integrable field theory can
formally be described in terms of a classical Lagrangian. If on the other
hand the conformal field theory is a priori known the TBA provides a
consistency check for the S-matrix and allows to remove the CDD ambiguities
inherent to the bootstrap construction (compare 3.1.2).

Both applications of the TBA will play a role in this chapter. To keep the
discussion self-contained some basic notions of conformal field theory are
introduced in the first section. Afterwards the thermodynamic Bethe Ansatz
in context of integrable relativistic field theories is presented. Central
issue is the derivation of a set of non-linear integral equations whose
solutions describe the thermodynamics of the system and allow to compute the
effective central charge. In general, the integral equations can only be
solved numerically by an iterative method and we will therefore comment on
existence and uniqueness of the solution as well as on the convergence
properties of the numerical procedure. However, the numerics can be
supplemented by various analytical approximation schemes which allow to
derive explicit formulas for the most important conformal data, such as the
effective central charge or the scaling dimension of the perturbing operator.

In a second step these approximation schemes are applied to affine Toda
field theories. Exploiting the Lie algebraic structure revealed in the last
chapter universal formulas are presented which describe the high energy
behaviour to lowest order for all ATFT models at once. These approximate
solutions are then checked for explicit examples, the Sinh-Gordon model and
the $(G_{2}^{(1)},D_{4}^{(3)})$-ATFT, against the numerical data. The
results will prove to be an additional consistency check for the ATFT
S-matrix constructed in Chapter 3.

In case of the colour valued S-matrices constructed in 3.3 emphasis will be
given first to the so-called Homogeneous Sine-Gordon models \cite{HSG},
which form a particular subclass of the $\frak{g}|\frak{\tilde{g}}$-theories
with $\frak{g}=A_{n}$ and $\frak{\tilde{g}}$ arbitrary but simply-laced. By
means of a semi-classical analysis these kind of integrable models have been
proposed as perturbations of WZNW coset models. In the TBA analysis several
of the semi-classical considerations like the breaking of parity invariance
and the existence of unstable bound states will be probed for consistency.
Especially, the $\frak{\tilde{g}}=A_{2}\equiv su(3)$ model is investigated
in detail numerically and analytically for several examples. In general, the
S-matrix will be shown to give rise to the correct central charge providing
strong evidence for the conjecture.

The TBA analysis of the Homogeneous Sine-Gordon models is then extended to
all possible $\frak{g}|\frak{\tilde{g}}$-theories by means of the underlying
Lie algebraic structure. The central charge calculation is performed and a
general formula derived depending only on the rank and on the Coxeter number
of the associated Lie algebras.

\section{Basic notions of conformal field theory}

The first considerations of conformally invariant quantum field theories
took place in the sixties, however, the subject started to flourish after
the recognition that local field theories which are scale invariant are also
conformally invariant \cite{Pol}. This motivated to use the notion of
conformal invariance in the study of statistical mechanics system at
criticality. The breakthrough in the understanding of these theories was the
appearance of the already mentioned pioneering paper by Belavin, Polyakov
and Zamolodchikov \cite{BPZ} in 1986. Today conformal field theory is an
established branch of research, which has recently experienced new interest
due to its applications in string theory. By now there are numerous review
articles and text books on the subject (see e.g. \cite{CFT}), whence in the
following only a survey of the most important results is given. Emphasis
lies on assembling the relevant notions for the subsequent discussion.

\subsection{Conformal coordinate transformations and Witt algebra}

Conformal symmetry is defined as the invariance under \textbf{conformal
coordinate transformations}. The latter are defined as such transformations $%
x^{\mu }\rightarrow y^{\mu }(x)$ which leave the metric tensor $g_{\mu \nu }$
invariant up to a positive scaling factor, i.e. 
\begin{equation}
g_{\mu \nu }^{\prime }(y)=\Omega (x)g_{\mu \nu }(x)\;.  \label{cmetric}
\end{equation}
In particular, the angle between two vectors is preserved. For an
infinitesimal conformal coordinate transformation $x^{\mu }\rightarrow
x^{\mu }+\epsilon ^{\mu }(x)$ the above requirement can be translated into
the so-called \textbf{conformal Killing equation} which for two dimensions
reads 
\begin{equation}
\partial _{\mu }\epsilon _{\nu }+\partial _{\mu }\epsilon _{\nu }=(\partial
_{\rho }\epsilon ^{\rho })\,g_{\mu \nu }  \label{Killing}
\end{equation}
The $\epsilon ^{\mu }$'s can be interpreted as the components of a conformal
Killing vector field, i.e. they determine the tangent vector to a conformal
coordinate transformation. \emph{Now, the crucial feature of two dimensions
is that the Killing equation specializes for Euclidean space }$g_{\mu \nu
}=\delta _{\mu \nu }$\emph{\ to the Cauchy-Riemann equations familiar from
complex analysis.} This motivates to introduce complex coordinates via the
relations 
\begin{gather}
z=x^{0}+ix^{1}\quad \quad \quad \bar{z}=x^{0}-ix^{1}  \label{comcor} \\
\partial =\frac{1}{2}\left( \partial _{0}-i\partial _{1}\right) \quad \quad
\quad \bar{\partial}=\frac{1}{2}\left( \partial _{0}+i\partial _{1}\right)
\;.  \notag
\end{gather}
The infinitesimal transformations $\epsilon =\epsilon ^{0}+i\epsilon ^{1},%
\bar{\epsilon}=\epsilon ^{0}-i\epsilon ^{1}$ solving (\ref{Killing}) are now
simply those which are holomorphic or anti-holomorphic, $\bar{\partial}%
\epsilon =0$ and $\partial \bar{\epsilon}=0$. Mathematically this implies
that they admit for a Laurent expansion 
\begin{equation*}
\epsilon (z)=\sum_{n=-\infty }^{\infty }c_{n}z^{n+1}\quad \quad \text{%
and\quad \quad }\bar{\epsilon}(\bar{z})=\sum_{n=-\infty }^{\infty
}c_{n}^{\prime }\bar{z}^{n+1}
\end{equation*}
showing that the infinitesimal conformal symmetry transformations are
locally generated by 
\begin{equation*}
l_{n}=-z^{n+1}\partial \quad \quad \text{and\quad \quad }\bar{l}_{n}=-\bar{z}%
^{n+1}\bar{\partial}\;.
\end{equation*}
These local generators form similar like the generators of the Poincare
group or the rotation group a Lie algebra. As one immediately deduces from
their definition they are subject to the commutation relations 
\begin{eqnarray}
\lbrack l_{m},l_{n}] &=&(m-n)l_{m+n}\;,  \label{Witt} \\
\lbrack \bar{l}_{m},\bar{l}_{n}] &=&(m-n)\bar{l}_{m+n}\;,  \notag \\
\lbrack l_{m},\bar{l}_{n}] &=&0\;.  \notag
\end{eqnarray}
In the literature the associated Lie algebra is known under the name \textbf{%
Witt algebra}. Note that in comparison to other space-time symmetries \emph{%
local} conformal transformations give rise to an infinite set of symmetry
generators what makes conformal invariance to an extremely powerful concept.
However, it should be pointed out that only a finite subalgebra $%
\{l_{-1},l_{0},l_{1}\}$ can be ``lifted'' to \emph{global} conformal
transformations, i.e. mappings which are defined everywhere and are
invertible. Explicitly, 
\begin{equation}
z\rightarrow \frac{a\,z+b}{c\,z+d}\;,\quad \quad ab-cd=1\;,  \label{Moebius}
\end{equation}
where $a,b,c,d$ are complex numbers. These coordinate transformations form a
group via composition and are known as \textbf{projective} or \textbf{%
M\"{o}bius group}. Interpreting the constraint on the complex constants in (%
\ref{Moebius}) as determinant it is easy to see that the projective
transformation can be parametrized by the complex $2\times 2$ matrices with
unit determinant modulo the negative unit matrix, i.e. $SL(2,\mathbb{C})/%
\mathbb{Z}_{2}$. The same holds true for the anti-holomorphic part of the
Witt algebra.

As is straightforward to verify the M\"{o}bius group contains for special
choices of the parameters translations, rotations and dilatations. In
particular, the generators which correspond to dilatations and rotations on
the real surface are 
\begin{equation*}
l_{0}+\bar{l}_{0}\quad \quad \text{and\quad }\quad i(l_{0}-\bar{l}_{0})\;,
\end{equation*}
respectively. As we will see below in conformally invariant field theories
the set of relevant operators will be assumed to be eigenstates to these
generators.

\subsection{The energy-momentum tensor and the Virasoro algebra}

Conformal invariance in terms of a field theory is equivalent to the
vanishing of the variation of the classical action functional under an
infinitesimal conformal transformation $x^{\mu }\rightarrow x^{\mu
}+\epsilon ^{\mu }(x)$, 
\begin{equation}
\delta S=\frac{1}{2}\int d^{2}x\,T^{\mu \nu }\left( \partial _{\mu }\epsilon
_{\nu }+\partial _{\mu }\epsilon _{\nu }\right) =\frac{1}{2}\int
d^{2}x\,T_{\;\;\mu }^{\mu }\cdot \partial _{\nu }\epsilon ^{\nu }\;.
\label{variation}
\end{equation}
Here $T^{\mu \nu }=T^{\nu \mu }$ is the canonical \textbf{energy-momentum
tensor}, which can always be made symmetric and we have used the determining
relation (\ref{Killing}) for the Killing field $\epsilon ^{\mu }$ in the
second step. Thus, conformal invariance is guaranteed if the trace of the
energy-momentum tensor is zero $T_{\;\;\mu }^{\mu }=0$, which in turn is the
requirement that the theory is scale invariant. Together with translation
and rotation invariance, $\partial _{\mu }T^{\mu \nu }=0\,$, this constraint
on the energy-momentum tensor can be transformed into complex coordinates (%
\ref{comcor}), 
\begin{equation}
\bar{\partial}T=0\quad \quad \text{and\quad \quad }\partial \bar{T}=0\;,
\label{holoT}
\end{equation}
where $T\equiv T_{zz}=\frac{1}{4}(T_{00}-T_{11}-2iT_{10}),\;\bar{T}\equiv T_{%
\bar{z}\bar{z}}=\frac{1}{4}(T_{00}-T_{11}+2iT_{10})$ are the only
non-vanishing complex components. Thus, the energy-momentum tensor splits
into a holomorphic and an anti-holomorphic part, a property which is assumed
to hold also true when the field theory is quantized. Similar as above,
Laurent expanding the components $T,\bar{T}$ gives then rise to the \emph{%
quantum} generators of the conformal space-time symmetry, 
\begin{equation}
T(z)=\sum_{n\in \mathbb{Z}}\frac{L_{n}}{z^{n+2}}\,,\quad \quad L_{-n}:=\frac{%
1}{2\pi i}\oint_{\mathbb{S}^{1}}\frac{dz}{z^{n+1}}\;T(z)\;.  \label{Virgen}
\end{equation}
An analogous relation is assumed for the anti-holomorphic part. However, in
contrast to the classical generators of conformal transformations the
coefficients in the Laurent expansion of the energy-momentum tensor are
assumed to obey commutation relations different from (\ref{Witt}) 
\begin{eqnarray}
\lbrack L_{n},L_{m}] &=&(n-m)L_{m+n}+\dfrac{c}{12}\,n(n^{2}-1)\delta _{m+n,0}
\label{Vir} \\
\lbrack \bar{L}_{n},\bar{L}_{m}] &=&(n-m)\bar{L}_{m+n}+\dfrac{c}{12}%
\,n(n^{2}-1)\delta _{m+n,0}  \notag \\
\lbrack L_{n},\bar{L}_{m}] &=&0\;.  \notag
\end{eqnarray}
Each set of the generators $\{L_{n}\},\{\bar{L}_{n}\}$ constitutes a copy of
the so-called \textbf{Virasoro algebra} which differs from the Witt algebra (%
\ref{Witt}) by the appearance of the term containing the \textbf{central
charge} $c$ \cite{cJoe}. As we will see below this additional term can be
understood as a soft breaking of conformal invariance when macroscopic
length scales are introduced into the system. One therefore refers to the
central charge also as \textbf{conformal anomaly}. Notice that the central
term is absent for the subalgebra $\{L_{-1},L_{0},L_{1}\}\cong sl(2,\mathbb{%
C)}$ belonging to the global conformal mappings.

Mathematically, the transition from the Witt to the Virasoro algebra can be
understood analogous to the Wigner-Bargmann theorem. Because of the
phase-freedom inherent to quantum theory classical symmetry groups are
reflected by projective representations on the quantum level, i.e. the
quantum symmetry operations form a representation of the classical group up
to a phase factor. In terms of the corresponding Lie algebra this amounts to
the allowance of central extensions. In fact, the Virasoro algebra is just
such a central extension of the Witt algebra.

\subsection{Primary fields and highest weight representations}

After having introduced the quantum generators of conformal transformations
it remains to construct a field theory invariant under their action. The
crucial step is the introduction of so-called \textbf{primary fields} $\phi
, $ which form the fundamental building constituents of the field content
and transform like covariant tensors under a change of variables 
\begin{equation}
\phi (z,\bar{z})\rightarrow \phi ^{\prime }(w,\bar{w})=\left( \frac{\partial
w}{\partial z}\right) ^{-\Delta }\left( \frac{\partial \bar{w}}{\partial 
\bar{z}}\right) ^{-\bar{\Delta}}\phi (z,\bar{z})\;.  \label{ctensor}
\end{equation}
The real exponents $(\Delta ,\bar{\Delta})$ are called \textbf{conformal
weights} of the primary field $\phi $. Their linear combinations 
\begin{equation}
d=\Delta +\bar{\Delta}\quad \quad \text{and}\quad \quad s=\Delta -\bar{\Delta%
}  \label{ds}
\end{equation}
determine the transformation behaviour of $\phi $ under dilatations and
rotations and are known as \textbf{anomalous scaling dimension} and \textbf{%
spin}, respectively. Clearly, the covariance property (\ref{ctensor}) is
only well defined for \emph{global} conformal transformations, but it might
be restated in terms of local generators and in this manner allows for a
generalization to \emph{local} conformal transformations, 
\begin{equation}
\lbrack L_{n},\phi (z,\bar{z})]=z^{n+1}\partial \phi (z,\bar{z})+\Delta
(n+1)z^{n}\phi (z,\bar{z})\;,\quad n\in \mathbb{Z}\,.  \label{primary}
\end{equation}
We shall take the above identity as defining relation for primary fields. A
similar relation is assumed to hold for the anti-holomorphic part. However,
since both parts commute we will henceforth only concentrate on the
holomorphic part and assume analogous algebraic relations to hold for the
anti-holomorphic one.

\subsubsection{Conformal families}

The primary fields are the central objects of a conformal field theory since
their linear combinations and derivatives generate the whole field content
of the algebra. In fact, introducing the operators 
\begin{equation}
L_{-n}(z)=\oint \frac{d\zeta }{2\pi i}\,\frac{T(\zeta )}{(\zeta -z)^{n+1}}
\label{Virz}
\end{equation}
one can assign to each primary field a \textbf{conformal family} $[\phi ]$
obtained by successive actions of $L_{n}(z)$ on it 
\begin{equation}
\lbrack \phi ]:=\{L_{n_{1}}(z)...L_{n_{N}}(z)\phi (z,\bar{z}):n_{1}\leq
...\leq n_{N}<0\}\;.  \label{family}
\end{equation}
The elements in $[\phi ]$ are called \textbf{secondary fields} or \textbf{%
descendants}. Notice the close connection between the operators (\ref{Virz})
and the Virasoro generators (\ref{Vir}) by means of the limit $%
\lim_{z\rightarrow 0}L_{n}(z)=L_{n}$. In fact, by exploiting this relation
one can easily show that each conformal family defines a highest weight
representation of the Virasoro algebra.

\subsubsection{Verma modules}

We introduce the Hilbert space of states by defining the \textbf{vacuum
vector} of the theory through the relation 
\begin{equation}
L_{n}\left| 0\right\rangle =0\;,\quad n\geq 0\;.  \label{vacuum}
\end{equation}
This in particular ensures the invariance of the vacuum sector under global
conformal transformations. To each primary field $\phi $ with conformal
weight $\Delta $ there exists a highest weight vector $\left| \Delta
\right\rangle :=\phi (0)\left| 0\right\rangle $ which by use of (\ref
{primary}) can be seen to satisfy 
\begin{equation}
L_{0}\left| \Delta \right\rangle =\Delta \left| \Delta \right\rangle \quad
\quad \text{and\quad }\quad L_{n}\left| \Delta \right\rangle =0\;,\quad
n>0\;.  \label{hweight}
\end{equation}
The remaining Virasoro generators with indices $n<0$ now produce a huge
state space upon acting on the highest weight vector, 
\begin{equation}
V_{c,\Delta }=\left\{ L_{n_{1}}\cdots L_{n_{k}}\left| \Delta \right\rangle
:n_{i}\leq 0,\;k\in \mathbb{N}\right\} \;.  \label{Verma}
\end{equation}
The space $V_{c,\Delta }$ is called a \textbf{Verma module} and is invariant
under the action of the Virasoro algebra. Clearly, from the defining
relation of the highest weight vector and the intimate relation between the
operators (\ref{Virz}) and the Virasoro generators one deduces that Verma
modules and conformal families $[\phi ]$ are in one-to-one correspondence.
Thus, the concept of a primary field allows to classify all possible CFT's,
namely via the highest weight representations of the Virasoro algebra.

\subsubsection{Singular vectors}

According to the above definitions the total state space may now be obtained
by summing over all tensor products $V_{c,\Delta }\otimes \bar{V}_{c,\bar{%
\Delta}}$, where the second factor represents the anti-holomorphic part and $%
\Delta ,\bar{\Delta}$ run over all conformal weights occurring in the
theory. However, in general the Verma modules will be reducible, i.e. it
contains a subspace invariant under the action of the Virasoro algebra. The
latter are generated from so-called \textbf{singular} or \textbf{null vectors%
} which are itself highest weight states. To see this more clearly, notice
first that all elements in a Verma module are eigenvectors of $L_{0}$. In
fact, exploiting the commutation relations (\ref{Vir}) one deduces that 
\begin{equation*}
L_{0}\,L_{n_{1}}\cdots L_{n_{k}}\left| \Delta \right\rangle =(\Delta +\ell
)\cdot L_{n_{1}}\cdots L_{n_{k}}\left| \Delta \right\rangle \;,\quad \ell
=-\sum_{i=1}^{k}n_{i}\;.
\end{equation*}
Here all $n_{i}<0$ and the integer $\ell $ is called the \textbf{level.} (It
is off course not related to the order of the Dynkin diagram automorphism
defined in Chapter 2 and 3). The states of a given level are in general
degenerate, in particular, they may contain a vector $\left| \upsilon
\right\rangle $ satisfying $L_{n}\left| \upsilon \right\rangle =0$, for all $%
n>0$. Thus, each singular vector defines another highest weight
representation $V_{c,\Delta +\ell },$ which lies inside the original one $%
V_{c,\Delta }$. Moreover, if one defines an inner product on $V_{c,\Delta }$
such that the Hermitian conjugate of a Virasoro generator is given by $%
L_{n}^{\ast }=L_{-n}$ one infers that singular vectors and all their
descendants have vanishing norm and are orthogonal to all other elements in
the Verma module. The physical state space is obtained by ''dividing'' out
all the possible null subspaces generated from singular vectors $\left|
\upsilon \right\rangle $ which can occur at different levels. The resulting
quotient space $V_{c,\Delta }^{\prime }$ forms then an irreducible
representation and the physical subspace has the structure 
\begin{equation}
\mathcal{H}=\tbigoplus_{\Delta ,\bar{\Delta}}m_{\Delta ,\bar{\Delta}%
}\,V_{c,\Delta }^{\prime }\otimes \bar{V}_{c,\bar{\Delta}}^{\prime }
\label{CFTspace}
\end{equation}
where $m_{\Delta ,\bar{\Delta}}$ is the multiplicity that the conformal
weights $\Delta ,\bar{\Delta}$ occur. The linear structure of a Verma module
is in general very intricate and complex and can be encoded in so-called 
\textbf{Virasoro characters,} which appear as natural mathematical objects, 
\begin{equation}
\chi _{c,\Delta }(q)=\limfunc{Tr}q^{L_{0}-c/24}=\sum_{\ell =0}^{\infty }\dim
(\ell )q^{\Delta +\ell -c/24}\;.  \label{Virch}
\end{equation}
Here $\dim (\ell )$ is the number of linear independent vectors at level $%
\ell $. The characters form important mathematical tools to elucidate the
structure of a conformal field theory, see \cite{CFT} for further details.
We comment on a possible connection between results from the TBA analysis of
integrable models and Virasoro characters in a subsequent section.

\subsubsection{The Kac determinant}

Another quantity to determine the number of linear independent vectors of a
Verma module $V_{c,\Delta }$ at a given level $\ell $ is the Gram matrix $%
M^{(\ell )}(c,\Delta )$ formed by all the inner products of the basis states 
\begin{equation*}
\left\langle \Delta |L_{m_{1}}\cdots L_{m_{l}}L_{-n_{1}}\cdots
L_{-n_{k}}|\Delta \right\rangle \;,\quad n_{i},m_{i}\geq
0,\;\sum_{i=1}^{k}n_{i}=\sum_{i=1}^{l}m_{i}=\ell \;.
\end{equation*}
Now, if the determinant $\det M^{(\ell )}(c,\Delta )$ vanishes one can
conclude that there are linear dependent or null vectors at level $\ell $.
Moreover, if the determinant is negative there must be states with negative
norm present and the representation is not unitary. Remarkably, $\det
M^{(\ell )}(c,\Delta )$ can be brought into the following universal form
found by Kac and proven by Feigin and Fuchs \cite{Kacdet}, 
\begin{equation}
\det M^{(\ell )}(c,\Delta )=a_{\ell }\prod_{rs\leq \ell }\left( \Delta
-\Delta _{r,s}\right) ^{P(\ell -rs)}  \label{Kdet}
\end{equation}
with the product running over all positive integers $r,s$ whose product is
smaller than or equal to the level. The constant $a_{\ell }$ does not depend
on the central charge or conformal weight and $P(\ell -rs)$ denotes the
number of partitions of the integer $\ell -rs$. The roots $\Delta _{r,s}$ of
the determinant can be parametrized as 
\begin{equation}
\Delta _{r,s}(m)=\frac{\left[ (m+1)r-ms\right] ^{2}-1}{4m(m+1)}\quad \text{%
with}\quad m=-\frac{1}{2}\pm \frac{1}{2}\sqrt{\frac{25-c}{1-c}}\;.
\label{par}
\end{equation}
Note that the auxiliary quantity has two branches and is in general complex.
One can now use this explicit formula to infer certain restriction on the
allowed central charges and conformal weights when the representation of the
Virasoro algebra ought to be unitary. Explicitly, one can show that all
representations with $c\geq 1,\Delta \geq 0$ are unitary, while for $c\geq
25 $ one deduces that $-1<m<0$ and therefore $\Delta _{r,s}(m)\leq 0$
implying that the associated representations are non-unitary. The structure
becomes especially restrictive for $0<c<1$, where only a discrete set of
possible unitary theories exist, 
\begin{equation}
c=1-\frac{6}{m(m+1)}\quad \text{with}\quad m=3,4,...\,,\;1\leq r<m,\,1\leq
s\leq r\;.  \label{min}
\end{equation}
The associated conformal weights are given by the parametrization (\ref{par}%
). The set of the above conformal models is known as the \textbf{minimal
unitary series }in the literature. They belong to the class of theories
possessing only a \emph{finite} number of primary fields associated with the
allowed conformal weights $\Delta _{r,s}(m)$ and are the best studied CFT's,
for further details see \cite{CFT}.

\subsection{Ward identities and elementary correlation functions}

After having analyzed the implications of conformal invariance on the
physical state space and the field content we now turn to more elementary
constraints on the measurable quantities of the theory, the correlation or $%
n $-point functions. In general, \textbf{Ward identities} for correlation
functions reflect symmetries possessed by a quantum field theory and can be
thought of as an infinitesimal version of covariance under a symmetry
operation. Implementation of covariance under global conformal
transformations with generators $\{L_{-1},L_{0},L_{1}\}$ leads to the
identities 
\begin{equation*}
\left\langle \lbrack L_{n},\Phi _{1}(z_{1},\bar{z}_{1})\cdots \Phi
_{n}(z_{n},\bar{z}_{n})]\right\rangle =0\;,\quad n=-1,0,1\;,
\end{equation*}
where the bracket indicates the vacuum expectation value of radially%
\footnote{%
Note that radial ordering in Euclidean space is the equivalent of
time-ordering in Minkowski space.} ordered fields, i.e. $|z_{1}|<...<|z_{n}|$%
. Again an analogous relation holds for the anti-holomorphic generators $\{%
\bar{L}_{-1},\bar{L}_{0},\bar{L}_{1}\}$. In case of primary fields the above
Ward identities take the more explicit form\emph{\ } 
\begin{eqnarray*}
\sum_{i}\partial _{z_{i}}\left\langle \phi _{1}\cdots \phi _{n}\right\rangle
&=&0 \\
\sum_{i}\left( z_{i}\partial _{z_{i}}+\Delta _{i}\right) \left\langle \phi
_{1}\cdots \phi _{n}\right\rangle &=&0 \\
\sum_{i}\left( z_{i}^{2}\partial _{z_{i}}+2z_{i}\Delta _{i}\right)
\left\langle \phi _{1}\cdots \phi _{n}\right\rangle &=&0\;
\end{eqnarray*}
where we have used the defining relation (\ref{primary}). These three
equations reflect invariance under translations, rotations and dilatations
as well as special conformal transformations and constitute the minimal
requirement on each conformal model. However, they are already sufficient to
determine the explicit form of the two and three-point functions, 
\begin{eqnarray}
\left\langle \phi _{1}(z_{1},\bar{z}_{1})\phi _{2}(z_{2},\bar{z}%
_{2})\right\rangle &=&\frac{C_{12}}{|z_{12}|^{2d}}  \label{cor2} \\
\left\langle \phi _{1}(z_{1},\bar{z}_{1})\phi _{2}(z_{2},\bar{z}_{2})\phi
_{3}(z_{3},\bar{z}_{3})\right\rangle &=&\frac{C_{123}}{%
|z_{12}|^{d_{1}+d_{2}-d_{3}}|z_{13}|^{d_{1}+d_{3}-d_{2}}|z_{23}|^{d_{2}+d_{3}-d_{1}}%
}  \label{cor3}
\end{eqnarray}
where $z_{ij}:=z_{i}-z_{j}$ and the coefficient $C_{12}$ of the two-point
function vanishes unless $d_{1}=d_{2}$. Thus, two and three point functions
are completely determined by the coefficients $C_{ij},C_{ijk}$. (Note that
the three point coefficient and the three-point coupling in context of ATFT
are different quantities. In particular, the \emph{fusion} of two primary
fields in the sense of the OPE discussed below is in general not linked to
the \emph{fusing} of quantum particles encountered before in context of
ATFT.) The latter carries further important information about the structure
of the field content via the operator product expansion and the algebra
hypothesis, which are additional ingredients to each CFT besides conformal
invariance.

\subsection{Operator product expansion}

From the explicit expressions (\ref{cor2}) and (\ref{cor3}) we infer that
the correlation functions possess singularities when the arguments of two
fields approach each other, $z_{ij}\rightarrow 0$. This holds true for all $%
n $-point functions and has its origin in the infinite quantum fluctuations
when a quantum field is taken at a precise position. The singular behaviour
of a correlation function can be extracted by a short-distance expansion of
operator products in a sum of terms involving only single operators.
Exploiting the decomposition (\ref{Virgen}) and the transformation
properties (\ref{primary}) one deduces for example the following \textbf{%
operator product expansion} (OPE) for the energy-momentum tensor and a
primary field, 
\begin{equation}
T(z)\phi (w,\bar{w})=\frac{\Delta \,\phi (w,\bar{w})}{(z-w)^{2}}+\frac{%
\partial _{w}\phi (w,\bar{w})}{z-w}+\,\text{regular terms}  \label{OPEp}
\end{equation}
where the regular terms vanish as $z\rightarrow w$. In fact, the above OPE
is equivalent to (\ref{primary}) and can be used as alternative definition
of primary fields. \emph{Note that the OPE is always understood to be well
defined only within correlation functions, i.e. in a weak sense.} Similar
one might now derive the OPE for the components of the energy-momentum
tensor by exploiting the commutation relations (\ref{Vir}) yielding 
\begin{equation}
T(z)T(w)=\frac{c/2}{(z-w)^{4}}+\frac{2T(w)}{(z-w)^{2}}+\frac{\partial
_{w}T(w)}{z-w}+\,\text{regular terms\ .}  \label{OPET}
\end{equation}
In comparison with (\ref{OPEp}) this shows that when the term involving the
central charge would be absent $T$ could be interpreted as a primary field
with conformal weight $(\Delta =2,\bar{\Delta}=0)$. In fact, $T$ is a
so-called \textbf{quasi-primary field}, i.e. it transforms covariantly under
global conformal mappings.

Now turning to the OPE of two primary fields one of the central assumptions
in context of CFT is the \textbf{operator algebra hypothesis} \cite{BPZ}
which states that the set of field forms a closed associative algebra such
that 
\begin{equation*}
\phi _{i}(z,\bar{z})\phi _{j}(w,\bar{w})=\sum_{k}\sum_{\varphi \in \lbrack
\phi _{k}]}C_{ijk}^{\varphi }(z-w)^{\Delta _{k}+\ell -\Delta _{i}-\Delta
_{j}}(\bar{z}-\bar{w})^{\bar{\Delta}_{k}+\bar{\ell}-\bar{\Delta}_{i}-\bar{%
\Delta}_{j}}\varphi (w,\bar{w})+...
\end{equation*}
Here the first sum runs over all primary fields $\phi _{k}$ and the second
about the descendants $\varphi \in \lbrack \phi _{k}]$ with $\ell ,\bar{\ell}
$ being its level w.r.t. the two copies of Virasoro algebras. By use of
conformal invariance the coefficients determining the expansion can be shown
to be of the form $C_{ijk}^{\varphi }=C_{ijk}\beta _{ijk}^{\varphi }\bar{%
\beta}_{ijk}^{\varphi }$, where the $C_{ijk}$'s are identical to the ones
occurring in the three-point function (\ref{cor3}) and the $\beta $%
-functions can be calculated from the conformal weights and central charge.
In principle, the above OPE allows to calculate all correlation functions
provided the structure constants $C_{ijk}^{\varphi }$ and the conformal
weights $\Delta _{i},\bar{\Delta}_{i}$ are known. From the correlation
functions the whole field theory can then be reconstructed. Thus, we can
summarize the central objects of interest in conformal field theory as
follows,

\begin{itemize}
\item  the central charge $c$ of the representation of the Virasoro algebra

\item  the conformal weights $\Delta ,\bar{\Delta}$ of the primary fields

\item  the OPE coefficients $C_{ijk}$ for primary fields

\item  the correlation functions.
\end{itemize}

\subsection{The central charge as Casimir energy}

We conclude this short survey of CFT by presenting an interpretation of the
central charge as Casimir energy which will play a central role throughout
the rest of this chapter.

From the OPE of the energy-momentum tensor we already pointed out that it
deviates from the one of primary fields. In fact, the term involving the
central charge leads to the following transformation behaviour under a local
conformal mapping \cite{CFT}, 
\begin{gather*}
T(z)\rightarrow T(w)=\left( \frac{\partial w}{\partial z}\right) ^{-2}T(z)+%
\frac{c}{12}\,s(z;w)\;,\quad \\
s(z;w):=\frac{z^{\prime }z^{\prime \prime \prime }-\frac{3}{2}(z^{\prime
\prime })^{2}}{(z^{\prime })^{2}}\;.
\end{gather*}
Here $s(z;w)$ is known as the Schwarzian derivative. Notice that it vanishes
for global conformal transformations. Consider the mapping from the plane
onto an infinite cylinder with circumference $R$ then the general formula
above yields 
\begin{equation*}
z\rightarrow w=\frac{R}{2\pi }\ln z\;,\quad T_{\text{cyl}}(w)=\left( \frac{%
2\pi }{R}\right) ^{2}\left( z^{2}T(z)-\frac{c}{24}\right) \;.
\end{equation*}
Thus, setting the vacuum energy density $\left\langle T(z)\right\rangle $ on
the plane to zero, we infer that due to the constant term it is nonzero on
the cylinder $\left\langle T_{\text{cyl}}(w)\right\rangle =-\pi ^{2}c/6R^{2}$%
. This can be understood in terms of a \textbf{Casimir energy} arising as
response of the system to the introduction of periodic boundary conditions
on the cylinder. In particular, this amounts to a shift of the Hamiltonian
on the cylinder which after integration of the energy density reads 
\begin{equation}
H_{\text{cyl}}=\frac{2\pi }{R}(L_{0}+\bar{L}_{0}-c/12)\;.  \label{CFTH}
\end{equation}
Notice that the central charge can therefore be recovered from the energy
spectrum. This observation will turn out to be crucial in context of the
thermodynamic Bethe ansatz which we are going to study in the next section.

There are more general scenarios which elucidate further the role of the
central charge as response to the introduction of macroscopic length scales.
For example, it was shown by Polyakov that in the context of curved
manifolds the following vacuum expectation value is in general non-zero, $%
\left\langle T_{\;\mu }^{\mu }\right\rangle =c/24\pi \,\mathcal{R},$ where $%
\mathcal{R}$ is the scalar curvature \cite{trace}. Since classically the
trace of the energy-momentum tensor ought to vanish, one refers to the
latter relation as \textbf{trace anomaly.} Additional examples also include
manifolds with boundaries as discussed in \cite{Peschel}.

\section{The thermodynamic Bethe ansatz}

The TBA-approach allows to extract various types of informations from a
massive integrable quantum field theory once its scattering matrix is known.
Most easily one obtains the central charge $c$ of the Virasoro algebra of
the underlying ultraviolet conformal field theory, the conformal dimension $%
d_{\Phi }$ and the factor of proportionality of the perturbing operator $%
\Phi $ in (\ref{perCFT}), the vacuum expectation value of the
energy-momentum tensor $\langle T_{\,\,\,\mu }^{\mu }\rangle $ and other
interesting quantities. In particular, the TBA provides a test laboratory in
which certain conjectured scattering matrices may be probed for consistency.

Moreover, the TBA is useful since it provides quantities which may be
employed in other contexts, like the computation of correlation functions.
For instance, the constant of proportionality, the dimension of the
perturbing field and $\langle T_{\,\,\,\mu }^{\mu }\rangle $ may be applied
in a perturbative approach around the operator product expansion of a two
point function within the conformal field theory \cite{Zamocorr}. On the
massive side, the vacuum expectation value $\langle T_{\,\,\,\mu }^{\mu
}\rangle $ may also be used as an initial value for the recursive system
between different $n$-particle form factors \cite{FF} when calculating the
correlation functions in the perturbed theory. Thus, the TBA plays not only
a key role in linking conformal and integrable field theories but also
serves as complementary approach to other techniques. The different aspects
of the TBA are shown in form of a diagram in Figure 4.1. 

\begin{center}
\includegraphics[width=5.2114in,height=6.6184in,angle=0]{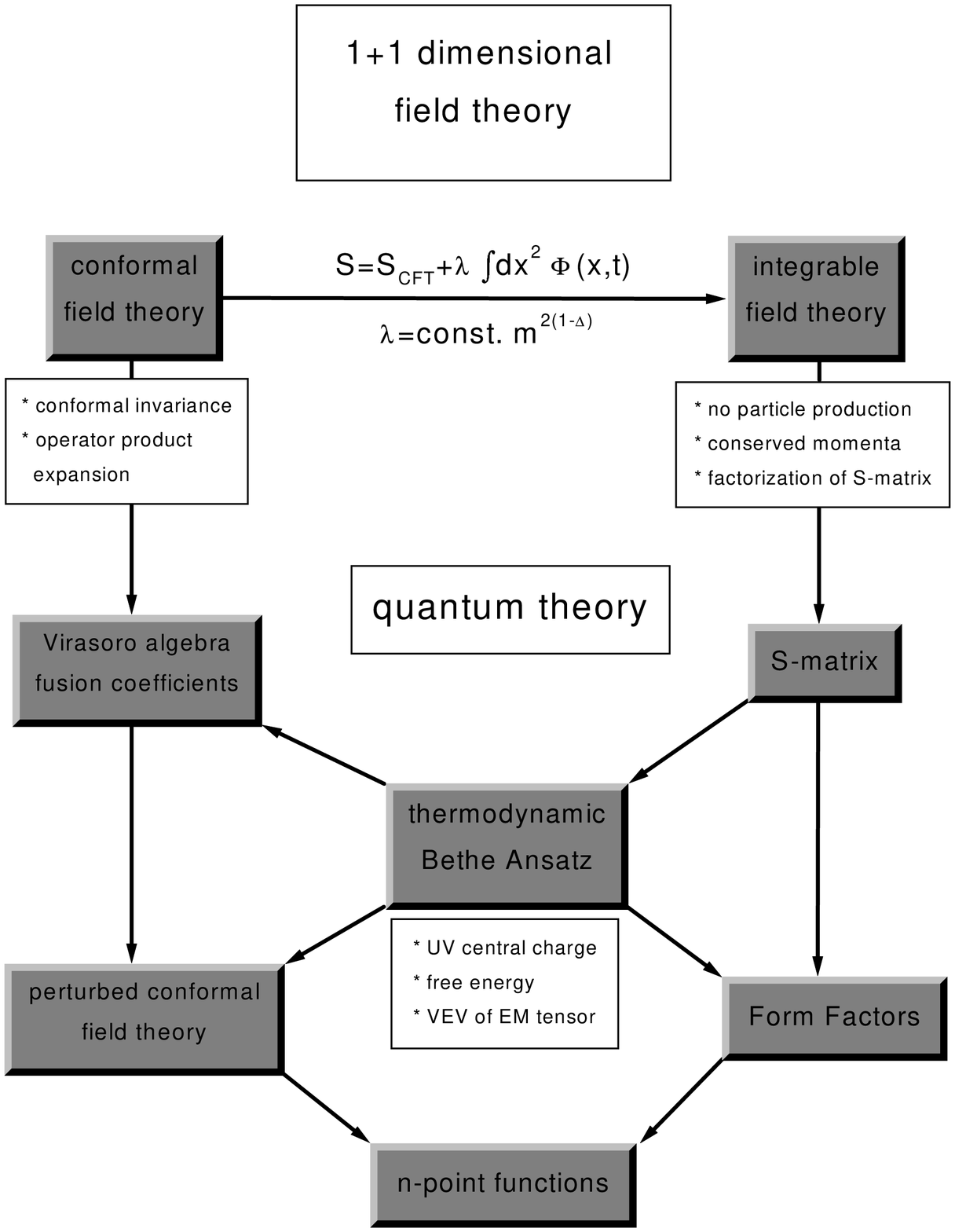}

{\small Figure 4.1: The role of the TBA in 1+1 dimensional field theory.}
\end{center}

\subsection{Quantum field theory on a torus}

As a preliminary step to the discussion of the TBA we present a simple
argument for the partition function of a Euclidean quantum field theory with
periodic boundary conditions.

Suppose first that we are dealing with a relativistic massive quantum field
theory in Minkowski space. Imposing periodic boundary conditions amounts to
a definition of the theory on a torus, see Figure 4.2. By performing a Wick
rotation we might change to Euclidean metric and one of the periods, say $R$%
, which specified the time direction before is now identified as inverse
temperature. The other period of the torus, say $L$, must then be the space
direction or so-called quantization axis of the system. However, due to the
Euclidean signature the choice of the quantization axis is arbitrary and
both periods play a symmetric role, whence we can write the associated
partition function in two different ways, 
\begin{equation}
Z(R,L)=\limfunc{Tr}e^{-LH_{R}}\quad \;\text{and\ \quad }Z(R,L)=\limfunc{Tr}%
e^{-RH_{L}}\;.  \label{ZQFT}
\end{equation}
Here $H_{R}$ and $H_{L}$ denote the Hamiltonians obtained when quantizing
the system along the $R$-axis and the $L$-axis, respectively. Later on in
the derivation we will perform the thermodynamic limit sending one of the
periods to infinity, $L\rightarrow \infty $, which in geometric terms
implies that the torus becomes an infinitely long cylinder. At the same time
the two expressions for the partition function in (\ref{ZQFT}) approach 
\begin{equation}
Z(R,L)\approx e^{-LE_{0}(R)}\quad \quad \text{and\quad \quad }Z(R,L)\approx
e^{-LR\,f(R)}\;,  \label{ZRL}
\end{equation}
where $E_{0}(R)$ denotes the non-degenerate ground state energy of $H_{R}$
and $f(R)$ is the bulk\ free energy per unit length at temperature $T=1/R$.
Comparing both equations gives the relation 
\begin{equation}
E_{0}(R)\approx Rf(R)\;.  \label{EF}
\end{equation}

\begin{center}
\includegraphics[width=4cm,height=14.5cm,angle=-90]{torus.epsi} \medskip 
{\small Figure 4.2: From the torus to the cylinder.\bigskip }
\end{center}

This simple observation will turn out to be crucial for the whole approach.
For large temperatures $T=1/R$ compared to the overall mass scale $m$ the
system will approach the UV limit $r:=mR\rightarrow 0,$ where the theory
becomes scale and therefore conformally invariant. Using now the expression (%
\ref{CFTH}) for the Hamiltonian of a CFT on a cylinder the above relation (%
\ref{EF}) specializes to 
\begin{equation}
r\rightarrow 0:\quad \quad Rf_{\text{CFT}}(R)\approx \frac{2\pi }{R}(\Delta
_{0}+\bar{\Delta}_{0})-\frac{\pi }{6R}c=:-\frac{\pi }{6R}\,c_{\text{eff}}\;.
\label{Aff}
\end{equation}
Here we have set the vacuum energy density in the plane to zero and used
that in the $L\rightarrow \infty $ limit only the lowest conformal weights $%
(\Delta _{0},\bar{\Delta}_{0})$ contribute to the partition function. The
quantity $c_{\text{eff}}=c-12d_{0}$ with $d_{0}=\Delta _{0}+\bar{\Delta}_{0}$
being the lowest scaling dimension is called \textbf{effective central charge%
}. In the case of a unitary CFT it coincides with the ordinary central
charge defined in (\ref{Vir}) since the lowest scaling dimension is then
given by the vacuum, $d_{0}=0$. The above result (\ref{Aff}) for the free
energy of a conformal model was derived independently by Affleck and Bloethe
et al. \cite{Aff}.

Thus, provided we are able to calculate the free energy of the massive
quantum field theory and to perform the high-energy limit we can recover the
central charge of the underlying UV conformal field theory. We will now show
how to achieve this in context of integrable models by means of the
thermodynamic Bethe ansatz.

\subsection{The thermodynamic Bethe ansatz equations}

In reference to the above geometry consider an integrable QFT on a circle of
circumference $L$, which we will take to infinity in due course. For
simplicity and to unburden formulas assume for the moment that there is only
one particle species of mass $m$. The latter determines the correlation
length of the system, $\xi =1/m$. Integrability ensures that the number of
particles and their individual momenta are conserved. Thus, the physical
state space decomposes into superselection sectors with fixed particle
number $N$. Restricting ourselves to one sector we can associate to each
state an asymptotic wave-function which is assumed to have similar
properties as known from the Bethe ansatz.

\subsubsection{The Bethe ansatz equations}

Intuitively the argumentation goes as follows \cite
{TBAZam1,TBAZam,TBAKM,Roltba}. Choose a configuration of particle positions $%
\{x_{1},\ldots ,x_{N}\}$ in real space such that $|x_{i}-x_{i+1}|\gg \xi
\,,\;i=1,...,N-1$. (Note that this is always possible, since the limit $%
L\rightarrow \infty $ will eventually be taken.) One therefore might neglect
off-shell contributions and the asymptotic wave function should be the one
of free particles in this region. Exchanging now two-particle positions, say 
$x_{k}$ and $x_{l}$, maps the original configuration to another satisfying
the same requirement. The corresponding asymptotic wave-function in the new
region is assumed to differ from the original one by the two-particle
scattering amplitude $S(\theta _{k}-\theta _{l})$. Hence, when translating
the $k^{\text{th}}$ particle by the period $L$ the wave function collects
all the scattering amplitudes with the other particles in the system. At the
same time we have imposed periodic boundary conditions by defining the
system on a circle, whence this contribution must cancel against the factor
acquired from translation leading to the following restriction on the
rapidities, 
\begin{equation}
e^{iLm\sinh \theta _{k}}\tprod_{l\neq k}^{N}S(\theta _{k}-\theta _{l})=1\;.
\label{BAE}
\end{equation}
This set of constraints is known as Bethe Ansatz equations and is believed
to be tightly linked to the integrable structure of the quantum field theory
under consideration. Taking the logarithm (\ref{BAE}) can be cast into the
equivalent form 
\begin{equation*}
Lm\sinh \theta _{k}+\tsum_{l\neq k}^{N}\delta (\theta _{k}-\theta _{l})=2\pi
n_{k}\;,\quad n_{k}\in \mathbb{Z}\,,
\end{equation*}
where $\delta =-i\ln S$ is the scattering phase. It is common in the
literature to call the allowed rapidities $\theta _{k}$ \emph{roots}. Notice
that for free particles $S=1$ we obtain the familiar quantization conditions
for non-interacting particles in a finite interval. Thus, the general case
can be interpreted as treating the system as a dilute quantum gas. The total
energy and momentum of the system can be deduced from (\ref{BAE}) to be
given by the formulas 
\begin{equation*}
E=m\sum_{k=1}^{N}\cosh \theta _{k}\quad \;\text{and\ \quad }%
P=m\sum_{k=1}^{N}\sinh \theta _{k}=\frac{2\pi }{L}\,\sum_{k=1}^{N}n_{k}\;
\end{equation*}
showing that the integers $n_{k}$ can be interpreted as pseudo-momenta. For
a typical factorizable S-matrix of an integrable model the Bethe ansatz
equations form a complicated system of transcendental equations, which in
general can not be solved directly. One therefore turns to the thermodynamic
limit, i.e. the period $L$ and the particle number $N$ tend to infinity such
that the density $N/L$ remains finite.

\subsubsection{The thermodynamic limit}

From (\ref{BAE}) one can deduce that the spacing between the rapidities $%
\theta _{k}$ is of order $1/L$ whence it is legitimate to introduce
continuous rapidity densities defined by 
\begin{eqnarray}
L\varrho _{p}(\theta )d\theta &=&\text{number of particles in }[\theta
,\theta +d\theta ]\;,  \notag \\
L\varrho _{h}(\theta )d\theta &=&\text{number of holes in }[\theta ,\theta
+d\theta ]\;,  \label{dens}
\end{eqnarray}
where the terminology of a 'hole' refers to a state which is allowed by the
Bethe ansatz equations but is not occupied. Notice that the rapidity
densities are assumed to be macroscopic variables, while the microscopic
structure is described by the solutions of (\ref{BAE}). Now the Bethe ansatz
equations can be re-written in terms of densities, 
\begin{equation}
m\sinh \theta _{k}+\dint_{-\infty }^{\infty }\delta (\theta _{k}-\theta
)\varrho _{p}(\theta )d\theta =\frac{2\pi }{L}\,n_{k}\;.  \label{Lieb}
\end{equation}
An analogue of the above equation was first derived in case of the
one-dimensional Bose gas \cite{ELieb}. Assuming the \textbf{Pauli exclusion
principle} to hold for the solutions the l.h.s. in (\ref{Lieb}) defines for
arbitrary rapidities $\theta $ a monotonic increasing function which upon
differentiation w.r.t. $\theta $ gives the total density $\varrho =\varrho
_{p}+\varrho _{h}$ of allowed states, 
\begin{equation}
\varrho (\theta )=\frac{m}{2\pi }\cosh \theta +\varphi \ast \varrho
_{p}(\theta ),\quad \quad \varphi =-i\frac{d}{d\theta }\ln S(\theta )\;,
\label{Lieb2}
\end{equation}
where the symbol $\ast $ stands for the convolution of two functions, 
\begin{equation}
f\ast g(\theta ):=\int \frac{d\theta ^{\prime }}{2\pi }f(\theta -\theta
^{\prime })g(\theta ^{\prime })\;.
\end{equation}
This integral equation is still not easy to solve. Instead, one discusses
the system in its thermal equilibrium, where the free energy is minimized.
This requires the notion of entropy first.

Since at finite temperature there should exist holes one obtains different
possibilities to arrange the particles leading to microscopic states of
approximately equal energy. Using the defining relation (\ref{dens}) and
exploiting again the exclusion principle the number of possibilities in an
infinitesimal rapidity interval is given by 
\begin{equation*}
\exp d\mathcal{S}=\frac{(L\varrho (\theta )d\theta )!}{\left( L\varrho
_{p}(\theta )d\theta \right) !\left( L\varrho _{h}(\theta )d\theta \right) !}%
\;,
\end{equation*}
whence the entropy per unit length $s=\mathcal{S}/L$ after integration
equals for large particle numbers 
\begin{equation*}
s[\varrho ,\varrho _{p}]\approx \int d\theta \left( \varrho \ln \varrho
-\varrho _{p}\ln \varrho _{p}-\varrho _{h}\ln \varrho _{h}\right) \;.
\end{equation*}
Notice that we used Stirling's formula $\ln k!\approx k\ln k,\;k\gg 1$ to
derive the above identity. The free energy of our QFT at temperature $T=1/R$
is then given by 
\begin{equation*}
f[\varrho ,\varrho _{p}]=e[\varrho _{p}]-1/R\;s[\varrho ,\varrho _{p}]\;,
\end{equation*}
where $e[\varrho _{p}]=\tint d\theta \;m\cosh \theta \,\varrho _{p}(\theta )$
is the internal energy of the system per unit length. Recall that $R$ is the
other period in the toroidal geometry used in Section 4.2.1. Minimizing the
free energy w.r.t. the densities $\varrho ,\varrho _{p}$ by exploiting the
constraint (\ref{Lieb2}) we obtain the properties of the system in thermal
equilibrium. Explicitly, the extremum condition $\delta f=0$ reads 
\begin{equation}
\ln \frac{\varrho _{h}}{\varrho _{p}}+\varphi \ast \ln \left( 1+\frac{%
\varrho _{p}}{\varrho _{h}}\right) =R\,m\cosh \theta \;.  \label{TBA0}
\end{equation}
The above constraint is known as thermodynamic Bethe ansatz equation. The
single quantity to be determined is the ratio of the densities $\varrho
_{p}/\varrho _{h}$ and the only input required is the logarithmic derivative
of the scattering matrix $\varphi =-i\frac{d}{d\theta }\ln S(\theta )$.
Since the Pauli exclusion principle was assumed to hold in the derivation
one usually rewrites the density ratio in form of a Fermi distribution 
\begin{equation}
\frac{\varrho _{p}}{\varrho _{p}+\varrho _{h}}=\frac{1}{e^{\varepsilon }+1}\;
\label{fermi}
\end{equation}
where the function $\varepsilon =\varepsilon (\theta )$ defined by the above
equation is referred to as \textbf{pseudo-energy} for obvious reasons.
However, there is a certain arbitrariness in introducing the distribution (%
\ref{fermi}) and the pseudo-energy $\varepsilon $ on which is discussed
below. The equilibrium condition may now be expressed in terms of the new
variable as 
\begin{equation}
\varepsilon +\varphi \ast L(\theta )=R\,m\cosh \theta \;,\quad L:=\ln \left(
1+e^{-\varepsilon }\right) \;.  \label{TBA1}
\end{equation}
This form of the TBA equations will turn out to be most convenient for
numerical purposes as will become apparent in the next section. Once this
integral equation is solved one is in the position to calculate the free
energy per unit length 
\begin{equation*}
f(R)=-\frac{m}{2\pi R}\int d\theta \,\cosh \theta \,L(\theta )
\end{equation*}
from which any further thermodynamic quantity can be derived. Now in view of
formula (\ref{Aff}) we introduce a normalized free energy which yields in
the UV regime directly the effective central charge, the so-called scaling
function 
\begin{equation}
c(r)=\frac{3\,r}{\pi ^{2}}\int d\theta \,\cosh \theta \,L(\theta )\;.
\label{Scale1}
\end{equation}
Here $r=mR=R/\xi $ is the scale parameter controlling the high energy limit.
Letting the latter tend to zero gives 
\begin{equation}
\lim_{r\rightarrow 0}c(r)=c_{\text{eff}}=c-12d_{0}\;.  \label{CEFF}
\end{equation}
Note that besides the explicit $r$ dependence of the scaling function there
is an additional hidden one entering through the solutions of the TBA
equations (\ref{TBA0}). With regard to our discussion in context of
conformal field theory concerning the physical meaning of the central charge
we might interpret the scaling function as off-critical measure of the
Casimir energy.

It is now straightforward to extend the discussion to the general case when $%
n$ particle species are present in the theory. Let the mass spectrum be $%
\{m_{1},...,m_{n}\}$ and assume that the particle content is non-degenerate,
i.e. the scattering matrix $S_{ij}(\theta )$ is diagonal. Then instead of a
single non-linear integral equation we now obtain a system of $n$ coupled
TBA equations 
\begin{equation}
\varepsilon _{i}(\theta )=r\,m_{i}\cosh \theta -\sum_{j=1}^{n}\varphi
_{ij}\ast L_{j}(\theta ),\quad L_{i}=\ln \left( 1+e^{-\varepsilon
_{i}}\right)  \label{TBAE}
\end{equation}
where $i=1,...,n$ and the pseudo-energies $\varepsilon _{i}$ are defined
analogous as in (\ref{fermi}) with $\varrho _{h},\varrho _{p}\rightarrow
\varrho _{h}^{(i)},\varrho _{p}^{(i)}$ being the densities belonging to the
particle species $i$. Analogously one obtains for the scaling function 
\begin{equation}
c(r)=\frac{3}{\pi ^{2}}\sum_{i=1}^{n}m_{i}^{\prime }\,r\int d\theta \,\cosh
\theta \,L_{i}(\theta )  \label{Scale}
\end{equation}
where the scale parameter is now defined as $r=m_{1}R$ with $m_{1}$ being
the lightest mass in the spectrum and $m_{i}^{\prime }:=m_{i}/m_{1}$.
However, henceforth this re-scaling of the mass spectrum is automatically
implied without changing the notation. Before we now start to discuss the
solutions of the TBA equations there is a different aspect which needs to be
discussed in more detail, statistics and the introduction of the
pseudo-energies.

\subsubsection{Statistical ambiguities}

When performing the thermodynamic limit and introducing the particle
densities we changed from microscopic to macroscopic variables (infinite
particle numbers). The thermodynamic analysis then followed the usual
classical procedure assuming the exclusion principle to hold when setting $%
\varrho =\varrho _{p}+\varrho _{h}$. On the microscopic level this should be
reflected by Fermi statistics which motivated the definition (\ref{fermi})
of the pseudo-energies. Alternatively, it might also occur that Bose
statistics governs the solutions of the TBA equations in which case the
density of holes just equals the density of all possible states, i.e. $%
\varrho =\varrho _{h}$. Following the analogous steps in the thermodynamic
analysis it would now be natural to define the pseudo-energies by $\varrho
_{p}/\varrho =1/(e^{\tilde{\varepsilon}}-1)$ leading to a TBA equation of
the form 
\begin{equation}
\tilde{\varepsilon}(\theta )+\varphi \ast \tilde{L}(\theta )=R\,m\cosh
\theta \;,\quad \tilde{L}:=-\ln (1-e^{-\tilde{\varepsilon}})\;.
\label{TBAbos}
\end{equation}
Inside the TBA approach the decision for either Fermi (\ref{TBA1}) or Bose
statistics (\ref{TBAbos}) is usually motivated by the symmetry property of
the asymptotic wave-function under an exchange of two identical particles 
\cite{TBAZam1}. The latter is assumed to be anti-symmetric for $S(0)=-1$ and
symmetric for $S(0)=1$ describing Fermi and Bose statistics, respectively.
However, this argument is rather ad hoc and should be viewed as a working
hypothesis.

An important observation in this context is that one might change from the
fermionic to the bosonic description by altering the S-matrix by a phase of
the following kind 
\begin{equation}
S(\theta )\rightarrow S^{\prime }(\theta )=e^{-2\pi i\,\Theta (\limfunc{Re}%
\theta )}S(\theta )\;.
\end{equation}
Here $\Theta $ denotes the Heavyside step function with $\Theta (0)=1/2$.
Notice that the phase factor leaves the bootstrap properties unchanged and
hence can be interpreted in terms of a CDD ambiguity \cite{BFs}. In terms of
the TBA-kernel this amounts to the replacement 
\begin{equation}
\varphi (\theta )\rightarrow \varphi ^{\prime }(\theta )=\varphi (\theta
)-2\pi \delta (\theta )  \label{BosFerm}
\end{equation}
in the TBA equation (\ref{TBA1}). Upon identifying $\tilde{\varepsilon}%
=\varepsilon +\ln (1+e^{-\varepsilon })$ one then obtains (\ref{TBAbos}).
Thus, we might always fix a different choice of statistics in terms of the
pseudo-energies by a re-definition of the TBA kernel. This important
observation that the TBA equations stay invariant under a simultaneous
change of the phase of the S-matrix and the introduction of a different
statistics was made in \cite{BFs}. Therein the more general case of Haldane
statistics \cite{Haldane} was also discussed.

\emph{When considering the S-matrices constructed in Chapter 3 below we will
always assume Fermi statistics to hold for the solutions of the TBA equations%
} due to the fact that they satisfy $S_{ii}(0)=S_{ii}^{aa}(0)=-1$ (compare
Chapter 3 for the notation) and that no other analytical and numerical
solutions with different statistics are known. Although the integrable
models under consideration, affine Toda field theories, are defined in terms
of bosonic fields via the action functional (\ref{ADEaction}) the
introduction of Fermi statistics at the TBA level does not lead to a
contradiction. For example we will recover Bose statistics from the
solutions of the TBA equations in certain limits, like the weak coupling in
affine Toda theory or the semiclassical limit in the homogeneous Sine-Gordon
models. \emph{Indeed, the statistics of the TBA solutions is not necessarily
the same as the statistics of the fields in which the classical action
functional is defined. }Recall in particular that the concept of bosons and
fermions is equivalent in 1+1 dimensions, i.e. bosonic fields might be
expressed in terms of fermionic ones and vice versa (see e.g. \cite{BosFer}).

\subsubsection{The central charge calculation and Roger's dilogarithm}

The most important quantity we are interested in the TBA analysis is the
effective central charge of the underlying UV conformal field theory. While
in general the TBA equations (\ref{TBAE}) cannot be solved analytically for
finite temperatures due to their non-linear nature and therefore also the
explicit form of the off-critical scaling function remains undetermined, the
situation improves in the UV regime, $r\rightarrow 0$. Following the
standard procedure developed in \cite{TBAZam1} we demonstrate how the
central charge might be extracted by analytical means specializing for
simplicity again to the case that only one particle species is present. The
first step is to set up a set of so-called ``massless'' TBA equations in the
rapidity range $\ln \frac{r}{2}\ll \theta \ll \ln \frac{2}{r}$ with $r\ll 1$%
. The nomenclature ``massless'' originates in the following approximative
replacement of the on-shell energies when performing the shift $\theta
\rightarrow \theta -\ln \frac{r}{2}$%
\begin{equation*}
r\,\cosh (\theta -\ln \tfrac{r}{2})\approx e^{\theta }\;.
\end{equation*}
Here we have assumed that $\theta \gg 2\ln \frac{r}{2}$ and neglected terms
of order $r^{2}$. Upon introducing the functions $\hat{\varepsilon}(\theta
)=\varepsilon (\theta -\ln \frac{r}{2}),\hat{L}(\theta )=L(\theta -\ln \frac{%
r}{2})$ the TBA equations (\ref{TBAE}) in the above mentioned rapidity range
are approximately given by 
\begin{equation}
r\ll 1:\quad e^{\theta }\approx \hat{\varepsilon}(\theta )+\varphi \ast \hat{%
L}(\theta )\;.  \label{uvTBA}
\end{equation}
Notice that the explicit dependence on the scale parameter $r$ has vanished.
We might now proceed similarly and perform also the shift $\theta
\rightarrow \theta -\ln \frac{2}{r}$ and obtain an analogous equation
involving on-shell energies of the kind $e^{-\theta }$ in the appropriate
range for the rapidity variable. However, at this point we make the
simplifying assumption that our theory is invariant under a parity
transformation, $\theta \rightarrow -\theta ,$ i.e. the solutions of the
original TBA equations are assumed to be symmetric. Then it is sufficient to
consider only the equation (\ref{uvTBA}) and the expression for the scaling
function simplifies to 
\begin{equation}
r\ll 1:\quad c(r)=\frac{6}{\pi ^{2}}\int\limits_{0}^{\infty }d\theta
\;L(\theta )\,\cosh \theta \approx \frac{6}{\pi ^{2}}\int\limits_{\ln \frac{r%
}{2}}^{\infty }d\theta \;\hat{L}(\theta )\,e^{\theta }\;,  \label{uvScale}
\end{equation}
where we have again discarded terms of order $r^{2}$ as in the derivation of
the approximate TBA equation (\ref{uvTBA}) in the UV regime. Now, the
crucial trick is to differentiate the latter and to replace the exponential
term in (\ref{uvScale}) by the r.h.s. which yields 
\begin{eqnarray*}
c(r) &\approx &\frac{6}{\pi ^{2}}\int\limits_{\hat{\varepsilon}(\ln \frac{r}{%
2})}^{\hat{\varepsilon}(\infty )}d\varepsilon \;\ln (1+e^{-\varepsilon })+%
\frac{6}{\pi ^{2}}\int\limits_{\ln \frac{r}{2}}^{\infty }d\theta \;\hat{L}%
(\theta )\,\varphi \ast \frac{d\hat{L}(\theta )}{d\theta } \\
&\approx &\frac{6}{\pi ^{2}}\int\limits_{\hat{\varepsilon}(\ln \frac{r}{2}%
)}^{\hat{\varepsilon}(\infty )}d\varepsilon \;\ln (1+e^{-\varepsilon })+%
\frac{6}{\pi ^{2}}\int\limits_{\ln \frac{r}{2}}^{\infty }d\theta \;\frac{d%
\hat{L}(\theta )}{d\theta }\,\varphi \ast \hat{L}(\theta )\;.
\end{eqnarray*}
The interchange of the two integrations in the second step ought to be exact
in the extreme UV limit $r\rightarrow 0$. Exploiting again the equation (\ref
{uvTBA}) in replacing the convolution term of the second integral and a
subsequent partial integration gives the final result 
\begin{eqnarray}
\lim_{r\rightarrow 0}c(r) &=&\frac{3}{\pi ^{2}}\int\limits_{\varepsilon
(0)}^{\varepsilon (\infty )}d\varepsilon \;\left[ \ln (1+e^{-\varepsilon })+%
\frac{\varepsilon }{1+e^{\varepsilon }}\right]  \label{ceff0} \\
&=&\frac{6}{\pi ^{2}}\mathcal{L}\left( \frac{1}{1+e^{\varepsilon (0)}}%
\right) -\frac{6}{\pi ^{2}}\mathcal{L}\left( \frac{1}{1+e^{\varepsilon
(\infty )}}\right) \;.  \notag
\end{eqnarray}
Here we have used the integral representation of Roger's dilogarithm
function \cite{Log} 
\begin{equation}
\mathcal{L}(x):=\sum_{k=1}^{\infty }\frac{x^{k}}{k^{2}}+\frac{1}{2}\ln x\ln
(1-x)=-\frac{1}{2}\int\limits_{0}^{x}dy\,\left[ \frac{\ln (1-y)}{y}+\frac{%
\ln y}{1-y}\right]  \label{roger}
\end{equation}
and employed the substitution $y=(1+e^{\varepsilon })^{-1}$. From the above
formula we infer that only the values of the TBA solutions $L(\theta )$ at
the origin and at infinity are required to compute the central charge. The
latter can be deduced to $\lim_{\theta \rightarrow \infty }L(\theta )=0$ on
physical grounds. At large rapidities the on-shell energies in (\ref{TBAE})
become dominant and the system becomes approximately free, i.e. $\varepsilon
(\theta )\approx r\cosh \theta $ for small but finite scale parameter. Thus,
the second term in (\ref{ceff0}) does not contribute. It is straightforward
to generalize this result to the general case with $n$ particle species
present, giving the compact formula 
\begin{equation}
c_{\text{eff}}=\frac{6}{\pi ^{2}}\sum_{i=1}^{n}\mathcal{L}\left( \frac{1}{%
1+e^{\varepsilon _{i}(0)}}\right) =\frac{6}{\pi ^{2}}\sum_{i=1}^{n}\mathcal{L%
}\left( 1-e^{-L_{i}(0)}\right) \;.  \label{ceffg}
\end{equation}
This formula is central in the TBA analysis and will be frequently used in
the following whence its derivation has been presented here in some detail.
To close the analytical calculation it remains to determine the value of the
TBA solutions at the origin as $r\rightarrow 0$. For the case when a
dynamical interaction between the particles is absent this is easily
achieved. As example consider systems with Fermi and Bose statistics. In the
first case we infer directly from (\ref{TBA1}) that the solution to the TBA
equation reads $L(\theta )=\ln (1+e^{-r\cosh \theta })$, whence we infer
from the general formula (\ref{ceffg}) that $c_{\text{eff}}=\frac{6}{\pi ^{2}%
}\mathcal{L}(\frac{1}{2})=1/2$. In accordance with the well known central
charge of the free fermion theory. For Bose statistics we must keep in mind
that the derivation of (\ref{ceffg}) was performed in context of Fermi
statistics. However, exploiting the statistical ambiguity explained in the
previous subsection we saw that (\ref{TBAbos}) is equivalent to (\ref{TBA1})
under the transformation (\ref{BosFerm}), whence we easily derive $L(\theta
)=-\ln (1-e^{-r\cosh \theta })$ and the effective central charge takes the
expected value $c_{\text{eff}}=\frac{6}{\pi ^{2}}\mathcal{L}(1)=1$.

The general case with a dynamical interaction present is more involved and
there exists various schemes which allow to determine the solution at the
origin. However, these depend crucially on the form of the TBA solution,
whence one refers usually to the outcome of numerical calculations before
one turns to analytical considerations. The numerical procedure applied to
solve (\ref{TBAE}) is described in the next section together with the
question whether the TBA equations admit more than one solution.

\subsection{Existence, uniqueness and the numerical procedure}

In this section we are going to investigate the existence and uniqueness
properties of the solutions of the TBA equations \cite{FKS1}. The main
physical motivation for this considerations is to clarify whether it is
possible to obtain different effective central charges for a fixed dynamical
interaction due to the existence of several different solutions. As a side
product we obtain useful estimates on the error and the rate of convergence
when solving the TBA equations numerically. Precise estimates of this kind
were not obtained previously in this context and convergence is simply
presumed. The procedure we are going to employ is the \textbf{contraction
principle} (or \textbf{Banach fixed point theorem}), see e.g. \cite{func}.
For the case of the non-relativistic one-dimensional Bose gas the uniqueness
question was already addressed by Yang and Yang \cite{Yang}, albeit with a
different method.

In order to keep the notation simple we commence our discussion for a system
with one particle only which is of fermionic type. Thereafter, we discuss
the straightforward generalization. The standard numerical procedure to
solve integral equations of the type (\ref{TBA1}) consists in evaluating the
original function at discrete points and a subsequent iteration. Introducing
the scale parameter $r=mR$, this means we consider (\ref{TBA1}) as recursive
equation 
\begin{equation}
\varepsilon _{n+1}(\theta ):=r\cosh \theta -\varphi \ast \ln
(1+e^{-\varepsilon _{n}})(\theta )  \label{sequence}
\end{equation}
and perform the iteration starting with $\varepsilon _{0}(\theta )=r\cosh
\theta $. The exact solution is then thought to be the limit $\varepsilon
=\lim_{n\rightarrow \infty }\varepsilon _{n}$. However, a priori it is not
clear whether this limit exists at all and how it depends on the initial
value $\varepsilon _{0}$. In particular, different initial values might lead
to different solutions.

The natural mathematical setup for this type of problem is to rewrite the
TBA-equation (\ref{TBA1}) as 
\begin{equation}
(A\xi )(\theta ):=\varphi \ast \ln (1+e^{\xi (\theta )-r\cosh \theta })=\xi
(\theta ),  \label{fixed}
\end{equation}
and treat it as a fixed point problem for the operator $A$ with $\xi (\theta
)=r\cosh \theta -\varepsilon (\theta )$.

In order to give meaning to the limit $\lim_{n\rightarrow \infty
}\varepsilon _{n}$ we have to specify a norm. Of course, it is natural to
assume that $\xi $ as function of $\theta $ is measurable, continuous and
essentially bounded on the whole real line. The latter assumption is
supported by all known numerical results. In fact, assuming on physical
grounds that $\varphi \sim e^{-|\theta |}$ for large rapidities, since the
particles become then approximately non-interacting, it follows from (\ref
{fixed}), that possible solutions $\xi $ vanish at infinity. This means
possible solutions of (\ref{fixed}) constitute a Banach space with respect
to the norm

\begin{equation}
\left\| f\right\| _{\infty }=\text{ess}\sup \left| f(\theta )\right| ,
\end{equation}
i.e. $L_{\infty }(\mathbb{R})$. In principle we are now in a position to
apply the Banach fixed point theorem\footnote{%
One may of course apply different types of fixed point theorems exploiting
different properties of the operator $A$. For instance if $A$ is shown to be
compact one can employ the Leray-Schauder fixed point theorem. In the second
refence of \cite{TBAKM} it is claimed that the problem at hand was treated
in this manner, albeit a proof was not provided.}, which states the
following:\medskip

\emph{Let }$D\subset L_{\infty }$\emph{\ be a non-empty set in a Banach
space and let }$A$\emph{\ be an operator which maps }$D$\emph{\ }$q$\emph{%
-contractively into itself, i.e. for all }$f,g\in D$\emph{\ and some fixed }$%
q$\emph{, }$0\leq q<1$\emph{\ } 
\begin{equation}
\Vert A(f)-A(g)\Vert _{\infty }\leq q\Vert f-g\Vert _{\infty }.
\label{contraction}
\end{equation}
\emph{Then the following statements hold:}

\begin{enumerate}
\item[i)]  \emph{There exists a unique fixed point }$\xi $\emph{\ in }$D$%
\emph{, i.e. equation (\ref{fixed}) has exactly one solution.}

\item[ii)]  \emph{The sequence constructed in (\ref{sequence}) by iteration
converges to the solution of (\ref{fixed}).}

\item[iii)]  \emph{The error of the iterative procedure may be estimated by }
\begin{equation*}
\Vert \xi -\xi _{n}\Vert _{\infty }\leq \frac{q^{n}}{1-q}\Vert \xi _{1}-\xi
_{0}\Vert _{\infty }\quad \text{and\quad }\left\| \xi -\xi _{n+1}\right\|
_{\infty }\leq \frac{q}{1-q}\left\| \xi _{n+1}-\xi _{n}\right\| _{\infty }.
\end{equation*}

\item[iv)]  \emph{The rate of convergence is determined by } 
\begin{equation*}
\left\| \xi -\xi _{n+1}\right\| _{\infty }\leq q\left\| \xi -\xi
_{n}\right\| _{\infty }.
\end{equation*}
\end{enumerate}

In order to be able to apply the theorem we first have to choose a suitable
set in the Banach space. We choose some $q\in \lbrack 0,1)$ such that $%
e^{-r}\leq q$ and take $D$ to be the convex\footnote{%
For $f,g\in D_{q,r}$ also $tf+(1-t)g\in D_{q,r}$, with $0\leq t\leq 1$.} set 
$D_{q,r}:=\left\{ f:\text{ }\Vert f\Vert _{\infty }\leq \ln \frac{q}{1-q}%
+r\right\} $. We may now apply the following estimate for the convolution
operator $\varphi \ast $ (which is a special case of Young's inequality) 
\begin{equation}
\Vert \varphi \ast f\Vert _{\infty }\leq \Vert \varphi \Vert _{1}\Vert
f\Vert _{\infty }.  \label{convolution}
\end{equation}
In the following we assume that $\Vert \varphi \Vert _{1}\leq 1$. In fact,
for the concrete one-particle theory we will consider below, the Sinh-Gordon
model, we have $\Vert \varphi \Vert _{1}:=\int \frac{d\theta }{2\pi }%
|\varphi (\theta )|=1$. Interpreting the function $L(\theta )=\ln
(1+e^{f(\theta )-r\cosh \theta })$ in (\ref{fixed}) as operator acting on $%
f\in D$ we have the estimate 
\begin{equation*}
L(f)\leq \ln \left[ 1+\exp \left( \left\| f\right\| _{\infty }-r\right) 
\right] \leq \ln \frac{1}{1-q}\leq \ln \frac{q}{1-q}+r\;.
\end{equation*}
The last inequality follows from our special choice of $q$. Thus, $A$ maps $%
D_{q,r\text{ }}$into itself.

In the final step we show that the contraction property (\ref{contraction})
is fulfilled on $D_{q,r\text{ }}$. It suffices to prove this for the map $L$%
, because of (\ref{convolution}) and the fact that $\Vert \varphi \Vert
_{1}\leq 1$. We have 
\begin{eqnarray*}
\Vert L(f)-L(g)\Vert _{\infty } &=&\left\| \tint_{0}^{1}dt\,\frac{d}{dt}%
L(g+t(f-g))\right\| _{\infty } \\
&=&\left\| \tint_{0}^{1}dt\,\frac{(f-g)}{1+\exp (-g-t(f-g)+r\cosh \theta )}%
\right\| _{\infty } \\
&\leq &\max_{0\leq t\leq 1}\left| \frac{1}{1+\exp (-g-t(f-g)+r)}\right|
\left\| f-g\right\| _{\infty } \\
&\leq &\max_{0\leq t\leq 1}\left| \frac{1}{1+\exp \left( -\left\|
g-t(f-g)\right\| _{\infty }+r\right) }\right| \left\| f-g\right\| _{\infty }
\\
&\leq &q\left\| f-g\right\| _{\infty }\,\,.
\end{eqnarray*}
In the last inequality we used the fact that $D_{q,r\text{ }}$ is a convex
set.

We may now safely apply the fixed point theorem. First of all we conclude
from i) and ii) that a solution of (\ref{fixed}) not only exists, but it is
also unique. In addition we can use iii) and iv) as a criterium for error
estimates. From our special choice of the closed set $D_{q,r}$ one sees that
the rate of convergence depends crucially on the parameter $r$, the smaller $%
r$ the greater $q$ is, whence the sequence $(\xi _{n})$ converges slower.

One could be mathematically more pedantic at this point and think about
different requirements on the function $\xi $. For instance one might allow
functions which are not bounded (there is no known example except when $r=0$
exactly) and then pursue similar arguments as before on $L_{p}$ rather than $%
L_{\infty }$.

The generalization of the presented arguments to a situation involving $n$
different types of particles may be carried out by the same arguments even
though it turns out to be more involved due to the coupling of the equations
belonging to different particle species.

\subsection{Approximate analytical solutions}

Having established how the TBA equations can be solved numerically by means
of the contraction principle we now turn to analytical considerations, see 
\cite{FKS1,FKtba}. The motivation is twofold. On the one hand the numerical
problem becomes quite complex when one increases the number of particle
species, on the other hand and more importantly one would like to gain a
deeper structural insight into the solutions of (\ref{TBAE}). Due to the
nonlinear nature of the TBA equations only few analytical solutions are
known. Nonetheless, one may obtain approximate analytical solutions when $r$
tends to zero, i.e. in the UV limit. These approximations depend on the form
of the solutions which vary dependent on the model.

In this section we shall assume that the particle densities become large in
the high energy regime, i.e. the solutions $L_{i}(\theta )$ of (\ref{TBAE})
should be much larger than one inside a rapidity range specified below. This
kind of behaviour will be found for all of the affine Toda field theories.

Nevertheless, most of the approximation scheme can be discussed in complete
generality and only in the last step when comparing against numerical data
one needs to specify a particular model. Therefore, we present the
derivation of the approximate formulas separately and justify the
assumptions made about the particular form of the solutions in retrospective
when discussing the concrete examples of affine Toda field theory.

We shall now generalize the method of \cite{ZamoR} which can be separated
into three distinct steps:

\begin{enumerate}
\item  Instead of regarding the TBA-equation as an integral equation, one
transforms it into an infinite order differential equation by means of the
convolution theorem.

\item  In the high-energy regime this infinite order differential equation
can for large densities be approximated by a second order differential
equation giving the leading contribution in the UV limit.

\item  After solving the finite order differential equations the solution is
used to compute an approximate scaling function.
\end{enumerate}

In this approach certain constants of integration are left undetermined,
whence in a second part we discuss further enhanced approximate solutions to
the TBA equation and provide a simple matching condition which allows to fix
the constants for a concrete model at hand.

\subsubsection{The TBA equation as infinite order differential equation}

The first step can be performed with the sole assumption that the Fourier
transform of the TBA kernel $\varphi _{ij}(\theta )=-i\frac{d}{d\theta }\ln
S_{ij}(\theta )$ can be expanded as a power series 
\begin{equation}
\tilde{\varphi}_{ij}(t)-2\pi \delta _{ij}:=\int\limits_{-\infty }^{\infty
}d\theta \;\varphi _{ij}(\theta )e^{it\theta }-2\pi \delta _{ij}=2\pi
\sum_{k=0}^{\infty }(-i)^{k}\eta _{ij}^{(k)}t^{k}\quad ,  \label{series}
\end{equation}
where for convenience we have subtracted a constant term which only effects
the definition of the zeroth order coefficient $\eta _{ij}^{(0)}$. In view
of the discussion in 4.2.2 this amounts to a reformulation of (\ref{TBAE})
in terms of Bose statistics, compare (\ref{BosFerm}). The coefficient $\eta
_{ij}^{(0)}$ will then turn out to vanish in context of ATFT and Fermi
statistics. Now, it is a simple consequence of the convolution theorem%
\footnote{$(f*g)(\theta )$ $=1/(2\pi )^{2}\int dk\widetilde{f}(k)\widetilde{g%
}(k)e^{-ik\theta }$} that the integral equations (\ref{TBAE}) may also be
written as a set of infinite order differential equations \cite
{ZamoR,MM,FKS1} 
\begin{equation}
rm_{i}\cosh \theta +\ln \left( 1-e^{-L_{i}(\theta )}\right)
=\sum\limits_{j=1}^{n}\sum\limits_{k=0}^{\infty }\eta
_{ij}^{(k)}L_{j}^{(k)}(\theta )\quad .  \label{TBADiff}
\end{equation}
Here we have expressed the pseudo-energies in terms of the $L$-function and
introduced the abbreviation $L_{i}^{(k)}(\theta )=(d/d\theta
)^{k}L_{i}(\theta )$. The whole dependence of the scattering matrix is now
incorporated in the coefficients $\eta _{ij}^{(k)}$. This alternative
formulation of the TBA-equations is most convenient for the analytical
considerations to follow.

\subsubsection{The ultraviolet regime}

In the second step we shall now pass on to the ultraviolet limit, i.e. the
scale parameter $r$ is going to zero. Similar to the general discussion of
the central charge calculation in 4.2.2 we introduce the quantity $\hat{L}%
_{i}(\theta ):=L_{i}(\theta -\ln \frac{r}{2})$ and perform the shift $\theta
\rightarrow \theta -\ln \frac{r}{2}.$ The TBA-equations (\ref{TBADiff}) then
acquire the form 
\begin{equation}
m_{i}e^{\theta }+\ln \left( 1-e^{-\hat{L}_{i}(\theta )}\right)
=\sum\limits_{j=1}^{n}\sum\limits_{k=0}^{\infty }\eta _{ij}^{(k)}\hat{L}%
_{j}^{(k)}(\theta )\,\,,  \label{TBAshift}
\end{equation}
where we have neglected the terms proportional to $e^{2\ln (r/2)-\theta }$,
under the assumption that $2\ln \frac{r}{2}\ll \theta $. Obviously the
dependence on the scale parameter $r$ has vanished, such that the $\hat{L}%
_{i}(\theta )$ are $r$-independent. Analogous manipulations can be applied
to the equation for the scaling function (\ref{Scale}). Assuming parity
invariance of the scattering matrix one easily verifies that the solutions
of (\ref{TBAE}) must be symmetric in the rapidity variable. Thus, rewriting
the integral such that the integration variable runs only over positive
values one obtains after a similar shift as in the derivation of (\ref
{TBAshift}) the approximate expression (compare also 4.2.2), 
\begin{equation}
r\rightarrow 0:\quad c(r)\approx \frac{6}{\pi ^{2}}\sum%
\limits_{i=1}^{n}m_{i}\int\limits_{\ln \frac{r}{2}}^{\infty }d\theta \,\hat{L%
}_{i}(\theta )e^{\theta }\;.  \label{cshift}
\end{equation}
Again terms proportional to $r^{2}$ have been neglected in the above
expression.

\subsubsection{Assumptions on the solutions and leading order behaviour}

At this point we make now several assumptions on the coefficients in the
series expansion (\ref{series}) and the behaviour of the solutions $\hat{L}%
_{i}$ which will be justified in the next section for the concrete models at
hand:

\begin{enumerate}
\item[i)]  The functions $\hat{L}_{i}(\theta )$ obey the equations (\ref
{TBAshift}).

\item[ii)]  In the power series expansion (\ref{series}) the coefficients
are symmetric in the particle type indices, i.e. $\eta _{ij}^{(k)}=\eta
_{ji}^{(k)}$.

\item[iii)]  All odd coefficients vanish in (\ref{series}), i.e. $\eta
_{ij}^{(2k+1)}=0$.

\item[iv)]  The asymptotic behaviour of the function $\hat{L}_{i}(\theta )$
and its derivatives read 
\begin{eqnarray}
\lim_{\theta \rightarrow \infty }\hat{L}_{i}^{(k)}(\theta ) &=&0\qquad \text{%
for }k\geq 1\,,1\leq i\leq n\quad ,  \label{a1} \\
\lim_{\theta \rightarrow \infty }e^{\theta }\hat{L}_{i}(\theta ) &=&0\qquad 
\text{for }1\leq i\leq n\quad .  \label{a2}
\end{eqnarray}
\end{enumerate}

Assumption ii) is guaranteed when the two-particle scattering matrix is
parity invariant. The requirement iii) will turn out to be satisfied for
Fermi statistics by all scattering matrices of interest to us. The
asymptotic behaviour iv) will be verified in retrospect, that is all known
numerical solutions exhibit this kind of asymptotics. Noting that the
conditions (\ref{a1}),(\ref{a2}) imply $\lim_{r,\theta \rightarrow
0}L_{i}^{(k)}(\theta )=0$ such that (\ref{TBADiff}) becomes a set of coupled
equations for $n$ constants $L_{i}(0)$, 
\begin{equation*}
\ln (1-e^{-L_{i}(0)})=\sum_{j=1}^{n}\eta _{ij}^{(0)}L_{j}(0)\;.
\end{equation*}
Solving this equation allows to determine the effective central charge via
the general formula (\ref{ceffg}). Later on we will see that in context of
affine Toda theories $\eta _{ij}^{(0)}=0$ for Fermi statistics whence the
above equation implies via (\ref{ceffg}) that $c_{\text{eff}}=n$. However,
our aim will be to go beyond the renormalization fixed point and calculate
the leading order perturbation term in the off-critical scaling function.

In order to deduce the leading order behaviour in the ultraviolet limit we
now proceed as follows: Under the assumption that the property $\hat{L}%
_{i}(\theta )=\hat{L}_{i}(2\ln \frac{r}{2}-\theta )$ originating in the
symmetry of the non-shifted solutions $L_{i}(\theta )$ still holds at this
point of the derivation, we may neglect in (\ref{TBAshift}) also the term $%
e^{\theta }$. \emph{Assuming that }$\hat{L}_{i}(\theta )$\emph{\ is large}
and (\ref{a1}) holds, the TBA-equation in the ultraviolet limit may be
approximated by 
\begin{equation}
\sum\limits_{j=1}^{n}(\eta _{ij}^{(2)}\hat{L}_{j}^{(2)}(\theta )+\eta
_{ij}^{(0)}\hat{L}_{j}(\theta ))+e^{-\hat{L}_{i}(\theta )}=0\quad .
\label{APPdiff}
\end{equation}
Unfortunately, this equation may not be solved analytically in its full
generality. However, in all cases we shall be considering in the following $%
\eta _{i}^{(0)}=\sum_{j=1}^{n}\eta _{ij}^{(0)}=0$ with fermionic type of
statistics. For this case the solution of (\ref{APPdiff}) reads 
\begin{equation}
\hat{L}_{i}(\theta )=\ln \left( \frac{\sin ^{2}\left( \kappa _{i}\left(
\theta -\delta _{i}\right) \right) }{2\kappa _{i}^{2}\eta _{i}^{(2)}}\right)
+\ln \left( \frac{\cos ^{2}\left( \tilde{\kappa}_{i}\left( \theta -\tilde{%
\delta}_{i}\right) \right) }{2\tilde{\kappa}_{i}^{2}\eta _{i}^{(2)}}\right)
\quad ,  \label{Soll}
\end{equation}
with $\kappa _{i},\delta _{i},\tilde{\kappa}_{i},\tilde{\delta}_{i}$ being
constants of integration and $\eta _{i}^{(2)}=\sum_{j=1}^{n}\eta _{ij}^{(2)}$%
. We will discard the second term in the following w.l.g. Invoking the
property $\hat{L}_{i}(\theta )=\hat{L}_{i}(2\ln \frac{r}{2}-\theta )$ of the
original function we obtain the following relation between the constants, 
\begin{equation}
\kappa _{i}=\frac{\pi }{2(\delta _{i}-\ln \frac{r}{2})}\;.  \label{ccc}
\end{equation}
Thus, we recovered the scale dependence of the solutions and only one
constant of integration remains undetermined. Not surprisingly it will be
characteristic for the specific model under consideration and we therefore
postpone the discussion of further restrictions on $\kappa _{i},\delta _{i}$%
. Now, removing the shift in the definition of the $\hat{L}_{i}$-functions
we have as approximate solutions to the TBA equations, 
\begin{equation}
L_{i}^{0}(\theta )=\ln \left( \frac{\cos ^{2}\left( \kappa _{i}\theta
\right) }{2\kappa _{i}^{2}\eta _{i}}\right) \quad \quad \quad \text{for }%
\left| \theta \right| \leq \frac{\arccos (\kappa _{i}\sqrt{2\eta _{i}})}{%
\kappa _{i}}.  \label{Lapp}
\end{equation}
Here the upper index shall indicate that we are dealing with an approximate
solution $L_{i}^{0}$ different form the exact one $L_{i}$. The restriction
on the range of the rapidity stems from the physical requirement $L_{i}\geq
0 $. In order to demonstrate consistency of the approximate solution with
regard to the assumption that the derivatives of $L_{i}(\theta )$ with
respect to $\theta $ are negligible as $r\rightarrow 0$, i.e. equation (\ref
{a1}), we report the following derivatives 
\begin{eqnarray}
(L_{i}^{0})^{(1)}(\theta ) &=&-2\kappa _{i}\tan (\kappa _{i}\theta )\quad
\quad  \label{L1} \\
(L_{i}^{0})^{(2)}(\theta ) &=&-2\kappa _{i}^{2}/\cos ^{2}(\kappa _{i}\theta )
\\
(L_{i}^{0})^{(3)}(\theta ) &=&-4\kappa _{i}^{3}\tan (\kappa _{i}\theta
)/\cos ^{2}(\kappa _{i}\theta ) \\
(L_{i}^{0})^{(k)}(\theta ) &\sim &\kappa _{i}^{k}\quad .  \label{L4}
\end{eqnarray}
Using the fact that $\kappa _{i}$ tends to zero for small $r$, the equations
(\ref{L1})-(\ref{L4}) are compatible with the assumption. Closer inspection
shows that for given $r$ the series build from the $L_{i}^{0}(\theta )$
starts to diverge at a certain value of $k$. Since (\ref{Lapp}) is not exact
this does not pose any problem, but one should be aware of it.

\subsubsection{The scaling function to leading order}

We now turn to the main quantity of interest in the TBA approach, the
scaling function. In order to apply the above approximation to its
computation one considers the so-called ``truncated scaling function''\cite
{ZamoR}, 
\begin{equation}
\hat{c}(r,r^{\prime })=\frac{6}{\pi ^{2}}\sum\limits_{i=1}^{n}m_{i}\int%
\limits_{r^{\prime }}^{\infty }d\theta \,\hat{L}_{i}(\theta )e^{\theta
}\quad ,  \label{truncc}
\end{equation}
which obviously coincides with (\ref{cshift}) for $r^{\prime }=$ $\ln \frac{r%
}{2}$. Recall that $\hat{L}_{i}$ are the solutions of (\ref{TBAshift}). The
reason of introducing the dummy variable $r^{\prime }$ is that we might
derive now a differential equation for the truncated version of the scaling
function, 
\begin{equation}
\frac{\partial \hat{c}(r,r^{\prime })}{\partial r^{\prime }}=-\frac{6}{\pi
^{2}}e^{r^{\prime }}\sum\limits_{i=1}^{n}m_{i}\hat{L}_{i}(r^{\prime })\quad .
\label{Diff}
\end{equation}
Using this differential equation together with the boundary condition $\hat{c%
}(r,\infty )=0$ one may now verify by direct substitution that under the
assumptions i) -- iv) on the solutions $\hat{L}_{i}$ the truncated scaling
function may also be written as 
\begin{eqnarray}
\hat{c}(r,r^{\prime }) &=&\frac{3}{\pi ^{2}}\sum\limits_{i,j=1}^{n}\left(
\sum\limits_{k=1}^{\infty }\eta
_{ij}^{(2k)}\sum\limits_{l=1}^{2k-1}(-1)^{l+1}\hat{L}_{i}^{(l)}(r^{\prime })%
\hat{L}_{j}^{(2k-l)}(r^{\prime })+\eta _{ij}^{(0)}\hat{L}_{i}(r^{\prime })%
\hat{L}_{j}(r^{\prime })\right)  \notag \\
&&\!\!\!-\frac{6}{\pi ^{2}}\sum\limits_{i=1}^{n}\left( \mathcal{L}(1-e^{-%
\hat{L}_{i}(r^{\prime })})+\frac{\hat{L}_{i}(r^{\prime })}{2}\ln (1-e^{-\hat{%
L}_{i}(r^{\prime })})+m_{i}e^{r^{\prime }}\hat{L}_{i}(r^{\prime })\right) .
\label{Solu}
\end{eqnarray}
Here $\mathcal{L}(x)$ denotes Rogers dilogarithm (\ref{roger})\footnote{%
The identity $\int\limits_{0}^{x}dy\ln (1-e^{-y})=\mathcal{L}%
(1-e^{-x})+x/2\ln (1-e^{-x})$ is useful in this context.}. Proceeding now
with the same assumptions which lead to the derivation of (\ref{APPdiff})
and (\ref{Lapp}) the leading order expression for the truncated scaling
function (\ref{Solu}) becomes 
\begin{equation}
\hat{c}(r,r^{\prime })=n+\frac{3}{\pi ^{2}}\sum\limits_{i=1}^{n}\left( \eta
_{i}^{(2)}\left( d\hat{L}_{i}^{0}/d\theta (r^{\prime })\right) ^{2}-2e^{-%
\hat{L}_{i}^{0}(r^{\prime })}\right) \quad .  \label{ctrunc}
\end{equation}
Substitution of the solution (\ref{Soll}) into (\ref{ctrunc}) yields 
\begin{equation}
\hat{c}(r,r^{\prime })=n-\frac{12\,}{\pi ^{2}}\sum\limits_{i=1}^{n}\eta
_{i}^{(2)}\kappa _{i}^{2}\quad .  \label{ctrun1}
\end{equation}
Notice that this expression for the truncated effective central charge is
independent of the dummy variable $r^{\prime }$. We can use the latter
property to argue that in fact the r.h.s. of (\ref{ctrun1}) corresponds to
the scaling function (\ref{cshift}) which upon exploiting (\ref{ccc})
becomes in this approximation 
\begin{equation}
c(r)=n-\,3\,\,\sum\limits_{i=1}^{n}\frac{\eta _{i}^{(2)}}{(\delta _{i}-\ln 
\frac{r}{2})^{2}}+...  \label{capp}
\end{equation}
where the whole dependence on the S-matrix is now reduced to the coefficient 
$\eta _{i}^{(2)}=\sum_{j=1}^{n}\eta _{ij}^{(2)}$ of the power series
expansion (\ref{series}). Additional information on the model under
investigation is encoded in the constant $\delta _{i}$. Hence, the remaining
step is now to deduce further restrictions on the constants of integration
besides (\ref{ccc}). This, can be done by going beyond the rapidity region
where the large density approximation (\ref{Lapp}) is valid.

\subsubsection{Enhanced approximate solutions and Y-systems}

The restriction on the range for the rapidities in (\ref{Lapp}), for which
the large density approximation $L_{i}^{0}(\theta )$ ceases to be valid,
makes it desirable to develop also an approximation outside the given
interval. For large rapidities we naturally expect that the solution will
tend to the one of a free theory and approaches small density values.
Solving (\ref{TBAE}) for vanishing kernel yields the free on-shell solution 
\begin{equation}
L_{i}^{f}(\theta )=\ln \left( 1+e^{-rm_{i}\cosh \theta }\right) \,\,.
\label{LF}
\end{equation}
Ideally we would like to have expressions for both regions which match at
some distinct rapidity value, say $\theta _{i}^{m}$, to be specified below.
At this point one can then determine the constant of integration $\delta
_{i} $. Since $L_{i}^{0}(\theta )$ and $L_{i}^{f}(\theta )$ become
relatively poor approximations in the transition region between large and
small densities, we first seek for improved analytical expressions. This is
easily achieved by expanding (\ref{TBAE}) around the ``zero order'' small
density approximations. In this case we obtain the integral representation 
\begin{equation}
L_{i}^{s}(\theta )=\exp \left( -rm_{i}\cosh \theta
+\sum\nolimits_{j=1}^{n}(\varphi _{ij}\ast L_{j}^{f})(\theta )\right) \,\,.\,
\end{equation}
For vanishing $\varphi _{ij}$ we may check for consistency and observe that
the functions $L_{i}^{s}(\theta )$ become the first term of the expansion in
(\ref{LF}). One could try to proceed similarly for the large density regime
and develop around $L_{i}^{0}$ instead of $L_{i}^{f}$. However, there is an
immediate problem resulting from the restriction on the range of rapidities
for the validity of $L_{i}^{0}$, which makes it problematic to compute the
convolution. We shall therefore proceed in a different manner for the large
density regime and employ so-called $Y$-systems for this purpose.

In many cases the TBA-equations may be expressed equivalently as a set of
functional relations referred to as $Y$-systems in the literature \cite
{TBAZamun}. Introducing the quantities $Y_{i}=\exp (-\varepsilon _{i})$, the
determining equations can always be cast into the general form 
\begin{equation}
Y_{i}(\theta +i\pi \omega _{i})Y_{i}(\theta -i\pi \omega _{i})=\exp
(g_{i}(\theta ))  \label{Yg}
\end{equation}
with $\omega _{i}$ being some real number and $g_{i}(\theta )$ being a
function whose precise form depends on the particular model. Below when
treating concrete examples we will see how these functional relations can be
explicitly derived from (\ref{TBAE}). We can formally solve the equation (%
\ref{Yg}) by Fourier transformation 
\begin{equation}
Y_{i}(\theta )=\exp \left[ (\Omega _{i}\ast g_{i})(\theta )\right]
\,,\,\qquad \Omega _{i}(\theta ):=\left[ 2\omega _{i}\cosh \tfrac{\theta }{%
2\omega _{i}}\right] ^{-1}  \label{soYg}
\end{equation}
i.e. substituting (\ref{soYg}) into the l.h.s. of (\ref{Yg}) yields $\exp
(g_{i}(\theta ))$. Of course this identification is not completely
compelling and we could have chosen also a different combination of $Y$'s.
However, in order to be able to evaluate the $g_{i}(\theta )$ we require a
concrete functional input for the function $Y_{i}(\theta )$ in form of an
approximated function. Choosing here the large density approximation $%
L_{i}^{0}$ makes the choice for $g_{i}(\theta )$ with hindsight somewhat
canonical, since other combinations lead generally to non-physical answers.

We replace now inside the defining relation of $g_{i}(\theta )$ the $Y$'s by 
$Y_{i}\left( \theta \right) \rightarrow \exp (L_{i}^{0}(\theta ))-1$.
Analogously to the approximating approach described in the previous
subsection, we can replace the convolution by an infinite series of
differentials 
\begin{equation}
\varepsilon _{i}(\theta )=-(g_{i}\ast \Omega _{i})(\theta
)\,=-\sum\limits_{m=0}^{\infty }\nu _{i}^{(m)}\frac{d^{m}}{d\theta ^{m}}%
g_{i}(\theta )\,\,\,,  \label{FTEg}
\end{equation}
where the $\nu $'s are defined by the power series expansion 
\begin{equation}
\int\limits_{-\infty }^{\infty }d\theta \;\Omega _{i}(\theta )\,e^{it\theta
}=2\pi \sum\limits_{m=0}^{\infty }(-i)^{m}\nu _{i}^{(m)}t^{m}=\pi
\sum\limits_{m=0}^{\infty }\frac{E_{2m}}{(2m)!}(\pi \omega
_{i})^{2m}\,t^{2m}\,\,.  \label{F2g}
\end{equation}
The $E_{m}$ denote the Euler numbers, which enter through the expansion $%
1/\cosh x=\sum_{m=0}^{\infty }x^{2m}E_{2m}/(2m)!$. In accordance with the
assumptions of our previous approximations for the solutions of the
TBA-equations in the large density regime, we can neglect all higher order
derivatives of the $L_{i}^{0}(\theta )$. Thus we only keep the zeroth order
in (\ref{FTEg}). From (\ref{F2g}) we read off the coefficient $\nu
_{i}^{(0)}=1/2$, such that we obtain a simple expression for an improved
large density approximation \cite{FKtba} 
\begin{equation}
L_{i}^{l}(\theta )=\ln [1+Y_{i}^{l}(\theta )]=\ln [1+\exp (g_{i}(\theta
)/2)]\,.  \label{LL}
\end{equation}
In principle we could proceed similarly for the small density approximation
and replace now $Y_{i}(\theta )\rightarrow \exp (L_{i}^{s}(\theta ))-1$ in
the defining relations for the $g_{i}$'s. However, in this situation we can
not neglect the higher order derivatives of the $L_{i}^{s}$ such that we
have to keep the convolution in (\ref{FTEg}) and end up with an integral
representation instead.

\subsubsection{The constant of integration and a matching condition}

We now wish to match $L_{i}^{s}$ and $L_{i}^{l}$ in the transition region
between the small and large density regimes at some distinct value of the
rapidity, say $\theta _{i}^{m}$. We select this point to be the value when
the following function proportional to the integrand in the scaling function
(\ref{Scale}) 
\begin{equation}
f_{i}(\theta )=(6/\pi ^{2})rm_{i}L_{i}(\theta )\cosh \theta \;,
\label{fdens}
\end{equation}
has its maximum in the small density approximation \cite{FKtba} 
\begin{equation}
\left. \frac{d}{d\theta }f_{i}^{s}(\theta )\right| _{\theta _{i}^{m}}=0\,\,.
\label{detttt1}
\end{equation}
In regard to the quantity we wish to compute, the scaling function, this is
the point in which we would like to have the highest degree of agreement
between the exact and approximated solution, since this will optimize the
outcome for $c(r)$. Having specified the $\theta _{i}^{m}$, the matching
condition provides a simple rational to fix the constant $\delta _{i}$ \cite
{FKtba}, 
\begin{equation}
L_{i}^{l}(\theta _{i}^{m})=L_{i}^{s}(\theta _{i}^{m})\qquad \quad
\Rightarrow \quad \delta _{i}^{m}\,\,\,.  \label{detttt}
\end{equation}

In general, we can not solve these equations analytically, but it is a
trivial numerical problem, which is by no means comparable with the one of
solving (\ref{TBAE}). Needless to say that the outcome of (\ref{detttt}) is
not to be considered as exact, but as our examples below demonstrate it will
lead to rather good approximations. One of the reasons why this procedure is
successful is that $L_{i}^{s}(\theta _{i}^{m})$ is still very close to the
precise solution, despite the fact that is at its worst in comparison with
the remaining rapidity range.

Combining the improved large and small density approximation we have the
following approximated analytical $L$-functions for the entire range of the
rapidity \cite{FKtba} 
\begin{equation}
L_{i}^{a}(\theta )=\left\{ 
\begin{array}{cc}
L_{i}^{l}(\theta ) & \qquad \text{for }\left| \theta \right| \leq \theta
_{i}^{m} \\ 
L_{i}^{s}(\theta ) & \qquad \text{for }\left| \theta \right| >\theta _{i}^{m}
\end{array}
\right. \;\,,
\end{equation}
such that the scaling function becomes well approximated by 
\begin{equation}
c(r)\simeq \sum\limits_{i=1}^{n}\dint_{0}^{\infty }d\theta
\;f_{i}^{a}(\theta )=\frac{6r}{\pi ^{2}}\sum\limits_{i=1}^{n}m_{i}%
\dint_{0}^{\infty }d\theta \,L_{i}^{a}(\theta )\cosh \theta \,\,\,.
\end{equation}

To develop matters further and report on the quality of $L^{0}$, $L^{f}$, $%
L^{s}$, $L^{l}$ we have to specify a particular theory at this point.

\section{The TBA analysis for affine Toda theory}

This section is concerned with the TBA analysis of affine Toda field
theories discussed in Chapter 3. In order to make contact with the general
considerations mentioned at the beginning of this chapter we re-write the
classical action functional (\ref{ADEaction}) of ATFT as the sum of two
terms 
\begin{equation}
S_{\text{ATFT}}(\frak{g})=S_{\text{TFT}}(\frak{g})-\frac{m^{2}}{\beta ^{2}}%
\int e^{\beta \left\langle \alpha _{0},\phi \right\rangle }d^{2}x\;.
\end{equation}
The implicitly defined action $S_{\text{TFT}}(\frak{g})$ describes Toda
field theory (TFT) which are known to be conformally invariant \cite{cToda}.
They differ in its definition from the affine theories by the above term
involving the affine or highest root $\alpha _{0}=-\theta $, compare Section
2.1.2. Thus, in accordance with the general picture of linking integrability
to broken conformal symmetry we might interpret ATFT as perturbed
conformally invariant Toda models.

To motivate the conformal invariance property of TFT at the classical level
we consider the equations of motion following from $S_{\text{TFT}}(\frak{g})$%
. The latter coincide with the ones of ATFT when omitting the affine
contribution of the potential term in (\ref{motion}). (Keep in mind that we
are dealing now with Euclidean geometry). Under a conformal coordinate
transformation the derivative term in (\ref{motion}) changes according to 
\begin{equation*}
\partial _{z}\partial _{\bar{z}}\phi \rightarrow \partial _{w}\partial _{%
\bar{w}}\phi =\frac{\partial z}{\partial w}\frac{\partial \bar{z}}{\partial 
\bar{w}}\partial _{z}\partial _{\bar{z}}\phi
\end{equation*}
where we have introduced as usual complex coordinates. Thus, the equations
of motion stay invariant if the potential term scales in an analogous
manner. This leads to the following requirement on the transformation
behaviour of the fields 
\begin{equation*}
\phi (z,\bar{z})\rightarrow \phi ^{\prime }(w,\bar{w})=\phi (z,\bar{z})+%
\frac{\rho ^{\vee }}{\beta }\ln \left| \frac{\partial z}{\partial w}\right|
^{2}\;,
\end{equation*}
where $\rho ^{\vee }$ is the dual Weyl vector (\ref{weylv}) satisfying $%
\left\langle \rho ^{\vee },\alpha _{i}\right\rangle =1$ for all simple roots 
$i=1,...,n$. One immediately verifies that this invariance is spoiled as
soon as the affine term is taken into account, since $\left\langle \rho
^{\vee },\alpha _{0}\right\rangle =1-h$. Alternatively, one might also
calculate the energy-momentum tensor from $S_{\text{TFT}}(\frak{g})$ and
show that it can be made traceless, see e.g. \cite{OTU,cToda}. This
classical conformal invariance of TFT was shown to survive quantization \cite
{cToda} and therein the conformal anomaly for the $ADE$ series was
determined to 
\begin{equation}
ADE:\quad c(\beta )=n+48\pi |\rho |^{2}\left( \frac{\beta }{4\pi }+\frac{1}{%
\beta }\right) ^{2}\;.  \label{cTFT}
\end{equation}
Notice that the central charge is coupling dependent and exhibits the same
self-duality behaviour w.r.t. the transformation $\beta \rightarrow 4\pi
/\beta $ as observed for simply-laced ATFT (compare Section 3.2.2). However,
much of the quantum structure of TFT still remains to be understood. Even
for the best known and simplest example, Liouville field theory with $\frak{g%
}=A_{1}$, operator algebra and structure constants are still subject to
investigations \cite{DO,Zamref}. As mentioned in the survey of CFT in
Section 4.1 the properties of conformal field theories with central charge $%
c\geq 1$ are much less understood than those of the minimal models with $%
c<1, $ see e.g. \cite{CFT} for further details. For the case at hand (\ref
{cTFT}) we realize by using the Freudenthal-de Vries strange formula $|\rho
|^{2}=h^{\vee }/12\dim X_{n}$ (see e.g. \cite{GO}) and (\ref{dim}) that 
\begin{equation*}
c(\beta )\geq n(1+4\,h(h+1))\geq 25\;.
\end{equation*}
Hence, Toda models belong to the class of conformal field theories for which
even unitarity can not be established, compare Subsection 4.1.3. (Note that
for Liouville theory the lower bound is just assumed at the self-dual point $%
\beta ^{2}=4\pi $.)

Recent advances in exploring the structure of the three and four point
functions in Liouville theory include the introduction of so-called
reflection amplitudes \cite{Zamref}. However, the origin of these reflection
amplitudes has not been rigorously derived so far \cite{Zamref,Ztalk},
whence semi-classical methods in connection with the thermodynamic Bethe
ansatz were used to provide consistency checks. Perturbing the conformal
field theory the scaling function for the Sinh-Gordon model was
approximately computed by considering only the zero mode dynamics which lead
to a re-formulation of the problem in the setting of quantum mechanics with
the reflection amplitude as crucial input \cite{Zamref, Arsch}. This
procedure was extended to all simply-laced ATFT in \cite{Fateev} and just
recently to the non simply-laced case in \cite{Fateev2} resulting in a
perturbative expansion of the scaling function of the kind 
\begin{equation*}
c(r)=n+\frac{a_{1}}{(\delta -\ln \frac{r}{2})^{2}}+\frac{a_{2}}{(\delta -\ln 
\frac{r}{2})^{5}}+\frac{a_{3}}{(\delta -\ln \frac{r}{2})^{7}}+...
\end{equation*}

The leading order term in this expansion matches the result found in \cite
{FKS1,FKtba} as described below. By direct comparison this allows now to fix
the constants of integration. However, one should keep in mind that the
semi-classical analysis based on reflection amplitudes in \cite{Zamref} and 
\cite{Fateev,Fateev2} relies on extra input information as for instance the
precise relationship between the coupling constant and the masses \cite{mum}
obtained by Bethe Ansatz methods. Therefore, it is certainly desirable to
determine the constant\ and the scaling function solely within the TBA
approach which uses the S-matrix as only input information. We already saw
how to derive the leading order behaviour in the previous subsection and we
will now provide in the context of ATFT a rigorous argument which
establishes that the constants of integration (\ref{ccc}) do not depend on
the particle type, such that we may replace $\kappa _{i}\rightarrow \kappa $
and $\delta _{i}\rightarrow \delta $. In addition, we determine them
approximately from within the TBA analysis by matching the large and small
density regimes. This allows then to compare against the semi-classical
approach \cite{Zamref,Fateev} and the agreement we find will provide an
additional consistency check for the ATFT S-matrix.

\subsection{The universal TBA-kernel}

Recall from our previous discussion that at the quantum level to each of the
affine Toda models a pair of dual affine Lie algebras $(X_{n}^{(1)},\hat{X}_{%
\hat{n}}^{(\ell )})$ is associated which describe the Lie algebraic
structure in the weak and strong coupling regime, respectively. As explained
in Chapter 2 $\hat{X}_{\hat{n}}^{(\ell )}$ denotes a twisted affine Lie
algebra w.r.t. a Dynkin diagram automorphism of order $\ell $, while $%
X_{n}^{(1)}$ is a non-twisted algebra of rank $n$. For $X_{n}^{(1)}$
simply-laced both algebras coincide, i.e. $X_{n}^{(1)}\cong \hat{X}^{(\ell
)},\,\ell =1$, which is reflected in the quantum theory by a strong-weak
self-duality in the coupling constant. In the following we will now exploit
the associated Lie algebraic structure to derive generic expressions for the
TBA analysis.

From the universal integral representation (\ref{SPPi}), we immediately
derive the Fourier transformed TBA-kernel (\ref{series}) for ATFT. However,
when taking the logarithmic derivative one has to be careful about
interchanging the derivative with the integral, since these two operations
do not commute. Comparison with the block representation of the S-matrix (%
\ref{blockS}) yields 
\begin{equation}
\varphi _{ij}(\theta )=\frac{1}{2\pi }\int_{-\infty }^{\infty }dt\,\left[
2\delta _{ij}-\phi _{ij}(t)\right] \,\exp \frac{t\theta }{i\pi },
\label{logder}
\end{equation}
such that the Fourier transformed universal TBA-kernel (\ref{series})
acquires the form 
\begin{eqnarray}
\widetilde{\varphi }_{ij}(t) &=&2\pi \delta _{ij}-\pi \phi _{ij}(\pi t) 
\notag \\
&=&2\pi \delta _{ij}-8\pi \sinh t\pi \vartheta _{h}\sinh t_{j}\pi \vartheta
_{H}\,\,A(e^{t\pi \vartheta _{h}},e^{t\pi \vartheta _{H}})_{ij\quad }^{-1}.
\label{uni2}
\end{eqnarray}
Recall that $A(q,\hat{q})$ is the $q$-deformed Cartan matrix (\ref{qA}) of
the non-twisted Lie algebra and that the integers $t_{i}=\ell \alpha
_{i}^{2}/2$ are the entries of the symmetrizer of the Cartan matrix, compare
(\ref{symm}) in Chapter 2. The deformation parameters inside the integral
representation are chosen in terms of the angles $\vartheta _{h}:=\frac{2-B}{%
2h}$ and $\vartheta _{H}:=\frac{B}{2H}$, previously defined in (\ref{Bangle}%
) of Chapter 3 together with the effective coupling constant $0\leq B\leq 2$%
. Remember further the definition of the Coxeter numbers $h,\,\hat{h}$ and
the dual Coxeter numbers $h^{\vee },\,\hat{h}^{\vee }$ of $X_{n}^{(1)}\frak{%
\ }$and $\hat{X}^{(\ell )}$, respectively, as well as the $\ell ^{\text{th}}$
Coxeter number $H=\ell \hat{h}$ of $\hat{X}^{(\ell )}$ in Chapter 2.

\subsubsection{The weak coupling limit}

Notice that the appearance of the extra term $\delta _{ij}$ in (\ref{logder}%
) is of particular importance in the weak or strong coupling limit, where
the particles become free, i.e. $S_{ij}=1$. Sending $\beta \rightarrow
0,\infty $ or equivalently $B\rightarrow 0,2$ we infer from (\ref{uni2})
that the TBA kernel approaches the $\delta $-function times the unit matrix,
whence the TBA equations (\ref{TBAE}) decouple and can be solved exactly, 
\begin{equation}
B=0:\quad L_{i}(\theta )=-\ln \left( 1-e^{-rm_{i}\cosh \theta }\right) \;.
\label{freeATFT}
\end{equation}
The above density distribution corresponds to the one of a free relativistic
Bose gas despite the fact that we used Fermi statistics in the formulation
of the TBA equations (\ref{TBAE}). This shows the statistical ambiguity in
the formulation of the TBA mentioned before, compare Section 4.2.2. Now, in
accordance with the physical picture suggested by the action functional (\ref
{ADEaction}) that the theory consists of $n$ free bosons at $\beta =0$, the
effective central charge deduced from (\ref{freeATFT}) in the extreme UV
limit is $c_{\text{eff}}=\lim_{r\rightarrow 0}c(r)=n$. The order in which
the weak coupling limit is to be performed, before or after taking the
logarithmic derivative $-i\frac{d}{d\theta }\ln S_{ij}(\theta ),$ is crucial
here, since in the former case the result would be a gas of $n$ free \emph{%
fermions} with central charge $c=n/2$. After this preliminary discussion of
vanishing coupling constant we now turn to the general case $B\neq 0$.

\subsubsection{Series expansion and the second order coefficient}

Based on the compact expression (\ref{uni2}) the discussion of the previous
section about approximate solutions can now be applied to all affine Toda
models at once. The only information we require to derive the leading order
behaviour in the UV limit is the zeroth and second order coefficient $\eta
_{ij}^{(2)}$ in the power series expansion (\ref{series}). From (\ref{uni2})
we read off directly $\eta _{ij}^{(0)}=0$ (which is in accordance with the
requirement imposed in 4.2.4) and 
\begin{equation}
\eta _{ij}^{(2)}=\frac{\pi ^{2}}{h\,H}\,B(2-B)A_{ij}^{-1}\,t_{j}=\frac{\pi
^{2}}{h\,h^{\vee }}\,B(2-B)\,\left\langle \lambda _{i},\lambda
_{j}\right\rangle \,\,\,\,.
\end{equation}
In the latter equality we used the fact that the inverse of the Cartan
matrix is related to the fundamental weights as $\lambda
_{i}=\sum_{j}A_{ij}^{-1}\,\alpha _{j}$, $t_{i}=\ell \alpha _{i}^{2}/2$ and $%
H=\ell \,\hat{h}=\ell \,h^{\vee }$. This implies on the other hand that 
\begin{equation}
\eta _{i}^{(2)}=\sum_{j=1}^{n}\eta _{ij}^{(2)}=\frac{\pi ^{2}}{h\,h^{\vee }}%
\,B(2-B)\,\left\langle \lambda _{i},\rho \right\rangle  \label{etta}
\end{equation}
with $\rho $ being the Weyl vector (\ref{weylv}). Therefore, 
\begin{equation}
\eta =\sum\limits_{i=1}^{n}\eta _{i}^{(2)}=B(2-B)\frac{\pi ^{2}|\rho |^{2}}{%
h\,h^{\vee }}\,=nB(2-B)\frac{\pi ^{2}(h+1)}{12h\,}\,\,.  \label{eta2}
\end{equation}
We used here again the Freudenthal-de Vries strange formula and $\dim
X_{n}^{(1)}=n(h+1)$, compare (\ref{dim}). Notice that in terms of the
quantities belonging to the untwisted Lie algebra $X_{n}^{(1)}$ the formula (%
\ref{eta2}) is identical for the simply-laced and the non simply-laced case.
Before we now use (\ref{etta}) and (\ref{eta2}) to apply the approximation
scheme of Section 4.2.4 we derive the set of functional equations or $Y$%
-systems needed to improve the leading order approximations.

\subsection{Universal TBA equations and Y-systems}

The universal expression for the kernel (\ref{uni2}) can be exploited in
order to derive universal TBA-equations for \emph{all} ATFT, which may be
expressed equivalently as a set of functional relations referred to as $Y$%
-systems. Fourier transforming (\ref{TBAE}) in a suitable manner and
invoking the convolution theorem we can manipulate the TBA equations by
using the expression (\ref{uni2}). After Fourier transforming back we obtain 
\begin{equation}
\varepsilon _{i}+\sum\limits_{j=1}^{n}\Delta _{ij}\ast
L_{j}=\sum\limits_{j=1}^{n}\,\Gamma _{ij}\ast (\varepsilon _{j}+L_{j})\,.
\label{uniTBA}
\end{equation}
The universal TBA kernels $\Delta $ and $\Gamma $ are then given by \qquad 
\begin{eqnarray}
\Omega _{i}(\theta ) &=&\left( 2(\vartheta _{h}+t_{i}\vartheta _{H})\cosh 
\tfrac{\theta }{2(\vartheta _{h}+t_{i}\vartheta _{H})}\right) ^{-1},\qquad \\
\Gamma _{ij}(\theta ) &=&\sum\limits_{k=1}^{I_{ij}}\Omega _{i}(\theta
+i(2k-1-I_{ij})\theta _{H}), \\
\Delta _{ij}(\theta ) &=&[\Omega _{i}(\theta +\theta _{h}-t_{i}\theta
_{H})+\Omega _{i}(\theta -\theta _{h}+t_{i}\theta _{H})]\,\delta _{ij}\,.
\label{kernel1}
\end{eqnarray}
The key point here is that the entire mass dependence, which enters through
the on-shell energies $m_{i}\cosh \theta $, has completely dropped out from
the equations due to the identity (\ref{ATFTmass}) for the mass spectrum.
Noting further that the identity 
\begin{equation}
\,[\,I_{ij}]_{\hat{q}(i\pi )}\,m_{j}\cosh \theta
\,=\sum_{k=1}^{I_{ij}}\,m_{j}\,\cosh \left[ \theta +(2k-1-I_{ij})\theta _{H}%
\right] \,
\end{equation}
can be employed to the l.h.s. of the mass formula, we have assembled all
ingredients to derive functional relations for the quantities $Y_{i}=\exp
(-\varepsilon _{i})$. For this purpose we may either shift the TBA equations
appropriately in the complex rapidity plane or use again Fourier
transformations, see \cite{FKS1}, 
\begin{multline}
Y_{i}(\theta +\theta _{h}+t_{i}\theta _{H})Y_{i}(\theta -\theta
_{h}-t_{i}\theta _{H})=\left[ 1+Y_{i}(\theta -\theta _{h}+t_{i}\theta _{H})%
\right] \left[ 1+Y_{i}(\theta +\theta _{h}-t_{i}\theta _{H})\right]
\label{YAT} \\
\times \prod_{j=1}^{n}\prod_{k=1}^{I_{ij}}\left[ 1+Y_{j}^{-1}(\theta
+(2k-1-I_{ij})\theta _{H})\right] ^{-1}\,\,.
\end{multline}
These equations are of the general form (\ref{Yg}) and specify concretely
the quantities $\omega _{i}$ and $g_{i}(\theta )$. We recover various
particular cases from (\ref{YAT}). In case the associated Lie algebra is
simply-laced, we have $\theta _{h}+t_{i}\theta _{H}\rightarrow i\pi /h$,$%
\quad \theta _{h}-t_{i}\theta _{H}\rightarrow i\pi /h(1-B)$ and $%
I_{ij}\rightarrow 0,1$, such that we recover the relations derived in \cite
{FKS1}, 
\begin{multline}
Y_{i}\left( {\theta +}\tfrac{i\pi }{h}\right) Y_{i}\left( {\theta }-\tfrac{%
i\pi }{h}\right) =\left[ 1+Y_{i}\left( \theta +\tfrac{i\pi }{h}(1-B)\right) %
\right] \left[ 1+Y_{i}\left( \theta -\tfrac{i\pi }{h}(1-B)\right) \right]
\label{YADE} \\
\times \prod\limits_{j=1}^{n}\,\left[ 1+Y_{j}^{\,-1}\left( \theta \right) %
\right] ^{-I_{ij}}\,\,.
\end{multline}
As stated therein we obtain the system for minimal ATFT \cite{TBAZamun} by
taking the limit $B\rightarrow i\infty $.

The concrete formula for the approximated solution of the Y-systems in the
large density regime, as defined in (\ref{LL}), reads 
\begin{equation}
Y_{i}^{l}(\theta )=\tfrac{\cos (2\theta \beta _{i})+\cos (2(\theta
_{h}-t_{i}\theta _{H})\beta _{i})}{4\eta _{i}\beta _{i}^{2}}%
\prod\limits_{j=1}^{n}\prod\limits_{m=1}^{I_{ij}}\left( 1-\tfrac{2\eta
_{j}\beta _{j}^{2}}{\cos ^{2}(\beta _{j}(\theta +(2m-1-I_{ij})\theta _{H})}%
\right) ^{\frac{1}{2}}\,\,.  \label{ih}
\end{equation}

Exploiting possible periodicities of the functional equations (\ref{YAT})
they may be utilized in the process of obtaining approximated analytical
solutions \cite{TBAZam}. As we demonstrated they can also be employed to
improve on approximated analytical solution in the large density regime. In
the following subsection we supply a further application and use them to put
constraints on the constant of integration $\delta _{i}$ in (\ref{ccc}).

\subsection{The constants of integration $\protect\kappa $ and $\protect%
\delta $}

There are various constraints we can put on the constants $\kappa _{i}$ and $%
\delta _{i}$ on general grounds, e.g. the lower bound already mentioned.
Having the numerical data at hand we can use them to approximate the
constant. In \cite{FKS1} this was done for the $ADE$ series by matching $%
L_{i}^{0}$ with the numerical data at $\theta =0$ and a simple analytical
approximation was provided 
\begin{equation}
\delta ^{\text{num}}=\ln [B(2-B)2^{1+B(2-B)}]\;.  \label{delnum}
\end{equation}
Of course the idea is to become entirely independent of the numerical
analysis. For this reason the argument which led to the matching condition (%
\ref{detttt}) was given. When we consider a concrete theory like ATFT, we
can exploit its particular structure and put additional constraints on the
constants from general properties. For instance, when we restrict ourselves
to the simply-laced case it is obvious to demand that the constants respect
also the strong-weak self-duality, i.e. $\kappa _{i}(B)=\kappa _{i}(2-B)$
and $\delta _{i}(B)=\delta _{i}(2-B)$.

Finally we present a brief argument which establishes that the constants $%
\kappa _{i}$ are in fact independent of the particle type $i$. We replace
for this purpose in the functional relations (\ref{YAT}) the $Y$-functions
by $Y_{i}^{0}(\theta )$ belonging to the solution (\ref{Lapp}) and consider
the equation at $\theta =0$, such that 
\begin{equation}
\frac{\cosh ^{2}[\pi \kappa _{i}(\vartheta _{h}+t_{i}\vartheta
_{H})]-2\kappa _{i}^{2}\eta _{i}^{(2)}}{\cosh ^{2}[\pi \kappa _{i}(\vartheta
_{h}+t_{i}\vartheta _{H})]}=\prod\limits_{j=1}^{n}\prod%
\limits_{m=1}^{I_{ij}}\,\tfrac{(\cosh ^{2}[\pi \kappa
_{i}(2m-1-I_{ij})\vartheta _{H}]-2\kappa _{i}^{2}\eta _{i}^{(2)})^{\frac{1}{2%
}}}{\cosh [\pi \kappa _{i}(2m-1-I_{ij})\vartheta _{H}]}\,\,.  \label{yy}
\end{equation}
Keeping in mind that $\kappa _{i}$ is a very small quantity in the
ultraviolet regime, we expand (\ref{yy}) up to second order in $\kappa _{i}$%
, which yields after cancellation 
\begin{equation}
4t_{i}\vartheta _{h}\vartheta _{H}=\frac{\alpha _{i}^{2}}{2}\frac{B(2-B)}{%
hh^{\vee }}=\sum_{j=1}A_{ij}\frac{\kappa _{j}^{2}}{\kappa _{i}^{2}}\frac{%
\eta _{j}^{(2)}}{\pi }=\frac{B(2-B)}{hh^{\vee }}\sum_{j=1}A_{ij}\frac{\kappa
_{j}^{2}}{\kappa _{i}^{2}}\left\langle \lambda _{j},\rho \right\rangle \,.
\label{bb}
\end{equation}
We substituted here the expression (\ref{etta}) for the constants $\eta _{j}$
in the last equality. Using once more the relation $\lambda
_{i}=\sum_{j}A_{ij}^{-1}\alpha _{j}$, we can evaluate the inner product such
that (\ref{bb}) reduces to 
\begin{equation}
\alpha _{i}^{2}=\sum_{j=1}^{n}\sum_{k=1}^{n}A_{ij}\left( \frac{\kappa
_{j}^{2}}{\kappa _{i}^{2}}\right) A_{jk}^{-1}\alpha _{k}^{2}\,\,.  \label{aa}
\end{equation}
Clearly this equation is satisfied if all the $\kappa _{i}$ are identical.
From the uniqueness of the solution of the TBA-equations follows then
immediately that we can always take $\kappa _{i}\rightarrow \kappa $. Since
the uniqueness discussion has only been presented for the one-particle case,
it is reassuring that we can obtain the same result also directly from (\ref
{aa}). From the fact that the $\kappa _{i}$ are real numbers and all entries
of the inverse Cartan matrix are positive follows that $\kappa
_{i}^{2}=\kappa _{j}^{2}$ for all $i$ and $j$. The ambiguity in the sign is
irrelevant for the use in $L_{i}^{0}(\theta )$.

\subsection{The scaling functions}

Exploiting the independence of $\kappa ,\delta $ of the particle index and
the relation (\ref{eta2}) the leading order behaviour for the scaling
function is now immediate to derive from the general expression (\ref{capp}%
), 
\begin{equation}
c(r)\approx n-\,\dfrac{3\eta }{(\delta -\ln \frac{r}{2})^{2}}=n\mathbf{\,}%
\left( 1-\dfrac{\pi ^{2}(h+1)B(2-B)}{4h(\delta -\ln \frac{r}{2})^{2}}\right)
\,\,.  \label{capp+}
\end{equation}
$\quad $ \noindent \noindent From the universal expression (\ref{uni2}) it
follows in fact that this expression holds for \emph{all} affine Toda field
theories related to a dual pair of simple affine Lie algebras $(X_{n}^{(1)}%
\frak{,}\hat{X}^{(\ell )})$. However, strong-weak duality is only guaranteed
for $\ell =1$, i.e. the $ADE$ series, since the constant $\delta $ is in
general coupling dependent.

Restricting ourselves to the simply-laced case, we can view the results of 
\cite{Zamref,Arsch,Fateev} obtained by means of a semi-classical treatment
for the scaling function as complementary to the one obtained from the TBA
analysis and compare directly with the expression (\ref{capp+}). Translating
the quantities in \cite{Zamref,Fateev} to our conventions, i.e. $%
R\rightarrow r$, $B\rightarrow B/2$, we observe that $c(r)$ becomes a power
series expansion in $\kappa $. We also observe that the second order
coefficients precisely coincide in their general form. Comparing the
expressions, we may read off directly 
\begin{equation}
\delta ^{\text{semi}}=\ln \left( \frac{4\pi \Gamma \left( \frac{1}{h}\right)
\left( \frac{2}{B}-1\right) ^{\frac{B}{2}-1}}{\varkappa \,\Gamma \left( 
\frac{1}{h}-\frac{B}{2h}\right) \Gamma \left( 1+\frac{B}{2h}\right) }\right)
-\gamma _{E}\,\,  \label{const}
\end{equation}
for all ATFT related to simply-laced Lie algebras\footnote{%
The expressions in \cite{Zamref} and \cite{Fateev} only coincide if in the
former case $m=1$ and in the latter $m=1/2$. In addition, we note a missing
bracket in equation (6.20) of \cite{Zamref}, which is needed for the
identification. Replace $C\rightarrow $ $-4QC$ therein.}. Here $\gamma _{E}$
denotes Euler's constant and $\varkappa =(\prod_{i=1}^{n}n_{i}^{n_{i}})^{%
\frac{1}{2h}}$ is a constant which can be computed from the Kac labels $%
n_{i} $ of the related Lie algebra. Contrary to the statement made in \cite
{Arsch}, this identification can be carried out effortlessly without the
need of higher order terms. Recalling the simple analytical expression (\ref
{delnum}) of \cite{FKS1} we may now compare. Figure 4.3 demonstrates
impressively that this working hypothesis shows exactly the same qualitative
behaviour as $\delta ^{\text{semi}}$ and also quantitatively the difference
is remarkably small. 

\begin{center}
\includegraphics[width=8cm,height=12cm,angle=-90]{const.epsi}

{\small Figure 4.3: The semi-classical constant }$\delta ^{\text{semi}}$%
{\small \ in comparison to the numerical fit }$\delta ^{\text{num}}${\small %
.\bigskip }
\end{center}

To illustrate the quality of our approximate solutions to the TBA-equations,
we shall now work out some explicit examples.

\subsection{Explicit examples}

To exhibit whether there are any qualitative differences between the
simply-laced and the non simply-laced case we consider the first examples of
these series.

\subsubsection{The Sinh-Gordon Model}

The Sinh-Gordon model is the easiest example in the simply-laced series and
therefore ideally suited as testing ground. The Coxeter number is $h=2$ in
this case. An efficient way to approximate the L-functions to a very high
accuracy is 
\begin{equation}
L^{a}(\theta )=\left\{ 
\begin{array}{cc}
\ln \left[ 1+\frac{\cos (2\kappa \theta )+\cosh (\pi \kappa (1-B))}{4\eta
\kappa ^{2}}\right] & \quad \text{for }\left| \theta \right| \leq \theta ^{m}
\\ 
\exp \left[ -rm\cosh \theta +(\varphi \ast L^{f})(\theta )\right] & \quad 
\text{for }\left| \theta \right| >\theta ^{m}
\end{array}
\right. \,\,
\end{equation}
with 
\begin{equation}
\varphi (\theta )=\frac{4\sin (\pi B/2)\cosh \theta }{\cosh 2\theta -\cos
\pi B},\qquad \eta =\frac{\pi ^{2}B(2-B)}{8}.
\end{equation}
The determining equation for the matching point reads 
\begin{equation}
\sinh \theta ^{m}-rm/2\sinh (2\theta ^{m})+\cosh (\theta ^{m})(\varphi
^{\prime }\ast L^{f})(\theta ^{m})=0\,\,\,.
\end{equation}
For instance at $B=0.4$ this equation yields $\theta ^{m}=11.9999$ such that
the matching condition (\ref{detttt}) gives $\delta ^{m}=0.4913$. Figure 4.4
(a) shows that the large and small density approximation $L^{0}$ and $L^{f}$
may be improved in a fairly easy way. In view of the simplicity of the
expression $L^{a}$ the agreement with the numerical solution is quite
remarkable. Figure 4.4(a) also illustrates that when using the constant $%
\delta ^{\text{semi}}$ instead of $\delta ^{m}$ the agreement with the
numerical solutions appears slightly better for small rapidities. When we
employ $\delta ^{\text{num}}$ instead of $\delta ^{\text{semi}}$ the
difference between the two approximated solutions is beyond resolution.
However, as may be deduced from Figure 4.4(b), with regard to the
computation of the scaling function the difference between using $\delta
^{m} $ instead of $\delta ^{\text{semi}}$ is almost negligible. Whereas in
the former case the resulting value for the scaling function is slightly
below the correct value, it is slightly above by almost the same amount in
the latter case. Finally we compute the scaling function (\ref{capp+}) up to
leading order and find \cite{FKS1} 
\begin{equation*}
c(r)=1-\frac{3\pi ^{2}B(2-B)}{8(\delta -\ln \frac{r}{2})^{2}}+...
\end{equation*}
which is in agreement with the leading order behaviour found in \cite{Zamref}%
.

\begin{center}
\includegraphics[width=12cm,height=16cm,angle=0]{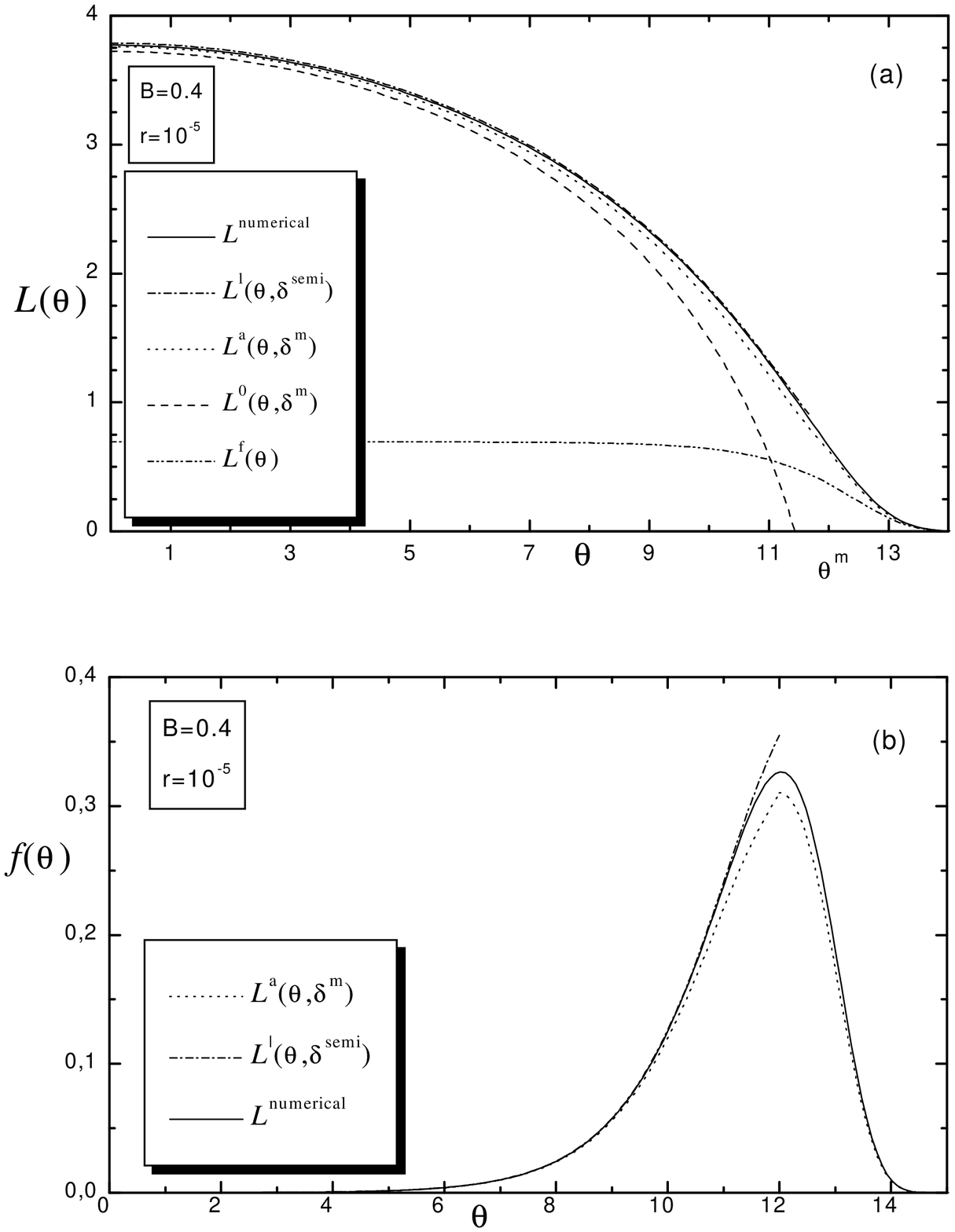}

{\small Figure 4.4: Numerical plots against the various analytical
approximations for the Sinh-Gordon model. The Graph (a) depicts the
solutions to the TBA equation, while (b) shows the free energy density (\ref
{fdens}).}
\end{center}

\subsubsection{$(G_{2}^{(1)},D_{4}^{(3)})$-ATFT}

In this case we have $h=6$ and $H=12$ for the related Coxeter numbers. The
two masses are $m_{1}=m\sin (\pi (1/6-B/24))$ and $m_{2}=m\sin (\pi
(1/3-B/12))$. The L-functions are well approximated by 
\begin{equation*}
L_{1}^{a}(\theta )=\left\{ 
\begin{array}{cc}
\ln [1+\tfrac{\cos (2\kappa \theta )+\cosh (\pi \kappa (\frac{1}{3}-\frac{B}{%
4}))}{4\eta _{1}^{(2)}\kappa ^{2}}\sqrt{1-\tfrac{2\eta _{2}^{(2)}\kappa ^{2}%
}{\cos ^{2}(\kappa \theta )}}\,\,\,] & \quad \;\text{for }\left| \theta
\right| \leq \theta _{1}^{m}\medskip \\ 
\exp [-rm_{1}\cosh \theta +(\varphi _{11}\ast L_{1}^{f}+\varphi _{12}\ast
L_{2}^{f})(\theta )] & \text{for }\left| \theta \right| >\theta _{1}^{m}
\end{array}
\right.
\end{equation*}
\begin{equation*}
\,\,L_{2}^{a}(\theta )=\left\{ 
\begin{array}{cc}
\ln [1+\tfrac{\cos (2\kappa \theta )+\cosh (\pi \kappa (\frac{1}{3}-\frac{5B%
}{24}))}{4\eta _{2}^{(2)}\kappa ^{2}}\prod\limits_{k=-1}^{1}\sqrt{1-\tfrac{%
2\eta _{1}^{(2)}\kappa ^{2}}{\cos ^{2}(\kappa (\theta +\frac{kB}{12}))}}\,]\,
& \text{for }\left| \theta \right| \leq \theta _{2}^{m}\medskip \\ 
\exp [-rm_{2}\cosh \theta +(\varphi _{21}\ast L_{1}^{f}+\varphi _{22}\ast
L_{2}^{f})(\theta )] & \text{for }\left| \theta \right| >\theta _{2}^{m}
\end{array}
\right. ,
\end{equation*}
with $\varphi $ given by (\ref{logder}) and 
\begin{equation}
\eta _{1}^{(2)}=\frac{5\pi ^{2}B(2-B)}{72},\quad \eta _{2}^{(2)}=\frac{\pi
^{2}B(2-B)}{8},\quad \eta =\frac{7\pi ^{2}B(2-B)}{36}\,\,.
\end{equation}
Using now the numerical data $L_{1}(0)=4.2524$ and $L_{2}(0)=3.67144$ as
benchmarks, we compute by matching them with $L_{1}^{a}(0)$ and $%
L_{2}^{a}(0) $ the constant to $\delta =1.1397$ in both cases. This confirms
our general result of Section 4.3.3. Evaluating the equations (\ref{detttt})
and (\ref{detttt1}) we obtain for $B=0.5$ the matching values for the
rapidities $\theta _{1}^{m}=12.744$ and $\theta _{2}^{m}=12.278$ such that $%
\delta _{1}^{m}=1.9539$ and $\delta _{2}^{m}=1.5572$. Figure 4.5(a) and
4.4(b) show a good agreement with the numerical outcome.

The approximated analytical expression for the scaling function reads

\begin{equation}
c(r)\simeq 2-\dfrac{7\,\pi ^{2}B(2-B)}{12(\delta -\ln \frac{r}{2})^{2}}\,.
\end{equation}
This expressions differs from the one quoted in \cite{FKS1}, since in there
the sign of some scattering matrices at zero rapidity was chosen differently.%
%
%
%

In conclusion, we have demonstrated that it is possible to find simple
analytical expressions which approximate the solutions of the TBA equations
in the large and small density regime to high accuracy. We derived the $Y$%
-systems for all ATFT and besides demonstrating how they can be utilized to
improve on the large density approximations we also showed how they can be
used to put constraints on the constant of integration (\ref{ccc}). By
matching the two solutions of the different regimes at the point in which
the particle density and the density of available states coincide, it is
possible to fix the constant of integration, which originated in the
approximation scheme of \cite{ZamoR,MM,FKS1} described in Subsection 4.2.4.
Moreover, the expression (\ref{capp}) for the scaling function has been
proven valid for \emph{all} ATFT which is in agreement with the leading
order term in the alternative approach \cite{Zamref,Fateev,Fateev2} using
reflection amplitudes originating in conformal TFT. In particular, this
yields sufficient evidence for the correctness of the universal ATFT
S-matrix constructed via the bootstrap approach.

\begin{center}
\includegraphics[width=12cm,height=16cm,angle=0]{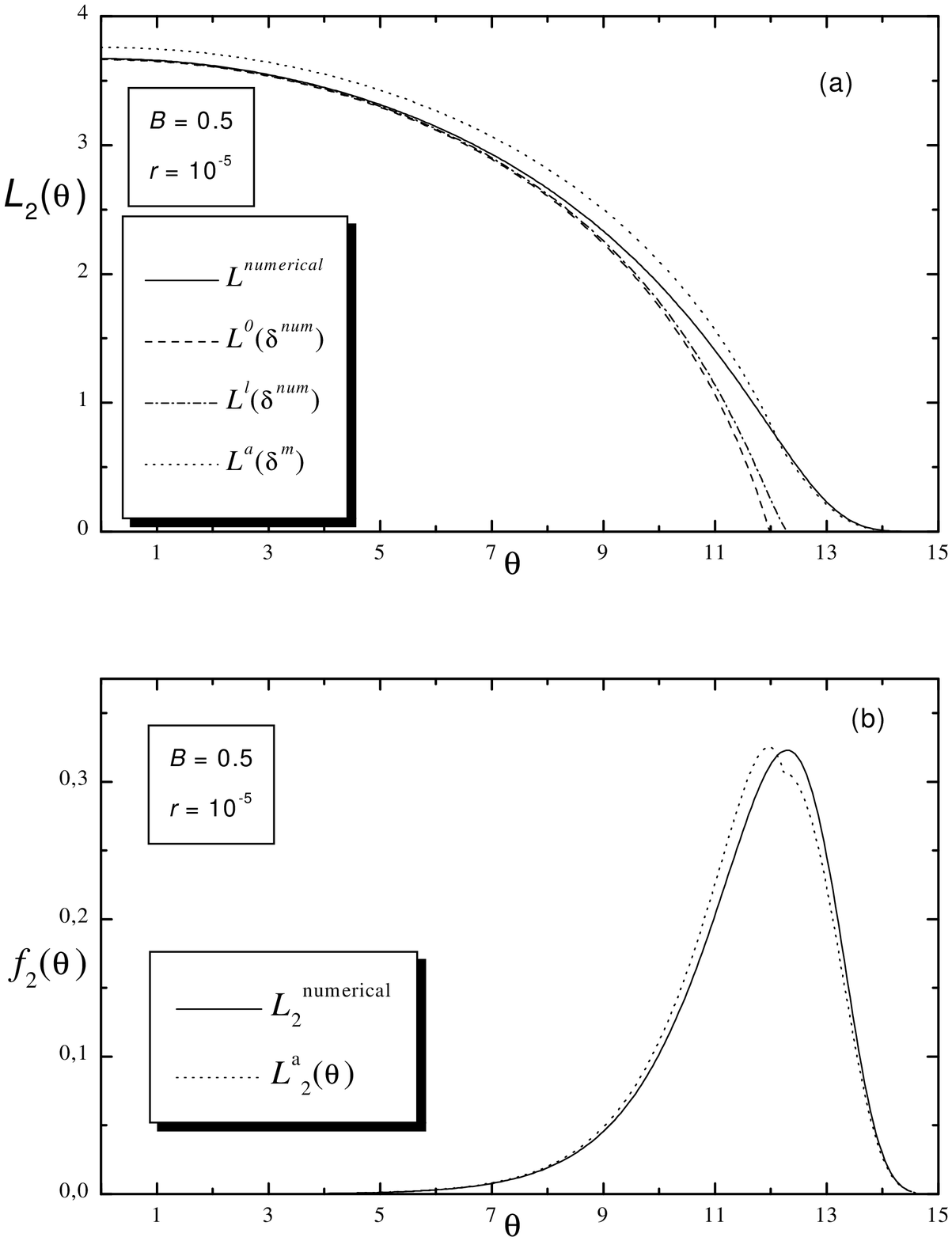}

{\small Figure 4.5: Numerical plots against the various analytical
approximations for the (G}$_{2}^{(1)}${\small ,D}$_{4}^{(3)}${\small %
)-theory. The Graph (a) depicts the solutions to the TBA equation, while (b)
shows the free energy density (\ref{fdens}).}\bigskip
\end{center}

\section{TBA for the Homogeneous Sine-Gordon models}

In this section we start with the analysis of the colour valued S-matrices
or the $\frak{g}|\frak{\tilde{g}}$-theories as we called them in Chapter 3.
Recall that both algebras are simply-laced and that one determines the main
quantum number, while the other specifies colour degrees of freedom. First
we start with a special subclass of these scattering matrices setting $\frak{%
g}=A_{n}$ while keeping the colour structure $\frak{\tilde{g}}=ADE$ generic.
This class has recently been proposed \cite{HSGS} to describe integrable
theories known as the \textbf{Homogeneous Sine-Gordon} (HSG) models \cite
{HSG,Park}.

For these models one has an explicit Lagrangian at hand whence many
informations of the field theory are accessible by a semi-classical
analysis. This makes the TBA analysis particularly interesting since one
might probe the different semi-classical predictions for consistency, as for
example parity violation and the appearance of unstable bound states. The
latter will be linked to resonance poles in the S-matrix leading to a
''staircase pattern'' in the scaling function, which has been observed
previously for similar models \cite{ZamoR,Stair}. However, in comparison
with the models studied so far, the HSG models are distinguished in two
aspects. First they break parity invariance and second some of the resonance
poles can be associated directly to unstable particles via a classical
Lagrangian. We present in particular a detailed numerical analysis for the $%
su(3)$-HSG\ model \cite{CFKM}, but expect that many of our findings for that
case are generalizable to other Lie groups. Especially, one of the main
outcomes of our TBA-analysis is that the suggested scattering matrix \cite
{HSGS} leads indeed to the expected coset central charge \cite{CFKM} (see
below), which gives strong support to the proposal. In addition, the
detailed analysis of the HSG theories will serve as a preliminary step for
the discussion of all $\frak{g}|\frak{\tilde{g}}$-theories.

\subsection{Classical Lagrangian and perturbed WZNW cosets}

The HSG models have been constructed as integrable perturbations of
WZNW-coset theories \cite{Witten} of the form $\frak{\tilde{g}}_{k}/u(1)^{%
\tilde{n}}$, where $\tilde{n}=\limfunc{rank}\frak{\tilde{g}}$ and $k=n+1$ is
an integer called the ``level'' with $n$ being the rank of $\frak{g}=A_{n}$.
The specific choice of the coset ensures that the perturbed theories possess
a mass gap \cite{HSG2}. The defining action of the HSG models reads 
\begin{equation}
S_{\text{HSG}}[g]=S_{\text{CFT}}[g]+\frac{m^{2}}{\pi \beta ^{2}}\,\int
d^{2}x\,\,\left\langle \Lambda _{+},g(x,t)^{-1}\Lambda
_{-}g(x,t)\right\rangle \ \,\;.  \label{HSGaction}
\end{equation}
Here $S_{\text{CFT}}$ denotes the coset action, $\left\langle
\,\,,\,\,\right\rangle $ the Killing form of $\frak{\tilde{g}}$ and $g(x,t)$
a bosonic scalar field taking values in the compact simple Lie group
associated with $\frak{\tilde{g}}$. $\Lambda _{\pm }$ are arbitrary
semi-simple elements of the Cartan subalgebra, which have to be chosen such
that they are not orthogonal to any root of $\frak{\tilde{g}}$. They play
the role of continuous vector coupling constants. The latter constraints do
not restrict the parameter choice in the quantum case with regard to the
proposed S-matrix which makes sense for every choice of $\Lambda _{\pm }$.
They determine the mass ratios of the particle spectrum as well as the
behaviour of the model under a parity transformation. The parameters $m$ and 
$\beta ^{2}=1/k+O(1/k^{2})$ are the bare mass scale and the coupling
constant, respectively. The non-perturbative definition of the theory is
achieved by identifying $\Phi (x,t)=\left\langle \Lambda
_{+},g(x,t)^{-1}\Lambda _{-}g(x,t)\right\rangle $ with a matrix element of
the WZNW-field $g(x,t)$ taken in the adjoint representation, which is a
spinless primary field of the coset-CFT and in addition exchanging $\beta
^{2}$ by $1/k$ and $m$ by the renormalized mass \cite{HSG2}. The conformal
data of $S_{\text{CFT}}[g]$, which are of interest for the TBA analysis are
the Virasoro central charge $c$ of the coset and the conformal dimensions $%
\Delta ,\bar{\Delta}$ of the perturbing operator in the massless limit 
\begin{equation}
c_{\frak{\tilde{g}}_{k}}=\frac{k\,\dim \frak{\tilde{g}}}{k+\tilde{h}}-\tilde{%
n}=\frac{k-1}{k+\tilde{h}}h\tilde{n}\,,\quad \quad \quad \quad \Delta =\bar{%
\Delta}=\frac{\tilde{h}}{k+\tilde{h}}\;.  \label{cdata}
\end{equation}
Here $\tilde{n}$ denotes the rank of $\frak{\tilde{g}}$ and $\tilde{h}$ its
Coxeter number\footnote{%
We slightly abuse here the notation and use $c_{\frak{\tilde{g}}_{k}}$
instead of $c_{\frak{\tilde{g}}_{k}/u(1)^{\tilde{n}}}$. Since we always
encounter these type of coset in our discussion, we can avoid bulky
expressions in this way.}. Since we have $\Delta <1$ for all allowed values
of $k$, the perturbation is always relevant in the sense of renormalization
showing the Lagrangian to be of the general form (\ref{perCFT}).

The classical equations of motion following from (\ref{HSGaction})
correspond to non-abelian affine Toda equations \cite{HSG,Nonab}, which are
known to be classically integrable and admit soliton solutions. Identifying
these solutions by a Noether charge allows for a semi-classical approach to
the quantum theory by applying the Bohr-Sommerfeld quantization rule. In 
\cite{HSGsol} the semi-classical mass for the soliton labelled by the
quantum numbers ($i,a$) was found to be 
\begin{equation}
M_{i}^{a}=\frac{m_{a}}{\pi \beta ^{2}}\,\sin \frac{\pi i}{k}\;,\quad \quad
1\leq i\leq k-1,\;1\leq a\leq \tilde{n},  \label{solmass}
\end{equation}
where $\beta $ is a coupling constant and the $m_{a}$ are $\tilde{n}$
different mass scales. We used here the same convention as in the
construction of the $\frak{g}|\frak{\tilde{g}}$-theories (compare 3.3.2),
namely that the lower index $i$ refers to a vertex in the Dynkin diagram of $%
\frak{g}=A_{n}$ determining the type of soliton solution while the upper
index $a$ specifies a vertex in the Dynkin diagram of $\frak{\tilde{g}}$
fixing the colour of the soliton. Thus, there are $\tilde{n}=\limfunc{rank}%
\frak{\tilde{g}}$ towers containing $n=k-1$ solitons each of which describes
the simplest example of a HSG theory, the complex Sine-Gordon model \cite
{Park,CSG,CSGS} associated with the coset $su(2)_{k}/u(1)$. Note that the
semi-classical HSG mass spectrum matches the more general structure of the $%
\frak{g}|\frak{\tilde{g}}$-theories, see Chapter 3. In fact, the definition
of the latter \cite{FKcol} was inspired by the analysis of the HSG-models 
\cite{HSGS,CFKM}. We postpone the discussion of further semi-classical
formulas concerning the mass spectrum to the next subsection when
introducing the S-matrix for the quantum theory.

In view of the discussion in Chapter 3 the latter should factorize. In fact,
the integrability on the quantum level was established in \cite{HSG2} by the
construction of higher spin conserved charges via the procedure of
Zamolodchikov described at the beginning of this chapter. Based on the
assumption that the semi-classical spectrum is exact, the S-matrix elements
have then been determined in \cite{HSGS} by means of the bootstrap program
for HSG-models related to simply-laced Lie algebras.

\subsection{The Homogeneous Sine-Gordon\ S-matrix and resonances}

With regard to the choice $\frak{g}=A_{n}$ the proposed scattering matrix
consists partially of $\tilde{n}$ copies of minimal $A_{n},k=n+1$ or
equivalently $su(k)$-affine Toda field theories (ATFT) found by K\"{o}berle
and Swieca \cite{Zper}, which describe the scattering of solitons with the
same colour. As a consequence, the anti-particles are constructed in analogy
to the ATFT, that is from the automorphism which leaves the $su(k)$-Dynkin
diagram invariant, such that $\overline{(i,a)}=(k-i,a)$. The colour of a
particle and its anti-particle is identical. The scattering between solitons
belonging to different copies is described by an CDD factor which violates
parity (compare 3.3.2 and 3.3.3). In the notation of the $\frak{g}|\frak{%
\tilde{g}}$-theories the S-matrix then reads 
\begin{equation}
S_{ij}^{ab}(\theta )=e^{i\pi \varepsilon _{ab}A_{ij}^{-1}}\exp \int \frac{dt%
}{t}\left( 2\cosh \frac{\pi t}{k}-\tilde{I}\right) _{ab}\left( 2\cosh \frac{%
\pi t}{k}-I^{su(k)}\right) _{ij}^{-1}e^{-it(\theta +\sigma _{ab})}\,
\label{Sint0}
\end{equation}
Here $I^{su(k)}=2-A^{su(k)}$ denotes the incidence matrix of the $su(k)$%
-Dynkin diagram and $\tilde{I}=2-\tilde{A}$ the one corresponding to the
Dynkin diagram of $\frak{\tilde{g}}$. A new feature in comparison with the
definition (\ref{iSnew}) is the shift $\sigma _{ab}$ in the rapidity
variable. Below we will see that these shifts are functions of the vector
couplings in (\ref{HSGaction}) and are anti-symmetric in the colour values $%
\sigma _{ab}=-\sigma _{ba}$. Thus, parity is broken not only by the phase
factors $e^{i\pi \varepsilon _{ab}A_{ij}^{-1}}$ as discussed in Chapter 3
but also by the shifts $\sigma _{ab}$. Due to the fact that these shifts are
real, the function $S_{ij}^{ab}(\theta )$ for $a\neq b$ will have poles
beyond the imaginary axis such that the parameters $\sigma _{ab}$
characterize resonance poles.

Analyzing the above S-matrix in the associated block form (\ref{Snew}) we
have therefore have the following picture concerning the formation of bound
states: Two solitons with the same colour value may form a bound state of
the same colour, whilst solitons of different colour with $\tilde{A}%
_{ab}\neq 0,2$, say $(i,a)$ and $(j,b)$, may only form an unstable state,
say $(\tilde{k},\tilde{c}),$ whose lifetime and energy scale are
characterized by the parameter $\sigma $ by means of the Breit-Wigner
formula, see e.g. \cite{BW}, in the form 
\begin{eqnarray}
(M_{\tilde{k}}^{\tilde{c}})^{2}-\frac{(\Gamma _{\tilde{k}}^{\tilde{c}})^{2}}{%
4} &=&(M_{i}^{a})^{2}+(M_{j}^{b})^{2}+2M_{i}^{a}M_{j}^{b}\cosh \sigma
_{ab}\cos \Theta  \label{BW1} \\
M_{\tilde{k}}^{\tilde{c}}\Gamma _{\tilde{k}}^{\tilde{c}}
&=&2M_{i}^{a}M_{j}^{b}\sinh |\sigma _{ab}|\sin \Theta \,\,,  \label{BW2}
\end{eqnarray}
where the resonance pole in $S_{ij}^{ab}(\theta )$ is situated at $\theta
_{R}=\sigma _{ab}-i\Theta $ and $\Gamma _{\tilde{c}}^{\tilde{k}}$ denotes
the decay width of the unstable particle with mass $M_{\tilde{k}}^{\tilde{c}%
} $. In the case $i=j$ these unstable states can be identified with solitons
in the semi-classical limit \cite{HSGS,HSGsol}. When $\sigma $ becomes zero,
(\ref{BW2}) shows that the unstable particles become stable, but are still
not at the same footing as the other asymptotically stable particles. They
become virtual states characterized by poles on the imaginary axis beyond
the physical sheet.

How many free parameters do we have in our model? Computing mass shifts from
renormalization, we only accumulate contributions from intermediate states
having the same colour as the two scattering solitons. Thus, making use of
the well known fact that the masses of the minimal $su(k)$-affine Toda
theory all renormalize with an overall factor (\ref{ADEmass}), i.e. for the
solitons $(i,a)$ we have that $\delta M_{i}^{a}/M_{i}^{a}$ equals a constant
for fixed colour value $a$ and all possible values of the main quantum
number $i$, we acquire in principle $\tilde{n}$ different mass scales $%
m_{1},\ldots ,m_{\tilde{n}}$ in the HSG-model, see (\ref{solmass}). In
addition there are $\tilde{n}-1$ independent parameters in the model in form
of the possible rapidity shifts $\sigma _{ab}=-\sigma _{ba}$ for each $a,b$
such that $\tilde{A}_{ab}\neq 0,2$. This means overall we have $2\tilde{n}-1$
independent parameters in the quantum theory. There is a precise
correspondence to the free parameters which one obtains from the classical
point of view. In the latter case we have the $2\tilde{n}$ independent
components of $\Lambda _{\pm }$ at our free disposal. This number is reduced
by $1$ as a result of the symmetry $\Lambda _{+}\rightarrow s\Lambda _{+}$
and $\Lambda _{-}\rightarrow s^{-1}\Lambda _{-}$ which introduces an
additional dependence as may be seen from the explicit expressions for the
classical mass ratios and the classical resonance shifts 
\begin{equation}
\frac{m_{a}}{m_{b}}=\frac{M_{i}^{a}}{M_{i}^{b}}=\sqrt{\frac{\left\langle
\alpha _{a},\Lambda _{+}\right\rangle \left\langle \alpha _{a},\Lambda
_{-}\right\rangle }{\left\langle \alpha _{b},\Lambda _{+}\right\rangle
\left\langle \alpha _{b},\Lambda _{-}\right\rangle }},\qquad \quad \sigma
_{ab}=\ln \sqrt{\frac{\left\langle \alpha _{a},\Lambda _{+}\right\rangle
\left\langle \alpha _{b},\Lambda _{-}\right\rangle }{\left\langle \alpha
_{a},\Lambda _{-}\right\rangle \left\langle \alpha _{b},\Lambda
_{+}\right\rangle }}\,.  \label{sig}
\end{equation}
Here the $\alpha _{a}$'s denote the simple roots of $\frak{\tilde{g}}$. Note
the anti-symmetry property $\sigma _{ab}=-\sigma _{ba}$ mentioned before. In 
\cite{HSGS} the parity violation to which this property gives rise was
motivated by the following semi-classical observation. The unstable bound
state formed by two solitons $(i,a)$ and $(i,b)$ is kinematically allowed to
decay in two different ways which are linked to each other by a parity
transformation. However, tree-level calculations in perturbation theory
showed that only the decay amplitude of one of these decay processes is
non-vanishing.

\subsection{Ultraviolet central charge for the HSG model}

In this section we are going to determine the conformal field theory which
governs the UV regime of the quantum field theory associated with the
S-matrix elements (\ref{Sint0}). According to the defining relation (\ref
{HSGaction}) and the discussion of the previous section, we expect to
recover the WZNW-coset theory with effective central charge (\ref{cdata}) in
the extreme ultraviolet limit. We start by stating the TBA-kernel $\varphi
_{ij}^{ab}(\theta )=-i\frac{d}{d\theta }\ln S_{ij}^{ab}(\theta )$\ for HSG
theories belonging to simply-laced algebras. From the integral
representation (\ref{Sint0}) it is calculated to 
\begin{equation}
\varphi _{ij}^{ab}(\theta )=\int\limits_{-\infty }^{\infty }dt\left[ \delta
_{ab}\delta _{ij}-\left( 2\cosh \frac{\pi t}{h}-\tilde{I}\right) _{ab}\left(
2\cosh \frac{\pi t}{h}-I^{su(k)}\right) _{ij}^{-1}\,\right]
\,\,e^{-it(\theta +\sigma _{ab})}\;,  \label{hsgk}
\end{equation}
where the extra term involving the two Kronecker $\delta $'s is due to the
fact that integration and differentiation do not commute analogous to the
discussion of affine Toda models. Careful comparison with the block
representation (\ref{Snew}) then yields the additional contribution.
Different form the discussion of affine Toda models we are not going to
apply the large density approximation since the numerical solution will show
a different behaviour. In fact, instead of growing over all bounds as $%
r\rightarrow 0$ we will find that the solutions to the TBA equations (\ref
{TBAE}) approach a finite constant value in the region $\ln \frac{r}{2}\ll
\theta \ll \ln \frac{2}{r}$. Thus, we need not to consider the series
expansion of the TBA-kernel as in Section 4.3.

\subsubsection{TBA with parity violation}

Since up to now the TBA analysis has only been carried out for parity
invariant S-matrices, a few preliminary comments are due to implement parity
violation. The starting point in the derivation of the key equations (\ref
{TBAE}) are the Bethe ansatz equations (\ref{BAE}), which are the outcome of
dragging one soliton, say of type $A=(i,a)$, along the world line, compare
4.2.2. For the time being we do not need the distinction between the two
quantum numbers. On this trip the formal wave-function of $A$ picks up the
corresponding S-matrix element as a phase factor when meeting another
soliton. Due to the parity violation it matters, whether the soliton is
moved clockwise or counter-clockwise along the world line, such that we end
up with two different sets of Bethe Ansatz equations 
\begin{equation}
e^{iLM_{A}\sinh \theta _{A}}\prod\limits_{B\neq A}S_{AB}(\theta _{A}-\theta
_{B})=1\quad \text{and\quad }e^{-iLM_{A}\sinh \theta
_{A}}\prod\limits_{B\neq A}S_{BA}(\theta _{B}-\theta _{A})=1\,,  \label{BA}
\end{equation}
with $L$ denoting the length of the compactified space direction, compare
4.2.1 and 4.2.2. These two sets of equations are of course not entirely
independent and may be obtained from each other by complex conjugation
exploiting the Hermitian analyticity (\ref{ha}) of the scattering matrix. We
may now carry out the thermodynamic limit of (\ref{BA}) as in 4.2.2 for the
parity preserving case and obtain the following sets of non-linear integral
equations 
\begin{eqnarray}
\epsilon _{A}^{+}(\theta )+\sum_{B}\,\varphi _{AB}\ast L_{B}^{+}(\theta )
&=&r\,M_{A}\cosh \theta \quad  \label{tba} \\
\epsilon _{A}^{-}(\theta )+\sum_{B}\,\varphi _{BA}\ast L_{B}^{-}(\theta )
&=&r\,M_{A}\cosh \theta \,\,\,.  \label{ptba}
\end{eqnarray}
Recall that '$\ast $' denotes the rapidity convolution of two functions and
that $r=m_{1}R$ is the inverse temperature times the overall mass scale $%
m_{1}$ of the lightest particle. We also re-defined the masses by $%
M_{i}^{a}\rightarrow M_{i}^{a}/m_{1}$ keeping, however, the same notation.
As in the case of ATFT we have chosen Fermi statistics by introducing the
so-called pseudo-energies $\epsilon _{A}^{+}(\theta )=\epsilon
_{A}^{-}(-\theta )$ and the related functions $L_{A}^{\pm }(\theta )=\ln
(1+e^{-\epsilon _{A}^{\pm }(\theta )})$. The TBA kernels $\varphi _{AB}$ in
the integrals carry the information of the dynamical interaction of the
system and are given by (\ref{hsgk}). Notice that (\ref{ptba}) may be
obtained from (\ref{tba}) simply by the parity transformation $\theta
\rightarrow -\theta $ and the equality $\varphi _{AB}(\theta )=\varphi
_{BA}(-\theta )$ following from Hermitian analyticity (\ref{ha}). The main
difference of these equations in comparison with the parity invariant case
is that we have lost the usual symmetry of the pseudo-energies as a function
of the rapidities, such that we have now in general $\epsilon
_{A}^{+}(\theta )\neq \epsilon _{A}^{-}(\theta )$. This symmetry may be
recovered when restoring parity.

The scaling function (\ref{Scale}) may be computed similar as in the usual
way 
\begin{equation}
c(r)=\frac{3\,r}{\pi ^{2}}\sum_{A}M_{A}\int\limits_{0}^{\infty }d\theta
\,\cosh \theta \,(L_{A}^{-}(\theta )+L_{A}^{+}(\theta ))\,,  \label{scale}
\end{equation}
once the equations (\ref{tba}) have been solved for the $\epsilon _{A}^{\pm
}(\theta )$. Of special interest is the deep UV limit, i.e. $r\rightarrow 0$%
, in which one recovers the effective central charge $c_{\text{eff}%
}=c-12(\Delta _{0}+\bar{\Delta}_{0})$. Recall that $c$ is the Virasoro
central charge and $\Delta _{0},\bar{\Delta}_{0}$ are the lowest conformal
dimensions related to the two chiral sectors of the conformal model. Since
the WZNW-coset is unitary, we expect that $\Delta _{0},\bar{\Delta}_{0}=0$
and $c_{\text{eff}}=c$. This assumption will turn out to be consistent with
the analytical and numerical results.

\subsubsection{The central charge calculation}

In this section we follow the usual argumentation of the TBA-analysis which
leads to the effective central charge, paying, however, attention to the
parity violation. We will recover indeed the value in (\ref{cdata}) as the
central charge of the HSG-models. Extending the general formula (\ref{ceffg}%
) to the case at hand we infer that only the value $\epsilon _{A}^{\pm }(0)$
of the solutions in the UV limit is required. Hence, we take the limits $%
r,\theta \rightarrow 0$ of (\ref{tba}) and (\ref{ptba}). When we assume that
the kernels $\varphi _{AB}(\theta )$ are strongly peaked\footnote[6]{%
That this assumption holds for the case at hand is most easily seen by
noting that the logarithmic derivative of a basic building block of the
S-matrix reads 
\begin{equation*}
-i\frac{d}{d\theta }\ln \frac{\sinh \frac{1}{2}(\theta +\frac{i\pi }{k})}{%
\sinh \frac{1}{2}(\theta +\frac{i\pi }{k})}=-\frac{\sin \left( \frac{\pi }{k}%
\,x\right) }{\cosh \theta -\cos \left( \frac{\pi }{k}\,x\right) }%
=-2\sum_{n=1}^{\infty }\sin \left( \frac{\pi }{k}\,x\right) e^{-n|\theta
|}\,.
\end{equation*}
From this we can read off directly the decay properties.} at $\theta =0$ and
that the solutions develop characteristic plateaus of constant height in the
region $\ln \frac{r}{2}\ll \theta \ll \ln \frac{2}{r}$ (as one observes for
the scaling models of minimal ATFT \cite{TBAZam1,TBAKM}\ for example), we
can take out the $L$-functions from the integral in the equations (\ref{tba}%
), (\ref{ptba}) and obtain the following determining equation 
\begin{equation}
\epsilon _{A}^{\pm }(0)+\sum_{B}\,N_{AB}L_{B}^{\pm }(0)=0\qquad \text{with
\quad }N_{AB}=\frac{1}{2\pi }\int\limits_{-\infty }^{\infty }d\theta
\,\,\varphi _{AB}(\theta )\,\,.  \label{consttba}
\end{equation}
The on-shell energies have been discarded since their contribution in the
interval $\ln \frac{r}{2}\ll \theta \ll \ln \frac{2}{r}$ is minute. In
particular, the above equation is assumed to become exact in the extreme
limit. Having the resonance parameter $\sigma $ present in our theory we may
also encounter a situation in which $\varphi _{AB}(\theta )$ is peaked at $%
\theta =\pm \sigma $ with $\sigma $ finite. This means in order for (\ref
{consttba}) to be valid, we have to assume $\epsilon _{A}^{\pm }(0)=\epsilon
_{A}^{\pm }(\pm \sigma )$ in the limit $r\rightarrow 0$ in addition to
accommodate that situation. \emph{The last assumption as well as the
characteristic plateau behaviour will be justified in retrospective for
particular cases from the numerical results (see e.g. Figure 4.6).} Note
that in (\ref{consttba}) we have recovered the parity invariance.

For small values of $r$ we may approximate, in analogy to the parity
invariant situation discussed in 4.2.2, $rM_{A}\cosh \theta $ by $\frac{r}{2}%
M_{A}\exp \theta $, such that taking the derivative of the relations (\ref
{tba}) and (\ref{ptba}) thereafter yields 
\begin{equation}
\frac{\epsilon _{A}^{\pm }(\theta )}{d\theta }+\frac{1}{2\pi }%
\sum_{B}\int\limits_{-\infty }^{\infty }d\theta ^{\prime }\frac{\varphi
_{AB}(\pm \theta \mp \theta ^{\prime })}{1+e^{\epsilon _{B}^{\pm }(\theta
^{\prime })}}\frac{d\epsilon _{B}^{\pm }(\theta ^{\prime })}{d\theta
^{\prime }}\,\approx \frac{r}{2}M_{A}\exp \theta \quad .  \label{der}
\end{equation}
The scaling function acquires the form 
\begin{equation}
c(r)\approx \frac{3\,r}{2\pi ^{2}}\sum_{A}M_{A}\int\limits_{0}^{\infty
}d\theta \,e^{\theta }(L_{A}^{-}(\theta )+L_{A}^{+}(\theta ))\,,\quad \quad 
\text{for }r\approx 0  \label{appc}
\end{equation}
in this approximation. Replacing in (\ref{appc}) the term $\frac{r}{2}%
M_{A}\exp \theta $ by the l.h.s. of (\ref{der}) a few manipulations similar
to those in Section 4.2.2 lead to 
\begin{equation}
\lim_{r\rightarrow 0}c(r)\simeq \frac{3}{2\pi ^{2}}\sum\limits_{p=+,-}%
\sum_{A}\int\limits_{\epsilon _{A}^{p}(0)}^{\epsilon _{A}^{p}(\infty
)}d\epsilon _{A}^{p}\left[ \ln (1+e^{-\epsilon _{A}^{p}})+\frac{\epsilon
_{A}^{p}}{1+e^{-\epsilon _{A}^{p}}}\right] \,\,\,.
\end{equation}
Upon the substitution $y_{A}^{p}=1/(1+\exp (\epsilon _{A}^{p}))$ we obtain
the well known expression (\ref{ceffg}) for the effective central charge 
\begin{equation}
c_{\text{eff}}=\frac{6}{\pi ^{2}}\sum_{A}\mathcal{L}\left( \frac{1}{%
1+e^{\epsilon _{A}^{\pm }(0)}}\right) \,\,.  \label{hsgceff}
\end{equation}
Here we used the integral representation for Roger's dilogarithm function (%
\ref{roger}) and the facts that $\epsilon _{A}^{+}(0)=\epsilon _{A}^{-}(0)$, 
$y_{A}^{+}(\infty )=y_{A}^{-}(\infty )=0$. This means we end up precisely
with the same situation as in the parity invariant case for scaling models 
\cite{TBAZam1}: Determining at first the phases of the scattering matrices
we have to solve the constant TBA-equation (\ref{consttba}) and may compute
the effective central charge in terms of Roger's dilogarithm thereafter.
Notice that in the ultraviolet limit we have recovered the parity invariance
and (\ref{hsgceff}) holds for all finite values of the resonance parameter.

For the case at hand we read off from the integral representation (\ref{hsgk}%
) of the TBA kernel 
\begin{equation}
N_{ij}^{ab}=\delta _{ij}\delta _{ab}-\tilde{A}_{ab}\,(A^{su(k)})_{\,\,\,%
\,ij}^{-1}\,\,.  \label{NN}
\end{equation}
With $N_{ij}^{ab}$ in the form (\ref{NN}) and the identifications $%
Q_{i}^{a}=\prod_{j=1}^{k-1}(1+e^{-\epsilon
_{j}^{a}(0)})^{(A^{su(k)})_{ij}^{-1}}$ the constant TBA-equations are
precisely the equations which occur in the context of the restricted
solid-on-solid models \cite{Kirillov,Kuniba}. It was noted in there that (%
\ref{consttba}) may be solved elegantly in terms of Weyl-characters and the
reported effective central charge coincides indeed with the one for the
HSG-models (\ref{cdata}).

It should be stressed that we understand the $N$-matrix here as defined in (%
\ref{consttba}) and not as the difference between the phases of the
S-matrix. In the latter case we encounter contributions from the non-trivial
constant phase factors in (\ref{Sint0}). Also in that case we may arrive at
the same answer by compensating them with a choice of a non-standard
statistical interaction as outlined in \cite{BFs}.

We would like to close this section with a comment which links our analysis
to structures which may be observed directly inside the conformal field
theory. When one carries out a saddle point analysis, see e.g. \cite{Rich},
on a Virasoro character (\ref{Virch}) which can be written in the particular
form 
\begin{equation}
\chi (q)=\sum\limits_{\vec{m}=0}^{\infty }\frac{q^{\frac{1}{2}\vec{m}(%
\mathbf{1}-N)\vec{m}^{t}+\vec{m}\cdot \vec{B}}}{(q)_{1}\ldots (q)_{(k-1)%
\text{$\tilde{n}$}}}\,\,,  \label{chi}
\end{equation}
with $(q)_{m}=\prod_{k=1}^{m}(1-q^{k})$, one recovers the set of coupled
equations as (\ref{consttba}) and the corresponding effective central charge
is expressible as a sum of Roger's dilogarithms as (\ref{ceff}). Note that
when we choose $\frak{g}\equiv A_{1}$ the HSG-model reduces to the minimal
ATFT and (\ref{chi}) reduces to the character formulae in \cite{KM}.

\emph{Thus, the equations (\ref{consttba}) and (\ref{ceff}) constitute an
interface between massive and massless theories, since they may be obtained
on one hand in the ultraviolet limit from a massive model and on the other
hand from a limit inside the conformal field theory.}

This means we can guess a new form of the coset character, by substituting (%
\ref{NN}) into (\ref{chi}). However, since the specific form of the vector $%
\vec{B}$ does not enter in this analysis (it distinguishes the different
highest weight representations) more work needs to be done in order to make
this more than a mere conjecture. This issue is left for future
investigations.

\subsection{The su(3)$_{k}$-HSG model}

We shall now focus our discussion on $\frak{\tilde{g}}=A_{2}\equiv su(3)$
and explore the structure of the scaling function in more detail
supplemented by a detailed numerical analysis at the end of this section.
Special emphasis will be given to the role of the resonance parameters.
Similar as for staircase models \cite{ZamoR,Stair} studied previously, they
allow to describe the ultraviolet limit of the HSG-model alternatively as
the flow between different conformal field theories in the ultraviolet and
infrared regime. Below we will find the following massless flow \cite{CFKM} 
\begin{equation}
\text{UV}\equiv su(3)_{k}/u(1)^{2}\leftrightarrow su(2)/u(1)\otimes
su(2)/u(1)\equiv \text{IR\ .}  \label{UVIR}
\end{equation}
We also observe the flow $(su(3)_{k}/u(1)^{2})/(su(2)_{k}/u(1))\rightarrow
su(2)_{k}/u(1)$ as a subsystem inside the HSG-model. For $k=2$ this
subsystem describes the flow between the tricritical Ising and the Ising
model previously studied in \cite{triZam}. In terms of the HSG-model we will
then obtain the following physical picture \cite{CFKM}: The resonance
parameter characterizes the mass scale of the unstable particles (\ref{BW1}%
). Approaching the extreme ultraviolet regime from the infrared we pass
various regions: At first all solitons are too heavy to contribute to the
off-critical central charge, then the two copies of the minimal ATFT will
set in, leading to the central charge corresponding to IR in (\ref{UVIR})
and finally the unstable bound states will start to contribute such that we
indeed obtain (\ref{cdata}) as the ultraviolet central charge of the
HSG-model in accordance with the result of the previous subsection.

First of all we need to establish how many free parameters we have at our
disposal in the case $\frak{\tilde{g}}=su(3)$. According to the discussion
in Section 4.4.1 we can tune the resonance parameter and the mass ratio 
\begin{equation}
\sigma :=\sigma _{21}=-\sigma _{12}\quad \text{and}\quad m_{1}/m_{2}\,\,.
\end{equation}
It will also be useful to exploit a symmetry present in the TBA-equations
related to $SU(3)_{k}$ by noting that the parity transformed equations (\ref
{ptba}) turn into the equations (\ref{tba}) when we exchange the masses of
the different types of solitons. This means the system remains invariant
under the simultaneous transformations 
\begin{equation}
\theta \rightarrow -\theta \quad \quad \text{and\qquad }m_{1}/m_{2}%
\rightarrow m_{2}/m_{1}\,\,.  \label{inv}
\end{equation}
For the special case $m_{1}/m_{2}=1$ we deduce therefore that $\epsilon
_{a}^{(1)}(\theta )=\epsilon _{a}^{(2)}(-\theta )$, meaning that a parity
transformation then amounts to an interchange of the colours. Furthermore,
we see from (\ref{ptba}) and the defining relation $\sigma =\sigma
_{21}=-\sigma _{12}$ that changing the sign of the rapidity variable is
equivalent to $\sigma \rightarrow -\sigma $. Therefore, we can restrict
ourselves to the choice $\sigma \geq 0$ without loss of generality.

\subsubsection{Staircase behaviour of the scaling function}

We will now come to the evaluation of the scaling function (\ref{scale}) for
finite and small scale parameter $r$. To do this we have to solve first the
TBA equations (\ref{tba}) for the pseudo-energies, which due to the
non-linear nature of the integral equations is hardly achieved analytically
as pointed out before. We therefore pass on to the UV regime with $r\ll 1$
where we can set up approximate TBA equations involving formally massless
particles\footnote[7]{%
The concept of massless scattering has been introduced originally in \cite
{triZam} as follows: The on-shell energy of a right and left moving particle
is given by $E_{\pm }=M/2e^{\pm \theta }$ which is formally obtained from
the on-shell energy of a massive particle $E=m\cosh \theta $ by the
replacement $\theta \rightarrow \theta \pm \sigma /2$ and taking the limit $%
m\rightarrow 0,\sigma \rightarrow \infty $ while keeping the expression $%
M=me^{\theta +\sigma /2}$ finite. It are these on-shell energies we will
encounter in our analysis.} analogous to the general discussion in 4.2.1.
Certain statements and approximation schemes we will use in order to extract
the staircase behaviour of the scaling function depend on the general form
of the $L$-functions. Since the latter is not known a priori, one may
justify ones assumptions in retrospective by referring to the numerics at
the end of this section, where we present numerical solutions for the
equations (\ref{tba}) for various levels $k$. The latter show that the
L-functions develop at most two (three if $m_{1}$ and $m_{2}$ are very
different) plateaus in the range $\ln \frac{r}{2}\ll \theta \ll \ln \frac{2}{%
r}$ and then fall off rapidly (see Figure 4.6). This type of behaviour is
similar to the one known from minimal ATFT \cite{TBAZam1,TBAKM}, and we can
therefore adopt various arguments presented in that context. The main
difficulty we have to deal with here is to find the appropriate ``massless''
TBA equations accommodating the dependence of the TBA equations on the
resonance shifts $\sigma $.

We start by separating the kernel (\ref{hsgk}) into two parts 
\begin{eqnarray}
\phi _{ij}(\theta ) &=&\Phi _{ij}^{aa}(\theta )=\int dt\,\left[ \delta
_{ij}-2\cosh \tfrac{\pi t}{k}\left( 2\cosh \tfrac{\pi t}{k}-I^{su(k)}\right)
_{ij}^{-1}\right] \,e^{-it\theta }\;,\quad  \label{kernelint} \\
\psi _{ij}(\theta ) &=&\Phi _{ij}^{ab}(\theta +\sigma _{ba})=\int dt\,\left(
2\cosh \tfrac{\pi t}{k}-I^{su(k)}\right) _{ij}^{-1}\,e^{-it\theta }\;\quad
,\;a\neq b\;.  \label{kernelint2}
\end{eqnarray}
Here $\phi _{ij}(\theta )$ is just the TBA kernel of the $su(k)$-minimal
ATFT and in the remaining kernels $\psi _{ij}(\theta )$ we have removed the
resonance shift. Note that $\phi ,\psi $ do not depend on the colour values $%
a,b$ which may therefore be dropped all together in the notation. Note that
the integral representations for these kernels are generically valid for all
values of the level $k$. The convolution term in (\ref{tba}) in terms of $%
\phi ,\psi $ is then re-written as 
\begin{equation}
\sum_{b=1}^{\tilde{n}}\sum_{j=1}^{k-1}\varphi _{ij}^{ab}\ast
L_{j}^{b}(\theta )=\sum_{j=1}^{k-1}\phi _{ij}\ast L_{j}^{a}(\theta )+\sum 
_{\substack{ b=1  \\ b\neq a}}^{\tilde{n}}\sum_{j=1}^{k-1}\psi _{ij}\ast
L_{j}^{b}(\theta -\sigma _{ba})\,.  \label{conv}
\end{equation}
These equations illustrate that whenever we are in a regime in which the
second term in (\ref{conv}) is negligible we are left with $\tilde{n}$
non-interacting copies of the $su(k)$-minimal ATFT.

We will now specialize the discussion on the $su(3)_{k}$-case for which we
can eliminate the dependence on $\sigma $ in the second convolution term by
performing the shifts $\theta \rightarrow \theta \pm \sigma /2$ in the TBA
equations. In the UV limit $r\rightarrow 0$ with $\sigma \gg 1$ the shifted
functions can be approximated by the solutions of the following sets of
integral equations 
\begin{eqnarray}
\varepsilon _{i}^{\pm }(\theta )+\sum_{j=1}^{k-1}\phi _{ij}\ast L_{j}^{\pm
}\left( \theta \right) +\sum_{j=1}^{k-1}\psi _{ij}\ast L_{j}^{\mp }\left(
\theta \right) &=&r^{\prime }\,M_{i}^{\pm }\,e^{\pm \theta }\quad \,
\label{uvTba} \\
\hat{\varepsilon}_{i}^{\pm }(\theta )+\sum_{j=1}^{k-1}\phi _{ij}\ast \hat{L}%
_{j}^{\pm }\left( \theta \right) &=&r^{\prime }\,M_{i}^{\mp }\,\,e^{\pm
\theta }\,,\quad  \label{kinktba}
\end{eqnarray}
where we have introduced the parameter $r^{\prime }=r\,e^{\frac{\sigma }{2}%
}/2$ familiar from the discussion of massless scattering and the masses $%
M_{i}^{+/-}=M_{i}^{(1)/(2)}$. The relationship between the solutions of the
massless system (\ref{uvTba}), (\ref{kinktba}) and those of the original
TBA-equations is given by 
\begin{eqnarray}
\epsilon _{i}^{(1)/(2)}(\theta ) &=&\varepsilon ^{+/-}(\theta \mp \sigma
/2)\qquad \text{for\quad }\ln \tfrac{r}{2}\ll \pm \theta \ll \ln \tfrac{r}{2}%
+\sigma  \label{e1} \\
\epsilon _{i}^{(1)/(2)}(\theta ) &=&\hat{\varepsilon}^{-/+}(\theta \pm
\sigma /2)\qquad \text{for\quad }\pm \theta \ll \min [\ln \tfrac{2}{r},\ln 
\tfrac{r}{2}+\sigma ]  \label{e2}
\end{eqnarray}
where we have assumed $m_{1}=m_{2}$. Similar equations may be written down
for the generic case. To derive (\ref{e2}) we have neglected here the
convolution terms $(\psi _{ij}\ast L_{j}^{(1)})(\theta +\sigma )$ and $(\psi
_{ij}\ast L_{j}^{(2)})(\theta -\sigma )$ which appear in the TBA-equations
for $\epsilon _{i}^{(2)}(\theta )$ and $\epsilon _{i}^{(1)}(\theta )$,
respectively. This is justified by the following argument: For $|\theta |\gg 
$ $\ln \frac{2}{r}$ the free on-shell energy term is dominant in the TBA
equations, i.e. $\epsilon _{i}^{a}(\theta )\approx r\,M_{i}^{a}\cosh \theta $
and the functions $L_{i}^{a}(\theta )$ are almost zero. The kernels $\psi
_{ij}$ are centered in a region around the origin\thinspace $\theta =0$
outside of which they exponentially decrease, see footnote in Subsection
4.4.3 for this. This means that the convolution terms in question can be
neglected safely if $\theta \ll \ln \frac{r}{2}+\sigma $ and $\theta \gg \ln 
\frac{2}{r}-\sigma $, respectively. Note that the massless system provides a
solution for the whole range of $\theta $ for the non-vanishing $L$-function
only if the ranges of validity in (\ref{e1}) and (\ref{e2}) overlap, i.e. if 
$\ln \frac{r}{2}\ll \min [\ln \frac{2}{r},\ln \frac{r}{2}+\sigma ],$ which
is always true in the limit $r\rightarrow 0$ when $\sigma \gg 0$. The
solution is uniquely defined in the overlapping region. These observations
are confirmed by our numerical analysis below.

The resulting equations (\ref{kinktba}) are therefore decoupled and we can
determine $\hat{L}^{+}$ and $\hat{L}^{-}$ individually. In contrast, the
equations (\ref{uvTba}) for $L_{i}^{\pm }$ are still coupled to each other
due to the presence of the resonance shift. Formally, the on-shell energies
for massive particles have been replaced by on-shell energies for massless
particles in the sense of \cite{triZam}, such that if we interpret $%
r^{\prime }$ as an independent new scale parameter the sets of equations (%
\ref{uvTba}) and (\ref{kinktba}) could be identified as massless TBA systems
in their own right.

Introducing then the scaling function associated with the system (\ref{uvTba}%
) as 
\begin{equation}
c_{\text{o}}(r^{\prime })=\frac{3\,r^{\prime }}{\pi ^{2}}\sum_{i=1}^{k-1}%
\int d\theta \,\,\left[ M_{i}^{+}\,e^{\theta }L_{i}^{+}(\theta
)+\,M_{i}^{-}\,e^{-\theta }L_{i}^{-}(\theta )\right]  \label{c0}
\end{equation}
and analogously the scaling function associated with (\ref{kinktba}) as 
\begin{equation}
\hat{c}_{\text{o}}(r^{\prime })=\frac{3\,r^{\prime }}{\pi ^{2}}%
\sum_{i=1}^{k-1}\int d\theta \,\left[ M_{i}^{+}\,e^{\theta }\hat{L}%
_{i}^{+}(\theta )+M_{i}^{-}\,e^{-\theta }\hat{L}_{i}^{-}(\theta )\right] \;
\label{ckink}
\end{equation}
we can express the scaling function (\ref{scale}) of the HSG model in the
parameter regime $r\rightarrow 0,\;\sigma \gg 1$ approximately by \cite{CFKM}
\begin{eqnarray}
c(r,\sigma ) &=&\frac{3\,r\,e^{\frac{\sigma }{2}}}{2\pi ^{2}}%
\sum_{a=1,2}\sum_{i=1}^{k-1}M_{i}^{a}\int d\theta \,\left[ \,e^{\theta
}L_{i}^{a}(\theta -\tfrac{\sigma }{2})+e^{-\theta }L_{i}^{a}(\theta +\tfrac{%
\sigma }{2})\right]  \notag \\
&\approx &c_{\text{o}}\left( r^{\prime }\right) +\hat{c}_{\text{o}}\left(
r^{\prime }\right) \,\;.  \label{uvscale}
\end{eqnarray}
\textbf{Remark}. \emph{Notice, that we have formally decomposed the massive }%
$su(3)_{k}$\emph{-HSG model in the UV regime into two massless TBA systems (%
\ref{uvTba}) and (\ref{kinktba}), reducing therefore the problem of
calculating the scaling function of the HSG model in the UV limit }$%
r\rightarrow 0$\emph{\ to the problem of evaluating the scaling functions (%
\ref{c0}) and (\ref{ckink}) for the scale parameter }$r^{\prime }$. \emph{%
The latter depends on the relative size of }$\ln \frac{2}{r}$\emph{\ and the
resonance shift }$\sigma /2.\medskip $

Keeping now $\sigma \gg 0$ fixed, and letting $r$ vary from finite values to
the deep UV regime, i.e. $r=0$, the scale parameter $r^{\prime }$ governing
the massless TBA systems will pass different regions. For the regime $\ln 
\frac{2}{r}<\sigma /2$ we see that the scaling functions (\ref{c0}) and (\ref
{ckink}) are evaluated at $r^{\prime }>1$, whereas for $\ln \frac{2}{r}%
>\sigma /2$ they are taken at $r^{\prime }<1$. Thus, when performing the UV
limit of the HSG model the massless TBA systems pass formally from the
``infrared'' to the ``ultraviolet'' regime with respect to the parameter $%
r^{\prime }$. We emphasize that the scaling parameter $r^{\prime }$ has only
a formal meaning and that the physical relevant limit we consider is still
the UV limit $r\rightarrow 0$ of the HSG model. However, proceeding this way
has the advantage that we can treat the scaling function of the HSG model by
the UV and IR central charges of the systems (\ref{uvTba}) and (\ref{kinktba}%
) as functions of $r^{\prime }$ \cite{CFKM} 
\begin{equation}
c(r,\sigma )\approx c_{\text{o}}\left( r^{\prime }\right) +\hat{c}_{\text{o}%
}\left( r^{\prime }\right) \approx \left\{ 
\begin{array}{ll}
\,c_{IR}+\hat{c}_{IR}\,, & 0\ll \ln \frac{2}{r}\ll \frac{\sigma }{2}\medskip
\\ 
c_{UV}+\hat{c}_{UV}\,, & \frac{\sigma }{2}\ll \ln \frac{2}{r}
\end{array}
\right. \,\,.  \label{step}
\end{equation}
Here we defined the quantities $c_{IR}:=\lim_{r^{\prime }\rightarrow \infty
}c_{\text{o}}(r^{\prime })$, $c_{UV}:=\lim_{r^{\prime }\rightarrow 0}c_{%
\text{o}}(r^{\prime })$ and $\hat{c}_{IR},\hat{c}_{UV}$ analogously in terms
of $\hat{c}_{\text{o}}(r^{\prime })$.

In the case of $c_{IR}+\hat{c}_{IR}\neq c_{UV}+\hat{c}_{UV}$, we infer from (%
\ref{step}) that the scaling function develops at least two plateaus at
different heights. A similar phenomenon was previously observed for theories
discussed in \cite{Stair}, where infinitely many plateaus occurred which
prompted to call them ``staircase models''. As a difference, however, the
TBA equations related to these models do not break parity. In the next
subsection we determine the central charges belonging to the different
regimes in (\ref{step}) by means of standard TBA central charge calculation,
setting up the so-called constant TBA equations.

\subsubsection{Central charges from constant TBA equations}

In this subsection we will perform the limits $r^{\prime }\rightarrow 0$ and 
$r^{\prime }\rightarrow \infty $ for the massless systems (\ref{uvTba}) and (%
\ref{kinktba}) referring to them formally as ``UV-'' and ``IR-limit'',
respectively, keeping however in mind that both limits are still linked to
the UV limit of the HSG model in the scale parameter $r$ as discussed in the
preceding subsection. We commence with the discussion of the ``UV limit'' $%
r^{\prime }\rightarrow 0$ for the subsystem (\ref{uvTba}). We then have
three different rapidity regions in which the pseudo-energies are
approximately given by 
\begin{equation}
\varepsilon _{i}^{\pm }(\theta )\approx \left\{ 
\begin{array}{ll}
r^{\prime }M_{i}^{\pm }\,e^{\pm \theta }, & \text{for }\pm \theta \gg -\ln
r^{\prime } \\ 
-\sum_{j}\phi _{ij}\ast L_{j}^{\pm }(\theta )-\sum_{j}\psi _{ij}\ast
L_{j}^{\mp }(\theta ), & \text{for }\ln r^{\prime }\ll \theta \ll -\ln
r^{\prime } \\ 
-\sum_{j}\phi _{ij}\ast L_{j}^{\pm }(\theta ), & \text{for }\pm \theta \ll
\ln r^{\prime }
\end{array}
\right. \;.
\end{equation}
We have only kept here the dominant terms up to exponentially small
corrections. We proceed analogously to the discussion as may be found in 
\cite{TBAZam1,TBAKM}. We see that in the first region the particles become
asymptotically free. For the other two regions the TBA equations can be
solved by assuming the L-functions to be constant. Exploiting once more that
the TBA kernels are centered at the origin and decay exponentially, we can
similar as in Subsection 4.4.3 take the $L$-functions outside of the
integrals and end up with the sets of equations 
\begin{eqnarray}
x_{i}^{\pm } &=&\prod_{j=1}^{k-1}(1+x_{j}^{\pm })^{\hat{N}%
_{ij}}(1+x_{j}^{\mp })^{N_{ij}^{\prime }}\quad \quad \text{for }\ln
r^{\prime }\ll \theta \ll -\ln r^{\prime }  \label{ctba1} \\
\hat{x}_{i} &=&\prod_{j=1}^{k-1}(1+\hat{x}_{j})^{\hat{N}_{ij}}\qquad \qquad
\qquad \quad \,\text{for }\pm \theta \ll \ln r^{\prime }  \label{ctba2}
\end{eqnarray}
for the constants $x_{i}^{\pm }=e^{-\varepsilon _{i}^{\pm }(0)}$ and $\hat{x}%
_{i}=e^{-\varepsilon _{i}^{\pm }(\mp \infty )}$. The N-matrices can be read
off directly from the integral representations (\ref{kernelint}) and (\ref
{kernelint2}) 
\begin{equation}
\quad \hat{N}:=\frac{1}{2\pi }\int \phi =1-2(A^{su(k)})^{-1}\qquad \text{%
and\qquad }\;N^{\prime }:=\frac{1}{2\pi }\int \psi =(A^{su(k)})^{-1}\,\,.
\end{equation}
Since the set of equations (\ref{ctba2}) has already been stated in the
context of minimal ATFT and its solutions may be found in \cite{TBAKM}, we
only need to investigate the equations (\ref{ctba1}). These equations are
simplified by the following observation. Sending $\theta $ to $-\theta $ the
constant L-functions must obey the same constant TBA equation (\ref{ctba1})
but with the role of $L_{i}^{+}$ and $L_{i}^{-}$ interchanged. The
difference in the masses $m_{1},m_{2}$ has no effect as long as $m_{1}\sim
m_{2}$ since the on-shell energies are negligible in the central region $\ln
r^{\prime }\ll \theta \ll -\ln r^{\prime }$. Thus, we deduce $%
x_{i}^{+}=x_{i}^{-}=:x_{i}$ and (\ref{ctba1}) reduces to 
\begin{equation}
x_{i}=\prod_{j=1}^{k-1}(1+x_{j})^{N_{ij}}\;\qquad \quad \text{with}\quad
N=1-(A^{su(k)})^{-1}\;.  \label{cTba2a}
\end{equation}
Remarkably, also these set of equations\ may be found in the literature in
the context of the restricted solid-on-solid models \cite{Kuniba} as already
has been pointed out in 4.4.3. Specializing some of the general
Weyl-character formulae in \cite{Kuniba} to the $su(3)_{k}$-case a
straightforward calculation leads to 
\begin{equation}
x_{i}=\frac{\sin \left( \frac{\pi \,(i+1)}{k+3}\right) \sin \left( \frac{\pi
\,(i+2)}{k+3}\right) }{\sin \left( \frac{\pi \,\,i}{k+3}\right) \sin \left( 
\frac{\pi \,(i+3)}{k+3}\right) }-1\quad \text{and\quad }\hat{x}_{i}=\frac{%
\sin ^{2}\left( \frac{\pi \,(i+1)}{k+2}\right) }{\sin \left( \frac{\pi \,\,i%
}{k+2}\right) \sin \left( \frac{\pi \,(i+2)}{k+2}\right) }-1\,\text{.}
\label{cccTBA}
\end{equation}
Having determined the solutions of the constant TBA equations (\ref{ctba1})
and (\ref{cTba2a}) one can proceed via the standard TBA calculation as
presented in 4.2.2, see also \cite{TBAZam1,triZam,TBAKM}, and compute the
central charges from (\ref{c0}), (\ref{ckink}) and express them in terms of
Roger's dilogarithm function (\ref{roger}) 
\begin{eqnarray}
c_{UV} &=&\lim_{r^{\prime }\rightarrow 0}c_{\text{o}}(r^{\prime })=\frac{6}{%
\pi ^{2}}\sum_{i=1}^{k-1}\left[ 2\mathcal{L}\left( \frac{x_{i}}{1+x_{i}}%
\right) -\mathcal{L}\left( \frac{\hat{x}_{i}}{1+\hat{x}_{i}}\right) \right]
\;,  \label{cuv} \\
\hat{c}_{UV} &=&\lim_{r^{\prime }\rightarrow 0}\hat{c}_{\text{o}}(r^{\prime
})=\frac{6}{\pi ^{2}}\sum_{a=1}^{k-1}\mathcal{L}\left( \frac{\hat{x}_{i}}{1+%
\hat{x}_{i}}\right) \,\,.
\end{eqnarray}
Using the non-trivial identities 
\begin{equation}
\frac{6}{\pi ^{2}}\sum_{i=1}^{k-1}\mathcal{L}\left( \frac{x_{i}}{1+x_{i}}%
\right) =3\,\frac{k-1}{k+3}\quad \text{and}\quad \frac{6}{\pi ^{2}}%
\sum_{i=1}^{k-1}\mathcal{L}\left( \frac{\hat{x}_{i}}{1+\hat{x}_{i}}\right)
=2\,\frac{k-1}{k+2}\;
\end{equation}
found in \cite{Log} and \cite{Kirillov}, we finally end up with \cite{CFKM} 
\begin{equation}
c_{UV}=\frac{\left( k-1\right) (4k+6)}{\left( k+3\right) \left( k+2\right) }%
\qquad \text{and\qquad }\hat{c}_{UV}=2\,\frac{k-1}{k+2}\,\,.  \label{cuvv}
\end{equation}
For the reasons mentioned above $\hat{c}_{UV}$ coincides with the effective
central charge obtained from $su(k)$ minimal ATFT describing parafermions 
\cite{Witten} in the conformal limit. Notice that $c_{UV}$ corresponds to
the coset $(su(3)_{k}/u(1)^{2})/(su(2)_{k}/u(1))$.

The discussion of the infrared limit may be carried out completely analogous
to the one performed for the UV limit. The only difference is that in case
of the system (\ref{uvTba}) the constant TBA equations (\ref{ctba1}) drop
out because in the central region $-\ln r^{\prime }\ll \theta \ll \ln
r^{\prime }$ the free energy terms becomes dominant when $r^{\prime
}\rightarrow \infty $. Thus, in the infrared regime the central charges of
both systems coincide with $\hat{c}_{UV}$, 
\begin{equation}
c_{IR}=\lim_{r^{\prime }\rightarrow \infty }c_{\text{o}}(r^{\prime })=\hat{c}%
_{IR}=\lim_{r^{\prime }\rightarrow \infty }\hat{c}_{\text{o}}(r^{\prime
})=2\,\frac{k-1}{k+2}\;.  \label{cir}
\end{equation}
In summary, collecting the results (\ref{cuvv}) and (\ref{cir}), we can
express equation (\ref{step}) explicitly in terms of the level $k$ \cite
{CFKM}, 
\begin{equation}
c(r,M_{\tilde{k}}^{\tilde{c}})\approx \left\{ 
\begin{array}{ll}
4\,\frac{k-1}{k+2}\,\,, & \qquad \text{for\quad }1\ll \frac{2}{r}\ll M_{%
\tilde{k}}^{\tilde{c}}\medskip \\ 
6\,\frac{k-1}{k+3}\,, & \qquad \text{for\quad }M_{\tilde{k}}^{\tilde{c}}\ll 
\frac{2}{r}
\end{array}
\right. \,\,.  \label{stepk}
\end{equation}
We have replaced the limits in (\ref{step}) by mass scales in order to
exhibit the underlying physical picture. Here $M_{\tilde{k}}^{\tilde{c}}$ is
the smallest mass of an unstable bound state which may be formed in the
process $(i,a)+(j,b)\rightarrow (\tilde{k},\tilde{c})$ for $\tilde{A}%
_{ab}\neq 0,2$. We also used that the Breit-Wigner formula (\ref{BW1})
implies that $M_{\tilde{k}}^{\tilde{c}}\sim e^{\sigma /2}$ for large
positive $\sigma $.

First one should note that in the deep UV limit we obtain the same effective
central charge as in Subsection 4.4.3 when discussing the general case,
albeit in a quite different manner. On the mathematical side this implies
some non-trivial identities for Rogers dilogarithm and on the physical (\ref
{stepk}) exhibits a more detailed behaviour than the analysis in Subsection
4.4.3.\medskip

\noindent \textbf{Summary}.\emph{\ In the first regime of (\ref{stepk}) the
lower limit indicates the onset of the lightest stable soliton in the two
copies of complex sine-Gordon model. The unstable particles are on an energy
scale much larger than the temperature of the system. Thus, the dynamical
interaction between solitons of different colours is ``frozen'' and we end
up with two copies of the }$su(2)_{k}/u(1)$\emph{\ coset which do not
interact with each other. As soon as the parameter }$r$\emph{\ reaches the
energy scale of the unstable solitons with mass }$M_{\tilde{k}}^{\tilde{c}}$%
\emph{, the solitons of different colours start to interact, being now
enabled to form bound states. This interaction breaks parity and forces the
system to approach the }$su(3)_{k}/u(1)^{2}$\emph{\ coset model with central
charge given by the formula in (\ref{cdata}).\medskip }

The case when $\sigma $ tends to infinity is special, since then the energy
scale on which unstable particles are formed is never reached. Therefore,
one needs to pay attention to the order in which the limits in $r,\sigma $
are taken, we have \cite{CFKM} 
\begin{equation}
4\,\frac{k-1}{k+2}=\lim_{r\rightarrow 0}\lim_{\sigma \rightarrow \infty
}c(r,\sigma )\neq \lim_{\sigma \rightarrow \infty }\lim_{r\rightarrow
0}c(r,\sigma )=6\,\frac{k-1}{k+3}\,.
\end{equation}

One might enforce an additional step in the scaling function by exploiting
the fact that the mass ratio $m_{1}/m_{2}$ is not fixed. So it may be chosen
to be very large or very small. This amounts to decouple the TBA systems for
solitons with different colour by shifting one system to the infrared with
respect to the scale parameter $r$. The plateau then has an approximate
width of $\sim \ln |m_{1}/m_{2}|$ (see Figure 4.7). However, as soon as $r$
becomes small enough the picture we discussed for $m_{1}\sim m_{2}$ is
recovered.

\subsubsection{Restoring parity and eliminating the resonances}

In this subsection we are going to investigate the special limit $\sigma
\rightarrow 0,$ which is equivalent to choosing the vector couplings $%
\Lambda _{\pm }$ in (\ref{HSGaction}) parallel or anti-parallel. For the
classical theory it was pointed out in \cite{HSG} that only then the
equations of motion are parity invariant. Also the TBA-equations become
parity invariant in the absence of the resonance shifts, albeit the S-matrix
still violates it through the phase factors in (\ref{Sint0}). Since in the
UV regime a small difference in the masses $m_{1}$ and $m_{2}$ does not
effect the outcome of the analysis, we can restrict ourselves to the special
situation $m_{1}=m_{2}$, in which case we obtain two identical copies of the
system 
\begin{equation}
\epsilon _{i}(\theta )+\sum_{j=1}^{k-1}(\phi _{ij}+\psi _{ij})\ast
L_{j}(\theta )=r\,M_{i}\cosh \theta \;.
\end{equation}
Then we have $\epsilon _{i}(\theta )=\epsilon _{i}^{(1)}(\theta )=\epsilon
_{i}^{(2)}(\theta )$, $M_{i}=M_{i}^{(1)}=M_{i}^{(2)}$ and the scaling
function is given by the expression 
\begin{equation}
c(r,\sigma =0)=\frac{12\,r}{\pi ^{2}}\sum_{i=1}^{k-1}M_{i}\int\limits_{0}^{%
\infty }d\theta \,\,L_{i}(\theta )\cosh \theta \;.
\end{equation}
The factor two in comparison with (\ref{scale}) takes the two copies for $%
a=1,2$ into account. The discussion of the high-energy limit follows now the
standard arguments similar to the one of the preceding subsection and as in
4.2.2. Instead of shifting by the resonance parameter $\sigma $, one now
shifts the TBA equations by $\ln \frac{r}{2}$. The constant TBA equation
which determines the UV central charge then just coincides with (\ref{ctba1}%
). We therefore obtain \cite{CFKM} 
\begin{equation}
\lim_{r\rightarrow 0}\lim_{\sigma \rightarrow 0}c(r,\sigma )=\frac{12}{\pi
^{2}}\sum_{i=1}^{k-1}\mathcal{L}\left( \frac{x_{i}}{1+x_{i}}\right) =6\,%
\frac{k-1}{k+3}\;.
\end{equation}
Thus, again we recover the coset central charge (\ref{cdata}) for $\frak{%
\tilde{g}}=su(3)$, but this time without breaking parity in the TBA
equations. This is in agreement with the results of Subsection 4.4.3, which
showed that we can obtain this limit for any finite value of the resonance
parameter $\sigma $.

\subsubsection{Universal TBA equations and Y-systems}

Before we turn to the discussion of specific examples for definite values of
the level $k$, we turn to the alternative formulation of the TBA equations (%
\ref{tba}) in terms of a single integral kernel and $Y$-systems analogous to
the discussion of affine Toda models in Section 4.3.2. In the present
context this variant of the TBA equations is of particular advantage when
one wants to discuss properties of the model and keep the level $k$ generic.
By means of the convolution theorem and the Fourier transforms of the TBA
kernels $\phi $ and $\psi $, which can be read off directly from (\ref
{kernelint}) and (\ref{kernelint2}), one derives the set of integral
equations \cite{CFKM} 
\begin{equation}
\epsilon _{i}^{a}(\theta )+\Omega _{k}\ast L_{i}^{b}(\theta -\sigma
_{ba})=\sum_{j=1}^{k-1}I_{ij}^{su(k)}\,\Omega _{k}\ast (\epsilon
_{j}^{a}+L_{j}^{a})(\theta )\;,\quad a\neq b\,.  \label{uTBA}
\end{equation}
We recall that $I^{su(k)}=2-A^{su(k)}$ denotes the incidence matrix of $%
su(k) $ and the kernel $\Omega _{k}$ is found to be 
\begin{equation}
\Omega _{k}(\theta )=\frac{k/2}{\cosh (k\theta /2)}\;.
\end{equation}
The on-shell energies have dropped out because of the crucial relation 
\begin{equation}
\sum_{j=1}^{k-1}I_{ij}M_{j}^{a}=2\cos \tfrac{\pi }{k}\,M_{i}^{a}\;,
\end{equation}
which is a property of the mass spectrum inherited from affine Toda field
theory, compare (\ref{ATFTmass}) in Chapter 3. Even though the explicit
dependence on the scale parameter has been lost, it is recovered from the
asymptotic condition 
\begin{equation}
\epsilon _{i}^{a}(\theta )\underset{\theta \rightarrow \pm \infty }{%
\longrightarrow }rM_{i}^{a}\,e^{\pm \theta }\,\,\,.
\end{equation}
The integral kernel present in (\ref{uTBA}) has now a very simple form and
the $k$ dependence is easily read off.

Closely related to the TBA equations in the form (\ref{uTBA}) are the
functional relations also referred to as $Y$-systems which we already
encountered in context of ATFT, compare in particular (\ref{YADE}). Using
complex continuation and defining the quantity $Y_{i}^{a}(\theta )=\exp
(-\epsilon _{i}^{a}(\theta ))$ the integral equations are replaced by \cite
{CFKM} 
\begin{equation}
Y_{i}^{a}(\theta +\tfrac{i\pi }{k})Y_{i}^{a}(\theta -\tfrac{i\pi }{k})=\left[
1+Y_{i}^{b}(\theta -\sigma _{ba})\right] \prod_{j=1}^{k-1}\left[
1+Y_{j}^{a}(\theta )^{-1}\right] ^{-I_{ij}},\quad a\neq b\,.  \label{Y}
\end{equation}
The $Y$-functions are assumed to be well defined on the whole complex
rapidity plane where they give rise to entire functions \cite{TBAZamun}. As
we already saw in Section 4.3 these systems are useful in many aspects, for
instance they may be exploited in order to establish periodicities in the $Y$%
-functions, which in turn can be used to provide approximate analytical
solutions of the TBA-equations. The scaling function can be expanded in
integer multiples of the period which is directly linked to the dimension of
the perturbing operator.

Noting that the asymptotic behaviour of the $Y$-functions is $\lim_{\theta
\rightarrow \infty }Y_{i}^{a}(\theta )\sim e^{-rM_{i}^{a}\cosh \theta }$, we
recover for $\sigma \rightarrow \infty $ the $Y$-systems of the $su(k)$%
-minimal ATFT derived originally in \cite{TBAZamun}. In this case the $Y$%
-systems were shown to have a period related to the dimension of the
perturbing operator (see (\ref{conj})). We found some explicit periods for
generic values of the resonance parameter $\sigma $ as we discuss in the
next section for some concrete examples.

\subsubsection{Explicit examples}

In this section we support our analytical discussion with some numerical
results and in particular justify various assumptions for which we have no
rigorous analytical argument so far. We numerically iterate the
TBA-equations (\ref{tba}) and have to choose specific values for the level $%
k $ for this purpose. As we pointed out in the introduction, quantum
integrability has only been established for the choice $k>h$. Since the
perturbation is relevant also for smaller values of $k$ and in addition the
S-matrix makes remains well defined, it will be interesting to see whether
the TBA-analysis in the case of $su(3)_{k}$ will exhibit any qualitative
differences for $k\leq 3$ and $k>3$. From our examples for the values $%
k=2,3,4$ the answer to this question is that there is no apparent
difference. For all cases we find the staircase pattern of the scaling
function predicted in the preceding section as the values of $\sigma $ and $%
x $ sweep through the different regimes. Besides presenting numerical plots
we also discuss some peculiarities of the systems at hand. We provide the
massless TBA equations (\ref{uvTba}) with their UV and IR central charges
and state the $Y$-systems together with their periodicities. Finally, we
also comment on the classical or weak coupling limit $k\rightarrow \infty $.

\paragraph{The su(3)$_{2}$-HSG model}

This is the simplest model for the $su(3)_{k}$ case, since it contains only
the two self-conjugate solitons (1,1) and (1,2). The formation of stable
particles via fusing is not possible and the only non-trivial S-matrix
elements are those between particles of different colour 
\begin{equation}
S_{11}^{11}=S_{11}^{22}=-1,\quad S_{11}^{12}(\theta -\sigma
)=-S_{11}^{21}(\theta +\sigma )=\tanh \frac{1}{2}\left( \theta -i\frac{\pi }{%
2}\right) \;.  \label{ZamS}
\end{equation}
Here we have chosen $\varepsilon _{12}=-\varepsilon _{21}=1$ in (\ref{Sint0}%
) and used that the inverse Cartan matrix of $su(2)$ is just $1/2$. This
scattering matrix may be related to various matrices which occurred before
in the literature. First of all when performing the limit $\sigma
\rightarrow \infty $ the scattering involving different colours becomes
trivial and the systems consists of two free fermions leading to the central
charge $c=1$. Taking instead the limit $\sigma \rightarrow 0$ the
expressions in (\ref{ZamS}) coincide precisely with a matrix which describes
the scattering of massless ``Goldstone fermions (Goldstinos)'' discussed in 
\cite{triZam}. Apart from a phase factor $\sqrt{-1}$, the matrix $%
S_{11}^{21}(\theta )|_{\sigma =0}$ was also proposed to describe the
scattering of a massive particle \cite{anticross}. Having only one colour
available one is not able to set up the usual crossing and unitarity
equations and in \cite{anticross} the authors therefore resorted to the
concept of ``anti-crossing''. As our analysis shows this may be consistently
overcome by breaking the parity invariance. The TBA-analysis is summarized
as follows \cite{CFKM}, 
\begin{eqnarray*}
\text{unstable particle formation} &:&\text{\qquad \quad }c_{su(3)_{2}}=%
\frac{6}{5}=c_{UV}+\hat{c}_{UV}=\frac{7}{10}+\frac{1}{2} \\
\text{no unstable particle formation} &:&\text{\qquad \quad }%
2c_{su(2)_{2}}=1=c_{IR}+\hat{c}_{IR}=\frac{1}{2}+\frac{1}{2}\,\,\,.
\end{eqnarray*}
It is interesting to note that the flow from the tricritical Ising to the
Ising model which was investigated in \cite{triZam}, emerges as a subsystem
in the HSG-model in the form $c_{UV}\rightarrow c_{IR}$. This suggests that
we could alternatively also view the HSG-system as consisting out of a
massive and a massless fermion, where the former is described by (\ref{c0}),(%
\ref{uvTba}) and the latter by (\ref{ckink}),(\ref{kinktba}), respectively.

Our numerical investigations of the model match the analytical discussion
and justify various assumptions in retrospect. Figure 4.6 exhibits various
plots of the $L$-functions in the different regimes. We observe that for $%
\ln \frac{2}{r}<\sigma /2,$ $\sigma \neq 0$ the solutions are symmetric in
the rapidity variable, since the contribution of the $\psi $ kernels
responsible for parity violation is negligible. The solution displayed is
just the free fermion $L$-function, $L^{a}(\theta )=\ln (1+e^{-rM^{a}\cosh
\theta })$. Advancing further into the ultraviolet regime, we observe that
the solutions $L^{a}$ cease to be symmetric signaling the violation of
parity invariance and the formation of unstable bound states. The second
plateau is then formed, which will extend beyond $\theta =0$ for the deep
ultraviolet (see Figure 4.6). The staircase pattern of the scaling function
is displayed in Figure 4.6 for the different cases discussed in the previous
section. We observe always the value $6/5$ in the deep ultraviolet regime,
but depending on the value of the resonance parameter and the mass ratio it
may be reached sooner or later. The plateau at $1$ corresponds to the
situation when the unstable particles can not be formed yet and we only have
two copies of $su(3)_{2}$ which do not interact. Choosing the mass ratios in
the two copies to be very different, we can also ``switch them on''
individually as the plateau at $1/2$ indicates. 

\begin{center}
\includegraphics[width=10cm,height=13.5cm,angle=-90]{HSGL.epsi}

{\small Figure 4.6: }$L${\small -function for the }$su(3)_{2}${\small \ HSG
model at various values of the scaling parameter }$r${\small .}
\end{center}


\paragraph{The su(3)$_{3}$-HSG model}

This model consists of two pairs of solitons $\overline{(1,1)}=(2,1)$ and $%
\overline{(1,2)}=(2,2)$. When the soliton $(1,a)$ scatters with itself it
may form $(2,a)$ for $a=1,2$ as a bound state. The two-particle S-matrix
elements in the general block representation (\ref{Snew}) of Chapter 3 read 
\begin{equation}
S^{ab}(\theta -\sigma _{ab})=\left( 
\begin{array}{ll}
\lbrack 1,\tilde{I}_{ab}]_{\theta ,ab}^{\tilde{A}_{ab}} & [2,\tilde{I}%
_{ab}]_{\theta ,ab}^{\tilde{A}_{ab}} \\ 
\lbrack 2,\tilde{I}_{ab}]_{\theta ,ab}^{\tilde{A}_{ab}} & [1,\tilde{I}%
_{ab}]_{\theta ,ab}^{\tilde{A}_{ab}}
\end{array}
\right) \quad \text{with\quad }\tilde{A}=\left( 
\begin{array}{cc}
2 & -1 \\ 
-1 & 2
\end{array}
\right) \;.
\end{equation}
Since soliton and anti-soliton of the same colour obey the same TBA
equations we can exploit charge conjugation symmetry to identify $\epsilon
^{a}(\theta ):=\epsilon _{1}^{a}(\theta )=\epsilon _{2}^{a}(\theta )$
leading to the reduced set of equations 
\begin{equation}
\epsilon ^{a}(\theta )+\varphi \ast L^{a}(\theta )-\varphi \ast L^{b}(\theta
-\sigma _{ba})=rM^{a}\cosh \theta ,\quad \varphi (\theta )=-\frac{4\sqrt{3}%
\cosh \theta }{1+2\cosh 2\theta }\;.
\end{equation}
The corresponding scaling function therefore acquires a factor two, 
\begin{equation}
c(r,\sigma )=\frac{6\,r}{\pi ^{2}}\sum_{a=1,2}M^{a}\int d\theta \,\cosh
\theta \,L^{a}(\theta )\;.
\end{equation}
This system exhibits remarkable symmetry properties. We consider first the
situation $\sigma =0$ with $m_{1}=m_{2}$ and note that the system becomes
free in this case 
\begin{equation}
M^{(1)}=M^{(2)}=:M\;\Rightarrow \;\epsilon ^{(1)}(\theta )=\epsilon
^{(2)}(\theta )=rM\cosh \theta \;.
\end{equation}
meaning that the theory falls apart into four free fermions whose central
charges add up to the expected coset central charge of $c=2$. Also for
unequal masses $m_{1}\neq m_{2}$ the system develops towards the free
fermion theory for high energies when the difference becomes negligible.
This is also seen numerically.

For $\sigma \neq 0$ the two copies of the minimal $A_{2}$-ATFT or
equivalently the scaling Potts model start to interact. The outcome of the
TBA-analysis in that case is summarized as \cite{CFKM} 
\begin{eqnarray*}
\text{unstable particle formation} &:&\text{\qquad \quad }%
c_{su(3)_{3}}=2=c_{UV}+\hat{c}_{UV}=\frac{6}{5}+\frac{4}{5} \\
\text{no unstable particle formation} &:&\text{\qquad \quad }2c_{su(2)_{3}}=%
\frac{8}{5}=c_{IR}+\hat{c}_{IR}=\frac{4}{5}+\frac{4}{5}\,\,\,.
\end{eqnarray*}
As discussed in the previous case for $k=2$ the $L$-functions develop an
additional plateau after passing the point $\ln \frac{2}{r}=\sigma /2$. This
plateau lies at $\ln 2$ which is the free fermion value signaling that the
system contains a free fermion contribution in the UV limit as soon as the
interaction between the solitons of different colours becomes relevant.
Figure 4.7 exhibits the same behaviour as the previous case, we clearly
observe the plateau at $8/5$ corresponding to the two non-interacting copies
of the minimal $A_{2}$-ATFT. As soon as the energy scale of the unstable
particles is reached the scaling function approaches the correct value of $%
c_{su(3)_{3}}=2$.

The Y-systems (\ref{Y}) for $k=3$ read 
\begin{equation}
Y_{1,2}^{a}\left( \theta +i\frac{\pi }{3}\right) Y_{1,2}^{a}\left( \theta -i%
\frac{\pi }{3}\right) =Y_{1,2}^{a}\left( \theta \right) \frac{%
1+Y_{1,2}^{b}(\theta +\sigma _{ab})}{1+Y_{1,2}^{a}\left( \theta \right) }%
\quad a,b=1,2,\;a\neq b\,\,.  \label{Y3}
\end{equation}
Once again we may derive a periodicity 
\begin{equation}
Y_{1,2}^{a}\left( \theta +2\pi i+\sigma _{ba}\right) =Y_{1,2}^{b}(\theta )
\end{equation}
by making the suitable shifts in (\ref{Y3}) and subsequent iteration.


\paragraph{The su(3)$_{4}$-HSG model}

This model involves 6 solitons, two of which are self-conjugate $\overline{%
(2,1)}=(2,1)$, $\overline{(2,2)}=(2,2)$ and two conjugate pairs $\overline{%
(1,1)}=(3,1)$, $\overline{(1,2)}=(3,2)$. The corresponding two-particle
S-matrix elements are obtained from the general formulae (\ref{Snew}) 
\begin{equation}
S^{ab}(\theta -\sigma _{ba})=\left( 
\begin{array}{ccc}
\lbrack 1,\tilde{I}_{ab}]_{\theta ,ab}^{\tilde{A}_{ab}} & [2,\tilde{I}%
_{ab}]_{\theta ,ab}^{\tilde{A}_{ab}} & [3,\tilde{I}_{ab}]_{\theta ,ab}^{%
\tilde{A}_{ab}} \\ 
\lbrack 2,\tilde{I}_{ab}]_{\theta ,ab}^{\tilde{A}_{ab}} & [3,\tilde{I}%
_{ab}]_{\theta ,ab}^{\tilde{A}_{ab}}[1,\tilde{I}_{ab}]_{\theta ,ab}^{\tilde{A%
}_{ab}} & [2,\tilde{I}_{ab}]_{\theta ,ab}^{\tilde{A}_{ab}} \\ 
\lbrack 3,\tilde{I}_{ab}]_{\theta ,ab}^{\tilde{A}_{ab}} & [2,\tilde{I}%
_{ab}]_{\theta ,ab}^{\tilde{A}_{ab}} & [1,\tilde{I}_{ab}]_{\theta ,ab}^{%
\tilde{A}_{ab}}
\end{array}
\right)
\end{equation}
In this case the numerics becomes more involved but for the special case $%
m_{1}=m_{2}$ one can reduce the set of six coupled integral equations to
only two by exploiting the symmetry $L_{i}^{(1)}(\theta
)=L_{i}^{(2)}(-\theta )$ and using charge conjugation symmetry, $%
L_{1}^{a}(\theta )=L_{3}^{a}(\theta )$. The numerical outcomes, shown in
Figure 4.7 again match, with the analytic expectations (\ref{stepk}) and
yield for $\ln \frac{2}{r}>\sigma /2$ the coset central charge of $18/7$. In
summary we obtain \cite{CFKM} 
\begin{eqnarray*}
\text{unstable particle formation} &:&\text{\qquad \quad }c_{su(3)_{4}}=%
\frac{18}{7}=c_{UV}+\hat{c}_{UV}=\frac{11}{7}+1 \\
\text{no unstable particle formation} &:&\text{\qquad \quad }%
2c_{su(2)_{4}}=2=c_{IR}+\hat{c}_{IR}=1+1\,\,\,,
\end{eqnarray*}
which matches precisely the numerical outcome in Figure 4.7, with the same
physical interpretation as already provided in the previous two examples.

\begin{center}
\includegraphics[width=12cm,height=16cm,angle=0]{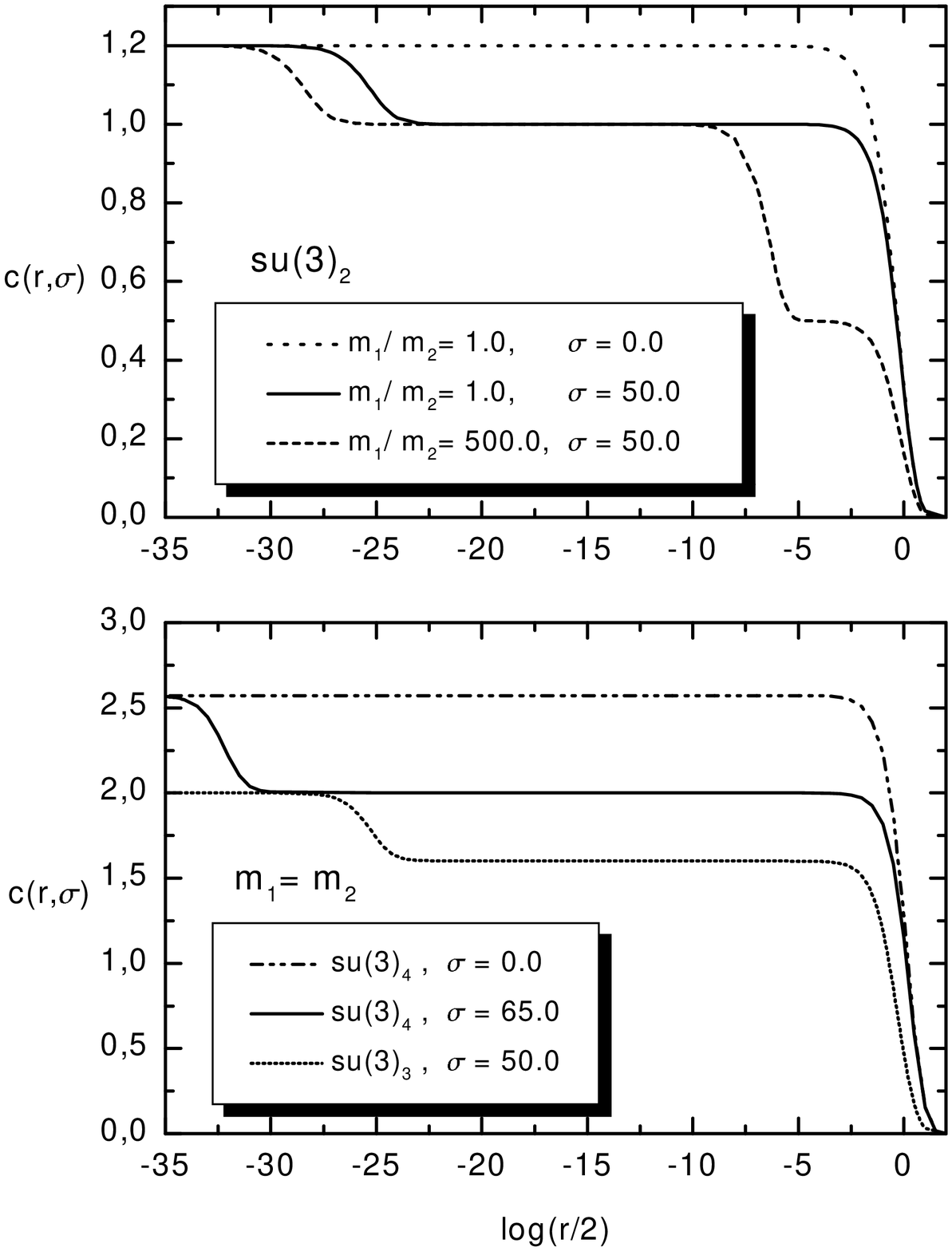}

{\small Figure 4.7: Scaling function for the }$su(3)_{k}${\small \ HSG
models at level }$k=2,3,4${\small \ with various choices of the parameters.}
\end{center}

\paragraph{The semi-classical limit $k\rightarrow \infty $}

As last example we carry out the limit $k\rightarrow \infty $, which is of
special physical interest, since it may be identified with the weak coupling
or equivalently the classical limit, as is seen from the relation $\hbar
\beta ^{2}=1/k+O(1/k^{2}).$ To illustrate this equivalence we have
temporarily re-introduced Planck's constant. It is clear from the
TBA-equations that this limit may not be taken in a straightforward manner.
However, we can take it in two steps, first for the on-shell energies and
the kernels and finally for the sum over all particle contributions. The
on-shell energies are easily computed by noting that the mass spectrum
becomes equally spaced for $k\rightarrow \infty $%
\begin{equation}
M_{i}^{a}=M_{k-i}^{a}=\frac{m_{a}}{\pi \beta ^{2}}\sin \frac{\pi \,i}{k}%
\approx i\,m_{a}\;\qquad ,\quad i<\frac{k}{2}\;.  \label{MM}
\end{equation}
For the TBA-kernels the limit may also be taken easily from their integral
representations 
\begin{equation}
\phi _{ij}(\theta )\underset{k\rightarrow \infty }{\longrightarrow }2\pi
\,\delta (\theta )\,\left( \delta _{ij}-2\left( A_{ij}^{su(k)}\right)
^{-1}\right) \quad \text{and\quad }\psi _{ij}(\theta )\underset{k\rightarrow
\infty }{\longrightarrow }2\pi \,\delta (\theta )\,\left(
A_{ij}^{su(k)}\right) ^{-1},\;
\end{equation}
when employing the usual integral representation of the delta-function.
Inserting these quantities into the TBA-equations yields 
\begin{equation}
\epsilon _{i}^{a}(\theta )\approx r\,i\,m_{a}\cosh \theta
-\sum_{j=1}^{k-1}\left( \delta _{ij}-2\left( A_{ij}^{su(k)}\right)
^{-1}\right) L_{j}^{a}(\theta )-\sum_{j=1}^{k-1}\left( A_{ij}^{su(k)}\right)
^{-1}L_{j}^{b}(\theta -\sigma )\;.  \label{TTT}
\end{equation}
We now have to solve these equations for the pseudo-energies. In principle,
we could proceed in the same way as in the case for finite $k$ by doing the
appropriate shifts in the rapidity. However, we will be content here to
discuss the cases $\sigma \rightarrow 0$ and $\sigma \rightarrow \infty $,
which as follows from our previous discussion correspond to the situation of
restored parity invariance and two non-interacting copies of the minimal
ATFT, respectively. The related constant TBA-equations (\ref{ctba2}) and (%
\ref{cTba2a}) become 
\begin{equation}
\sigma \rightarrow \infty :\;\hat{x}_{j}\underset{k\rightarrow \infty }{%
\longrightarrow }\frac{(j+1)^{2}}{j(j+2)}-1\quad \text{and\quad }\sigma
\rightarrow 0:\;x_{j}\underset{k\rightarrow \infty }{\longrightarrow }\frac{%
(j+1)(j+2)}{j(j+3)}-1\;.  \label{l1}
\end{equation}
The other information we may exploit about the solutions of (\ref{TTT}) is
that for large rapidities they tend asymptotically to the free solution,
meaning that 
\begin{equation}
\sigma \rightarrow 0,\infty \;:\;\;\;\;\;L_{j}^{a}(\theta )\underset{\theta
\rightarrow \pm \infty }{\longrightarrow }\ln (1+e^{-r\,j\,m_{a}\cosh \theta
})\;.  \label{l2}
\end{equation}
We are left with the task to seek functions which interpolate between the
properties (\ref{l1}) and (\ref{l2}). Inspired by the analysis in \cite
{Fowler} we take these functions to be 
\begin{eqnarray}
\sigma &\rightarrow &\infty :\quad L_{j}^{a}(\theta )=\ln \left[ \frac{\sinh
^{2}\left( \frac{j+1}{2}\,rm_{a}\cosh \theta \right) }{\sinh \left( \frac{j\,%
}{2}\,rm_{a}\cosh \theta \right) \sinh \left( \frac{j+2}{2}\,rm_{a}\cosh
\theta \right) }\right]  \label{p1} \\
\sigma &\rightarrow &0:\quad L_{j}^{a}(\theta )=\ln \left[ \frac{\sinh
\left( \frac{j+1}{2}\,rm_{a}\cosh \theta \right) \sinh \left( \frac{j+2}{2}%
\,rm_{a}\cosh \theta \right) }{\sinh \left( \frac{j}{2}\,rm_{a}\cosh \theta
\right) \sinh \left( \frac{j+3}{2}\,rm_{a}\cosh \theta \right) }\right] \,\,.
\label{p2}
\end{eqnarray}
The expression (\ref{p1}) coincides with the expressions discussed in the
context of the breather spectrum of the sine-Gordon model \cite{Fowler} and (%
\ref{p2}) is constructed in analogy. We are now equipped to compute the
scaling function in the limit $k\rightarrow \infty ,$%
\begin{equation}
c(r,\sigma )=\lim_{k\rightarrow \infty }\frac{3\,r}{\pi ^{2}}%
\sum_{a=1}^{2}\int d\theta \,\cosh \theta
\sum_{i=1}^{k-1}M_{i}^{a}L_{i}^{a}(\theta )\,\,.
\end{equation}
Using (\ref{MM}), (\ref{p1}) and (\ref{p2}) the sum over the main quantum
number may be computed directly by expanding the logarithm. We obtain for $%
k\rightarrow \infty $ \cite{CFKM} 
\begin{eqnarray}
c(r)|_{\sigma =\infty }\!\! &=&\!\!\dfrac{-6r}{\pi ^{2}}\!\!\sum_{a=1}^{2}\!%
\!\int \!\!d\theta \,m_{a}\cosh \theta \ln \left( 1-e^{-r\,m_{a}\cosh \theta
}\right)  \label{in1} \\
c(r)|_{\sigma =0}\!\! &=&\!\!\dfrac{-6\,r}{\pi ^{2}}\!\!\sum_{a=1}^{2}\!\!%
\int \!\!d\theta \,\,m_{a}\cosh \theta \lbrack \ln \left(
1\!-\!e^{-r\,m_{a}\cosh \theta }\right) +\ln (1\!-\!e^{-r\,2m_{a}\cosh
\theta })].\,\,\quad \,  \label{in2}
\end{eqnarray}
Here we have acquired an additional factor of 2, resulting from the
identification of particles and anti-particles which is needed when one
linearizes the masses in (\ref{MM}). Taking now the limit $r\rightarrow 0$
we obtain \cite{CFKM} 
\begin{eqnarray}
\text{no unstable particle formation} &:&\text{\quad\ }2\,c_{su(2)_{\infty
}}=4\;  \label{in11} \\
\text{unstable particle formation} &:&\text{\quad\ }c_{su(3)_{\infty
}}=6\,\,.  \label{in22}
\end{eqnarray}

The results (\ref{in1}), (\ref{in11}) and (\ref{in2}), (\ref{in22}) allow a
nice physical interpretation. We notice that for the case $\sigma
\rightarrow \infty $ we obtain four times the scaling function of a free
boson. This means in the classical limit we obtain twice the contribution of
the non-interacting copies of $SU(2)_{\infty }/U(1)$, whose particle content
reduces to two free bosons each of them contributing $1$ to the effective
central charge which is in agreement with (\ref{cdata}). For the case $%
\sigma \rightarrow 0$ we obtain the same contribution, but in addition the
one from the unstable particles, which are two free bosons of mass $2m_{a}$.
This is also in agreement with (\ref{cdata}).

Finally it is interesting to observe that when taking the resonance poles to
be $\theta _{R}=\sigma -i\pi /k$ the semi-classical limit taken in the
Breit-Wigner formula (\ref{BW1}) leads to $m_{\tilde{c}%
}^{2}=(m_{a}+m_{b})^{2}$. On the other hand (\ref{in2}) seems to suggest
that $m_{\tilde{c}}=2m_{a}$, which implies that the mass scales should be
the same. However, since our analysis is mainly based on exploiting the
asymptotics we have to be cautious about this conclusion.\bigskip

We can summarize that the main outcome of our the TBA-analysis indeed
confirms the consistency of the scattering matrix proposed in \cite{HSGS}.
In the deep ultraviolet limit we recover the $\frak{\tilde{g}}_{k}/u(1)^{%
\tilde{n}}$-coset central charge for any value of the $2\tilde{n}-1$ free
parameters entering the S-matrix, including the choice when the resonance
parameters vanish and parity invariance is restored on the level of the
TBA-equations. This is in contrast to the properties of the S-matrix, which
is still not parity invariant due to the occurrence of the phase factors,
which are required to close the bootstrap equations, compare Section 3.3.
However, they do not contribute to our TBA-analysis, which means that so far
we can not make any definite statement concerning the necessity of the
parity breaking in the quantum theory of HSG theories (\ref{HSGaction}),
since the same value for the central charge is recovered irrespective of the
value of the $\sigma $'s. The underlying physical behaviour is, however,
quite different as our numerical analysis demonstrates. For vanishing
resonance parameter the deep ultraviolet coset central charge is reached
straight away, whereas for non-trivial resonance parameter one passes the
different regions in the energy scale. Also the choice of different mass
scales leads to a theory with a different physical content, but still
possessing the same central charge.

To settle this issue, it would therefore be highly desirable to carry out
the series expansion of the scaling function in $r$ analogous to the
discussion in \cite{TBAZam1} and determine the dimension $d_{\Phi }$ of the
perturbing operator. It will be useful for this to know the periodicities of
the Y-functions. We conjecture that they will be \cite{CFKM} 
\begin{equation}
Y_{i}^{a}\left( \theta +i\pi (1-d_{\Phi })^{-1}+\sigma _{ba}\right) =Y_{\bar{%
\imath}}^{b}(\theta ),  \label{conj}
\end{equation}
which is confirmed by our $su(3)$-examples for various levels. For vanishing
resonance parameter and the choice $k=2$, this behaviour coincides with the
one obtained in \cite{TBAZamun}. This means the form in (\ref{conj}) is of a
very universal nature beyond the models discussed here.

We also observe from our $su(3)$-example that the different regions, i.e. $k>%
\tilde{h}$ and $k\leq \tilde{h}$, for which quantum integrability was shown
and for which not, respectively, do not show up in our analysis. It would be
very interesting to extend the case-by-case analysis to other algebras. The
first challenge in these cases is to incorporate the different resonance
parameters.

\section{The central charge calculation for $\frak{g}|\frak{\tilde{g}}$
theories}

This section is devoted to an extension of the previous TBA analysis for the
HSG models to all $\frak{g}|\frak{\tilde{g}}$-theories as defined in Chapter
3. This means, the two algebras involved are now arbitrary but simply-laced.
Since this gives a total of $ADE\times ADE$ possibilities we restrict
ourselves to the central charge calculation analogous to the one performed
for HSG models in Section 4.4.3 but without resonance poles, i.e. we set $%
\sigma =0$. Emphasis in the following considerations will be given to the
powerful Lie algebraic structure behind these theories. The latter will lead
to a generic formula for the central charge and suggests an intimate
relation between the models $\frak{g}|\frak{\tilde{g}}$ and $\frak{\tilde{g}}%
|\frak{g}$ in the UV regime via the constant TBA equations encountered
before.

Similar to the discussion of the previous section TBA-equations with Fermi
statistics and absence of resonances read 
\begin{equation}
rm_{i}^{a}\cosh \theta =\varepsilon _{i}^{a}(\theta )+\sum\limits_{b=1}^{%
\tilde{n}}\sum\limits_{j=1}^{n}\int\limits_{-\infty }^{\infty }d\theta
^{\prime }\varphi _{ij}^{ab}(\theta -\theta ^{\prime })\ln \left(
1+e^{-\varepsilon _{b}^{j}(\theta ^{\prime })}\right) \,\,  \label{TBA}
\end{equation}
where $r$ is again the scale parameter, $m_{i}^{a}$ the mass of the particle 
$(i,a)$, $\varepsilon _{i}^{a}(\theta )$ the pseudo-energies and kernels are
obtained from (\ref{iSnew}) 
\begin{equation}
\varphi _{ij}^{ab}(\theta )=\int\limits_{-\infty }^{\infty }dt\left[ \delta
_{ab}\delta _{ij}-\left( 2\cosh \frac{\pi t}{h}-\tilde{I}\right) _{ab}\left(
2\cosh \frac{\pi t}{h}-I\right) _{ij}^{-1}\,\right] \,\,e^{-it\theta }\,\,.
\label{9}
\end{equation}
Note that in contrast to the HSG models both incidence matrices may now
belong to arbitrary simply-laced algebras. The effective central charge $c_{%
\text{eff}}=c-24\Delta _{0}$ of the underlying ultraviolet conformal field
theory will now be calculated by the standard procedure (compare 4.2.2)
under the assumption that the solutions of the TBA-equations develop the
same ``plateau behaviour'' as we encountered for the HSG models. Notice that
also for the class of scaling or minimal ATFT models \cite{TBAZam1} this
behaviour holds true and that the latter are obtained by setting $\frak{%
\tilde{g}}=A_{1}$. Thus, one may again approximate $\varepsilon
_{i}^{a}(\theta )=\varepsilon _{i}^{a}=$ $const$ in the large region $\ln 
\frac{r}{2}\ll \theta \ll \ln \frac{2}{r}$ when $r$ is small. By the same
standard TBA arguments used also in the discussion of the previous section
it follows that the effective central charge is expressible as 
\begin{equation}
c_{\text{eff}}=\frac{6}{\pi ^{2}}\sum\limits_{a=1}^{\tilde{n}%
}\sum\limits_{i=1}^{n}\mathcal{L}\left( \frac{x_{i}^{a}}{1+x_{i}^{a}}\right)
\label{ceff}
\end{equation}
with $\mathcal{L}(x)$ denoting Rogers dilogarithm and where the $%
x_{i}^{a}=\exp (-\varepsilon _{i}^{a})$ are obtained as solutions from the
constant TBA-equations in the form 
\begin{equation}
x_{i}^{a}=\prod\limits_{j=1}^{n}\prod\limits_{b=1}^{\tilde{n}%
}(1+x_{j}^{b})^{N_{ij}^{ab}}\,\,.  \label{cTBA}
\end{equation}
The matrix $N_{ij}^{ab}$ is defined analogously to (\ref{NN}) via the
asymptotic behaviour of the scattering matrix which for the more general
case at hand reads \cite{FKcol} 
\begin{equation}
N_{ij}^{ab}=\frac{1}{2\pi }\int\limits_{-\infty }^{\infty }d\theta
\,\,\varphi _{ij}^{ab}(\theta )=\,\delta _{ab}\delta _{ij}-A_{ij}^{-1}\tilde{%
A}_{ab}\,.
\end{equation}
In regard to finding explicit solutions for the set of coupled equations (%
\ref{cTBA}) and to exhibit the Lie algebraic structure present, it turns out
to be convenient to introduce new variables because they may be related to
Weyl characters of the Lie algebra $\frak{g}$\textbf{\ }or $\frak{\tilde{g}}$%
. Following \cite{Resh,Kun} we define 
\begin{equation}
Q_{i}^{a}=\prod_{j=1}^{n}(1+x_{j}^{a})^{A_{ij}^{-1}}\qquad \Leftrightarrow
\qquad x_{i}^{a}=\prod_{j=1}^{n}\left( Q_{j}^{a}\right) ^{A_{ij}}-1
\end{equation}
such that the constant TBA-equations (\ref{cTBA}) acquire the more symmetric
form 
\begin{equation}
\prod_{j=1}^{n}\left( Q_{j}^{a}\right) ^{I_{ij}}+\prod_{b=1}^{\tilde{n}%
}\left( Q_{i}^{b}\right) ^{\tilde{I}_{ab}}=\left( Q_{i}^{a}\right) ^{2}\,\,
\label{QQ}
\end{equation}
exhibiting a ''dual'' relation between the two algebras. The effective
central charge (\ref{ceff}) is then expressible in various equivalent ways 
\cite{FKcol} 
\begin{eqnarray}
\hspace{-0.5cm}\!\!\!c_{\text{eff}}^{\frak{g}|\frak{\tilde{g}}}\!\!\!
&=&\!\!\!\tfrac{6}{\pi ^{2}}\sum_{i=1}^{n}\sum_{a=1}^{\tilde{n}}\mathcal{L}%
\left( 1-\prod_{j=1}^{n}\left( Q_{j}^{a}\right) ^{-A_{ij}}\right)
\!=\!\medskip n\tilde{n}-\tfrac{6}{\pi ^{2}}\sum_{i=1}^{n}\sum_{a=1}^{\tilde{%
n}}\mathcal{L}\left( \prod_{b=1}^{\tilde{n}}\left( Q_{i}^{b}\right) ^{-%
\tilde{A}_{ab}}\right)  \label{cc} \\
\hspace{-0.5cm}\!\!\! &=&\!\!\!n\tilde{n}-\tfrac{6}{\pi ^{2}}%
\sum_{i=1}^{n}\sum_{a=1}^{\tilde{n}}\mathcal{L}\left( \prod_{j=1}^{n}\left(
Q_{j}^{a}\right) ^{-A_{ij}}\right) \!=\!\tfrac{6}{\pi ^{2}}\sum_{a=1}^{%
\tilde{n}}\sum_{i=1}^{n}\mathcal{L}\left( 1-\prod_{b=1}^{\tilde{n}}\left(
Q_{i}^{b}\right) ^{-\tilde{A}_{ab}}\right)  \label{cc2}
\end{eqnarray}
where we used the well-known identity $\mathcal{L}(x)+\mathcal{L}(1-x)=\pi
^{2}/6$, see e.g. \cite{Log}. It is also clear that having solved the
equations (\ref{QQ}) for the case $\frak{g}|\frak{\tilde{g}}$\textbf{\ }we
have exploiting the symmetry present immediately a solution for the case $%
\frak{\tilde{g}}|\frak{g}$\textbf{\ }simply by interchanging the roles of
the two algebras, i.e. interchanging main quantum numbers and colour degrees
of freedom. Supposing now that $c_{\text{eff}}^{\frak{g}|\frak{\tilde{g}}%
}=\lambda \,c_{\text{eff}}^{\frak{\tilde{g}}|\frak{g}}$ for some unknown
constant $\lambda $, it follows directly from (\ref{cc}) and (\ref{cc2})
that $c_{\text{eff}}^{\frak{g}|\frak{\tilde{g}}}=\lambda n\tilde{n}%
/(1+\lambda )$. We conjecture now this constant to be $\lambda =\tilde{h}/h$
such that \cite{FKcol} 
\begin{equation}
c_{\text{eff}}^{\frak{g}|\frak{\tilde{g}}}=\frac{n\tilde{n}\,\tilde{h}}{h+%
\tilde{h}}\,\,\,.  \label{ceffn}
\end{equation}
As expected from the identification of the scattering matrices for several
special cases we must recover the results of the already known cases when we
fix the algebras appropriately. For instance we obtain $c_{\text{eff}}^{%
\frak{g}\mathbf{|}A_{1}}=2n/(h+2)$ which is the well known formula for the
effective central charge of the minimal affine Toda theories. Furthermore,
we recover the effective central charge of the Homogeneous sine-Gordon
models $c_{\text{eff}}^{A_{n}|\frak{\tilde{g}}}=n\tilde{n}\tilde{h}/(n+1+%
\tilde{h})$, compare (\ref{cdata}). It should be noted that this is
independent of whether a resonance parameter is present or not despite the
fact that the TBA-equation are not parity invariant in that case, see our
discussion of the previous section. Numerically we also solved (\ref{QQ})
explicitly for numerous examples with $\frak{g}\neq A_{n}$ and confirmed (%
\ref{ceffn}).

The knowledge of the ultraviolet central charge (\ref{ceffn}) will certainly
be useful in identifying the Lagrangian corresponding to the $\frak{g}|\frak{%
\tilde{g}}$-theories, since it provides the renormalization fixed point. As
we know from the Homogeneous Sine-Gordon models the $A_{n}|\frak{\tilde{g}}$%
--theory may be viewed as perturbed $\frak{\tilde{g}}_{n+1}/u(1)^{\otimes 
\tilde{n}}$-coset WZNW theories. In analogy, we could view for instance the
``dual'' theory of this, i.e. the $\frak{\tilde{g}}\mathbf{|}A_{n}$--theory,
formally as perturbed $\frak{\tilde{g}}_{1}^{\otimes (n+1)}/\frak{\tilde{g}}%
_{n+1}$-coset WZNW theory as one infers from the above table \cite{FKcol}.
Besides the identification of the fixed point theory for the situation in
which $\frak{g}\neq \!\!A_{n}$, it remains open to find the precise form of
the perturbing operators. Also the Lie algebraic structure present in the
constant TBA equations (\ref{QQ}) deserves further investigations, since
these provide an additional link to the conformal models and their spectra
as mentioned in 4.4.3.

\begin{center}
\begin{tabular}{|c||c|c|c|c|c|}
\hline
$\frak{g}|\frak{\tilde{g}}$ & $A_{m}$ & $D_{m}$ & $E_{6}$ & $E_{7}$ & $E_{8}$
\\ \hline\hline
$A_{n}$ & $\frac{nm\,(m+1)}{n+m+2}$ & $\frac{nm\,(2m-2)}{n+2m-1}$ & $\frac{%
72\,n}{n+13}$ & $\frac{126\,n}{n+19}$ & $\frac{240\,n}{n+31}$ \\ \hline
$D_{n}$ & $\frac{nm\,(m+1)}{2n+m-1}$ & $\frac{nm\,(m-1)}{n+m-2}$ & $\frac{%
36\,n\,}{n+5}$ & $\frac{63\,n\,}{n+8}$ & $\frac{120\,\,n\,}{n+14}$ \\ \hline
$E_{6}$ & $\frac{6\,m\,(m+1)}{m+13}$ & $\frac{6\,m\,(m-1)}{m+5}$ & $18$ & $%
\frac{126}{5}$ & $\frac{240}{7}$ \\ \hline
$E_{7}$ & $\frac{7\,m\,(m+1)}{m+19}$ & \multicolumn{1}{|l|}{$\frac{%
7\,m\,(m-1)}{m+8}$} & $\frac{84}{5}$ & $\frac{49}{2}$ & $35$ \\ \hline
$E_{8}$ & $\frac{8\,m\,(m+1)}{m+31}$ & $\frac{8\,m\,(m-1)}{m+14}$ & $\frac{96%
}{7}$ & $21$ & $32$ \\ \hline
\end{tabular}

\vspace{0.2cm} {\small Table 4.1: Effective central charges }$c_{\text{eff}%
}^{\frak{g}|\frak{\tilde{g}}}$ {\small of the}\textbf{\ }$\frak{g}$%
\TEXTsymbol{\vert}$\frak{\tilde{g}}${\small -theories.}
\end{center}

\subsection{An explicit example: $D_{4}|D_{4}$}

In order to illustrate the working of our general formulae beyond the
already treated case it is instructive to evaluate them for a concrete model
with $\frak{g}\neq A_{n}$ and $\frak{\tilde{g}}\neq A_{1}$. We chose the $%
D_{4}|D_{4}$-model which is an example for a theory hitherto unknown. The
model contains 16 different particles labeled by $(i,a)$ with $1\leq a,i\leq
4$. The Coxeter number is $h=\tilde{h}=6$ for $D_{4}$. Naming the central
particle in the $D_{4}$--Dynkin diagram by $1$ the S-matrix elements
according to (\ref{Snew}) are computed to \cite{FKcol} 
\begin{equation*}
\begin{tabular}{ll}
$S_{22}^{aa}(\theta )=S_{33}^{aa}(\theta )=S_{44}^{aa}(\theta
)=[1,0]_{\theta ,aa}^{2}[5,0]_{\theta ,aa}^{2}$ & $\text{for }%
a=1,2,3,4,\smallskip $ \\ 
$S_{12}^{aa}(\theta )=S_{13}^{aa}(\theta )=S_{14}^{aa}(\theta
)=[2,0]_{\theta ,aa}^{2}[4,0]_{\theta ,aa}^{2}$ & $\text{for }%
a=1,2,3,4,\smallskip $ \\ 
$S_{23}^{aa}(\theta )=S_{24}^{aa}(\theta )=S_{34}^{aa}(\theta
)=[3,0]_{\theta ,aa}^{2}$ & $\text{for }a=1,2,3,4,\smallskip $ \\ 
$S_{11}^{aa}(\theta )=[1,0]_{\theta ,aa}^{2}[3,0]_{\theta
,aa}^{4}[5,0]_{\theta ,aa}^{2}$ & $\text{for }a=1,2,3,4,\smallskip $ \\ 
$S_{22}^{1b}(\theta )=S_{33}^{1b}(\theta )=S_{44}^{1b}(\theta
)=[1,1]_{\theta ,1b}^{-1}[5,1]_{\theta ,1b}^{-1}$ & $\text{for }%
b=2,3,4,\smallskip $ \\ 
$S_{12}^{2b}(\theta )=S_{23}^{2b}(\theta )=S_{24}^{2b}(\theta
)=[2,1]_{\theta ,2b}^{-1}[4,1]_{\theta ,2b}^{-1}\;$ & $\text{for }%
b=2,3,4,\smallskip $ \\ 
$S_{23}^{1b}(\theta )=S_{24}^{1b}(\theta )=S_{34}^{1b}(\theta
)=[3,1]_{\theta ,1b}^{-1}$ & $\text{for }b=2,3,4,\smallskip $ \\ 
$S_{11}^{1b}(\theta )=[1,1]_{\theta ,1b}^{-1}[3,1]_{\theta
,1b}^{-2}[5,1]_{\theta ,1b}^{-1}$ & $\text{for }b=2,3,4,\smallskip $ \\ 
$S_{ij}^{ab}(\theta )=1$ & $\text{for }i,j=1,2,3,4;\,\,a\neq b;\,a,b\neq
1\,. $%
\end{tabular}
\end{equation*}
The solutions to the constant TBA-equations (\ref{cTBA}) read \cite{FKcol} 
\begin{eqnarray}
x_{1}^{1}
&=&x_{3}^{2}=x_{4}^{2}=x_{2}^{3}=x_{2}^{4}=x_{3}^{3}=x_{3}^{4}=x_{3}^{4}=x_{4}^{4}=x_{2}^{2}=1
\\
x_{1}^{2} &=&x_{1}^{3}=x_{1}^{4}=1/2 \\
x_{2}^{1} &=&x_{3}^{1}=x_{4}^{1}=2
\end{eqnarray}
such that the effective central charge according to (\ref{ceff}) is 
\begin{equation}
c_{\text{eff}}=\frac{6}{\pi ^{2}}\left( 10\mathcal{L}\left( \frac{1}{2}%
\right) +3\mathcal{L}\left( \frac{2}{3}\right) +3\mathcal{L}\left( \frac{1}{3%
}\right) \right) =8\,\,.
\end{equation}
This result confirms the general formula (\ref{ceffn}).

\chapter{Conclusions}

... {\small \emph{When your mind is going hither and thither, discrimination
will never be brought to a conclusion. With an intense, fresh and undelaying
spirit, one will make his judgements within the space of seven breaths. It
is a matter of being determined and having the spirit to break right through
to the other side.}}

\qquad \qquad \qquad \qquad \qquad \qquad {\small From 'The Book of the
Samurai, Hagakure'\bigskip }

It has been demonstrated that Lie algebraic structures play an essential
role in affine Toda models, which form a vast class of integrable quantum
field theories. Coxeter geometry and its $q$-deformed extension
accommodating the renormalization flow have been proved to provide a concise
mathematical framework for the treatment of the quantum aspects of ATFT. The
strategy to express various physical quantities like the mass spectrum, the
fusing rules, and the S-matrices in generic and universal Lie algebraic
formulas leads not only to a great simplification in the treatment of the
on-shell structure of these models, but has also been shown to carry through
to the level of the thermodynamic Bethe ansatz when investigating the
underlying conformal field theories in the ultraviolet limit. Indeed, all
relevant calculations have been performed in a complete generic Lie
algebraic setting, treating all models at once.

In addition, the Lie algebraic structures have been explicitly exploited to
formulate hitherto unknown exact S-matrices via the introduction of colour
degrees of freedom, the so-called $\frak{g}|\frak{\tilde{g}}$-models. By
means of a detailed TBA analysis the universal formula (\ref{ceffn}) giving
the effective central charges of the associated conformal models in the high
energy regime has been derived. For particular subclasses of $\frak{g}|\frak{%
\tilde{g}}$-scattering matrices this result could be directly linked to the
central charges of WZNW coset models, see Table 5.1.$\medskip $

\begin{center}
$
\begin{tabular}{|c|c|c|c|}
\hline\hline
$\frak{g}$ & $\frak{\tilde{g}}$ & integrable model & WZNW coset \\ 
\hline\hline
$A_{n}$ & $ADE$ & HSG models & $\frak{g}_{n+1}/u(1)^{\times \limfunc{rank}%
\frak{g}}$ \\ \hline\hline
$ADE$ & $A_{1}$ & minimal ATFT & $\frak{g}_{1}\otimes \frak{g}_{1}/\frak{g}%
_{2}$ \\ \hline\hline
$ADE$ & $A_{n}$ & ``dual'' HSG models & $\frak{g}_{1}^{\otimes (n+1)}/\frak{g%
}_{n+1}$ \\ \hline\hline
\end{tabular}
\medskip $

\vspace{0.4cm} {\small Table 5.1: }$\frak{g}|\frak{\tilde{g}}${\small %
-theories related to WZNW cosets.}
\end{center}

\noindent This demonstrates that in many cases the Lie algebraic structures
occurring in S-matrices of affine Toda type can be traced back to the ones
appearing in WZNW models. They are therefore particularly interesting
candidates for further investigations relating massive and conformal
spectra, since the Lie algebraic framework might serve as a guiding line in
this task. The motivation to explore the interplay between conformal and
integrable models in greater depth is manifold. For example, it might shed
light on the origin of mass as already mentioned in the introduction. From a
more technical point of view it will certainly help in the construction of
correlation functions. The latter are to a much deeper extent explored in
conformal theories than in massive ones. In fact, two and three-point
functions can immediately be written down in presence of conformal symmetry.
Moreover, in the case of WZNW models the higher $n$-point functions are
accessible by differential equations, the so-called Knizhnik-Zamolodchikov
equations. By means of conformal perturbation theory one might then check
against outcomes of the form factor program. At the moment such relations
are still a far posted goal, but the techniques of the bootstrap approach
and the TBA analysis performed in this thesis are first steps in this
direction.

As outlined in the introduction the development of efficient techniques to
calculate correlation functions are of use in the area of quantum field
theory as well as of two-dimensional statistical mechanics and condensed
matter systems with second order phase transitions. In particular, it should
be emphasized once more that the interpretation of integrable field theories
as perturbed conformal models and the related TBA analysis are conceptually
closely related to the ideas of the renormalization group originating in the
study of critical phenomena. For example, the central object of interest in
the TBA considerations, the scaling function, has been conjectured to be
tightly linked to the $\beta $-function appearing in the renormalization
group equations \cite{ZamoR}. Exploring this connection in more detail one
might learn from 1+1 dimensional theories about higher dimensional ones,
since the concept of the renormalization group is applicable in any
dimension.

These general remarks are now supplemented by a more detailed presentation
of the results in order to pinpoint concrete starting points for further
investigations.

\section{The affine Toda S-matrix}

A systematic treatment has been given for the application of geometrical
arguments in context of ATFT. Starting with the theory of ordinary Coxeter
elements their $q$-deformation has been motivated: first on an abstract
mathematical level by generalizing numerous formulas and identities from the
non-deformed case, second by exhibiting how $q$-deformation matches with the
renormalization picture obtained from perturbative calculations, which have
already been performed in the literature. Recall that the latter indicated
that the renormalized particle masses flow between the classical values of
two different affine Toda models belonging to a pair of algebras related by
Langlands duality ($\alpha \leftrightarrow \alpha ^{\vee }$). The crucial
property of the $q$-deformed Coxeter element, twisted or untwisted, turned
out to be the merging of the data of these two dual algebras in a consistent
manner.

The first physical application where this property has been used was the
formulation of the quantum fusing rules in ATFT indicating the vanishing of
the three-point coupling, which in turn is associated with the fusing of two
quantum particles to a third. While the fusing rules found by Chari and
Pressley have to be formulated in terms of both the non-deformed Coxeter 
\emph{and} the non-deformed twisted Coxeter element, $q$-deformation allows
to state them either in terms of the orbits $\Omega ^{q}$ \emph{or} $\hat{%
\Omega}^{q}$ solely. The latter are generated by the $q$-deformed Coxeter
and the $q$-deformed twisted Coxeter element, respectively. One of the
central results presented in this thesis is the derivation of the precise
relation between these different versions of the fusing rules (\ref{3eta})
and the proof of their equivalence. This clarifies the interplay between the
two dual Lie algebraic structures and shows their equivalence. In this
context, it would be particularly interesting to investigate, whether for
the $q$-deformed fusing rules a similar connection to representation theory
exists as for the non-deformed ones \cite{CP}. This might give rise to new
''quantum symmetries''. However, the link to representation theory is so far
only partially understood and the formalism presented here is by now the
most restrictive one.

The $q$-deformed fusing rules were then shown to be consistent with the
description of the mass spectrum and other conserved quantities as
nullvector of the $q$-deformed Cartan matrix (\ref{qA}). Because of its
compactness and mathematical beauty this characterization of the quantum
masses is recalled here (compare (\ref{ATFTmass})), 
\begin{equation*}
\sum_{j=1}^{n}[I_{ij}]_{\hat{q}}\,m_{j}=2\cos \pi \left( \frac{2-B}{2h}+%
\frac{t_{i}B}{2H}\right) \,m_{i}\;,\quad \hat{q}=e^{i\pi \frac{sB}{2H}}\;.
\end{equation*}
From the above formula the renormalized mass flow of all affine Toda
theories can be directly read off. It would be desirable to construct the
corresponding ''quantum'' Lagrangian from which this mass spectrum can be
deduced in terms of a mass matrix, as it was done for the $ADE$ case w.r.t.
the classical Lagrangian. It is most likely that $q$-deformation will also
play a vital role in this construction.

Furthermore, the $q$-deformed fusing rules have been directly employed in
the construction of the S-matrix by means of the bootstrap equations. Their
generic formulation allowed to write down a universal formula for the
two-particle scattering amplitude in terms of hyperbolic functions whose
powers can be directly inferred from the $q$-deformed Coxeter orbits. These
have been explicitly worked out for the first time in \cite{FKS2} (see also
the appendix). Alternatively, one might use the $M$-matrix (\ref{M2}), which
is a slightly modified version of the inverse $q$-deformed Cartan matrix. It
was argued that the matrix elements of the latter always simplify to
polynomials in the two deformation parameters involved. These polynomials
reflect the structure of the building blocks of hyperbolic functions and
encode the complete information about the S-matrix. Using the structure of
the $M$-matrix, it has been systematically demonstrated that the
two-particle scattering amplitude fulfills all required bootstrap
properties. Equivalently, this also followed from the inner product
identities for the $q$-deformed Coxeter elements derived in Section 2.4.4
and 2.4.7. Moreover, it is intriguing that the combined bootstrap equation (%
\ref{cb}), which is closely linked to the structure of the $M$-matrix,
incorporates the information of \emph{all }individual fusing processes. The
discussion of the analytic properties of the S-matrix has been completed by
providing a simple criterion which enables one to exclude unphysical poles
from the participation in the bootstrap.

The matrix structure was also exploited in the rigorous derivation of the
universal integral representation of the ATFT S-matrix (\ref{SPPi}) 
\begin{equation*}
S_{ij}(\theta )=\,\exp 8\int\limits_{0}^{\infty }\frac{dt}{t}\,\sinh
(\vartheta _{h}t)\sinh (t_{j}\vartheta _{H}t)A(e^{t\vartheta
_{h}},e^{t\vartheta _{H}})^{-1}\sinh \left( \frac{\theta t}{i\pi }\right)
\,\,.
\end{equation*}
This formula is not only a neat and compact expression for all ATFT
S-matrices, but also of direct use in several applications. For instance,
the two-particle form factor can be immediately extracted from the above
identity. The discovery of similar powerful Lie algebraic structures might
then yield a valuable advantage in the calculation of higher particle form
factors. In this thesis the above integral representation has been
explicitly applied when discussing the thermodynamic Bethe ansatz for ATFT,
where it lead to additional universal formulas.

\section{The TBA analysis of ATFT}

In Section 4.3 it has been demonstrated that it is possible to extract the
leading order behaviour of the scaling function for \emph{all} ATFT by
simple analytical approximations schemes in the large and small density
regime. By matching the approximate solutions of the TBA equations in the
two different regimes at the point in which the particle density and the
density of available states coincide, it is possible to fix the constant of
integration, which originated in the approximation scheme of \cite
{ZamoR,MM,FKS1} and was left undetermined therein. Since the leading order
behaviour derived for the scaling function is in agreement with the
semi-classical results found in \cite{Zamref} and also \cite{Fateev,Fateev2}%
, one has an alternative method to fix the constant and to compare directly
the different approaches. Thus, it is not necessary for this to proceed to
higher order differential equations as was claimed in \cite{Arsch}. Since
the solutions to the higher order differential equations may only be
obtained approximately one does not gain any further structural insight this
way and, moreover, one has lost the virtue of the leading order
approximation, namely its simplicity.

With regard to future investigations it would be desirable to extend the
analysis presented here and to find analytical expressions for the constant
of integration totally within the TBA approach. This would allow to verify
the semi-classical results in the literature on the basis of the ``pure
quantum'' S-matrix. For this purpose further exact insight on the solutions
of the TBA equations is required. This motivated the derivation of the
Y-systems, which were also presented in a unique formula for all ATFT. As
demonstrated they can be utilized to improve the large density
approximations and can also be applied to put constraints on the constant of
integration. Future considerations about possible periodicities and their
analytic features might lead to additional requirements on the constant
allowing to extract it analytically. There also exist interesting links
between the Y-systems and spectral functions in quantum mechanics \cite
{spectral}, such that one can expect more exact and universal results to
follow.

The TBA results are also of interest in comparison to alternative methods
which allow to calculate the ultraviolet central charge, such as the \textbf{%
c-theorem} \cite{ZamoC}. The latter allows to compute the difference of the
central charges recovered in the ultraviolet and infrared limit of the
massive theory from the Schwinger function, 
\begin{equation}
\Delta c=c_{\text{UV}}-c_{\text{IR}}=-\frac{3}{2}\int_{0}^{\infty
}dr\,r^{3}\left\langle 0|T_{\,\,\,\mu }^{\mu }(r)T_{\,\,\,\mu }^{\mu
}(0)|0\right\rangle \;.
\end{equation}
The correlation function might be calculated via the form factor approach
upon inserting a complete set of states as explained in the introduction. In
fact, in most cases the two-particle form factor already yields an excellent
approximation. Since $c_{\text{IR}}=0$ for purely massive theories one might
then compare with the outcome from the TBA. Closely related to this
observation is the problem to compute the vacuum expectation value of the
energy momentum tensor. Unlike in the situation of conformal invariance, in
which the trace vanishes, this tensor is not unique and acquires some
scaling behaviour for massive theories. One therefore needs an external
input for the recursive form factor equations removing this ambiguity. The
latter is provided by the TBA, where the vacuum expectation value can be
directly obtained from the scaling function \cite{TBAZam1}, 
\begin{equation*}
\left\langle 0|T_{\,\,\,\mu }^{\mu }(r)|0\right\rangle =-\frac{\pi ^{2}}{3r}%
\frac{d}{dr}c(r)\;.
\end{equation*}
However, in case of ATFT one first needs a deeper physical understanding of
the logarithmic corrections appearing in the leading order behaviour of the
scaling function before one may proceed this way. Notice that in the context
of fixing the vacuum expectation value by means of the TBA also the
investigation of existence and uniqueness of the solution in Section 4.2.3
plays an important role. Restricting the number of possible solutions to
one, this rules out any ambiguity inside this approach, except for different
possibilities in choosing the statistics. The impact of statistics is
another issue which needs to be clarified in more detail, especially in its
relevance for relating conformal and massive spectra.

\section{Colour valued S-matrices, HSG models and WZNW cosets}

Suggesting a general construction principle a new class of exact scattering
matrices exhibiting Lie algebraic structures very alike to those in affine
Toda models has been generated. Extending the techniques originally
developed in context of ATFT to these integrable systems, it has been shown
that the proposed S-matrices for the $\frak{g}|\frak{\tilde{g}}$-theories
provide consistent solutions of the bootstrap equations (\ref{boot}). The
special feature of parity violation has been shown to enter this
construction as a consequence of taking the square root of the CDD-factor
occurring in ATFT. The motivation for this particular choice was to recover
S-matrices already known in the literature, especially the ones proposed in
context of the Homogeneous Sine-Gordon models.

For these integrable theories strong evidence has been presented that the
proposal originally made in \cite{HSGS} matches with the semi-classical
picture obtained when perturbing the WZNW cosets $\frak{\tilde{g}}/u(1)^{%
\limfunc{rank}\frak{\tilde{g}}}$, see Table 5.1. In particular, the detailed
analytical and numerical TBA analysis of the $su(3)$ model supported several
assumptions made about unstable particles in the spectrum. They led to a
direct physical interpretation of the staircase pattern in the scaling
function resulting from the resonance poles in the S-matrix. The staircase
pattern has then been interpreted as the sign for a more sophisticated
renormalization group flow of these theories which dependent on the choice
of external parameters might end in different fixed points, i.e. different
conformal field theories. In particular, massless subflows might occur which
have a non-trivial UV as well as a non-trivial IR fixed point. As one
particular example the flow between the tricritical and the critical Ising
model was provided. Similar findings are to be expected when generalizing to
other HSG models.

However, the subject is still far from being closed. The exact nature of the
unstable particles has to be explored in more detail, particularly in
hindsight to a complete explanation of the resonance poles of the S-matrix.
Also their relevance for recovering the correct conformal spectrum is not
settled, since the expected coset central charge is reached irrespective
whether or not the resonance parameter is chosen to be zero. In particular,
on the level of the TBA equations the latter decides about the violation of
parity. Thus, so far a definite statement concerning the loss of parity
invariance can only be made w.r.t. the S-matrix which due to the phase
factors breaks parity also when the resonances are removed. Further
investigations of the quantum theory of these integrable models are
necessary in order to settle the issues mentioned.

A first step towards this direction was performed in \cite{CFK} where the
complete set of form factors for the $su(3)_{2}$-HSG theory has been
calculated exhibiting powerful determinant structures similar to those found
in the context of the Yang-Lee \cite{Zamocorr} and the Sinh-Gordon model
(see Fring et al. in \cite{FFex}). The knowledge about the form factors of
the energy-momentum tensor allowed to compare the UV central charge obtained
in the TBA approach of Section 4.4.4 against the one obtained from the
c-theorem mentioned above. Taking the form factors up to the six particle
contribution into account one ends up with \cite{CFK}, 
\begin{equation*}
\text{TBA:\quad }c_{su(3)_{2}}=1.2\quad \quad \quad \quad \text{%
c-theorem:\quad }c_{su(3)_{2}}^{(6)}=1.199...\;.
\end{equation*}
This demonstrates perfect agreement between both methods and supports the
findings presented here. Again it would be very interesting to extend this
analysis to higher levels $k>2$ and other algebras.

A further open question is to identify the corresponding Lagrangian for
general $\frak{g}|\frak{\tilde{g}}$-theories. The knowledge of the
ultraviolet central charge (\ref{ceffn}) 
\begin{equation*}
c_{\text{eff}}^{\frak{g}|\frak{\tilde{g}}}=\frac{\tilde{h}}{h+\tilde{h}}\,n\,%
\tilde{n}
\end{equation*}
obtained by extending the TBA analysis of the HSG models to the general case
will certainly be helpful in this search. It provides the renormalization
group fixed point and the perturbing operator might then be identified among
the spinless relevant fields in the conformal spectrum. For the minimal
affine Toda or scaling models this has already been achieved in the
literature. Exploiting periodicities present in the Y-systems a series
expansion of the scaling function of the following form has been derived
from which the dimension $d_{\Phi }$ of the perturbing operator can be
directly read off \cite{TBAZam1}, 
\begin{equation*}
c(r)=c_{\text{eff}}+f_{0}\,r^{2}+\sum_{n}c_{n}r^{2n(1-d_{\Phi })}\;.
\end{equation*}
Here $f_{0}$ is a constant related to the bulk free energy of the massive
model. One might now proceed similar for the more complex $\frak{g}|\frak{%
\tilde{g}}$-theories paying attention to the parity violation. For the HSG
models it has already been demonstrated for the $su(3)$ model that
periodicities in the Y-systems occur, which are consistent with the expected
dimension of the perturbing field.

The general Lie algebraic classification of these new integrable models
might also lead to the discovery of new ''duality'' relations, an
intensively discussed issue in string theory. Analyzing the structure of the
constant TBA equations (\ref{QQ}) in Section 4.5, it became apparent that
theories linked to each other by an exchange of the kind $\frak{g}%
\leftrightarrow \frak{\tilde{g}}$ share the same set of constant TBA
solutions. Since structures analogous to the constant TBA equations have
also been found in Virasoro characters associated to conformal models \cite
{Rich} one might conjecture on new Lie algebraic identities connecting
different conformal field theories. For instance, the ''dual'' relation
between the $A_{n}|\frak{\tilde{g}}$ and $\frak{\tilde{g}}|A_{n}$-theory
would relate the WZNW cosets $\frak{\tilde{g}}/u(1)^{\times \limfunc{rank}%
\frak{\tilde{g}}}$ and $\frak{\tilde{g}}_{1}^{\otimes (n+1)}/\frak{\tilde{g}}%
_{n+1}$ with each other. However, this issue has to be understood in more
detail before definite conclusions can be drawn.\newpage

\noindent \textbf{Acknowledgments}: I would especially like to thank Dr.
Andreas Fring, who introduced me to the area of integrable quantum field
theory and from whom I have learned about the various techniques applied in
this thesis. I have greatly benefited from his insightful explanations and
our intense discussions on the subject. His friendly good advice as well as
his critical comments have been of great help in our collaborations and in
writing this thesis.

I am also very grateful to Professor Robert Schrader for his continuing
constant support throughout my doctoral studies and providing the
opportunity to participate in several workshops and conferences, which have
deepened my understanding of the research area. I owe also thanks for
numerous valuable suggestions and fruitful discussions.

I would like to thank Professor David I. Olive for agreeing to be referee
and his interest in this work.

I would like to express my gratitude to Professor J. Luis Miramontes for
making my stay at the Departamento de F\'{i}sica de Part\'{i}culas,
Universidad de Santiago de Compostela possible. I am indebted to him for his
explanations concerning the Homogeneous Sine-Gordon models, numerous
profound discussions and friendly support. I deeply enjoyed our
collaboration. I would also like to thank his student O.A. Castro-Alvaredo
for her kindness during my visit and our cooperations on the HSG models. I
must also acknowledge my gratitude to Professor Joaquin
S\'{a}nchez-Guill\'{e}n for his help in preparing the stay at the
Universidad de Santiago and valuable comments. Special thanks go to Dr.
Jos\'{e} Edelstein for being a kind and understanding office mate during my
visit and his wife Mariella for her warm hospitality.

Thanks are also due to Dr. Martin Schmidt for helpful discussions and Svante
Wellershoff in helping to prepare the title page. Last but not least I would
like to thank B. Jesko Schulz for our joint work on ATFT. His good sense of
humor and cheerfulness made it an enjoyable experience.\medskip

The financial support of the DFG Sfb 288 is gratefully acknowledged.

\appendix

\chapter{The affine Toda S-matrix case-by-case}

In order to illustrate the working of the general formulae derived for the
affine Toda S-matrix it is useful to work them out explicitly for some
concrete examples. We concentrate here on the non-simply laced case, since
the simply laced case is covered extensively in the literature \cite{TodaS}.
We will be most detailed for the ($G_{2}^{(1)},D_{4}^{(3)}$)-case. The
conventions with regard to numbering and colouring may be read off from the
Dynkin diagrams. As usual the arrow points towards the short roots. A black
and white vertex corresponds to the colour value $c_{i}=-1$ and $c_{i}=1$,
respectively.

\subsection{$(G_{2}^{(1)},D_{4}^{(3)})$}

\unitlength=0.780000pt 
\begin{picture}(223.00,105.00)(-120.00,200.00)
\put(130.00,0.00){\line(0,1){0.00}}
\qbezier(160.00,223.00)(176.00,213.00)(191.00,223.00)
\put(223.00,218.00){\makebox(0.00,0.00){$\alpha_3$}}
\put(175.00,316.01){\makebox(0.00,0.00){$\alpha_4$}}
\put(127.00,218.00){\makebox(0.00,0.00){$\hat{\alpha}_2$}}
\put(193.00,266.00){\makebox(0.00,0.00){$\hat{\alpha}_1$}}
\put(46.67,274.17){\makebox(0.00,0.00){$\alpha_2$}}
\put(6.00,274.01){\makebox(0.00,0.00){$\alpha_1$}}
\qbezier(189.33,298.34)(227.67,279.33)(213.33,242.00)
\qbezier(160.67,297.67)(126.00,281.01)(137.67,242.33)
\put(178.67,256.33){\line(1,-1){23.33}}
\put(171.67,256.33){\line(-1,-1){23.00}}
\put(175.00,294.68){\line(0,-1){31.33}}
\put(175.00,300.00){\circle{10.00}}
\put(205.00,230.00){\circle{10.00}}
\put(145.00,230.00){\circle{10.00}}
\put(175.00,260.00){\circle*{10.00}}
\put(20.00,260.00){\line(1,-2){7.00}}
\put(20.00,260.00){\line(1,2){7.00}}
\put(5.00,255.00){\line(1,0){40.00}}
\put(5.33,265.00){\line(1,0){39.67}}
\put(10.00,260.00){\line(1,0){30.00}}
\put(5.00,260.00){\circle*{10.00}}
\put(45.00,260.00){\circle{10.00}}
\end{picture}

\noindent The S-matrices of the theory read \cite{G2} 
\begin{eqnarray}
S_{11}(\theta ) &=&\{\overset{2}{\overbrace{1,1}};\overset{1}{\overbrace{%
3,5_{2}}};5,11\}_{\theta }  \label{11} \\
S_{12}(\theta ) &=&\{2,2_{3};\overset{1}{\overbrace{4,6_{3}}\}}_{\theta }
\label{22} \\
S_{22}(\theta ) &=&\{1,1_{3};\overset{2}{\overbrace{3,3_{3}}}%
;3,5_{3};5,7_{3}\}_{\theta }\,\,.  \label{33}
\end{eqnarray}
Here we indicated which block is responsible for which type of fusing
process. We have $h=6$ and $H=12$ for the Coxeter numbers. With the help of (%
\ref{shif}), we easily verify that for (\ref{11}), (\ref{22}) and (\ref{33})
the following bootstrap identities hold 
\begin{eqnarray}
S_{1l}\left( \theta +\theta _{h}+\theta _{H}\right) S_{1l}\left( \theta
-\theta _{h}-\theta _{H}\right) &=&S_{2l}\left( \theta \right) \quad \quad
l=1,2  \label{b2} \\
S_{1l}\left( \theta +2\theta _{h}+4\theta _{H}\right) S_{1l}\left( \theta
-2\theta _{h}-4\theta _{H}\right) &=&S_{1l}\left( \theta \right) \,\quad
\quad l=1,2\,  \label{b4} \\
S_{2l}\left( \theta +2\theta _{h}+4\theta _{H}\right) S_{2l}\left( \theta
-2\theta _{h}-4\theta _{H}\right) &=&S_{2l}\left( \theta \right) \,\quad
\quad l=1,2\,\,\,.  \label{b5}
\end{eqnarray}
As an example for the working of the generalized bootstrap and our criterion
(\ref{res1}), (\ref{kii}) provided in Section 3.2.8, we plotted the
imaginary part of the residues of $S_{22}(\theta )$ in Figure A.1 for
several poles. We observe that the sign changes throughout the range for
poles resulting from $\{1,1_{3}\}$ and $\{3,5_{3}\}$. Only the poles
responsible for the self-coupling of particle $2$ has a positive imaginary
part of the residue throughout the range of the coupling constant $\beta $.
Except at $B=4/3$ where it is zero, such that this fusing process decouples.%
%
%
%

\begin{center}
\includegraphics[width=10cm,height=13.5cm,angle=-90]{Fig1v2.epsi}

\vspace*{0.5cm} \noindent {\small Figure A.1: The imaginary part of several
residues of $S_{22}(\theta )$ as a function of the effective coupling
constant.} \vspace*{1.2mm}
\end{center}

\noindent Besides (\ref{b2}) the combined bootstrap identities (\ref{cb})
also yield 
\begin{equation}
S_{l2}\left( \theta +\theta _{h}+3\theta _{H}\right) S_{l2}\left( \theta
-\theta _{h}-3\theta _{H}\right) \!\!=\!\!S_{l1}\left( \theta \right)
S_{l1}\left( \theta +2\theta _{H}\right) S_{l1}\left( \theta -2\theta
_{H}\right) ,  \label{g2boot}
\end{equation}
for $l=1,2$. These equations may be derived from (\ref{b2}) and (\ref{b4})
or verified directly for (\ref{11}), (\ref{22}) and (\ref{33}), with the
help of (\ref{shif}). The process corresponding to the combined bootstrap
identity (\ref{g2boot}) is depicted in Figure A.2.

\noindent Reading off the fusing angles from the bootstrap equations we
obtain the mass ratios according to (\ref{mratio}) 
\begin{equation}
\frac{m_{1}}{m_{2}}=\frac{\sinh \left( \theta _{h}+\theta _{H}\right) }{%
\sinh \left( 2\theta _{h}+2\theta _{H}\right) }\,\,.
\end{equation}

We may construct all these formulae from the Lie algebraic data in
alternative ways.

\medskip

\begin{center}
\medskip 
\begin{picture}(299.67,150.00)(-57.00,0.00)
\put(281.33,95.00){\makebox(0.00,0.00){$1$}}
\put(245.67,75.00){\makebox(0.00,0.00){$1$}}
\put(234.67,75.00){\makebox(0.00,0.00){$1$}}
\put(76.33,74.67){\makebox(0.00,0.00){$1$}}
\put(248.67,130.00){\makebox(0.00,0.00){$1$}}
\put(197.67,94.67){\makebox(0.00,0.00){$1$}}
\put(259.67,0.00){\makebox(0.00,0.00){$2$}}
\put(219.67,0.00){\makebox(0.00,0.00){$2$}}
\put(170.00,109.33){\makebox(0.00,0.00){$l$}}
\put(125.00,130.00){\makebox(0.00,0.00){$1$}}
\put(79.00,129.67){\makebox(0.00,0.00){$1$}}
\put(35.33,129.67){\makebox(0.00,0.00){$1$}}
\put(63.33,74.67){\makebox(0.00,0.00){$1$}}
\put(0.00,29.67){\makebox(0.00,0.00){$l$}}
\put(89.67,0.00){\makebox(0.00,0.00){$2$}}
\put(49.67,0.00){\makebox(0.00,0.00){$2$}}
\put(139.67,86.00){\line(1,0){21.00}}
\put(140.67,90.00){\line(1,0){19.00}}
\put(179.67,110.00){\line(6,1){120.00}}
\put(259.67,60.00){\line(0,-1){50.00}}
\put(239.67,110.00){\line(0,1){40.00}}
\put(259.67,60.00){\line(1,3){30.00}}
\put(239.67,110.00){\line(2,-5){20.00}}
\put(219.67,60.00){\line(2,5){20.00}}
\put(219.67,60.00){\line(0,-1){50.00}}
\put(189.67,150.00){\line(1,-3){30.00}}
\put(9.67,30.00){\line(6,1){120.00}}
\put(89.67,60.00){\line(0,-1){50.00}}
\put(49.67,60.00){\line(0,-1){50.00}}
\put(69.67,150.00){\line(0,-1){40.00}}
\put(89.67,60.00){\line(1,3){30.00}}
\put(69.67,110.00){\line(2,-5){20.00}}
\put(49.67,60.00){\line(2,5){20.00}}
\put(19.67,150.00){\line(1,-3){30.00}}
\end{picture}
\bigskip

\noindent {\small Figure A.2: }$(G_{2}^{(1)},D_{4}^{(3)})${\small -combined
bootstrap identities (\ref{g2boot}). }
\end{center}

\medskip

\subsubsection{$S_{ij}\left( \protect\theta \right) $ from $G_{2}^{(1)}$}

We start by exploiting the properties of $G_{2}^{(1)}$. The non-vanishing
entries of the incidence matrix are $I_{12}=1$ and $I_{21}=3$. Consequently
equation (\ref{symm}) yields $t_{1}=1$ and $t_{2}=3$. As indicated in the
Dynkin diagram we choose $c_{1}=-1$ and $c_{2}=1$, such that the $q$%
-deformed Coxeter element reads $\sigma _{q}=\sigma _{1}^{q}\tau \sigma
_{2}^{q}\tau $. The result of successive actions of this element on the
simple roots is reported in Table A.1. Here and in all further tables we
choose the following conventions: To each $\gamma _{i}$ we associate a
column in which we report the powers of the $q$ of the coefficients of the
simple roots. We abbreviate 
\begin{equation}
\pm (q^{\mu _{1}^{1}}+\ldots +q^{\mu _{1}^{l_{1}}})\alpha _{1}\pm \ldots \pm
(q^{\mu _{n}^{1}}+\ldots +q^{\mu _{n}^{l_{n}}})\alpha _{n}\rightarrow \pm
\mu _{1}^{1},\ldots ,\mu _{l_{1}}^{1};\ldots ;\mu _{n}^{1},\ldots ,\mu
_{l_{n}}^{1}\,\,,  \label{notat}
\end{equation}
with $n=\limfunc{rank}\frak{g}$. When $q^{\mu }$ occurs $x$-times we denote
this by $\mu ^{x}$. Like in the undeformed case the overall sign of any
element in $\Omega _{i}^{q}$ is definite. Therefore it suffices to report
the sign only once as stated in (\ref{notat}). In the complete orbit we
always have an equal number of plus and minus signs. When we do not report
any signs in the column at all, the signs of the column to the left are
adopted. In case the coefficient of the root is zero, we indicate this by a $%
\ast $. For instance from Table A.1 we read off : $\sigma _{q}\gamma
_{1}=-(q^{4}+q^{6})\alpha _{1}-q^{4}\alpha _{2}$.

\begin{center}
\begin{tabular}{|c||c|c|}
\hline\hline
$\sigma _{q}^{x}$ & $\alpha _{1}=-\gamma _{1}$ & $\alpha _{2}=\gamma _{2}$
\\ \hline\hline
$1$ & $4,6;4$ & $-4,6,8;6$ \\ \hline
$2$ & $10;8$ & $-8,10,12;8,10$ \\ \hline
$3$ & $-12;\ast $ & $-\ast ;12$ \\ \hline
$4$ & $-16,18;16$ & $16,18,20;18$ \\ \hline
$5$ & $-22;20$ & $20,22,24;20,22$ \\ \hline
$6$ & $24;\ast $ & $\ast ;24$ \\ \hline\hline
\end{tabular}
\medskip

\noindent {\small Table A.1: The orbits $\Omega _{i}^{q}$ created by the
action of $\sigma _{q}^{x}$ on $\gamma _{i}$} \medskip
\end{center}

\noindent For the conventions chosen the generating functions (\ref{muCox})
for the powers of the building blocks are obtainable from the generating
functions 
\begin{eqnarray}
\sum\limits_{y}\mu _{11}\left( 2x+1,y\right) q^{y} &=&-q^{1}(\,\lambda
_{1}^{\vee }\cdot (\sigma _{q})^{x}\gamma _{1})/2\,\, \\
\sum\limits_{y}\mu _{21}\left( 2x,y\right) q^{y} &=&-q^{-2}(\,\lambda
_{1}^{\vee }\cdot (\sigma _{q})^{x}\gamma _{2})/2 \\
\sum\limits_{y}\mu _{22}\left( 2x-1,y\right) q^{y} &=&-q^{-3}\left[ 3\right]
_{q}(\,\lambda _{2}^{\vee }\cdot (\sigma _{q})^{x}\gamma _{2})/2\,\,.
\end{eqnarray}
We may now read off the Lie algebraic data from Table A.1 and we can
construct the scattering matrices (\ref{11}), (\ref{22}) and (\ref{33})
according to formula (\ref{blockS}).

The two non-equivalent solutions to (\ref{Ootafuse1}) corresponding to the
S-matrix bootstrap equations (\ref{b2}), (\ref{b4}) and (\ref{b5}) read 
\begin{equation}
q\sigma _{q}^{-1}\gamma _{1}+q^{-1}\gamma _{1}=q^{-3}\gamma _{2},\text{%
\qquad }q^{-1}\gamma _{1}+q\sigma _{q}^{-1}\gamma _{1}=q^{-3}\gamma _{2}\,,
\label{67}
\end{equation}
\begin{equation}
q^{3}\sigma _{q}^{-1}\gamma _{1}+q^{-5}\sigma _{q}\gamma _{1}=q^{-1}\gamma
_{1},\text{\qquad }q^{-3}\gamma _{1}+q^{5}\sigma _{q}^{-2}\gamma
_{1}=q\sigma _{q}^{-1}\gamma _{1}\,\,,  \label{68}
\end{equation}
\begin{equation}
q^{16}\sigma _{q}\gamma _{2}+\sigma _{q}^{5}\gamma _{2}=q^{20}\gamma _{2},%
\text{\qquad }q^{4}\sigma _{q}^{4}\gamma _{2}+q^{20}\gamma _{2}=\sigma
_{q}^{5}\gamma _{2}\,\,,  \label{69}
\end{equation}
respectively. These relations may be obtained either from (\ref{b2}), (\ref
{b4}) and (\ref{b5}) together with the formulae (\ref{1eta}) which relate
the fusing angles to the solution of the fusing rules in terms of the $q$%
-deformed Coxeter element or alternatively they may be read off directly
from Table A.1. For a direct comparison with (\ref{1eta}) one should cross
all term to one side of the equation by means of (\ref{anti}).

It is also instructive to consider explicitly the matrix representation and
verify the general formulae (\ref{M2}), (\ref{qA}) and (\ref{muM}) of
Section 2.4.3 and 3.2.7. The $q$--deformed Cartan matrix for generic $q$ and 
$\hat{q}$ reads 
\begin{equation}
A(q,\hat{q})=\left( 
\begin{array}{ll}
q\hat{q}+q^{-1}\hat{q}^{-1} & -1 \\ 
-(1+\hat{q}^{2}+\hat{q}^{-2}) & q\hat{q}^{3}+q^{-1}\hat{q}^{-3}
\end{array}
\right)  \label{KG}
\end{equation}
with determinant $\det A(q,\hat{q})=q^{2}\hat{q}^{4}+q^{-2}\hat{q}^{-4}-1$.
The right nullvectors are evaluated to 
\begin{eqnarray}
y(1) &=&(\sinh (\theta _{h}+\theta _{H}),\sinh (2\theta _{h}+2\theta _{H}))
\\
y(2) &=&(\sinh (5\theta _{h}+5\theta _{H}),\sinh (10\theta _{h}+10\theta
_{H}))\,\,.
\end{eqnarray}
From (\ref{KG}) we compute the $M$-matrix according to (\ref{M2}) 
\begin{equation}
M(q,\hat{q})=\frac{1-q^{12}\hat{q}^{24}}{2}\left( 
\begin{array}{ll}
\frac{q\hat{q}+q^{3}\hat{q}^{7}}{1-q^{2}\hat{q}^{4}+q^{4}\hat{q}^{8}} & 
\frac{1+\hat{q}^{2}+\hat{q}^{-2}}{q^{2}\hat{q}^{4}+q^{-2}\hat{q}^{-4}-1} \\ 
\frac{1+\hat{q}^{2}+\hat{q}^{-2}}{q^{2}\hat{q}^{4}+q^{-2}\hat{q}^{-4}-1} & 
\frac{(q\hat{q}+q^{3}\hat{q}^{3})(1+\hat{q}^{2}+\hat{q}^{4})}{1+q^{2}\hat{q}%
^{4}+q^{4}\hat{q}^{8}}
\end{array}
\right) \;.  \label{MAA}
\end{equation}

\noindent Now, careful cancellation against the prefactor produces the
polynomials

\begin{center}
\begin{tabular}{ll}
$M_{11}(q,\hat{q})=$ & $\frac{1}{2}(1+q^{2}\hat{q}^{4}-q^{6}\hat{q}%
^{12}-q^{8}\hat{q}^{16})(q\hat{q}+q^{3}\hat{q}^{7})$ \\ 
$M_{12}(q,\hat{q})=$ & $\frac{1}{2}(1+\hat{q}^{2}+\hat{q}^{-2})(q^{2}\hat{q}%
^{4}+q^{4}\hat{q}^{8}-q^{10}\hat{q}^{20}-q^{8}\hat{q}^{16})$ \\ 
$M_{22}(q,\hat{q})=$ & $\frac{1}{2}(1+\hat{q}^{2}+\hat{q}^{4})(q\hat{q}+q^{3}%
\hat{q}^{3})(1+q^{6}\hat{q}^{12}-q^{2}\hat{q}^{4}-q^{8}\hat{q}^{16})$%
\end{tabular}
\end{center}

\noindent which after expansion give rise to the S-matrix elements via (\ref
{blockS}) and (\ref{muCox}). Evaluating the $M$-matrix at roots of unity, $%
M(e^{s_{k}\theta _{h}},e^{s_{k}\theta _{H}})$ with the exponents $%
s_{1}=1=h-s_{2},$ leads to 
\begin{eqnarray}
M_{ij}(e^{\theta _{h}},e^{\theta _{H}})\! &=&\!\!\frac{2i\sqrt{3}(1+2\cosh
\theta _{H})}{\sinh (\theta _{h}+\theta _{H})\sinh (2\theta _{h}+2\theta
_{H})}\,\,y_{i}(1)y_{j}(1)\,\, \\
M_{ij}(e^{5\theta _{h}},e^{5\theta _{H}})\! &=&\!\!\!\frac{-2i\sqrt{3}%
(1+2\cosh (5\theta _{H}))}{\sinh (5\theta _{h}+5\theta _{H})\sinh (10\theta
_{h}+10\theta _{H})}\,\,y_{i}(2)y_{j}(2)\,,\,\,\,\,\,\,\,\,
\end{eqnarray}
which confirms equation (\ref{Mcharge}) including also the precise factor of
proportionality.

\subsubsection{$S_{ij}\left( \protect\theta \right) $ from $D_{4}^{(3)}$}

Instead of using the data from $G_{2}^{(1)}$, we can also employ the
properties of $D_{4}^{(3)}$. As indicated in the Dynkin diagram, we choose
the values of the bi-colouration to be $c_{1}=-1$ and $c_{2}=c_{3}=c_{4}=1$.
Our conventions for the incidence matrix $I$, the action of $\hat{\tau}$ on
the simple roots and the action of the automorphism $\omega $ on the simple
roots are 
\begin{equation}
I=\left( 
\begin{array}{llll}
0 & 1 & 1 & 1 \\ 
1 & 0 & 0 & 0 \\ 
1 & 0 & 0 & 0 \\ 
1 & 0 & 0 & 0
\end{array}
\right) ,\quad \quad \hat{\tau}(\vec{\alpha})=\left( 
\begin{array}{l}
q^{2}\alpha _{1} \\ 
q^{2}\alpha _{2} \\ 
\alpha _{3} \\ 
\alpha _{4}
\end{array}
\right) ,\quad \quad \omega (\vec{\alpha})=\left( 
\begin{array}{l}
\alpha _{1} \\ 
\alpha _{4} \\ 
\alpha _{2} \\ 
\alpha _{3}
\end{array}
\right) \,.  \label{ig2}
\end{equation}
The lengths of the orbits are $l_{1}=1$, $l_{2}=l_{3}=l_{4}=3$ and the $q$%
-deformed twisted Coxeter element reads therefore $\hat{\sigma}_{q}=\omega
^{-1}\hat{\sigma}_{1}^{q}\hat{\tau}\hat{\sigma}_{2}^{q}$. Successive actions
of this element on the representatives of $\Omega _{i}^{\omega }$ are
reported in Table A.2.\bigskip

\begin{center}
\begin{tabular}{|c||c|c|}
\hline\hline
$\hat{\sigma}_{q}^{x}$ & $\hat{\alpha}_{1}=-\hat{\gamma}_{1}^{\omega }$ & $%
\hat{\alpha}_{2}=\hat{\gamma}_{2}^{\omega }$ \\ \hline\hline
$1$ & $\ast ;\ast ;2;\ast $ & $-2;\ast ;2;\ast $ \\ \hline
$2$ & $2;\ast ;\ast ;2$ & $-2;\ast ;4;2$ \\ \hline
$3$ & $2;2;4;\ast $ & $-2,4;2;4;4$ \\ \hline
$4$ & $\ast ;\ast ;\ast ;4$ & $-4;4;6;4$ \\ \hline
$5$ & $4;4;\ast ;\ast $ & $-4;4;\ast ;6$ \\ \hline
$6$ & $-6;\ast ;\ast ;\ast $ & $-\ast ;6;\ast ;\ast $ \\ \hline
$7$ & $-\ast ;\ast ;8;\ast $ & $8;\ast ;8;\ast $ \\ \hline
$8$ & $-8;\ast ;\ast ;8$ & $8;\ast ;10;8$ \\ \hline
$9$ & $-8;8;10;\ast $ & $8,10;8;10;10$ \\ \hline
$10$ & $-\ast ;\ast ;\ast ;10$ & $10;10;12;10$ \\ \hline
$11$ & $-10;10;\ast ;\ast $ & $10;10;\ast ;12$ \\ \hline
$12$ & $12;\ast ;\ast ;\ast $ & $\ast ;12;\ast ;\ast $ \\ \hline\hline
\end{tabular}
\medskip

\noindent {\small Table A.2: The orbits $\hat{\Omega}_{i}^{q}$ created by
the action of $\hat{\sigma}_{q}^{x}$ on $\gamma _{i}$} \medskip
\end{center}

\noindent For the generating functions (\ref{mutCox}) we obtain 
\begin{eqnarray}
\sum\limits_{x}\mu _{11}\left( x,2y+1\right) q^{x} &=&-q(\,\hat{\lambda}%
_{1}\cdot (\hat{\sigma}_{q})^{y}\hat{\gamma}_{1}^{\omega })/2 \\
\,\,\sum\limits_{x}\mu _{12}\left( x,2y\right) q^{x} &=&-(\,\hat{\lambda}%
_{2}\cdot (\hat{\sigma}_{q})^{y}\hat{\gamma}_{1}^{\omega })/2 \\
\sum\limits_{x}\mu _{22}\left( x,2y-1\right) q^{x} &=&-q^{-1}(\,\hat{\lambda}%
_{2}\cdot (\hat{\sigma}_{q})^{y}\hat{\gamma}_{2}^{\omega })/2
\end{eqnarray}
which yield the scattering matrices (\ref{11}), (\ref{22}) and (\ref{33})
with the help of table 2.

The two non-equivalent solutions to (\ref{Ootafuse2}) corresponding to (\ref
{b2}), (\ref{b4}) and (\ref{b5}) read 
\begin{equation}
q^{2}\hat{\gamma}_{1}^{\omega }+\hat{\sigma}_{q}\hat{\gamma}_{1}^{\omega }=%
\hat{\sigma}_{q}\hat{\gamma}_{2}^{\omega },\qquad \hat{\sigma}_{q}\hat{\gamma%
}_{1}^{\omega }+q^{2}\hat{\gamma}_{1}^{\omega }=\hat{\sigma}_{q}\hat{\gamma}%
_{2}^{\omega },  \label{77}
\end{equation}
\begin{equation}
q\hat{\sigma}_{q}^{-1}\hat{\gamma}_{1}^{\omega }+q^{-3}\hat{\sigma}_{q}^{3}%
\hat{\gamma}_{1}^{\omega }=q^{-1}\hat{\sigma}_{q}\hat{\gamma}_{1}^{\omega
},\qquad q^{-3}\hat{\sigma}_{q}^{3}\hat{\gamma}_{1}^{\omega }+\,q\hat{\sigma}%
_{q}^{-1}\hat{\gamma}_{1}^{\omega }=q^{-1}\hat{\sigma}_{q}\hat{\gamma}%
_{1}^{\omega },  \label{78}
\end{equation}
\begin{equation}
q^{-2}\hat{\sigma}_{q}^{6}\hat{\gamma}_{2}^{\omega }+q^{2}\hat{\sigma}%
_{q}^{2}\hat{\gamma}_{2}^{\omega }=\hat{\sigma}_{q}^{4}\hat{\gamma}%
_{2}^{\omega },\qquad q^{2}\hat{\sigma}_{q}^{2}\hat{\gamma}_{2}^{\omega
}+q^{-2}\hat{\sigma}_{q}^{6}\hat{\gamma}_{2}^{\omega }=\hat{\sigma}_{q}^{4}%
\hat{\gamma}_{2}^{\omega },  \label{79}
\end{equation}
respectively. These relations may be obtained either from (\ref{b2}), (\ref
{b4}) and (\ref{b5}) together with the relation (\ref{2eta}) which relates
the fusing angles to the solution of the fusing rules in terms of the $q$%
-deformed twisted Coxeter element or alternatively they may be read off
directly from Table A.2. Exploiting the relationship between the different
versions of the fusing rules (\ref{fus}), we may also obtain (\ref{77}), (%
\ref{78}) and (\ref{79}) from (\ref{67}), (\ref{68}) and (\ref{69}).

\subsection{$(F_{4}^{(1)},E_{6}^{(2)})$}

\unitlength=0.780000pt 
\begin{picture}(370.00,107.58)(-100.00,0.00)
\put(70.00,38.00){\line(-1,2){7.00}}
\put(70.00,38.00){\line(-1,-2){7.00}}
\qbezier(211.00,28.00)(285.00,-28.00)(360.00,28.00)
\qbezier(251.00,28.00)(285.00,7.01)(320.00,28.00)
\put(365.00,52.00){\makebox(0.00,0.00){$\alpha_6$}}
\put(325.00,52.00){\makebox(0.00,0.00){$\alpha_5$}}
\put(296.25,52.17){\makebox(0.00,0.00){$\hat{\alpha}_3$}}
\put(285.00,93.00){\makebox(0.00,0.00){$\hat{\alpha}_4$}}
\put(245.00,52.00){\makebox(0.00,0.00){$\hat{\alpha}_2$}}
\put(206.00,52.00){\makebox(0.00,0.00){$\hat{\alpha}_1$}}
\put(125.00,53.00){\makebox(0.00,0.00){$\alpha_4$}}
\put(85.83,53.00){\makebox(0.00,0.00){$\alpha_3$}}
\put(45.00,53.00){\makebox(0.00,0.00){$\alpha_2$}}
\put(5.00,53.00){\makebox(0.00,0.00){$\alpha_1$}}
\put(330.00,38.00){\line(1,0){30.00}}
\put(285.33,72.67){\line(0,-1){31.00}}
\put(290.00,38.00){\line(1,0){30.00}}
\put(250.00,38.00){\line(1,0){30.00}}
\put(210.00,38.00){\line(1,0){30.00}}
\put(45.33,33.00){\line(1,0){40.33}}
\put(45.33,43.00){\line(1,0){40.33}}
\put(90.00,38.00){\line(1,0){30.00}}
\put(10.00,38.00){\line(1,0){30.00}}
\put(285.00,78.00){\circle{10.00}}
\put(245.00,38.00){\circle{10.00}}
\put(325.00,38.00){\circle{10.00}}
\put(365.00,38.00){\circle*{10.00}}
\put(285.00,38.00){\circle*{10.00}}
\put(45.00,38.00){\circle{10.00}}
\put(125.00,38.00){\circle{10.00}}
\put(5.00,38.00){\circle*{10.00}}
\put(85.00,38.00){\circle*{10.00}}
\put(205.00,38.00){\circle*{10.00}}
\end{picture}

\noindent The S-matrices of the theory read \cite{PD} 
\begin{eqnarray*}
S_{11}(\theta ) &=&\left\{ 1,1_{2};5,7_{2};7,9_{2};11,15_{2}\right\}
_{\theta } \\
S_{12}(\theta ) &=&\left\{
2,3_{2};4,5_{2};6,7_{2};6,9_{2};8,11_{2};10,13_{2}\right\} _{\theta } \\
S_{13}(\theta ) &=&\left\{ 3,4_{2};5,6_{2};7,10_{2};9,12_{2}\right\}
_{\theta } \\
S_{14}(\theta ) &=&\left\{ 4,5_{2};8,11_{2}\right\} _{\theta } \\
S_{22}(\theta )
&=&%
\{1,1_{2};3,3_{2};3,5_{2};5,5_{2};5,7_{2}^{2};7,9_{2}^{2};7,11_{2};9,11_{2};9,13_{2};11,15_{2}\}_{\theta }
\\
S_{23}(\theta )
&=&\{2,2_{2};4,4_{2};4,6_{2};6,8_{2}^{2};8,10_{2};8,12_{2};10,14_{2}\}_{%
\theta } \\
S_{24}(\theta ) &=&\left\{ 3,3_{2};5,7_{2};7,9_{2};9,13_{2}\right\} _{\theta
} \\
S_{33}(\theta ) &=&\left\{
1,1;3,3_{2};5,7;5,7_{2};7,9_{2};7,11;9,13_{2};11,17\right\} _{\theta } \\
S_{34}(\theta ) &=&\left\{ 2,2;4,6;6,8_{2};8,12;10,16\right\} _{\theta } \\
S_{44}(\theta ) &=&\{1,1;5,7;7,11;11,17\}_{\theta }\,\,.
\end{eqnarray*}
We have $h=12$ and $H=18$ for the Coxeter numbers. We will not report here
all boostrap identities, but we state the combined bootstrap identities (\ref
{cb}) 
\begin{eqnarray}
S_{1l}(\theta +\theta _{h}+2\theta _{H})S_{1l}(\theta -\theta _{h}-2\theta
_{H}) &=&S_{l2}(\theta ) \\
S_{2l}(\theta +\theta _{h}+2\theta _{H})S_{2l}(\theta -\theta _{h}-2\theta
_{H}) &=&S_{l1}(\theta )S_{l3}(\theta -\theta _{H})S_{l3}(\theta +\theta
_{H})\,\,\,\,\,\,\,\,\,\,  \label{f4boot} \\
S_{3l}(\theta +\theta _{h}+\theta _{H})S_{3l}(\theta -\theta _{h}-\theta
_{H}) &=&S_{l2}(\theta )S_{l4}(\theta ) \\
S_{4l}(\theta +\theta _{h}+\theta _{H})S_{4l}(\theta -\theta _{h}-\theta
_{H}) &=&S_{l3}(\theta )
\end{eqnarray}
for $l=1,2,3,4$. Once again there occurs one equation which is more involved
than the usual bootstrap which we depict in Figure A.3. \bigskip

\begin{center}
\unitlength=0.780000pt 
\begin{picture}(299.67,150.00)(-58.00,0.00)
\put(281.33,95.00){\makebox(0.00,0.00){$3$}}
\put(245.67,75.00){\makebox(0.00,0.00){$4$}}
\put(234.67,75.00){\makebox(0.00,0.00){$1$}}
\put(76.33,74.67){\makebox(0.00,0.00){$4$}}
\put(248.67,130.00){\makebox(0.00,0.00){$3$}}
\put(197.67,94.67){\makebox(0.00,0.00){$1$}}
\put(259.67,0.00){\makebox(0.00,0.00){$2$}}
\put(219.67,0.00){\makebox(0.00,0.00){$2$}}
\put(170.00,109.33){\makebox(0.00,0.00){$l$}}
\put(125.00,130.00){\makebox(0.00,0.00){$3$}}
\put(79.00,129.67){\makebox(0.00,0.00){$3$}}
\put(35.33,129.67){\makebox(0.00,0.00){$1$}}
\put(63.33,74.67){\makebox(0.00,0.00){$1$}}
\put(0.00,29.67){\makebox(0.00,0.00){$l$}}
\put(89.67,0.00){\makebox(0.00,0.00){$2$}}
\put(49.67,0.00){\makebox(0.00,0.00){$2$}}
\put(140.67,86.00){\line(1,0){20.00}}
\put(140.67,90.00){\line(1,0){20.00}}
\put(179.67,110.00){\line(6,1){120.00}}
\put(259.67,60.00){\line(0,-1){50.00}}
\put(239.67,110.00){\line(0,1){40.00}}
\put(259.67,60.00){\line(1,3){30.00}}
\put(239.67,110.00){\line(2,-5){20.00}}
\put(219.67,60.00){\line(2,5){20.00}}
\put(219.67,60.00){\line(0,-1){50.00}}
\put(189.67,150.00){\line(1,-3){30.00}}
\put(9.67,30.00){\line(6,1){120.00}}
\put(89.67,60.00){\line(0,-1){50.00}}
\put(49.67,60.00){\line(0,-1){50.00}}
\put(69.67,150.00){\line(0,-1){40.00}}
\put(89.67,60.00){\line(1,3){30.00}}
\put(69.67,110.00){\line(2,-5){20.00}}
\put(49.67,60.00){\line(2,5){20.00}}
\put(19.67,150.00){\line(1,-3){30.00}}
\end{picture}
\medskip

\noindent {\small Figure A.3: $(F_{4}^{(1)},E_{6}^{(2)})$-combined bootstrap
identities \ (\ref{f4boot}).}\medskip
\end{center}

\noindent Reading off the fusing angles from the bootstrap equations we
obtain the mass ratios from (\ref{mratio})

\begin{eqnarray}
\frac{m_{1}}{m_{2}} &=&\frac{\sinh (\theta _{h}+2\theta _{H})}{\sinh
(10\theta _{h}+14\theta _{H})}\qquad \qquad \frac{m_{1}}{m_{3}}=\frac{\sinh
(3\theta _{h}+5\theta _{H})}{\sinh (7\theta _{h}+10\theta _{H})} \\
\frac{m_{1}}{m_{4}} &=&\frac{\sinh (3\theta _{h}+5\theta _{H})}{\sinh
(2\theta _{h}+3\theta _{H})}\qquad \quad \qquad \frac{m_{2}}{m_{3}}=\frac{%
\sinh (9\theta _{h}+15\theta _{H})}{\sinh (2\theta _{h}+2\theta _{H})} \\
\frac{m_{2}}{m_{4}} &=&\frac{\sinh (9\theta _{h}+15\theta _{H})}{\sinh
(\theta _{h}+\theta _{H})}\qquad \qquad \,\,\,\frac{m_{3}}{m_{4}}=\frac{%
\sinh (2\theta _{h}+2\theta _{H})}{\sinh (\theta _{h}+\theta _{H})}.
\label{mrf4_end}
\end{eqnarray}
As in the previous case these formulae can be re-constructed from the
twisted as well as the untwisted Lie algebra.

\subsubsection{$S_{ij}\left( \protect\theta \right) $ from $F_{4}^{(1)}$}

According to our conventions the q-deformed Coxeter element reads in terms
of simple Weyl reflections $\sigma _{q}=\sigma _{1}^{q}\sigma _{3}^{q}\tau
\sigma _{2}^{q}\sigma _{4}^{q}\tau $. The result of successive actions of
this element on the simple roots is reported in Table A.3.\medskip

\hspace{-1.5cm} {\footnotesize 
\begin{tabular}{|c||c|c|c|c|}
\hline\hline
$\sigma _{q}^{x}$ & $\alpha _{1}=-\gamma _{1}$ & $\alpha _{3}=-\gamma _{3}$
& $\alpha _{2}=\gamma _{2}$ & $\alpha _{4}=\gamma _{4}$ \\ \hline\hline
$1$ & $\ast ;4;3,5;\ast $ & $3;3;2,4;2$ & $-4;4;3,5;\ast $ & $\ast ;\ast
;2;2 $ \\ \hline
$2$ & $6;6;5,7;5,7$ & $5;5,7;6^{2},8;6$ & $-6;6,8;5,7^{2},9;5,7$ & $%
5;5;6;\ast $ \\ \hline
$3$ & $8;8,10;9,11;\ast $ & $9;9^{2};8,10^{2};8,10$ & $%
-8,10;8,10^{2};9^{2},11^{2};9,11$ & $\ast ;9;8,10;8$ \\ \hline
$4$ & $\ast ;12;11,13;11,13$ & $11;11,13;12,14;12$ & $%
-12;12^{2},14;11,13^{2},15;11,13$ & $11;11;12;12$ \\ \hline
$5$ & $14;14;\ast ;\ast $ & $\ast ;15;16;16$ & $-14;14,16;15,17;15,17$ & $%
\ast ;15;16;\ast $ \\ \hline
$6$ & $-18;\ast ;\ast ;\ast $ & $\ast ;\ast ;18;\ast $ & $-\ast ;18;\ast
;\ast $ & $\ast ;\ast ;\ast ;18$ \\ \hline
$7$ & $-\ast ;22;21,23;\ast $ & $21;21;20,22;20$ & $22;22;21,23;\ast $ & $%
\ast ;\ast ;20;20$ \\ \hline
$8$ & $-\!24;\!24;\!23,\!25;\!23,\!25$ & $23;23,25;24^{2},26;24$ & $%
24;24,26;23,25^{2},27;23,25$ & $23;23;24;\ast $ \\ \hline
$9$ & $-26;26,28;27,29;\ast $ & $\!27;\!27^{2};\!26,\!28^{2};\!26,\!28$ & $%
26,\!28;\!26,\!28^{2};\!27^{2},\!29^{2};\!27,\!29$ & $\ast
;\!27;\!26,\!28;\!26$ \\ \hline
$10$ & $-\ast ;30;29,31;29,31$ & $29;29,31;30,32;30$ & $%
30;30^{2},32;29,31^{2},33;29,31$ & $29;29;30;30$ \\ \hline
$11$ & $-32;32;\ast ;\ast $ & $\ast ;33;34;34$ & $32;32,34;33,35;33,35$ & $%
\ast ;33;34;\ast $ \\ \hline
$12$ & $36;\ast ;\ast ;\ast $ & $\ast ;\ast ;36;\ast $ & $\ast ;36;\ast
;\ast $ & $\ast ;\ast ;\ast ;36$ \\ \hline\hline
\end{tabular}
}

\begin{center}
{\small Table A.3: The orbits $\Omega _{i}^{q}$ created by the action of $%
\sigma _{q}^{x}$ on $\gamma _{i}.$} \medskip
\end{center}

\noindent By using Table A.3 we may recover the {\small $%
(F_{4}^{(1)},E_{6}^{(2)})-$}S-matrices with the help of generating functions
(\ref{muCox}). The two non-equivalent solutions of the fusing rule in $%
\Omega ^{q}$ are 
\begin{eqnarray*}
\gamma _{l}+q^{-12}\sigma _{q}^{4}\gamma _{l} &=&q^{-6}\sigma _{q}^{2}\gamma
_{l},\;\;\;\sigma _{q}^{-1}\gamma _{l}+q^{12}\sigma _{q}^{-5}\gamma
_{l}=q^{6}\sigma _{q}^{-3}\gamma _{l},\;\;l=1,2,3,4 \\
\gamma _{1}+q^{-4}\sigma _{q}\gamma _{1} &=&q^{-4}\sigma _{q}\gamma
_{2},\;\;\;\sigma _{q}^{-1}\gamma _{1}+q^{4}\sigma _{q}^{-2}\gamma
_{1}=\sigma _{q}^{-1}\gamma _{2}, \\
\gamma _{2}+q^{-14}\sigma _{q}^{5}\gamma _{1} &=&\gamma
_{1},\;\;\;q^{-4}\gamma _{2}+q^{14}\sigma _{q}^{-6}\gamma _{1}=\sigma
_{q}^{-1}\gamma _{1}, \\
\gamma _{4}+q^{-2}\sigma _{q}\gamma _{4} &=&\gamma _{3},\;\;\;q^{-2}\gamma
_{4}+\sigma _{q}^{-1}\gamma _{4}=\sigma _{q}^{-1}\gamma _{3}, \\
\gamma _{4}+q^{-16}\sigma _{q}^{5}\gamma _{3} &=&q^{-16}\sigma
_{q}^{5}\gamma _{4},\;\;\;q^{-2}\gamma _{4}+q^{16}\sigma _{q}^{-6}\gamma
_{3}=q^{14}\sigma _{q}^{-5}\gamma _{4}, \\
\gamma _{1}+q^{-15}\sigma _{q}^{5}\gamma _{3} &=&q^{-11}\sigma
_{q}^{4}\gamma _{4},\;\;\;\sigma _{q}^{-1}\gamma _{1}+q^{15}\sigma
_{q}^{-6}\gamma _{3}=q^{9}\sigma _{q}^{-4}\gamma _{4}, \\
\gamma _{1}+q^{-9}\sigma _{q}^{3}\gamma _{4} &=&q^{-3}\sigma _{q}\gamma
_{3},\;\;\;\sigma _{q}^{-1}\gamma _{1}+q^{7}\sigma _{q}^{-3}\gamma
_{4}=q^{3}\sigma _{q}^{-2}\gamma _{3}, \\
\gamma _{3}+q^{-14}\sigma _{q}^{5}\gamma _{4} &=&q^{-3}\sigma _{q}\gamma
_{1},\;\;\;\sigma _{q}^{-1}\gamma _{3}+q^{12}\sigma _{q}^{-5}\gamma
_{4}=q^{3}\sigma _{q}^{-2}\gamma _{1}, \\
\gamma _{2}+q^{-15}\sigma _{q}^{5}\gamma _{3} &=&q^{-1}\sigma _{q}\gamma
_{4},\;\;\;q^{-4}\gamma _{2}+q^{15}\sigma _{q}^{-6}\gamma _{3}=q^{-1}\sigma
_{q}^{-1}\gamma _{4}, \\
\gamma _{2}+q^{-15}\sigma _{q}^{5}\gamma _{4} &=&q\gamma
_{3},\;\;\;q^{-4}\gamma _{2}+q^{13}\sigma _{q}^{-5}\gamma _{4}=q^{-1}\sigma
_{q}^{-1}\gamma _{3}, \\
\gamma _{3}+q^{-4}\sigma _{q}^{2}\gamma _{4} &=&q^{-3}\sigma _{q}\gamma
_{2},\;\;\;\sigma _{q}^{-1}\gamma _{3}+q^{2}\sigma _{q}^{-2}\gamma
_{4}=q^{-1}\sigma _{q}^{-1}\gamma _{2}, \\
\gamma _{4}+q^{-8}\sigma _{q}^{3}\gamma _{4} &=&q^{-3}\sigma _{q}\gamma
_{1},\;\;\;q^{-2}\gamma _{4}+q^{6}\sigma _{q}^{-3}\gamma _{4}=q^{3}\sigma
_{q}^{-2}\gamma _{1}, \\
\gamma _{4}+q^{-13}\sigma _{q}^{4}\gamma _{1} &=&q^{-10}\sigma
_{q}^{3}\gamma _{4},\;\;\;q^{-2}\gamma _{4}+q^{13}\sigma _{q}^{-5}\gamma
_{1}=q^{8}\sigma _{q}^{-3}\gamma _{4}.
\end{eqnarray*}
Once again we can confirm from these solution the equivalence of the
bootstrap equations and the fusing rules by means of (\ref{1eta}) and also
verify the relation for the mass ratios (\ref{mratio}).

\subsubsection{$S_{ij}\left( \protect\theta \right) $ from $\hat{E}%
_{6}^{(2)} $}

The $q$-deformed twisted Coxeter element in the conventions stated reads
explicitly $\hat{\sigma}_{q}=\omega ^{-1}\hat{\sigma}_{1}^{q}\hat{\sigma}%
_{3}^{q}\hat{\tau}\hat{\sigma}_{2}^{q}\hat{\sigma}_{4}^{q}$. The successive
actions of this element on the representatives of $\Omega _{i}^{\omega }$
are reported in Table A.4.\bigskip

\hspace{-2cm} {\footnotesize 
\begin{tabular}{|c||c|c|c|c|}
\hline\hline
$\hat{\sigma}_{q}^{x}$ & $\alpha _{6}=-\hat{\gamma}_{1}^{\omega }$ & $\hat{%
\alpha}_{3}=-\hat{\gamma}_{3}^{\omega }$ & $\hat{\alpha}_{2}=\hat{\gamma}%
_{2}^{\omega }$ & $\hat{\alpha}_{4}=\hat{\gamma}_{4}^{\omega }$ \\ 
\hline\hline
$1$ & $0;\ast ;\ast ;\ast ;\ast ;\ast $ & $\ast ;\ast ;2;2;2;2$ & $-\ast
;\ast ;2;\ast ;2;2$ & $\ast ;\ast ;2;2;\ast ;\ast $ \\ \hline
$2$ & $\ast ;\ast ;2;\ast ;2;\ast $ & $2;2;2;\ast ;4;4$ & $-2;2;2,4;4;4;4$ & 
$\ast ;\ast ;\ast ;\ast ;4;4$ \\ \hline
$3$ & $\ast ;2;2,4;4;4;4$ & $4;4;4^{2};4;4;\ast $ & $-4;4;4^{2};4;4,6;6$ & $%
4;4;4;\ast ;\ast ;\ast $ \\ \hline
$4$ & $4;4;4;4;6;6$ & $\ast ;4;4,6;6;6^{2};6$ & $-6;4,6;4,6^{2};6;6^{2};6$ & 
$\ast ;\ast ;6;6;6;\ast $ \\ \hline
$5$ & $6;6;6;\ast ;6;\ast $ & $6;6^{2};6^{2};6;8;8$ & $%
-6;6^{2};6^{2},8;6,8;8^{2};8$ & $\ast ;6;6;\ast ;8;8$ \\ \hline
$6$ & $\ast ;6;6,8;8;8;\ast $ & $8;8;8;8;8;\ast $ & $%
-8;8^{2};8^{2};8;8,10;10 $ & $8;8;8;8;\ast ;\ast $ \\ \hline
$7$ & $\ast ;8;8;8;10;10$ & $\ast ;8;8;\ast ;10;\ast $ & $%
-10;8,10;8,10;10;10;\ast $ & $\ast ;\ast ;\ast ;\ast ;10;\ast $ \\ \hline
$8$ & $10;10;\ast ;\ast ;\ast ;\ast $ & $\ast ;10;10;10;\ast ;\ast $ & $%
-\ast ;10;10;10;12;\ast $ & $\ast ;10;10;\ast ;\ast ;\ast $ \\ \hline
$9$ & $-\ast ;\ast ;\ast ;\ast ;\ast ;12$ & $\ast ;\ast ;12;\ast ;\ast ;\ast 
$ & $-\ast ;12;\ast ;\ast ;\ast ;\ast $ & $\ast ;\ast ;\ast ;12;\ast ;\ast $
\\ \hline
$10$ & $-12;\ast ;\ast ;\ast ;\ast ;\ast $ & $\ast ;\ast ;14;14;14;14$ & $%
\ast ;\ast ;14;\ast ;14;14$ & $\ast ;\ast ;14;14\ast ;\ast $ \\ \hline
$11$ & $-\ast ;\ast ;14;\ast ;14;\ast $ & $14;14;14;\ast ;16;16$ & $%
14;14;14,16;16;16;16$ & $\ast ;\ast ;\ast ;\ast ;16;16$ \\ \hline
$12$ & $-\!\ast ;14;\!14,\!16;\!16;\!16;\!16$ & $16;16;16^{2};16;16;\ast $ & 
$16;16;16^{2};16;16,18;18$ & $16;16;16;\ast ;\ast ;\ast $ \\ \hline
$13$ & $-16;16;16;16;18;18$ & $\ast ;\!16;\!16,\!18;\!18;\!18^{2};\!18$ & $%
18;\!16,\!18;\!16,\!18^{2};\!18;\!18^{2};\!18$ & $\ast ;\ast ;18;18;18;\ast $
\\ \hline
$14$ & $-18;18;18;\ast ;18;\ast $ & $18;18^{2};18^{2};18;20;20$ & $%
18;\!18^{2};\!18^{2},\!20;\!18,\!20;\!20^{2};\!20$ & $\ast ;18;18;\ast
;20;20 $ \\ \hline
$15$ & $-\ast ;18;18,20;20;20;\ast $ & $20;20;20;20;20;\ast $ & $%
20;20^{2};20^{2};20;20,22;22$ & $20;20;20;20;\ast ;\ast $ \\ \hline
$16$ & $-\ast ;20;20;20;22;22$ & $\ast ;20;20;\ast ;22;\ast $ & $%
22;20,22;20,22;22;22;\ast $ & $\ast ;\ast ;\ast ;\ast ;22;\ast $ \\ \hline
$17$ & $-22;22;\ast ;\ast ;\ast ;\ast $ & $\ast ;22;22;22;\ast ;\ast $ & $%
\ast ;22;22;22;24;\ast $ & $\ast ;22;22;\ast ;\ast ;\ast $ \\ \hline
$18$ & $\ast ;\ast ;\ast ;\ast ;\ast ;24$ & $\ast ;\ast ;24;\ast ;\ast ;\ast 
$ & $\ast ;24;\ast ;\ast ;\ast ;\ast $ & $\ast ;\ast ;\ast ;24;\ast ;\ast $
\\ \hline\hline
\end{tabular}
}\medskip

\begin{center}
{\small Table A.4: The orbits $\hat{\Omega}_{i}^{q}$ created by the action
of $\hat{\sigma}_{q}^{x}$ on $\gamma _{i}$.} \medskip
\end{center}

\noindent Using the orbits $\hat{\Omega}_{i}^{q}$ listed in Table A.4 we
recover with help of the generating functions (\ref{mutCox}) the {\small $%
(F_{4}^{(1)},E_{6}^{(2)})$-}S-matrices. The two non-equivalent solutions to
the fusing rule in $\hat{\Omega}_{q}$ read 
\begin{eqnarray*}
\hat{\gamma}_{l}^{\omega }+q^{-8}\hat{\sigma}_{q}^{6}\hat{\gamma}%
_{l}^{\omega } &=&q^{-4}\hat{\sigma}_{q}^{3}\hat{\gamma}_{l}^{\omega
},\;\;\;q^{2}\hat{\sigma}_{q}^{2}\hat{\gamma}_{l}^{\omega }+q^{10}\hat{\sigma%
}_{q}^{-4}\hat{\gamma}_{l}^{\omega }=q^{6}\hat{\sigma}_{q}^{-1}\hat{\gamma}%
_{l}^{\omega },\;\;l=1,2,3,4 \\
\hat{\gamma}_{1}^{\omega }+q^{-2}\hat{\sigma}_{q}^{2}\hat{\gamma}%
_{1}^{\omega } &=&q^{-2}\hat{\sigma}_{q}\hat{\gamma}_{2}^{\omega
},\;\;\;q^{2}\hat{\sigma}_{q}^{2}\hat{\gamma}_{1}^{\omega }+q^{4}\hat{\gamma}%
_{1}^{\omega }=q^{2}\hat{\sigma}_{q}\hat{\gamma}_{2}^{\omega }, \\
\hat{\gamma}_{2}^{\omega }+q^{-10}\hat{\sigma}_{q}^{8}\hat{\gamma}%
_{1}^{\omega } &=&\hat{\sigma}_{q}\hat{\gamma}_{1}^{\omega },\;\;\;\hat{%
\sigma}_{q}^{2}\hat{\gamma}_{2}^{\omega }+q^{12}\hat{\sigma}_{q}^{-6}\hat{%
\gamma}_{1}^{\omega }=q^{2}\hat{\sigma}_{q}\hat{\gamma}_{1}^{\omega }, \\
\hat{\gamma}_{4}^{\omega }+q^{-2}\hat{\sigma}_{q}\hat{\gamma}_{4}^{\omega }
&=&\hat{\gamma}_{3}^{\omega },\;\;\;\hat{\sigma}_{q}^{2}\hat{\gamma}%
_{4}^{\omega }+q^{2}\hat{\sigma}_{q}\hat{\gamma}_{4}^{\omega }=q^{2}\hat{%
\sigma}_{q}\hat{\gamma}_{3}^{\omega }, \\
\hat{\gamma}_{4}^{\omega }+q^{-10}\hat{\sigma}_{q}^{8}\hat{\gamma}%
_{3}^{\omega } &=&q^{-10}\hat{\sigma}_{q}^{8}\hat{\gamma}_{4}^{\omega
},\;\;\;\hat{\sigma}_{q}^{2}\hat{\gamma}_{4}^{\omega }+q^{12}\hat{\sigma}%
_{q}^{-7}\hat{\gamma}_{3}^{\omega }=q^{10}\hat{\sigma}_{q}^{-6}\hat{\gamma}%
_{4}^{\omega }, \\
\hat{\gamma}_{1}^{\omega }+q^{-10}\hat{\sigma}_{q}^{7}\hat{\gamma}%
_{3}^{\omega } &=&q^{-8}\hat{\sigma}_{q}^{5}\hat{\gamma}_{4}^{\omega
},\;\;\;q^{2}\hat{\sigma}_{q}^{2}\hat{\gamma}_{1}^{\omega }+q^{12}\hat{\sigma%
}_{q}^{-6}\hat{\gamma}_{3}^{\omega }=q^{8}\hat{\sigma}_{q}^{-3}\hat{\gamma}%
_{4}^{\omega }, \\
\hat{\gamma}_{1}^{\omega }+q^{-6}\hat{\sigma}_{q}^{4}\hat{\gamma}%
_{4}^{\omega } &=&q^{-2}\hat{\sigma}_{q}\hat{\gamma}_{3}^{\omega
},\;\;\;q^{2}\hat{\sigma}_{q}^{2}\hat{\gamma}_{1}^{\omega }+q^{6}\hat{\sigma}%
_{q}^{-2}\hat{\gamma}_{4}^{\omega }=q^{4}\hat{\gamma}_{3}^{\omega }, \\
\hat{\gamma}_{3}^{\omega }+q^{-10}\hat{\sigma}_{q}^{7}\hat{\gamma}%
_{4}^{\omega } &=&q^{-2}\hat{\sigma}_{q}^{2}\hat{\gamma}_{1}^{\omega
},\;\;\;q^{2}\hat{\sigma}_{q}\hat{\gamma}_{3}^{\omega }+q^{10}\hat{\sigma}%
_{q}^{-5}\hat{\gamma}_{4}^{\omega }=q^{4}\hat{\gamma}_{1}^{\omega }, \\
\hat{\gamma}_{2}^{\omega }+q^{-10}\hat{\sigma}_{q}^{8}\hat{\gamma}%
_{3}^{\omega } &=&q^{-2}\hat{\sigma}_{q}\hat{\gamma}_{4}^{\omega },\;\;\;%
\hat{\sigma}_{q}^{2}\hat{\gamma}_{2}^{\omega }+q^{12}\hat{\sigma}_{q}^{-7}%
\hat{\gamma}_{3}^{\omega }=q^{2}\hat{\sigma}_{q}\hat{\gamma}_{4}^{\omega },
\\
\hat{\gamma}_{2}^{\omega }+q^{-10}\hat{\sigma}_{q}^{8}\hat{\gamma}%
_{4}^{\omega } &=&\hat{\gamma}_{3}^{\omega },\;\;\;\hat{\sigma}_{q}^{2}\hat{%
\gamma}_{2}^{\omega }+q^{10}\hat{\sigma}_{q}^{-6}\hat{\gamma}_{4}^{\omega
}=q^{2}\hat{\sigma}_{q}\hat{\gamma}_{3}^{\omega }, \\
\hat{\gamma}_{3}^{\omega }+q^{-4}\hat{\sigma}_{q}^{2}\hat{\gamma}%
_{4}^{\omega } &=&q^{-2}\hat{\sigma}_{q}\hat{\gamma}_{2}^{\omega
},\;\;\;q^{2}\hat{\sigma}_{q}\hat{\gamma}_{3}^{\omega }+q^{4}\hat{\gamma}%
_{4}^{\omega }=q^{2}\hat{\sigma}_{q}\hat{\gamma}_{2}^{\omega }, \\
\hat{\gamma}_{4}^{\omega }+q^{-6}\hat{\sigma}_{q}^{4}\hat{\gamma}%
_{4}^{\omega } &=&q^{-2}\hat{\sigma}_{q}^{2}\hat{\gamma}_{1}^{\omega },\;\;\;%
\hat{\sigma}_{q}^{2}\hat{\gamma}_{4}^{\omega }+q^{6}\hat{\sigma}_{q}^{-2}%
\hat{\gamma}_{4}^{\omega }=q^{4}\hat{\gamma}_{1}^{\omega }, \\
\hat{\gamma}_{4}^{\omega }+q^{-8}\hat{\sigma}_{q}^{7}\hat{\gamma}%
_{1}^{\omega } &=&q^{-6}\hat{\sigma}_{q}^{5}\hat{\gamma}_{4}^{\omega },\;\;\;%
\hat{\sigma}_{q}^{2}\hat{\gamma}_{4}^{\omega }+q^{10}\hat{\sigma}_{q}^{-5}%
\hat{\gamma}_{1}^{\omega }=q^{6}\hat{\sigma}_{q}^{-3}\hat{\gamma}%
_{4}^{\omega }.
\end{eqnarray*}

Again we confirm from these solution the equivalence between the bootstrap
equations and the fusing rules by means of (\ref{2eta}) and also verify the
relation for the mass ratios (\ref{mratio}).

\subsection{$(C_{2}^{(1)},D_{3}^{(2)})$}

\unitlength=0.780000pt 
\begin{picture}(437.92,149.59)(-40.00,50.00)
\qbezier(420.00,170.00)(440.00,150.00)(420.00,130.00)
\put(57.00,125.00){\makebox(0.00,0.00){$c_1=-1$ if $N$ even}}
\put(437.50,104.17){\makebox(0.00,0.00){$\alpha_{N+1}$}}
\put(437.92,184.17){\makebox(0.00,0.00){$\hat{\alpha}_N$}}
\put(411.17,149.59){\makebox(0.00,0.00){$\hat{\alpha}_{N-1}$}}
\put(347.50,165.00){\makebox(0.00,0.00){$\hat{\alpha}_{N-2}$}}
\put(296.00,165.00){\makebox(0.00,0.00){$\hat{\alpha}_2$}}
\put(255.00,165.00){\makebox(0.00,0.00){$\hat{\alpha}_1$}}
\put(177.92,165.42){\makebox(0.00,0.00){$\alpha_N$}}
\put(137.92,165.42){\makebox(0.00,0.00){$\alpha_{N-1}$}}
\put(97.50,165.00){\makebox(0.00,0.00){$\alpha_{N-2}$}}
\put(45.00,165.00){\makebox(0.00,0.00){$\alpha_2$}}
\put(5.00,165.00){\makebox(0.00,0.00){$\alpha_1$}}
\put(389.00,146.33){\line(1,-1){22.67}}
\put(411.67,176.67){\line(-1,-1){23.00}}
\put(350.00,150.00){\line(1,0){30.00}}
\put(330.00,150.00){\line(1,0){10.00}}
\put(300.00,150.00){\line(1,0){10.00}}
\put(260.00,150.00){\line(1,0){30.00}}
\put(150.00,150.00){\line(1,-2){7.00}}
\put(150.00,150.00){\line(1,2){7.00}}
\put(135.33,145.00){\line(1,0){39.67}}
\put(135.33,155.00){\line(1,0){39.67}}
\put(100.00,150.00){\line(1,0){30.00}}
\put(80.00,150.00){\line(1,0){10.00}}
\put(50.00,150.00){\line(1,0){10.00}}
\put(10.00,150.00){\line(1,0){30.00}}
\put(415.00,120.00){\circle{10.00}}
\put(415.00,180.00){\circle{10.00}}
\put(385.00,150.00){\circle*{10.00}}
\put(345.00,150.00){\circle{10.00}}
\put(135.00,150.00){\circle*{10.00}}
\put(175.00,150.00){\circle{10.00}}
\put(255.00,150.00){\circle{10.00}}
\put(295.00,150.00){\circle{10.00}}
\put(95.00,150.00){\circle{10.00}}
\put(45.00,150.00){\circle{10.00}}
\put(5.00,150.00){\circle{10.00}}
\end{picture}

\noindent The S-matrices are given as 
\begin{equation*}
S_{11}(\theta )=\{1,1;3,5\}_{\theta }\quad S_{12}(\theta
)=\{2,2_{2}\}_{\theta }\quad S_{22}(\theta )=\{1,1_{2};3,3_{2}\}_{\theta }.
\end{equation*}
We have $h=4$ and $H=6$ for the Coxeter numbers. The combined bootstrap
equations (\ref{cb}) yield 
\begin{eqnarray}
S_{1l}(\theta +\theta _{h}+\theta _{H})S_{1l}(\theta -\theta _{h}-\theta
_{H}) &=&S_{l2}(\theta ) \\
S_{2l}(\theta +\theta _{h}+2\theta _{H})S_{2l}(\theta -\theta _{h}-2\theta
_{H}) &=&S_{l1}(\theta -\theta _{H})S_{l1}(\theta +\theta _{H})  \label{xxx}
\end{eqnarray}
for $l=1,2$.

\begin{center}
\unitlength=0.780000pt 
\begin{picture}(260.00,100.00)(-65.00,0.00)
\put(210.00,40.00){\makebox(0.00,0.00){$1$}}
\put(60.00,60.00){\makebox(0.00,0.00){$1$}}
\put(250.00,100.00){\makebox(0.00,0.00){$1$}}
\put(170.00,100.00){\makebox(0.00,0.00){$1$}}
\put(100.00,100.00){\makebox(0.00,0.00){$1$}}
\put(20.00,100.00){\makebox(0.00,0.00){$1$}}
\put(250.00,0.00){\makebox(0.00,0.00){$2$}}
\put(170.00,0.00){\makebox(0.00,0.00){$2$}}
\put(100.00,0.00){\makebox(0.00,0.00){$2$}}
\put(20.00,0.00){\makebox(0.00,0.00){$2$}}
\put(160.00,60.00){\makebox(0.00,0.00){$l$}}
\put(0.00,20.00){\makebox(0.00,0.00){$l$}}
\put(170.00,60.00){\line(5,1){90.00}}
\put(230.00,50.00){\line(1,-2){20.00}}
\put(230.00,50.00){\line(1,2){20.00}}
\put(190.00,50.00){\line(1,0){40.00}}
\put(190.00,50.00){\line(-1,-2){20.00}}
\put(170.00,90.00){\line(1,-2){20.00}}
\put(126.00,50.00){\line(1,0){20.00}}
\put(126.00,54.00){\line(1,0){20.00}}
\put(10.00,20.00){\line(5,1){100.00}}
\put(80.00,50.00){\line(1,-2){20.00}}
\put(80.00,50.00){\line(1,2){20.00}}
\put(40.00,50.00){\line(1,0){40.00}}
\put(40.00,50.00){\line(-1,2){20.00}}
\put(40.00,50.00){\line(-1,-2){20.00}}
\end{picture}
\medskip

\noindent {\small Figure A.4: $(C_{2}^{(1)},D_{3}^{(2)})$-combined bootstrap
identities (\ref{xxx}).}\medskip
\end{center}

\noindent The mass ratio according to (\ref{ATFTmass}) are

\begin{equation}
\frac{m_{1}}{m_{2}}=\frac{\sinh (\theta _{h}+\theta _{H})}{\sinh (2\theta
_{h}+4\theta _{H})}.
\end{equation}

\subsubsection{$S_{ij}(\protect\theta )$ from $C_{2}^{(1)}$:}

The result of successive actions of the $q$-deformed Coxeter element on the
simple roots is reported in Table A.5.

\begin{center}
\begin{tabular}{|c||c|c|}
\hline\hline
$\sigma _{q}^{x}$ & $\alpha _{1}=-\gamma _{1}$ & $\alpha _{2}=\gamma _{2}$
\\ \hline\hline
$1$ & $4;3$ & $-3,5;4$ \\ \hline
$2$ & $-6;\ast $ & $-\ast ;6$ \\ \hline
$3$ & $-10;9$ & $9,11;10$ \\ \hline
$4$ & $12;\ast $ & $\ast ;12$ \\ \hline\hline
\end{tabular}
\medskip

\noindent {\small Table A.5: The orbits }$\ \Omega _{i}^{q}${\small \
created by the action of }$\sigma _{q}^{x}$ {\small on }$\gamma _{i}$\medskip
\end{center}

\noindent The two non-equivalent solutions to the fusing rule in $\Omega
_{q} $ read 
\begin{eqnarray*}
\gamma _{1}+q^{-2}\sigma _{q}\gamma _{1} &=&q^{-3}\sigma _{q}\gamma
_{2},\;\;\;\sigma _{q}^{-1}\gamma _{1}+q^{2}\sigma _{q}^{-2}\gamma
_{1}=q^{-1}\sigma _{q}^{-1}\gamma _{2}, \\
\gamma _{1}+q^{-7}\sigma _{q}^{2}\gamma _{2} &=&q^{-4}\sigma _{q}\gamma
_{1},\;\;\;\sigma _{q}^{-1}\gamma _{1}+q^{3}\sigma _{q}^{-2}\gamma
_{2}=q^{4}\sigma _{q}^{-2}\gamma _{1}.
\end{eqnarray*}

\subsubsection{$S_{ij}(\protect\theta )$ from $\hat{D}_{3}^{(2)}$:}

The result of successive actions of the q-deformed twisted Coxeter element
on the simple roots is reported in Table A.6.

\begin{center}
\begin{tabular}{|c||c|c|}
\hline\hline
$\hat{\sigma}_{q}^{x}$ & $\hat{\alpha}_{1}=-\hat{\gamma}_{1}^{\omega }$ & $%
\hat{\alpha}_{2}=\hat{\gamma}_{2}^{\omega }$ \\ \hline\hline
$1$ & $\ast ;\ast ;2$ & $-2;\ast ;2$ \\ \hline
$2$ & $2;2;\ast $ & $-2;2;4$ \\ \hline
$3$ & $-4;\ast ;\ast $ & $-\ast ;4;\ast $ \\ \hline
$4$ & $-\ast ;\ast ;6$ & $6;\ast ;6$ \\ \hline
$5$ & $-6;6;\ast $ & $6;6;8$ \\ \hline
$6$ & $8;\ast ;\ast $ & $\ast ;8;\ast $ \\ \hline\hline
\end{tabular}
\medskip

\noindent {\small Table A.6: The orbits }$\ \hat{\Omega}_{i}^{q}${\small \
created by the action of }$\hat{\sigma}_{q}^{x}$ {\small on }$\hat{\gamma}%
_{i}^{\omega }$\medskip
\end{center}

\noindent The two non-equivalent solutions to the fusing rule in $\hat{\Omega%
}_{q}$ read 
\begin{eqnarray*}
\hat{\gamma}_{1}^{\omega }+q^{-2}\hat{\sigma}_{q}\hat{\gamma}_{1}^{\omega }
&=&q^{-2}\hat{\sigma}_{q}\hat{\gamma}_{2}^{\omega },\;\;\;q^{2}\hat{\sigma}%
_{q}\hat{\gamma}_{1}^{\omega }+q^{4}\hat{\gamma}_{1}^{\omega }=q^{2}\hat{%
\sigma}_{q}\hat{\gamma}_{2}^{\omega }, \\
\hat{\gamma}_{1}^{\omega }+q^{-4}\hat{\sigma}_{q}^{3}\hat{\gamma}%
_{2}^{\omega } &=&q^{-2}\hat{\sigma}_{q}^{2}\hat{\gamma}_{1}^{\omega
},\;\;\;q^{2}\hat{\sigma}_{q}\hat{\gamma}_{1}^{\omega }+q^{4}\hat{\sigma}%
_{q}^{-1}\hat{\gamma}_{2}^{\omega }=q^{4}\hat{\sigma}_{q}^{-1}\hat{\gamma}%
_{1}^{\omega }.
\end{eqnarray*}

\subsection{$(C_{3}^{(1)},D_{4}^{(2)})$}

The S-matrices are 
\begin{eqnarray*}
S_{11}(\theta ) &=&\{1,1;5,7\}_{\theta }\quad S_{12}(\theta
)=\{2,2;4,6\}_{\theta }\quad S_{33}(\theta
)=\{1,1_{2};3,3_{2};5,5_{2}\}_{\theta } \\
S_{22}(\theta ) &=&\{1,1;3,3_{2};5,7\}_{\theta }\quad S_{23}(\theta
)=\{2,2_{2};4,4_{2}\}_{\theta }\quad S_{13}(\theta )=\{3,3_{2}\}_{\theta }.
\end{eqnarray*}
We have $h=6$ and $H=8$ for the Coxeter numbers. The combined bootstrap
identities read 
\begin{eqnarray}
S_{1l}(\theta +\theta _{h}+\theta _{H})S_{1l}(\theta -\theta _{h}-\theta
_{H}) &=&S_{l2}(\theta ) \\
S_{2l}(\theta +\theta _{h}+\theta _{H})S_{2l}(\theta -\theta _{h}-\theta
_{H}) &=&S_{l1}(\theta )S_{l3}(\theta ) \\
S_{3l}(\theta +\theta _{h}+2\theta _{H})S_{3l}(\theta -\theta _{h}-2\theta
_{H}) &=&S_{l2}(\theta -\theta _{H})S_{l2}(\theta +\theta _{H}).
\end{eqnarray}

\noindent The mass ratios turn out to be 
\begin{equation}
\frac{m_{1}}{m_{2}}=\frac{\sinh (\theta _{h}+\theta _{H})}{\sinh (4\theta
_{h}+6\theta _{H})}\quad \frac{m_{1}}{m_{3}}=\frac{\sinh (\theta _{h}+\theta
_{H})}{\sinh (3\theta _{h}+5\theta _{H})}\quad \frac{m_{2}}{m_{3}}=\frac{%
\sinh (2\theta _{h}+2\theta _{H})}{\sinh (3\theta _{h}+5\theta _{H})}.
\end{equation}

\subsubsection{$S_{ij}(\protect\theta )$ from $C_{3}^{(1)}$}

The result of successive actions of the q-deformed Coxeter element on the
simple roots is reported in Table A.7.

\begin{center}
\begin{tabular}{|c||c|c|c|}
\hline\hline
$\sigma _{q}^{x}$ & $\alpha _{1}=\gamma _{1}$ & $\alpha _{3}=\gamma _{3}$ & $%
\alpha _{2}=-\gamma _{2}$ \\ \hline\hline
$1$ & $-2;2;\ast $ & $\ast ;3,5;4$ & $2;2,4;3$ \\ \hline
$2$ & $-\ast ;6;5$ & $5,7;5,7;6$ & $6;6;5$ \\ \hline
$3$ & $-8;\ast ;\ast $ & $\ast ;\ast ;8$ & $-\ast ;8;\ast $ \\ \hline
$4$ & $10;10;\ast $ & $\ast ;11,13;12$ & $-10;10,12;11$ \\ \hline
$5$ & $\ast ;14;13$ & $13,15;13,15;14$ & $-14;14;13$ \\ \hline
$6$ & $16;\ast ;\ast $ & $\ast ;\ast ;16$ & $\ast ;16;\ast $ \\ \hline\hline
\end{tabular}
\medskip

\noindent {\small Table A.7 : The orbits }$\ \Omega _{i}^{q}${\small \
created by the action of }$\sigma _{q}^{x}$ {\small on }$\gamma _{i}$\medskip
\end{center}

\noindent The solutions of the fusing rule in $\Omega ^{q}$ are{\small \ } 
\begin{eqnarray*}
\gamma _{1}+q^{-2}\sigma _{q}\gamma _{1} &=&\gamma _{2},\;\;\;q^{-2}\gamma
_{1}+\sigma _{q}^{-1}\gamma _{1}=\sigma _{q}^{-1}\gamma _{2}, \\
\gamma _{1}+q^{-6}\sigma _{q}^{2}\gamma _{2} &=&q^{-6}\sigma _{q}^{2}\gamma
_{1},\;\;\;q^{-2}\gamma _{1}+q^{6}\sigma _{q}^{-3}\gamma _{2}=q^{4}\sigma
_{q}^{-2}\gamma _{1}, \\
\gamma _{1}+q^{-2}\sigma _{q}\gamma _{2} &=&q^{-3}\sigma _{q}\gamma
_{3},\;\;\;q^{-2}\gamma _{1}+q^{2}\sigma _{q}^{-2}\gamma _{2}=q^{-1}\sigma
_{q}^{-1}\gamma _{3}, \\
\gamma _{1}+q^{-7}\sigma _{q}^{2}\gamma _{3} &=&q^{-4}\sigma _{q}\gamma
_{2},\;\;\;q^{-2}\gamma _{1}+q^{3}\sigma _{q}^{-2}\gamma _{3}=q^{4}\sigma
_{q}^{-2}\gamma _{2}, \\
\gamma _{2}+q^{-9}\sigma _{q}^{3}\gamma _{3} &=&q^{-6}\sigma _{q}^{2}\gamma
_{1},\;\;\;\sigma _{q}^{-1}\gamma _{2}+q^{5}\sigma _{q}^{-3}\gamma
_{3}=q^{4}\sigma _{q}^{-2}\gamma _{1}.
\end{eqnarray*}

\subsubsection{$S_{ij}(\protect\theta )$ from $\hat{D}_{4}^{(2)}$}

The result of successive actions of the q-deformed twisted Coxeter element
on the simple roots is reported in Table A.8.

\begin{center}
\begin{tabular}{|c||c|c|c|}
\hline\hline
$\hat{\sigma}_{q}^{x}$ & $\hat{\alpha}_{1}=\hat{\gamma}_{1}^{\omega }$ & $%
\hat{\alpha}_{3}=\hat{\gamma}_{3}^{\omega }$ & $\hat{\alpha}_{2}=-\hat{\gamma%
}_{2}^{\omega }$ \\ \hline\hline
$1$ & $-2;2;\ast ;\ast $ & $\ast ;2;\ast ;2$ & $2;2;\ast ;2$ \\ \hline
$2$ & $-\ast ;\ast ;\ast ;4$ & $4;2,4;2;4$ & $\ast ;2;2;4$ \\ \hline
$3$ & $-\ast ;4;4;\ast $ & $4;4;4;6$ & $4;4;4;\ast $ \\ \hline
$4$ & $-6;\ast ;\ast ;\ast $ & $\ast ;\ast ;6;\ast $ & $-\ast ;6;\ast ;\ast $
\\ \hline
$5$ & $8;8;\ast ;\ast $ & $\ast ;8;\ast ;8$ & $-8;8;\ast ;8$ \\ \hline
$6$ & $\ast ;\ast ;\ast ;10$ & $10;8,10;8;10$ & $-\ast ;8;8;10$ \\ \hline
$7$ & $\ast ;10;10;\ast $ & $10;10;10;12$ & $-10;10;10;\ast $ \\ \hline
$8$ & $12;\ast ;\ast ;\ast $ & $\ast ;\ast ;12;\ast $ & $\ast ;12;\ast ;\ast 
$ \\ \hline\hline
\end{tabular}
\medskip

\noindent {\small Table A.8 : The orbits }$\ \hat{\Omega}_{i}^{q}${\small \
created by the action of }$\hat{\sigma}_{q}^{x}$ {\small on }$\hat{\gamma}%
_{i}^{\omega }$\medskip
\end{center}

\noindent The solutions of the fusing rule in $\hat{\Omega}^{q}$ are{\small %
\ } 
\begin{eqnarray*}
\hat{\gamma}_{1}^{\omega }+q^{-2}\hat{\sigma}_{q}\hat{\gamma}_{1}^{\omega }
&=&\hat{\gamma}_{2}^{\omega },\;\;\;\hat{\sigma}_{q}^{2}\hat{\gamma}%
_{1}^{\omega }+q^{2}\hat{\sigma}_{q}\hat{\gamma}_{1}^{\omega }=q^{2}\hat{%
\sigma}_{q}\hat{\gamma}_{2}^{\omega }, \\
\hat{\gamma}_{1}^{\omega }+q^{-4}\hat{\sigma}_{q}^{3}\hat{\gamma}%
_{2}^{\omega } &=&q^{-4}\hat{\sigma}_{q}^{3}\hat{\gamma}_{1}^{\omega },\;\;\;%
\hat{\sigma}_{q}^{2}\hat{\gamma}_{1}^{\omega }+q^{6}\hat{\sigma}_{q}^{-2}%
\hat{\gamma}_{2}^{\omega }=q^{4}\hat{\sigma}_{q}^{-1}\hat{\gamma}%
_{1}^{\omega }, \\
\hat{\gamma}_{1}^{\omega }+q^{-2}\hat{\sigma}_{q}\hat{\gamma}_{2}^{\omega }
&=&q^{-2}\hat{\sigma}_{q}\hat{\gamma}_{3}^{\omega },\;\;\;\hat{\sigma}%
_{q}^{2}\hat{\gamma}_{1}^{\omega }+q^{4}\hat{\gamma}_{2}^{\omega }=q^{2}\hat{%
\sigma}_{q}\hat{\gamma}_{3}^{\omega }, \\
\hat{\gamma}_{1}^{\omega }+q^{-4}\hat{\sigma}_{q}^{3}\hat{\gamma}%
_{3}^{\omega } &=&q^{-2}\hat{\sigma}_{q}^{2}\hat{\gamma}_{2}^{\omega },\;\;\;%
\hat{\sigma}_{q}^{2}\hat{\gamma}_{1}^{\omega }+q^{4}\hat{\sigma}_{q}^{-1}%
\hat{\gamma}_{3}^{\omega }=q^{4}\hat{\sigma}_{q}^{-1}\hat{\gamma}%
_{2}^{\omega }, \\
\hat{\gamma}_{2}^{\omega }+q^{-6}\hat{\sigma}_{q}^{4}\hat{\gamma}%
_{3}^{\omega } &=&q^{-4}\hat{\sigma}_{q}^{3}\hat{\gamma}_{1}^{\omega
},\;\;\;q^{2}\hat{\sigma}_{q}\hat{\gamma}_{2}^{\omega }+q^{6}\hat{\sigma}%
_{q}^{-2}\hat{\gamma}_{3}^{\omega }=q^{4}\hat{\sigma}_{q}^{-1}\hat{\gamma}%
_{1}^{\omega }.
\end{eqnarray*}

\subsection{$(B_{2}^{(1)},A_{3}^{(2)})$}

\begin{center}
{
\unitlength=0.780000pt 
\begin{picture}(430.00,78.12)(0.00,0.00)
\qbezier(270.00,38.12)(330.00,18.13)(380.00,38.12)
\qbezier(230.00,38.12)(326.25,-13.12)(420.00,38.12)
\put(390.00,48.12){\line(1,0){30.00}}
\put(270.00,48.12){\line(1,0){10.00}}
\put(230.00,48.12){\line(1,0){30.00}}
\put(430.25,62.71){\makebox(0.00,0.00){$\alpha_{2N-1}$}}
\put(386.67,62.71){\makebox(0.00,0.00){$\alpha_{2N-2}$}}
\put(326.00,63.12){\makebox(0.00,0.00){$\hat{\alpha}_N$}}
\put(266.25,62.29){\makebox(0.00,0.00){$\hat{\alpha}_2$}}
\put(226.00,62.12){\makebox(0.00,0.00){$\hat{\alpha}_1$}}
\put(175.42,63.54){\makebox(0.00,0.00){$\alpha_N$}}
\put(136.00,63.12){\makebox(0.00,0.00){$\alpha_{N-1}$}}
\put(95.00,63.12){\makebox(0.00,0.00){$\alpha_{N-2}$}}
\put(46.00,63.12){\makebox(0.00,0.00){$\alpha_2$}}
\put(5.00,64.12){\makebox(0.00,0.00){$\alpha_1$}}
\put(278.75,0.00){\makebox(0.00,0.00){$c_1=-1$ if $N$ odd}}
\put(57.50,26.88){\makebox(0.00,0.00){$c_1=-1$ if $N$ odd}}
\put(160.00,48.12){\line(-1,-2){7.00}}
\put(160.00,48.12){\line(-1,2){7.00}}
\put(135.00,42.79){\line(1,0){40.33}}
\put(135.00,53.12){\line(1,0){39.67}}
\put(330.00,48.12){\line(1,0){10.00}}
\put(310.00,48.12){\line(1,0){10.00}}
\put(100.00,48.12){\line(1,0){30.00}}
\put(80.00,48.12){\line(1,0){10.00}}
\put(50.00,48.12){\line(1,0){10.00}}
\put(10.00,48.12){\line(1,0){30.00}}
\put(425.00,48.12){\circle{10.00}}
\put(385.00,48.12){\circle{10.00}}
\put(325.00,48.12){\circle*{10.00}}
\put(265.00,48.12){\circle{10.00}}
\put(225.00,48.12){\circle{10.00}}
\put(175.00,48.12){\circle*{10.00}}
\put(135.00,48.12){\circle{10.00}}
\put(95.00,48.12){\circle*{10.00}}
\put(45.00,48.12){\circle{10.00}}
\put(5.00,48.12){\circle{10.00}}
\end{picture}\medskip }
\end{center}

\noindent The S-matrices read 
\begin{equation*}
S_{11}(\theta )=\{1,1_{2};3,3_{2}\}_{\theta }\quad S_{12}(\theta
)=\{2,2_{2}\}_{\theta }\quad S_{22}(\theta )=\{1,1;3,5\}_{\theta }.
\end{equation*}
We have $h=4$ and $H=6$ for the Coxeter numbers. The combined bootstrap
identities are

\begin{eqnarray}
S_{1l}(\theta +\theta _{h}+2\theta _{H})S_{1l}(\theta -\theta _{h}-2\theta
_{H}) &=&S_{l2}(\theta -\theta _{H})S_{l2}(\theta +\theta _{H}) \\
S_{2l}(\theta +\theta _{h}+\theta _{H})S_{2l}(\theta -\theta _{h}-\theta
_{H}) &=&S_{l1}(\theta ).
\end{eqnarray}
The mass ratio is 
\begin{equation}
\frac{m_{1}}{m_{2}}=\frac{\sinh (2\theta _{h}+4\theta _{H})}{\sinh (\theta
_{h}+\theta _{H})}.
\end{equation}

\subsubsection{$S_{ij}(\protect\theta )$ from $B_{2}^{(1)}$}

The result of successive actions of the $q$-deformed Coxeter element on the
simple roots is reported in Table A.9.

\begin{center}
$
\begin{tabular}{|c||c|c|}
\hline\hline
$\sigma _{q}^{x}$ & $\alpha _{1}=\gamma _{1}$ & $\alpha _{2}=-\gamma _{2}$
\\ \hline\hline
$1$ & $-4;3,5$ & $3;4$ \\ \hline
$2$ & $-6;\ast $ & $-\ast ;6$ \\ \hline
$3$ & $10;9,11$ & $-9;10$ \\ \hline
$4$ & $12;\ast $ & $\ast ;12$ \\ \hline\hline
\end{tabular}
$\medskip

\noindent {\small Table A.9: The orbits }$\ \Omega _{i}^{q}${\small \
created by the action of }$\sigma _{q}^{x}$ {\small on }$\gamma _{i}$\medskip
\end{center}

\noindent Solutions of the fusing rule in $\Omega ^{q}${\small \ } 
\begin{eqnarray*}
\gamma _{1}+q^{-3}\sigma _{q}\gamma _{2} &=&q\gamma _{2},\;\;\;q^{-4}\gamma
_{1}+q^{3}\sigma _{q}^{-2}\gamma _{2}=q^{-1}\sigma _{q}^{-1}\gamma _{2}, \\
\gamma _{2}+q^{-2}\sigma _{q}\gamma _{2} &=&q^{-3}\sigma _{q}\gamma
_{1},\;\;\;\sigma _{q}^{-1}\gamma _{2}+q^{2}\sigma _{q}^{-2}\gamma
_{2}=q^{-1}\sigma _{q}^{-1}\gamma _{1}.
\end{eqnarray*}

\subsubsection{$S_{ij}(\protect\theta )$ from $\hat{A}_{3}^{(2)}$}

The result of successive actions of the q-deformed twisted Coxeter element
on the simple roots is reported in Table A.10.

\begin{center}
\begin{tabular}{|c||c|c|}
\hline\hline
$\hat{\sigma}_{q}^{x}$ & $\hat{\alpha}_{1}=\hat{\gamma}_{1}^{\omega }$ & $%
\hat{\alpha}_{2}=-\hat{\gamma}_{2}^{\omega }$ \\ \hline\hline
$1$ & $-\ast ;2;2$ & $\ast ;\ast ;2$ \\ \hline
$2$ & $-2;2;4$ & $2;2;\ast $ \\ \hline
$3$ & $-4;\ast ;\ast $ & $-\ast ;4;\ast $ \\ \hline
$4$ & $\ast ;6;6$ & $-\ast ;\ast ;6$ \\ \hline
$5$ & $6;6;8$ & $-6;6;\ast $ \\ \hline
$6$ & $8;\ast ;\ast $ & $\ast ;8;\ast $ \\ \hline\hline
\end{tabular}
\medskip

\noindent {\small Table A.10 : The orbits }$\ \hat{\Omega}_{i}^{q}${\small \
created by the action of }$\hat{\sigma}_{q}^{x}$ {\small on }$\hat{\gamma}%
_{i}^{\omega }$\medskip
\end{center}

\noindent The solutions to the fusing rule in $\hat{\Omega}^{q}${\small \ } 
\begin{eqnarray*}
\hat{\gamma}_{1}^{\omega }+q^{-2}\hat{\sigma}_{q}^{2}\hat{\gamma}%
_{2}^{\omega } &=&\hat{\gamma}_{2}^{\omega },\;\;\;\hat{\sigma}_{q}^{2}\hat{%
\gamma}_{1}^{\omega }+q^{4}\hat{\sigma}_{q}^{-1}\hat{\gamma}_{2}^{\omega
}=q^{2}\hat{\sigma}_{q}\hat{\gamma}_{2}^{\omega }, \\
\hat{\gamma}_{2}^{\omega }+q^{-2}\hat{\sigma}_{q}\hat{\gamma}_{2}^{\omega }
&=&q^{-2}\hat{\sigma}_{q}\hat{\gamma}_{1}^{\omega },\;\;\;q^{2}\hat{\sigma}%
_{q}\hat{\gamma}_{2}^{\omega }+q^{4}\hat{\gamma}_{2}^{\omega }=q^{2}\hat{%
\sigma}_{q}\hat{\gamma}_{1}^{\omega }.
\end{eqnarray*}

\subsection{$(B_{3}^{(1)},A_{5}^{(2)})$}

The S-matrices read 
\begin{eqnarray*}
S_{11}(\theta ) &=&\{1,1_{2};5,7_{2}\}_{\theta }\quad S_{12}(\theta
)=\{2,3_{2};4,5_{2}\}_{\theta }\quad S_{33}(\theta )=\{1,1;3,5;5,9\}_{\theta
} \\
S_{22}(\theta ) &=&\{1,1_{2};3,3_{2};3,5_{2};5,7_{2}\}_{\theta }\quad
S_{23}(\theta )=\{2,2_{2};4,6_{2}\}_{\theta }\quad S_{13}(\theta
)=\{3,4_{2}\}_{\theta }.
\end{eqnarray*}
We have $h=6$ and $H=10$ for the Coxeter numbers. The combined bootstrap
identities read 
\begin{eqnarray}
S_{1l}(\theta +\theta _{h}+2\theta _{H})S_{1l}(\theta -\theta _{h}-2\theta
_{H}) &=&S_{l2}(\theta ) \\
S_{2l}(\theta +\theta _{h}+2\theta _{H})S_{2l}(\theta -\theta _{h}-2\theta
_{H}) &=&S_{l1}(\theta )S_{l3}(\theta -\theta _{H})S_{l3}(\theta +\theta
_{H})\,\,\,\,\,\,\,\,\, \\
S_{3l}(\theta +\theta _{h}+\theta _{H})S_{3l}(\theta -\theta _{h}-\theta
_{H}) &=&S_{l2}(\theta ).
\end{eqnarray}
The mass ratios are 
\begin{equation}
\frac{m_{1}}{m_{2}}=\frac{\sinh (\theta _{h}+2\theta _{H})}{\sinh (4\theta
_{h}+6\theta _{H})}\quad \frac{m_{1}}{m_{3}}=\frac{\sinh (2\theta
_{h}+4\theta _{H})}{\sinh (2\theta _{h}+3\theta _{H})}\quad \frac{m_{2}}{%
m_{3}}=\frac{\sinh (4\theta _{h}+8\theta _{H})}{\sinh (\theta _{h}+\theta
_{H})}.
\end{equation}

\subsubsection{$S_{ij}(\protect\theta )$ from $B_{3}^{(1)}$}

The result of successive actions of the q-deformed Coxeter element on the
simple roots is reported in Table A.11.

\begin{center}
\begin{tabular}{|c||c|c|c|}
\hline\hline
$\sigma _{q}^{x}$ & $\alpha _{1}=-\gamma _{1}$ & $\alpha _{3}=-\gamma _{3}$
& $\alpha _{2}=\gamma _{2}$ \\ \hline\hline
$1$ & $\ast ;4;3,5$ & $3;3;4$ & $-4;4;3,5$ \\ \hline
$2$ & $6;6;\ast $ & $\ast ;7;8$ & $-6;6,8;7,9$ \\ \hline
$3$ & $-10;\ast ;\ast $ & $\ast ;\ast ;10$ & $-\ast ;10;\ast $ \\ \hline
$4$ & $-\ast ;14;13,15$ & $13;13;14$ & $14;14;13,15$ \\ \hline
$5$ & $-16;16;\ast $ & $\ast ;17;18$ & $16;16,18;17,19$ \\ \hline
$6$ & $20;\ast ;\ast $ & $\ast ;\ast ;20$ & $\ast ;20;\ast $ \\ \hline\hline
\end{tabular}
\medskip

\noindent {\small Table A.11: The orbits }$\ \Omega _{i}^{q}${\small \
created by the action of }$\sigma _{q}^{x}$ {\small on }$\gamma _{i}$\medskip
\end{center}

\noindent The solutions of the fusing rule in $\Omega ^{q}$ are 
\begin{eqnarray*}
\gamma _{1}+q^{-4}\sigma _{q}\gamma _{1} &=&q^{-4}\sigma _{q}\gamma
_{2},\;\;\;\sigma _{q}^{-1}\gamma _{1}+q^{4}\sigma _{q}^{-2}\gamma
_{1}=\sigma _{q}^{-1}\gamma _{2}, \\
\gamma _{1}+q^{-10}\sigma _{q}^{3}\gamma _{2} &=&q^{-6}\sigma _{q}^{2}\gamma
_{1},\;\;\;\sigma _{q}^{-1}\gamma _{1}+q^{6}\sigma _{q}^{-3}\gamma
_{2}=q^{6}\sigma _{q}^{-3}\gamma _{1}, \\
\gamma _{1}+q^{-7}\sigma _{q}^{2}\gamma _{3} &=&q^{-3}\sigma _{q}\gamma
_{3},\;\;\;\sigma _{q}^{-1}\gamma _{1}+q^{7}\sigma _{q}^{-3}\gamma
_{3}=q^{3}\sigma _{q}^{-2}\gamma _{3}, \\
\gamma _{2}+q^{-7}\sigma _{q}^{2}\gamma _{3} &=&q\gamma
_{3},\;\;\;q^{-4}\gamma _{2}+q^{7}\sigma _{q}^{-3}\gamma _{3}=q^{-1}\sigma
_{q}^{-1}\gamma _{3}, \\
\gamma _{3}+q^{-6}\sigma _{q}^{2}\gamma _{3} &=&q^{-3}\sigma _{q}\gamma
_{1},\;\;\;\sigma _{q}^{-1}\gamma _{3}+q^{6}\sigma _{q}^{-3}\gamma
_{3}=q^{3}\sigma _{q}^{-2}\gamma _{1}, \\
\gamma _{3}+q^{-2}\sigma _{q}\gamma _{3} &=&q^{-3}\sigma _{q}\gamma
_{2},\;\;\;\sigma _{q}^{-1}\gamma _{3}+q^{2}\sigma _{q}^{-2}\gamma
_{3}=q^{-1}\sigma _{q}^{-1}\gamma _{2}.
\end{eqnarray*}

\subsubsection{$S_{ij}(\protect\theta )$ from $\hat{A}_{5}^{(2)}$:}

The result of successive actions of the q-deformed twisted Coxeter element
on the simple roots is reported in Table A.12.

\begin{center}
\begin{tabular}{|c||c|c|c|}
\hline\hline
$\hat{\sigma}_{q}^{x}$ & $\alpha _{5}=-\hat{\gamma}_{1}^{\omega }$ & $\hat{%
\alpha}_{3}=-\hat{\gamma}_{3}^{\omega }$ & $\hat{\alpha}_{2}=\hat{\gamma}%
_{2}^{\omega }$ \\ \hline\hline
$1$ & $0;\ast ;\ast ;\ast ;\ast $ & $\ast ;\ast ;\ast ;2;2\ $ & $-\ast ;\ast
;2;2;2$ \\ \hline
$2$ & $\ \ast ;\ast ;2;2;\ast $ & $\ 2;2;2;\ast ;\ast $ & $-2;2;2;4;4$ \\ 
\hline
$3$ & $\ \ast ;2;2;4;4$ & $\ \ast ;\ast ;\ast ;4;\ast $ & $\ -4;4;4;4;\ast $
\\ \hline
$4$ & $\ 4;4;\ast ;\ast ;\ast $ & $\ \ast ;4;4;\ast ;\ast $ & $\ -\ast
;4;4;6;\ast $ \\ \hline
$5$ & $\ \ -\ast ;\ast ;\ast ;\ast ;6$ & $\ \ast ;\ast ;6;\ast ;\ast $ & $\
-\ast ;6;\ast ;\ast ;\ast $ \\ \hline
$6$ & $-6;\ast ;\ast ;\ast ;\ast $ & $\ \ast ;\ast ;\ast ;8;8$ & $\ \ast
;\ast ;8;8;8$ \\ \hline
$7$ & $\ -\ast ;\ast ;8;8;\ast $ & $\ 8;8;8;\ast ;\ast $ & $\ 8;8;8;10;10$
\\ \hline
$8$ & $\ -\ast ;8;8;10;10$ & $\ast ;\ast ;\ast ;10;\ast $ & $\
10;10;10;10;\ast $ \\ \hline
$9$ & $-10;10;\ast ;\ast ;\ast $ & $\ast ;10;10;\ast ;\ast $ & $\ast
;10;10;12;\ast $ \\ \hline
$10$ & $\ast ;\ast ;\ast ;\ast ;12$ & $\ast ;\ast ;12;\ast ;\ast $ & $\ast
;12;\ast ;\ast ,\ast $ \\ \hline\hline
\end{tabular}
\medskip

\noindent {\small Table A.12 : The orbits }$\ \hat{\Omega}_{i}^{q}${\small \
created by the action of }$\hat{\sigma}_{q}^{x}$ {\small on }$\hat{\gamma}%
_{i}^{\omega }$\medskip
\end{center}

\noindent The solutions of the fusing rule in $\ \hat{\Omega}^{q}$%
\begin{eqnarray*}
\hat{\gamma}_{1}^{\omega }+q^{-2}\hat{\sigma}_{q}^{2}\hat{\gamma}%
_{1}^{\omega } &=&q^{-2}\hat{\sigma}_{q}\hat{\gamma}_{2}^{\omega
},\;\;\;q^{2}\hat{\sigma}_{q}^{2}\hat{\gamma}_{1}^{\omega }+q^{4}\hat{\gamma}%
_{1}^{\omega }=q^{2}\hat{\sigma}_{q}\hat{\gamma}_{2}^{\omega }, \\
\hat{\gamma}_{1}^{\omega }+q^{-6}\hat{\sigma}_{q}^{4}\hat{\gamma}%
_{2}^{\omega } &=&q^{-4}\hat{\sigma}_{q}^{3}\hat{\gamma}_{1}^{\omega
},\;\;\;q^{2}\hat{\sigma}_{q}^{2}\hat{\gamma}_{1}^{\omega }+q^{6}\hat{\sigma}%
_{q}^{-2}\hat{\gamma}_{2}^{\omega }=q^{6}\hat{\sigma}_{q}^{-1}\hat{\gamma}%
_{1}^{\omega }, \\
\hat{\gamma}_{1}^{\omega }+q^{-4}\hat{\sigma}_{q}^{3}\hat{\gamma}%
_{3}^{\omega } &=&q^{-2}\hat{\sigma}_{q}\hat{\gamma}_{3}^{\omega
},\;\;\;q^{2}\hat{\sigma}_{q}^{2}\hat{\gamma}_{1}^{\omega }+q^{6}\hat{\sigma}%
_{q}^{-2}\hat{\gamma}_{3}^{\omega }=q^{4}\hat{\gamma}_{3}^{\omega }, \\
\hat{\gamma}_{2}^{\omega }+q^{-4}\hat{\sigma}_{q}^{4}\hat{\gamma}%
_{3}^{\omega } &=&\hat{\gamma}_{3}^{\omega },\;\;\;\hat{\sigma}_{q}^{2}\hat{%
\gamma}_{2}^{\omega }+q^{6}\hat{\sigma}_{q}^{-3}\hat{\gamma}_{3}^{\omega
}=q^{2}\hat{\sigma}_{q}\hat{\gamma}_{3}^{\omega }, \\
\hat{\gamma}_{3}^{\omega }+q^{-4}\hat{\sigma}_{q}^{3}\hat{\gamma}%
_{3}^{\omega } &=&q^{-2}\hat{\sigma}_{q}^{2}\hat{\gamma}_{1}^{\omega
},\;\;\;q^{2}\hat{\sigma}_{q}\hat{\gamma}_{3}^{\omega }+q^{6}\hat{\sigma}%
_{q}^{-2}\hat{\gamma}_{3}^{\omega }=q^{4}\hat{\gamma}_{1}^{\omega }, \\
\hat{\gamma}_{3}^{\omega }+q^{-2}\hat{\sigma}_{q}\hat{\gamma}_{3}^{\omega }
&=&q^{-2}\hat{\sigma}_{q}\hat{\gamma}_{2}^{\omega },\;\;\;q^{2}\hat{\sigma}%
_{q}\hat{\gamma}_{3}^{\omega }+q^{4}\hat{\gamma}_{3}^{\omega }=q^{2}\hat{%
\sigma}_{q}\hat{\gamma}_{2}^{\omega }.
\end{eqnarray*}

\end{document}